\documentclass[aps,showpacs,showkeywords, preprint]{revtex4-1}%
\usepackage{amsmath}
\usepackage{amssymb}
\usepackage{graphicx}%

\providecommand{\U}[1]{\protect\rule{.1in}{.1in}}
\newtheorem{theorem}{Theorem}

\begin{document}
\preprint{.}
\title[Two null gravitational cones in GPS-intersatellite communications]{Two null gravitational cones in the theory of GPS-intersatellite communications between two moving satellites. I. Physical and mathematical theory of the space-time interval and the geodesic distance on intersecting null cones.}

\author{Bogdan G. Dimitrov}
\email{dimitrov.bogdan.bogdan@gmail.com $^1$}
\affiliation{Institute for Nuclear Research and Nuclear Energetics $^{1}$, Bulgarian
Academy of Sciences, $72$ Tzarigradsko Chaussee Blvd., \ $1784$ Sofia,
Bulgaria}
\email{bogdan.dimitrov@iaps.institute $^2$}
\affiliation{Institute for Advanced Physical Studies $^2$, New Bulgarian University, 21 Montevideo str., Sofia 1618, Bulgaria }

\begin{abstract}
Several space missions such as GRACE, GRAIL, ACES and others rely on
intersatellite communications (ISC) between two satellites at a large distance
one from another. The main goal of the theory is to formulate all the navigation
observables within the General Relativity Theory (GRT). The same approach should be applied
also to the intersatellite GPS-communications (in perspective also between the GPS, GLONASS and
Galileo satellite constellations). In this paper a theoretical approach has
been developed  for ISC between two satellites moving on (one-plane)
elliptical orbits, based on the introduction of two gravity null cones with
origins at the emitting-signal and receiving-signal satellites. The two null
cones  account for the variable distance between the satellites during their
uncorrelated motion. The intersection of the two null cones defines a distance, which can be found from a differential equation in full derivatives. This distance is the space-time interval in GRT. Applying some theorems from higher algebra, it was proved that this space-time distance can become zero, consequently it can be also negative and positive. But in order to represent the geodesic distance travelled by the signal, the space-time interval has to be "compatible" with the Euclidean distance. So this "compatibility condition", conditionally called "condition for ISC" is the most important consequence of the theory. The other important consequence is that the geodesic distance turns out to be the space-time interval, but with account also of the "condition for ISC". This interpretation enables the strict mathematical proof that the geodesic distance is greater than the Euclidean distance - a result, entirely based on the "two null cones approach" and moreover, without any use of the Shapiro delay formulae. The theory places also a restriction on the ellipticity of the orbit ($e\leq0.816496580927726$). For the typical GPS
orbital parameters, the condition for ISC gives a value $E=45.00251$ $[\deg]$,
which is surprisingly close to the value for the true anomaly angle
\ $f=45.54143$ \ \ \ $[\deg]$ and also to the angle of disposition of the
satellites in the GLONASS satellite constellation (the Russian analogue of
the American GPS) - $8$ satellites within one and the same plane equally
spaced at $45$ $\deg$. Consistency between several other newly derived numerical parameters is noted.
 The paper is the first step towards constructing a new
and consistent relativistic physical theory of ISC between moving
(non-stationary) satellites on space-distributed Kepler orbits, which further will include the mathematical realization of the
s.c. concept of "multi-ranging" in a curved space-time - the transmission of signals by means of a "chain" of satellites on different space orbits and from different satellite constellations. For the case of two satellites, the corresponding equations
have been analyzed and the possibility was pointed out to extend the approach of intersecting cones and hyperplanes to the problem
about the change of the atomic time under the transportation of the atomic clock on the satellite. Under some specific restrictions
and for the case of plane motion of the satellites, the analytical formulae was derived for the propagation time of the signal,
emitted by a moving along an elliptical orbit satellite. This represents the first part of the problem for finding the (second)
propagation time of reception of the signal for the case of moving satellites.

\end{abstract}
\volumeyear{year}
\volumenumber{number}
\issuenumber{number}
\eid{identifier}
\date[Date text]{date}
\received[Received text]{date}

\revised[Revised text]{date}

\accepted[Accepted text]{date}

\published[Published text]{date}

\startpage{1}
\endpage{ }
\maketitle

\section{INTRODUCTION}
\label{sec:intro}

\subsection{General idea about intersatellite communications and autonomous navigation}
\label{sec:intersat commun}

Currently GPS technologies have developed rapidly due to the wide
implementation of atomic clocks \cite{AAB1} (particularly optical atomic
clocks based on atomic optical transitions and precise frequency standards),
which have found numerous applications in navigation, satellite
communications, frequency and time transfer over optical fibres \cite{AAB2},
time-variable gravity potential components, induced by tides and non-tidal mass
redistributions on the Earth \cite{AAB3}.

The main stream of research in the last $20$ years was concentrated mainly on
the problem about the communications between satellites and ground stations.
The central issue in this communication is the broadcast message
\cite{AAB22}, which contains information about the orbit of the satellite in
the form of Kepler elements and also determines the accuracy of the
navigation and point positioning. This message contains information about the
perturbation - dependent deviation from the two-body ellipse.

In the past $10$ years the problem about GPS satellite-ground station
communications has been replaced by the problem about autonomous navigation
and intersatellite communications (ISC) (links), which has been mentioned yet
in 2005 in the monograph \cite{C27}. Autonomous navigation means that
generations of satellite Block II F (replenishment) and Block III
satellites have the capability to transmit data between them via
intersatellite cross-link ranging and thus they will essentially position
themselves without extensive ground tracking. Consequently, autonomous navigation
is achieved by means of exchanging time signals and other information among the satellites
through ISL (Inter Satellite Links) for ranging and calculating clock offsets \cite{G1}. In such way,
navigation accuracy can be maintained for six months \cite{AAB16} without ground support and
control. However, one of the serious problems is that the accuracy of the
navigation message degrades over time such that the user range in
satellite-to-satellite tracking is bounded by $10000$ $m$ after $180$ days
\cite{AAB16A}.

    In fact, autonomous navigation should not obligatory be based on intersatellite
communication links - in the sense of the definition about "trajectory determination and guidance" in the
monograph \cite{HofmanNavig}, it can include also inertial navigation (gyroscopes and accelerometers,
providing orientation and position, respectively) \cite{Noureldin} and also astronomical observations. If however
autonomous navigation is realized for spacecrafts in large elliptical orbits \cite{Liu}, at low orbit the spacecrafts
can receive much signal from the navigation stars, while at high Earth orbit the valid stars decrease sharply due to
the Earth-shielding factors and others. Also, while in inertial navigation the measurement accumulates over time, the navigation by means of
the Global Positioning System does not lead to accumulation of the error with time. Consequently, the establishment of intersatellite links (ISL) and measurement communications is of primary importance for the relative ranging and relative velocity determination between the satellites
from one constellation or from different constellations such as the GPS system, the Russian GLONASS, the European Galileo and
the China Beidou second-generation system, all of them considered to be interoperable with each other. This also means that
the relative ranging and relative velocity model should account also for the bending of the transmitting path of the
signal, which is significant for such large distances between the satellites due to the action of the gravitational field.

This paper will propose a new theoretical approach for intersatellite
communications between satellites moving on one-plane Kepler elliptical orbits.
In principle, the precision measurement of the propagation time of the signal
(this is mostly performed for the signal between station on the Earth and a
satellite) is of key importance for Satellite Laser Ranging, which measures
relativistic effects on the light propagation between station and satellite
with the accuracy of $1$ micrometer ($\sim0.01$ $ps$ - picoseconds)
\cite{AAB1}. But this accuracy ($1$ micrometer is diameter of the blood
cell) can be attained also in measuring the distance between the two
spacecrafts (about $220$ $km$) in the GRACE (Gravity Recovery and Climate
Experiment) space mission \cite{AAB50} on a low-Earth orbit by means of the
microwave ranging (MWR) system \cite{AAB59}.

It is important that next generation space missions will attain sub-millimeter
precision of measuring distances beyond $10^{6}$ meters by means of ultrashort
femtosecond pulse lasers. A new technology based on comparison of the phase of
the laser pulses is proposed currently by the Jet Propulsion Laboratory (JPL).
So the precision measurement of distances is the first important moment.

The second important moment is that the theoretical description of such
measurements is inevitably related with General Relativity Theory (GRT). As an
example one can point out the space mission GRAIL (Gravity Recovery and
Interior Laboratory), also comprised of two spacecrafts, launched on September
10, 2011 and with data acquisition from March 1, 2012. This mission together
with the one- and two- way Doppler observations from the NASA Deep Space
Network (DSN) allows to recover the lunar gravitational field with the purpose
to investigate the interior structure of the Moon from crust to core
\cite{AAB61}. The peculiar and essential fact in the theoretical formalism
is that all the observables for the two radio links at the K- and Ka-
bands ($26$ GHz and $32$ GHz) for inter-satellite ranging, also for the
second inter-spacecraft link at the S-band ($\sim2.3$ GHz) for the Time
Transfer System (TTS) and the one-way X - band link should be formulated
within the GRT.

\subsection{Intersatellite communications and the space experiments
GRACE, GRAIL, ACES and the RadioAstron
 ground-space VLBRI project}
 \label{sec:VLBRI}

The theory of intersatellite communications (ISC) is developed in the series
of papers by S. Turyshev, V. Toth, M. Sazhin \cite{AAB62,AAB63}
and S. Turyshev, N. Yu, V. Toth \cite{AAB64} and concerns the space missions
GRAIL (Gravity Recovery and Interior Laboratory), GRACE- FOLLOW-ON
(GRACE-FO - Gravity Recovery and Climate Experiment - Follow On) mission and
\ the Atomic Clock Ensemble in Space (ACES) \ experiment \cite{ACES1,
AAB51, ACES2}  on the International Space Station (ISS).
It should be stressed that the theory in these papers is developed for
low-orbit satellites when in the theoretical description of the gravitational
field the multipoles of the Earth as a massive celestial body should be taken
into account. In the case currently investigated, GPS satellites are on a more distant orbit of
$26560$ $km$, so the gravitational field at such height will not be influenced
by such multipoles. Nevertheless, many features of the theory may be applied
also to the intersatellite GPS-communications theory. For example, a key
property of the theory for the exchange of signals between two non-moving
satellites is that if the first spacecraft A is sending $dn_{A_{0}}$ cycles
(number of pulses), then they should be equal to the number of pulses
$dn_{A_{0}}^{B}$ received by the spacecraft $B$ \cite{AAB62}. The question
which arises in reference to the problem treated in this paper is: will this
equality be preserved in the case when the satellites are moving? In this paper,
this problem will be treated from a different perspective:  if a signal is being sent from a
signal-emitting satellite, then will this signal be received by a second (signal-receiving) satellite, provided
that 1. the curved trajectory of the signal propagation has to be taken into account and also 2. the distance,
travelled by the second satellite from the moment of emission of the signal from the first satellite to the moment
of reception of this signal by the second satellite. Consequently, the successful reception of the signal is ensured
by the fulfillment of both conditions, not only the first one, which is sufficient for determining the propagation time
for the case of stationary (non-moving) satellites.
\\
\   There is also a concrete experimental situation, related to the RadioAstron
interferometric project, where the baseline distance (which is in fact $R_{AB}%
$) is changing. RadioAstron is a ground-space interferometer \cite{AAB41},
consisting of a space radio telescope (SRT) with a diameter $10$ meters,
launched into a highly elongated and perturbed orbit \cite{AAB42}, and a
ground radio telescope (GRT) with a diameter larger than $60$ $m$. The
baseline between the SRT and the GRT is changing its length due to the
variable parameters of the orbit - the perigee varies from $7065$ $km$ to
$81500$ $km$, the apogee varies from $280$ $000$ $km$ to $353$ $000$ $km$ and
for a period of $100$ days the eccentricity of the orbit changes from $0.59$
to $0.96$. In this conjunction of SRT and a GRT, commonly called VLBRI
(Very Large Baseline Radio Interferometry) \cite{AAB43}, the SRT time
turns out to be undefined due to the changing delay time, which is a
difference between the time of the SRT and the time of the Terrestrial
Station (TS). This might mean that the commonly accepted formulae for the
Shapiro time delay might not account for the relative motion between the SRT
and the GRT. This will be explained in the next section \ and represents one
of the main motivation for the search for a formulae, which would account for a
variable baseline distance between the SRT and the GRT.

\subsection{General idea about algebraic geometry approach and the multi-range model in a
curved space-time}
\label{sec:multi-range curved}

The purpose of this paper is to construct a theoretical model for intersatellite
communications and relative ranging between satellites moving on one-plane Kepler elliptical orbits (with small ellipticities), based on the
null cone equation in General Relativity Theory. The main peculiar moment in the novel formalism is that in order to account for
the relative motion between two satellites, two intersecting null cones will be introduced. Now it can be guessed why the theoretical model
for the two moving one with respect to another satellites changes for this case. The situation is similar to the one, when relativistic
effects in the signal propagation between an near-Earth satellite and a station on the rotating Earth have to be taken into account. In such a
theoretical model, as noted in the monograph \cite{HandbookGNSS}, the displacement of the receiver on the Earth surface relative an inertial frame
during the time of flight of the signal must be included. It is known that in the Earth rotating frame of reference, this property is called
the Earth rotation correction or the Sagnac effect. For example, for a receiver on the rotating geoid, observing a GPS satellite,
this maximum correction is about $133$ nanoseconds. In the case, the influence of the relative displacement between the two satellites on the signal propagation is modelled by means of two intersecting null cones. This formalism is of interest also from a mathematical point of view, because for the general case, when the orbit motion is characterized by the full set of six Keplerian parameters, it is required to find the intersection of two four-dimensional null cones with a six-dimensional hyperplane equation. In the present investigation, the case will be simplified, because the plane motion will require to solve the problem for the intersection of two three-dimensional null cones with a four-dimensional hyperplane equation.

In this paper, the case of small ellipticities of the orbits is considered. However, the current technologies may propose another options. For example, the proposed in the monograph \cite{Liu} experimental set-up for high-ellipticity orbits might require the treatment of the ellipticity as a variable parameter, signifying a transition from a low orbit (where the gravitational potential of the Earth as a massive body is influenced by the harmonics decomposition) to a high orbit (where the gravitational potential is in its standard form). In all cases, the model will not concern the case about signal transmission between an Earth-based station and a satellite, when the pseudorange equation for the distance station-satellite
\begin{equation}
\rho _{i}=\sqrt{(x_{i}-x_{u})^{2}+(y_{i}-y_{u})^{2}+(z_{i}-z_{u})^{2}}+ct_{u}%
  \label{MMM1}
\end{equation}%
is defined with $\rho _{i}$ - the pseudorange between the station and the $%
i- $th satellite, the indice $i=1,2,3,4$ enumerates the satellites, $(x_{i},y_{i},z_{i})$ are the changing coordinates of each
of the four satellites, $(x_{u},y_{u},z_{u})$ represent the fixed coordinates of the
user (the station) on the Earth in the defined geocentric coordinate system
and $t_{u}$ is the offset between the atomic clocks on the Earth station and
on the satellite (see the monographs \cite{AAB16A} and \cite{Grewal} for the
standard definition of the pseudorange). It is natural to think that in the
case of large distances between the station and the satellite (or between
four satellites), the system of equations for the four straight lines should
be replaced by four non-straight (bending) trajectories, each one of which
in the mathematical sense is determined by a pair of intersecting null cones
and also by the hyperplane equation for each pair of satellites in the
configuration - this can be a configuration of satellites on one (plane)
orbit or a constellation of satellites from different, space-distributed
orbits. In such a way, the intersection of each pair of the null cones and
the hyperplane is the mathematical condition for  the transmission and
reception of the signal, because the intersection will enable to find, in a consecutive order, the
space-time interval on the intersection of the null cones (and the hyperplane) and afterwards - the geodesic distance on the
intersecting null cones, which is the distance, travelled by the light or radio signal. One of the main achievements of this paper is the
algorithm for finding the geodesic distance from the space-time interval by means of the s.c. "compatibility condition for
intersatellite communications". This condition is in fact found from the "compatibility" (i.e. equality) of the space-time distance with the
large-scale, Euclidean distance. Further, the fact that the geodesic distance is found by substituting the compatibility condition into the expression for the space-time interval turns out to be fully consistent with the physical interpretation about the geodesic distance as the large-scale
distance, travelled by the signal.
\\  A more concise exposition of the new approach in this paper, concerning the space-time interval, the "condition for intersatellite communications" and the geodesic distance for the case of signal transmission between satellites, moving in a two-dimensional plane, can be found in the conference proceedings \cite{BOGCONF}.

\subsection{Multi-path ranging and the algebraic geometry approach of intersecting null cones and hyperplanes}
\label{sec:ranging algebr}

Further the formalism in this paper will concern only two satellites, exchanging signals between each
other. But if the whole configuration of $n$ satellites is taken into
consideration (this can be also a station on the ground and $(n-1)$
satellites or two stations and $(n-2)$ satellites), there will be $\ $a
total of $\binom{n}{2}$ pairs of intersatellite links, or $\binom{n}{2}-1$
links for the case of $(n-2)$ satellites - the case about multi-path ranging
or two-dimensional position fixing, mentioned in \cite{HofmanNavig}. Due to
the signal bending between each pair of satellites, moving with respect to
each other, there will be a number of $\binom{n}{3}$ possible spherical
triangles with endpoints at the corresponding satellites. They can be
conditionally called "possible dynamical spherical triangulations" of the
satellites from one or several constellations, because due to the permanent
relative motion between the satellites, these endpoints will change along
each satellite orbit. However, if the mathematical conditions for the
intersection of all possible pairs of null cones and the corresponding
hyperplanes within a given dynamic satellite configuration are fulfilled,
then the satellites will be able to communicate, while moving with respect to each other. This is the
interesting dynamical model of "remote ranging" between many satellites in a
curved space-time, if the (relatively) simple geometrical model about the
two intersecting four-dimensional null cones with a hyperplane (defined by
differentiating the expression for the Euclidean distance in the
three-dimensional space) is worked out.

\subsection{The necessity to use the gravitational null equation instead the Shapiro delay formulae}
\label{sec:null Shapiro}

In this paper, which is the starting point for many further theoretical
developments with interesting experimental applications, a model has been
proposed, in which due to the curved space-time, the Euclidean distance
between two satellites (i.e. between the emitter and the receptor) has been
replaced by the geodesic distance.  It is defined as the distance along the
curved line, travelled by radio or light signals. However,this definition is applicable
when the emitter and the receiver are non-moving. The calculation of the propagation time of the
signal for this case is performed on the base of the well-known Shapiro delay formulae. It should be stressed that
the Shapiro delay formulae is also derived from the gravitational null cone equation. This is and will be the guiding
principle for the derivation of the propagation time of the signal for several other cases. Some authors claim
that this formulae can be used also for the case of moving satellites. It should be distinguished, however, when
the motion is arbitrary and when there is a restriction on the motion due to  some peculiar assumptions -
for example, a Taylor decomposition of the velocity up to the first or the second order. But in any case, the structure of
the Shapiro delay formulae obviously suggests that it is not applicable for any movement - the first term in this formulae
is the geometric distance, divided by the velocity of light, and the second term represents a logarithmic correction,
accounting for the delay of the signal under the action of the gravitational field.

In this paper, the case when the emitter is moving along an elliptical orbit will be considered. The calculation is performed not by means of the Shapiro delay formulae,but by means of the null cone equation, taking into account the Kepler parametrization of the coordinates. This is equivalent to a transformation to the moving system of the satellite elliptic orbit. The analytical expressions, which will be obtained confirm the conclusion that the propagation time for a signal, emitted by a satellite, moving along an elliptical orbit will be different from the propagation time for a signal, emitted by a satellite on a geocentric, circular orbit. This will become clear if one sets up the eccentricity of the orbit equal to zero in the final expression  for the propagation time. It will be proved in future publications that the ellipticity of the orbit influences the propagation time (calculated for a given numerical value of the eccentric anomaly angle) after the fifth digit after the decimal dot. Also, since in the formulae for the propagation time the integration is along the eccentric anomaly angle, it will be inapplicable for the case of stationary (non-moving) satellites.

\subsection{Signal emitted by one moving satellite or signal transmission between two moving one with respect to
another satellites - how does the theoretical formalism change?}
\label{sec:onesateltwosatel}

 Following the same line of reasoning, it will be natural to ask: what will change if not only the signal-emitting satellite is moving,
 but also the signal-receiving satellite? At this point, one should remember that the signal is propagating on the gravitational null cone,
 no matter whether this propagation is related with the emission or the reception of the signal. Consequently, it naturally follows that in the case of moving signal-emitting and signal receiving satellites, two gravitational null cones have to be introduced, and since the Euclidean distance
 between the satellites is changing with time, in the strict mathematical sense these two cones have to be intersected by a hyperplane equation, obtained after taking the differential of the expression for the Euclidean distance. In this paper, we have attacked the problem from two perspectives.

 The first perspective is related to the hyperplane equation and the two null cones in their general form, meaning the case of space-distributed orbits. This method is developed in Section \ref{sec:propag space0} and only the general formulaes (\ref{DOPC1}) for the first and for the (differential) of the second (\ref{DOPC3}) propagation times have been shown. If the concrete expressions have to be found, it is clear that the calculations here are rather complicated and it is not the purpose of this paper to resolve the case to the end. However, an important feature will become evident - the combination of the formulae for the Euclidean distance with the formulae for the second propagation time will change the formulae for the second propagation time. Consequently, the motion of both satellites have to be accounted in order to calculate the correct expression for the propagation time. There is one more important peculiarity in the formalism - in this paper for the first time it is proposed to express the propagation time as a difference between the time of reception of the signal (counted from some initial moment of perigee passing for the second satellite) and the time of emission of the signal from the emitter of the first satellite (also counted from the time of perigee passage). It will be shown in the next Section \ref{sec:atomic} that this definition gives the opportunity to find an important algebraic relation between the propagation times and the two atomic times of the corresponding atomic clocks on the two satellites. In future research, this more general method will be suitable for the application of variational procedures for the optimization (minimization) of the atomic time and the propagation time, when the process of successive transmission of signals will be achieved by means of a chain of satellites (belonging to different satellite configurations such as GPS, GLONASS, Galileo, BeiDou) on different space-distributed orbits.
     \\  The second perspective is based on the method of comparing the differentials $(dx_{1})^{2}+(dy_{1})^{2}$ and
$d(x_{1}^{2}+y_{1}^{2})$ and also using the two gravitational null cone equations (\ref{ABC1}) and (\ref{ABC2}). The method is developed in Section \ref{sec:Two diff}. The first differential is related to the expression for the given diagonal metric (\ref{DOP25}) and the second differential - to the expression (\ref{ABC2A2}) for the square of the differential of the formulae for the Euclidean distance. This method turned out to be suitable for the particular case of two-dimensional plane parametrization of the orbit, but could be developed also for the general case of space-distributed orbits. A disadvantage of the second method is that it might be not so easy to apply it for the case of some other, non-diagonal metric. At the same time, an apparent advantage of the second method of the two-dimensional method of plane parametrization is the possibility to obtain the space-time interval after integrating the s.c. differential equation in full derivatives (\ref{ABC36A}) in Section \ref{sec:Deriv form}. For the general case of space orbit, a linear integral equation (\ref{DOPC13}) with respect to the square of the Euclidean distance has been obtained in Section \ref{sec:integral}, which is supposed to be the equivalent to the differential equation in full derivatives (\ref{ABC36A}). Yet, the solution of the integral equation requires the application of the mathematical theory of integral equations, which is based on approximate methods and is much more complicated in comparison with the exact methods for solution of equations in full derivatives.
\\  Consequently, it can naturally be concluded that it will be recommended to apply both methods in a research, dedicated to optimization of signal transmission in the gravitational field.

\subsection{Change of the atomic time of two satellites, moving one with respect to another}
\label{sec:atomtime movsatel}

 At this step, the paper goes a little further, and explicitly proves that the same approach of intersecting null cones and a hyperplane may be applied with respect to the atomic time, displayed by the atomic clocks on the satellites. The  atomic time changes with the transportation of the atomic clocks, while the satellites are moving along the elliptical orbit. The problem here is still unexplored both from the fundamental (physical) point of view and also from the technical (mathematical) aspect, since the concrete expressions for the atomic time represent complicated integrals. The fundamental aspect is related to the fact that the usual definition for atomic time, given for example in the known monograph by Fock \cite{Fock}, is related to non-accelerated, inertial motion. In the case of plane motion along an elliptic orbit and also satellite motion along space-distributed orbits, the motion is accelerating, since the acceleration of the orbital motion is non-zero. So for the case of the atomic time, the full definition for the atomic time is applied, which includes all the metric tensor components. Concerning the atomic time definition, an important fact is proved in this paper, valid for the chosen metric of the Geocentric Celestial System - the Geocentric Coordinate Time (TCG) can be identified with the celestial time of motion for the satellite, which is determined from the  Kepler equation. This fact enables the concrete calculation of the change of the atomic time under transportation of the atomic clock along the orbit of the satellites, based on the defining formulae (\ref{DOP23}) for the atomic time.

 As for the propagation time, in view of the fact that after the emission of the signal,it decouples from the motion of the satellite, the acceleration and the velocity of the signal-emission satellite turn out to be the initial conditions for the propagation of the signal, mathematically expressed by the null cone gravitational equation, which in fact is a first-order differential equation. In case of two moving one with respect to another satellites, there are two null equations and the second one  enables to express the differential of the second propagation time as a complicated function of the differential of the first propagation time, taking also into account the change of the Euclidean distance between the satellites. Thus, it becomes clear that the solution of the problem about the propagation time of a signal, transmitted between moving satellites in fact consists of two mutually related problems: 1. Finding the propagation time for a signal, emitted by a  moving satellite. In this problem, one is not interested in the interception of the signal. 2. Solving the differential equation for the second propagation time, which takes into account the changing Euclidean distance between the satellites. Due to the mathematical structure of this equation, it is clear also that the problem can be generalized to many satellites in the framework of the proposed "multi-ranging model". This means that the $n$-th propagation time shall be expressed as a complicated recurrent relation between the preceding  propagation times. The same refers also to the atomic times, which in fact will provide the opportunity for synchronization  of a "chain" of the atomic times of atomic clocks, situated on moving with respect one to another satellites. The important conceptual moment here is that the "relative motion" of the satellites will result also in an atomic time, different from the case of two non-moving satellites or for the case of one moving and one non-moving satellite.
\subsection{The space-time and geodesic distances and the "multi-ranging model" in a curved space-time}
\label{sec:geodmultirange}

 However, in this paper the problems about the propagation and atomic times (which are mutually related) are not the central problems. There is one more problem, related to the propagation of the signal in a curved space-time, where the signal trajectory is a curved one and has a greater length than the Euclidean distance, because of the action of the gravitational field. The length of this signal trajectory is in fact the geodesic distance. If it is possible to construct the mathematical algorithm for finding the geodesic distance, travelled by the light or radio signal, then this gives hope that the "ranging" model in a curved space-time for the case of many satellites is possible to be worked out, and then the total length of the geodesic distance for the signal transmission between all the chosen (moving) satellites in the satellite configurations can be minimized.  There are two supporting arguments in favour of such a conclusion: 1. The geodesic distance for this
particular case of intersecting null cones and a hyperplane for the case of plane motion on the orbit is constructed
from the space-time distance, which in turn is derived on the base of the Euclidean distance. The Euclidean distance is the basic ingredient of the
classical "ranging" model, which neglects the curvature of space-time. 2. The geodesic and the space-time distances preserve their properties, when
they are determined for the case of intersecting null cones. This means that the space-time interval  preserves its property of being positive, null or negative, while the geodesic distance for this case continues to be only positive. The interesting problems are 1. Whether these properties will be
simultaneously fulfilled also for the case of three and even more intersecting null cones and hyperplanes? 2. Whether these properties will be valid also for the general case of space-distributed orbits, described by all the six Kepler elements? Therefore, the model about three
intersecting null cones with three hyperplanes (both for the cases of plane motion or satellite motion on space-distributed orbits) will be the starting point for creating a combinatorial model from graph theory \cite{Combinatorial}, in which the vertices $V(G)$ of the graph will be the satellites (the emitters and receivers of the signal) and the edges $E(G)$ will be the
curved trajectories of the signals between the moving satellites. In the
case, the graph is the ordered pair $(V(G),E(G))$, in which the edges are
dynamically changing their configurations and lengths. This means that the
transmission of a signal from a satellite in a given constellation to
another, distant satellite has to be performed by means of a series of
transmissions between a "chain" of different satellites from one or from
different constellations, so that the total propagation time of the signal is
optimized. In other words, this is an optimization (minimization) problem in
a dynamically changing graph configuration.

An extension of the multi-ranging model in a curved space-time can be
related for example with planetary spacecraft navigation \cite{Miller},
which will require the introduction of hyperbolic (Earth departure)
trajectories or interplanetary trajectories for transferring a spacecraft
from one planet to another.

\subsection{Objectives of this paper}
\label{sec:objectives}

\subsubsection{Two intersecting null gravitational cones and the algebraic geometry problem}
\label{sec:inters cones}

The key objective of the paper is to construct a mathematical formalism for the exchange of signals in the gravitational field of the Earth between satellites, which are not stationary with time, but are moving on one-plane elliptic orbits. A new approach here is the introduction of two gravitational null cones with origins at the signal-emitting and signal receiving satellites. The two null cones signify a transition to the moving reference systems of the two satellites, parametrized by the Kepler parameters (the eccentricities, the semi-major axis and the eccentric anomaly angles) for the case of plane elliptic motions. The eccentricity angles can be found as solutions of the Kepler equation for a given value of the mean anomaly angles (determined for one full revolution along the Kepler orbit), but generally, they may be considered changing with time. For this general case, the distance between the satellites will be a variable quantity, which means that the distance will not be expressed by a number, but may depend on the space coordinates. Thus, the known formulae for the Shapiro time delay cannot be applied, because it presumes non-moving emitters and receivers in the gravitational field around a massive body. The most peculiar moment of the approach in this paper is that the Euclidean distance can be expressed from the two null cone equations - this variable distance will turn out to be a solution of a differential equation in full derivatives. The distance (note that again, it is a macroscopic quantity) will then be the distance between two points on the corresponding null cones. Consequently, since the null gravitational cones are an essential ingredient of General Relativity Theory (GRT), this distance will represent in fact the space-time interval, determined on the intersection of the two four - dimensional null cones. In other words, the initial function, denoting the Euclidean distance shall be expressed by another formulae, which will give the space-time interval.

\subsubsection{The concept about the space-time interval of two intersecting gravitational null cones}
\label{sec:concept spacetime}

   In the first place, let us clarify  why in the present case the two gravitational null cones are intersecting. This is so because for a given (variable) Euclidean distance between two space points on the corresponding four - dimensional null cones, there will be a relation between the space-time coordinates on the two null cones. In other words, the two four-dimensional null cones will be intersecting along some three-dimensional hypersurface, which will be a function also of the variable Euclidean distance (depending on the space coordinates). Therefore, the problem about finding the propagation times $T_{1}$ and $T_{2}$ in fact is equivalent to the algebraic geometry problem about finding the intersection variety \cite{AAB43B} of the two gravitational null cones, written in terms of the corresponding variables  $dT_{1}$, $dx_{1}$, $dy_{1}$, $dz_{1}$ and $dT_{2}$, $dx_{2}$, $dy_{2}$, $dz_{2}$ with the six-dimensional hyperplane in terms of the variables $dx_{1}$, $dy_{1}$, $dz_{1}$, $dx_{2}$, $dy_{2}$, $dz_{2}$. For the investigated case of plane Keplerian motion, since there will be no dependence on the $z_{1}$ and $z_{2}$ coordinates, the null cones will be three-dimensional instead and the hyperplane - a four-dimensional one.

    In the second place, following the GRT concepts, this space-time interval can be positive, negative or zero.  Further, the availability of all these options will be confirmed by the concrete calculations. Let us clarify this unusual moment: in the standard literature no proof is given whether the intersection of the two null cones will give again a space-time distance with the property of being null, positive or negative. In this paper, this fact will be proved for several partial cases (for example, equal eccentricities, semi-major axis but different eccentric anomaly angles or the other case, when the eccentric anomaly angles are also equal), but also for the general case, when the space-time interval represents a fourth-degree polynomial with respect to the square of the sine of the eccentric anomaly angle. By applying the Schur theorem from higher algebra, it will be proved not only that the polynomial  has  roots, but also the interval of values for the eccentric anomaly will be found, for which this polynomial might have zeroes.

     \subsubsection{How does the concept about the geodesic distance appear based on the notion of space-time interval}
     \label{sec:concept geodesic}

       This problem was mentioned also in the Introduction, but now a more concrete argumentation shall be given. Propagation of signals is a macroscopic process, because the signal (light or electromagnetic) has to travel a certain macroscopic distance,  which is of course positive. Therefore, this space-time interval has to be compatible with the large-scale, Euclidean distance. Thus, the equality between the space-time interval and the Euclidean distance will give the s.c. "condition for intersatellite communications" (CISC)(further in the text - eq. (\ref{ABC45}). Interestingly, this condition can be obtained also (but only for a certain partial case) without comparing with the Euclidean distance, only by means of setting up equal to zero the space-time interval. Again, for coinciding points on the orbit (equal semi-major axis, eccentricities and eccentric anomaly angles), this is a consistent result, since it does not change the physical essence about zero Euclidean distance and zero space-time interval for coinciding points.  If the CISC is substituted into the equation (\ref{ABC43}) for the space-time interval, the obtained expression will be called the geodesic distance (equation (\ref{ABC46})). This physical interpretation will be correct, because by using some properties of the condition for intersatellite communications, it will be proved that the geodesic distance is greater than the Euclidean distance. Again, this was established for certain partial cases, but it turned out that a simple proof can be made also for the general case. This should be so, because from the Shapiro formulae it follows that due to the action of the gravitational field, a signal travels a greater time (the sum of the Euclidean time and the logarithmic correction). The curious and very interesting fact in the present case is that this result has been confirmed without the use of the Shapiro delay formulae and in the framework of the approach of the two gravitational null cones. The other curious moment is that the algebraic treatment of the fourth-order algebraic equation for the geodesic distance fully complies with its positivity. The Schur theorem applied to this equation proves that it does not possess any roots. This is a substantial difference from the previous case with the algebraic equation for the space-time interval. Therefore, in spite of the relative motion between the satellites and the emission and reception of signals by the moving satellites, some basic facts about the geodesic distance and the delay of the signal in the gravitational field still remain.

    In view of the above considerations,the paper will have the following major objectives:

    1. Deriving the expression for the space-time interval as an intersection of the two null cone equations, relating this interval also to the Euclidean distance. Proving that the space-time interval for certain partial cases can be positive, negative and also zero. Presenting a complicated mathematical proof (without solving the equation) that the fourth-order algebraic equation for the space-time interval for the general case of different eccentricities, semi-major axis and eccentric anomaly angles possesses roots (zeroes). The mathematical proof is valid however for the case of small eccentricities, which is the case for GPS orbits.

    2. Deriving the s.c. "condition for intersatellite communications" and by means of it, clarifying the physical meaning of the space-time interval and the geodesic distance, by considering also the limiting cases of equal eccentricities and semi-major axis, but different eccentric anomaly angles and also another case - equal eccentricities, semi-major axis and eccentric anomaly angles.

    3. A mathematical proof is given that the square of the geodesic distance is greater than the square of the Euclidean distance, based on the "two gravitational null cones approach" and not on the Shapiro delay formulae. Based on the proof that the geodesic distance is only positive and greater than the Euclidean distance, a new physical interpretation is proposed for the Euclidean distance as the positive space-time distance on the intersection of two gravitational null cones. However, the requirement for positive distance may not be taken into account, if the condition for intersatellite communications is taken into consideration, since it is derived from the equality of the space-time distance and the Euclidean distance. Then the distance measured on the intersection of the null cones will be positive.

    4. Some numerical restrictions are found on the eccentricity of the orbit (valid for any eccentric anomaly angles and any semi-major axis) and on the eccentric anomaly angle (valid only for the typical eccentricity of the GPS orbit). In is very interesting to note that the first restriction is closely related to the fact that the geodesic distance is greater than the Euclidean one.

    5. The equations for the two gravitational null cones and the hyperplane have been analyzed in the general case of space-distributed orbits. It was demonstrated that the second propagation time of reception of the signal can be expressed through the first propagation time of reception of the signal. The applicability of the algebraic geometry approach of intersecting cones has been demonstrated with respect to the atomic time, displayed by the atomic clocks on the satellites.

    6. For the partial case of elliptic motion of the satellite on a plane orbit, the analytical expression for the propagation time of the signal, emitted by the emitter of the satellite, has been found. This expression, found from the gravitational null cone equation,  has the dimension of seconds, which proves the mathematical correctness of the approach.

\subsection{Organization of this paper}
\label{sec:organization}

This paper is organized as follows:

In section \ref{sec:intro} it has been pointed out that contemporary experiments such as GRACE, GRAIL, RadioAstron
and others perform precision measurements of the distance between the satellites, which requires
the formulation of all the observables for the radio links within the GRT. So the main prerequisite for
this investigation comes from an experimental point of view and the necessity to establish the s.c. "intersatellite
communications" between moving satellites (see Section \ref{sec:intersat commun} and also Section \ref{sec:VLBRI} ). However, there is also a serious theoretical motivation for this, presented in section \ref{sec:inters cones}, which is based on the
introduced new concept in this paper about the "intersecting null four-dimensional gravitational null cones". This concept, allowing to create a theory for exchange of signals between moving satellites, inevitably leads to two other
concepts,  the physical and mathematical aspects of which will be investigated in details in this paper: the concept in section \ref{sec:concept spacetime} about the space-time distance on intersecting null cones and the other, closely related to the first one, but yet different
concept in section \ref{sec:concept geodesic} about the geodesic distance on intersecting null cones. For each concept, several partial cases will be worked out before investigating the general case since the partial cases will "suggest" the ideas about the space-time distance being positive, negative or null and the geodesic distance-being only positive.

But firstly, some initial and important notions will be reminded in section \ref{sec:world null}, namely the relation between the null cone equation and the world function, which has been defined in General Relativity by Synge. The next section \ref{sec:Shapiro delay} reminds the well known and currently widely used Shapiro time delay formulae for the theoretical modelling of the mentioned in section \ref{sec:intro} experiments . It enables to find explicitly the propagation time of a signal, sent from one satellite to another but this formulae presupposes that the emitter and receiver of the signal are non-moving. But if they are moving, it cannot be expected that the distance travelled by the signal (this is in fact the geodesic distance) can be represented as a sum of the geometric distance and the logarithmic term in the Shapiro formulae multiplied by the velocity of light $c$. The formulae for the case of moving emitters and receivers will be another, and this will become evident from formulae (\ref{ABC47}) in section \ref{sec:compat intersat} and formulae (\ref{ABC55A1}) in section \ref{sec:all equal2}.

   Further in section \ref{sec:celest eccentric} the correspondence celestial time-eccentric anomaly angle is discussed. The reason is that the celestial time is proportional to the mean anomaly $M$, which is a numerical characteristics of the orbit. However, the really important characteristics for the elliptic motion is the eccentric anomaly angle $E$, which for known $M$ is found as solution of the transcendental Kepler equation. But at the same time, the null cone equation establishes a correspondence between the eccentric anomaly angle and the propagation time (see \ref{sec:propag eccentr}). This means that the propagation time is a solution of the null cone equation (see section \ref{sec:propag null}), but the peculiar moment in the present investigation is that there are two propagation times - the propagation time $T_{1}$ of emission of the signal by the first satellite and a propagation time $T_2$ of reception of the signal, and these propagation times are the solutions of two null cone equations. In Section \ref{sec:null vardist} a motivation is presented why a variable distance between an emitter and a receiver (or a variable baseline of a space interferometer such as RadioAstron) should be accounted by two gravitational null cones.

   Section \ref{sec:atomic} has the purpose to show that the algebraic geometry approach of intersecting cones with the hyperplane equation (derived after taking the differential of the Euclidean distance) can be applied also with respect to the atomic time - this is the time of the atomic clocks on each satellite, defined in a non-moving reference frame. The readings of these clocks are influenced by their transportation along the elliptic orbit. In \ref{sec:propag space} the expression for the dependence of the second propagation time on the initial one will be presented for the general case of space-distributed orbits. Since further in the text a differential equation in full derivatives will be derived with respect to the Euclidean distance for the partial case of plane motion of the satellites, the problem is will there be an analogous equation for the general case of space-distributed orbits. In Section \ref{sec:integral} a linear integral equation (\ref{DOPC13}) will be derived again with respect to the Euclidean distance, which might be considered to be an analogue to the differential equation in full derivatives (\ref{ABC36A}) in Section \ref{sec:Deriv form}. In Section \ref{sec:interscones atomic} it has been noted that the same approach of intersecting cones and a hyperplane can be applied also with respect to the defining equations for the atomic time at two different space points.  Further in Sections \ref{sec:celest propag1} and \ref{sec:celest propag2} two possible conditions are given for the reception and emission of a signal by moving satellites. The second condition is particularly important and useful especially for the case of signal transmission between moving satellites, since it is based on a newly proposed and non-contradictory concept in this paper (outlined further in Section \ref{sec:proptime 1satel}) about "initial" and "final" propagation times. These times are really propagation times, because are counted from some "fictitious" moment until the moment of emission or reception of the signal. The real propagation time of the signal between the emitter of the first satellite and the receiver of the second satellite is the difference between these two propagation times. Then the condition for reception of the signal is the equality of the propagation time of the signal to the celestial time of motion for the second satellite from some initial moment of time (corresponding to the emission of the signal from the first satellite) to the final moment of time (corresponding to the reception of the signal by the second satellite). The second definition, the usefulness of which will be explained in Section  \ref{sec:proptime 1satel}, enables to obtain a four-dimensional cubic algebraic surface (\ref{DOP3B25}) in terms of the differentials of both the atomic and the propagation times. The next Section \ref{sec:atom prop} deals with the ratio of the atomic time interval and the propagation time interval. In Section \ref{sec:equal geoccelest} the equality of the geocentric time in the defining equation for the atomic time with the celestial time is proved. This important result gives the opportunity to calculate the change of the atomic time under transportation of the atomic clock along the plane elliptic orbit. Section \ref{sec:atom geoc} deals with the rate of change of the atomic time with respect to the rate of change of the Geocentric Coordinate Time (TCG). The result is an imaginary number (equation (\ref{5B.12})), but this is a consistent and correct result, because the space-time interval turns out to be a real-valued function with respect to the Geocentric Coordinate Time (in view of the definition $ds=icd\tau $ in (\ref{DOP23})), as it should be. Based on the equivalence of all the representations for both the propagation and atomic times and for the case of a diagonal metric, in Section \ref{sec:phys interpr} the ratio of two atomic time differentials at two different space points is calculated to be equal to the ratio of the gravitational frequency shifts for a photon, assuming the source and the receiver at rest. The interesting fact here is that for lines of constant Geocentric Time, this ratio is not modified by any velocity terms, because the ratio (\ref{5B.14}) of the differential atomic time to the differential Geocentric Coordinate Time (TCG) does not contain velocity terms. However, in case of using the ratio (\ref{KLM14}) of the differential atomic and celestial times, the under-square expression will be modified by the term $\frac{g_{kk}(v^{k})^{2}}{c^{2}}$, which is second inverse powers in the velocity of light. This again confirms the physical fact that the rate of change of the atomic time is related to the motion of the satellite (i.e.transportation of the atomic clock). Section \ref{sec:eccentranom} has the purpose to find the two eccentric anomaly angles of the orbits of the satellites from the condition for equality of the corresponding celestial times with the propagation time of the signal.

In Section \ref{sec:one satellite} the standard case for the propagation time of a signal, emitted by a moving along an elliptical orbit satellite is considered. Since only the emission of the satellite is taken into account, only one null cone equation is needed for the calculation of the propagation time. Before presenting the theoretical formalism for the cases of one satellite and two moving satellites (considered in the next sections), the physical argumentation for choosing the gravitational potential with a positive sign is presented in Section \ref{sec:Convent GravPotential}. The guiding principle here is that if a body is lifted at a distance $h$ above the Earth surface, the acting force has to overcome the force of gravitational attraction, consequently the acting force should be negative. This in turn means that the potential difference should be negative, which is fulfilled if the potential is taken with a positive sign. Next, Section \ref{sec:proptime dimens} is the most important one for the case of one satellite, because the analytical expression for the propagation time, depending on the eccentric anomaly angle is presented. The most important conclusion from the expression for the propagation time is that the two constituent terms have coefficients with the correct dimensions of seconds. It can be expected that this conclusion will be valid also for the next case of two null cone equations. In Section \ref{sec:proptime 1satel} it is explained why it is natural to introduce two propagation times $T_{1}$ and $T_{2}$ for the case of two moving satellites. One of the reasons is that since the two propagation times enter the equations for the two null gravitational null cones, each propagation time can be represented as a solution of the corresponding null cone equation and consequently - the difference of the two propagation times will represent the propagation time of the signal.

    The next Section \ref{sec:Signal propag} has the purpose to find explicitly the formulae for the propagation time for the case of two moving
   one with respect to another satellites as a function of the eccentric anomaly angles of the two satellites. This is equivalent to solving the algebraic geometry problem about the intersection of two null gravitational cones with a six-dimensional hyperplane (Section \ref{sec:Gen consid}), derived from the variable Euclidean distance. Section \ref{sec:Sign ellip} deals with the partial case of equal eccentric anomaly angles as characteristics of the orbits, after first clarifying what is the meaning of this notion.

The following two sections \ref{sec:SpaceTime Int} and \ref{sec:geod dist} are the most important contributions in this paper. Based on the initial physical concept about intersecting null cones, the two sections have the purpose to build up a detailed physical and mathematical theory of the space-time interval and of the geodesic distance. Since the geodesic distance is the distance travelled by light (or electromagnetic signal), it might seem that it should have a more important physical meaning in comparison with the space-time distance. But in fact, the space-time distance also has an important meaning, proposed for the first time in section \ref{sec:Dist cones} in this paper. Namely, the Euclidean distance may be considered as the positive distance measured on the intersection of two null four-dimensional null cones. This is a new and rather non-trivial moment, because a large-scale notion from celestial mechanics - the Euclidean distance, turns out to have another meaning and representation in terms of a notion with a "broader" physical meaning - the space-time interval, which is related to General Relativity Theory. The explicit derivation of the expression for the space-time interval for the case of two-dimensional Keplerian (elliptic) motion has been performed in Section \ref{sec:Deriv form}. It should be kept in mind that the space interval will be possible to be found also in the general case of space-distributed orbits, because the derivation will be based again on the relation (\ref{ABC6}) in Section \ref{sec:Two diff}, but this time taken for the three-dimensional parametrization (with six Keplerian parameters) of the orbital motion. By "broader" meaning it is meant also that (positive) Euclidean distance is just one option for the space-time distance - besides positive, it can be also zero and negative (section \ref{sec:Phys spacetime}). The emergence of a negative (macroscopic) distance is not prohibited by geometry - these are the s.c. Lobachevsky geometries with a negative scalar curvature. On the other hand, the derivation of the formulae (\ref{ABC43}) for the space-time distance in Section \ref{sec:Compat cond} clearly suggests that there should be some compatibility between the space-time distance and the Euclidean one, especially when light or electromagnetic signals propagate a macroscopic distance. The mathematical expression of this compatibility is the "condition for intersatellite communications". It is important to stress that formulae (\ref{ABC43}) is fully legitimate and can be used independently from the compatibility condition (\ref{ABC45}). Then, for certain partial cases in Section \ref{sec:Posit negat} and in Section \ref{sec:Nonzero Euclid}, it can easily be established that the space-time interval can be of any signs. Particularly interesting is the simple proof in Section \ref{sec:Nonzero Euclid} that even for non-zero Euclidean distance, the space-time interval can also be negative. However, for the general case of non-equal eccentricities, semi-major axis and eccentric anomaly angles, it cannot become evident whether or not the space-time distance can become zero, because the formulae represents a complicated polynomial of fourth degree. One of the main achievements of this paper is that even for such a complicated case and without solving the algebraic equation, it is possible to establish that the polynomial has roots. A general overview of the theorems from higher algebra is given in  Section \ref{sec:High algeb} and particularly in Section \ref{sec:Unit circle}. Among the several theorems from higher algebra,  dedicated to polynomials with roots within the unit circle (see the monograph by Obreshkoff on higher algebra \cite{AAB16AB}), only two of the theorems are most appropriate to be applied - the Schur theorem in Section \ref{sec:Shur theor} and the so called "substitution theorem". Since the reader might not be familiar with these theorems, their mathematical proofs are given in Section \ref{sec:AppB}. The Schur theorem is applied to the fourth-order algebraic equation (\ref{F16}) for the space-time distance in Section \ref{sec:AppC} and also the substitution theorem is applied to the same equation in Section \ref{sec:AppD}. Both algebraic methods confirm that the space-time algebraic equation really does have roots within the unit circle, related to the chosen variable. So it is amazing that the results from the two higher algebra theorems are fully consistent with the simple algebraic analysis performed in Section \ref{sec:Nonzero Euclid} for the case of different eccentric anomaly angles, when the Euclidean distance is non-zero.

The next Section \ref{sec:Phys conseq} gives the restriction on the eccentric anomaly angle for the assumed value of the eccentricity of the GPS orbit (see Section \ref{sec:orbit anomaly}) and on the numerical value of the eccentricity of the orbit (see Section \ref{sec:restr ellip}). The importance of these restrictions from a physical point of view will be discussed in the Discussion part of the paper.

Section \ref{sec:Spat orient} does not propose a full solution to any problem and only discusses the perspectives for applying the mathematical formalism in this paper to the more complicated case of exchange of signals between satellites on different space-oriented orbits, which differ by the numerical values of the Keplerian orbital parameters. Surely there is a mathematical consistency for developing such an approach, but the main motivation outlined in Section \ref{sec:GPS GLONASS} comes from the necessity for operational interaction between the satellites on different satellite constellations such as GPS, GLONASS and Galileo. As pointed out in the monograph \cite{C26} by Xu, a combined GNSS (Global Navigation Satellite System) of $75$ satellites from the GPS, GLONASS and the Galileo constellations may increase greatly the visibility of the satellites, especially in critical areas such as urban canyons. However, the theoretical investigation (with account of General Relativity Theory) of the process of propagation of signals between satellites on different, space-oriented orbits contains a number of peculiar moments, which are clarified in Sections \ref{sec:Runge Lentz}, \ref{sec:Gener Kepler},  \ref{sec:Modif Kepler} and  \ref{sec:small eccentr}.

Further in Section \ref{sec:geod dist} the theory of the geodesic distance is exposed, again starting from the partial cases and afterwards treating the general case after applying a complicated higher algebra technique, which indeed confirms the main conclusion that the geodesic distance does not have the property to be negative or zero. In fact, confirming this important property entirely different from the space-time interval, is the key moment in this investigation. Firstly, the expression for the geodesic distance (\ref{ABC46}) is derived in Section \ref{sec:compat intersat} by means of substituting the derived compatibility "condition for intersatellite communications" (\ref{ABC45}) in the space-time distance formulae (\ref{ABC43}). So the geodesic distance is a "further step" in the theory after finding the space-time interval, and the relation between these two notions is very important. In Sections \ref{sec:all equal1} and \ref{sec:all equal2} some subcases are investigated for the geodesic distance. Firstly, the subcase of equal eccentricities, semi major axis and eccentric anomaly angles is considered, when the geodesic distance is zero as it should be, because for zero Euclidean distance (coinciding points on the orbit), the geodesic distance should also be zero (i.e. no propagation of any signals between coinciding points). Although this case is trivial, it serves as a consistency check of the correctness of the calculations. The second case in Section \ref{sec:all equal2} is more interesting and corresponds to the case of different eccentric anomaly angles (non-zero Euclidean distance). Most remarkable is formulae (\ref{ABC55A1}), showing that the geodesic distance is greater than the Euclidean distance. This is a result similar to formulae (\ref{AA19}) for the Shapiro time delay, where the time travelled by the signal (consequently the distance) is greater than the geometric time (and also the "geometric", Euclidean distance). But since formulae (\ref{ABC55A1}) is derived in the framework of the formalism of "two null intersecting gravitational cones", the result can be interpreted as a clear evidence about the physical consistency of this formalism. It is interesting to note that the additional (second) term under the square root in this formulae is positive due to the found in Section \ref{sec:restr ellip} restriction $e\leq0.816496580927726$ (formulae (\ref{ABC54A})), which follows from the condition for intersatellite communications. Therefore, this restriction on the ellipticity plays an important role for the positivity of the geodesic distance.
\\ The greatness of the geodesic distance in comparison with the Euclidean one is proved also in Section \ref{sec:posit gener} in the general case by substituting  inequality (\ref{ABC55A8}) from the condition for intersatellite communications (\ref{ABC55A3}) into expression (\ref{ABC47}) for the differences between the squares of the geodesic distance and the Euclidean one. In Section \ref{sec:geod anom} a numerical value for the lower bound of the eccentricity anomaly angle is obtained, but the condition for intersatellite communications gives a higher bound.
\\ In the general case of different eccentricities, semi-major axis and eccentric anomaly angles, the geodesic distance assumes the form of a fourth-degree algebraic equation (\ref{ABC55A19}), which was obtained in Section \ref{sec:restr lower}. The equation has been analyzed in Section \ref{sec:AppE} again by applying the Schur theorem and afterwards - the substitution theorem in Section \ref{sec:AppF}. Both higher algebra methods confirm that the fourth-order algebraic equation for the geodesic distance (\ref{E1}) does not have any roots, which is fully consistent with the previous considerations in Section \ref{sec:all equal2} and in Section \ref{sec:posit gener} about the positivity of the geodesic distance.
\\  The results from the application of the higher algebra theorems to the space-time equation in Section \ref{sec:SpaceTime Int} and to the geodesic equation in Section \ref{sec:geod dist} are valid only under the assumption of the smallness of the eccentricity of the GPS orbit, but this fully corresponds to the real small value of the eccentricity.
\\  In the Discussion section \ref{sec:Discus} some of the obtained results are summarized, but the emphasis is on the importance and consistency between the different numerical parameters, obtained as a result of the proposed new theoretical formalism.

\section{PHYSICAL\ ARGUMENTATION FOR\ THE\ NEW\ APPROACH
OF\ TWO\ GRAVITATIONAL\ NULL\ CONE\ EQUATIONS}
\label{sec:Phys argum}

\subsection{World function in GPS theory and relation to
the null-cone equation}
\label{sec:world null}

The  "point positioning problem" in GPS theory means that if $\left\{
t_{j},r_{j}\right\}  $ ($j=1,2,3,4$) are respectively the time of the
transmission events and the positions of the four satellites \cite{C5B},
then the position of the station on the ground and the time can be found from
the s.c. "navigation equations" (see also the review article by Neil
Ashby \cite{AAB23}). All other issues of GPS satellite-ground station
communications - Earths rotation, determination of the geoid, the
gravitational frequency shifts and the second-order Doppler shifts are treated
in the contemporary review articles \cite{C4, C5, C5A,
C5B, CA8, AAB23, AAB27}.
\\  For high accuracy measurements and determination of coordinate positions and
time, performed over large distances (long baselines of at least $1000$
kilometers) for the purposes of space-based interferometers, the navigation
equations in the framework of the GRT are modified and replaced by the
two-point world function $\Omega(P_{1},P_{2})$ (initially determined in Synge
monograph \cite{AAB29}), accounting for the delay of the electromagnetic
signals due to the presence of the gravitational field. The physical meaning
of the world function is that the flat - space null cones are replaced by the
null geodesic equations \cite{AAB28}. In other words, from a mathematical
point of view navigation in a curved space-time means that a set of four
unique null geodesics connecting four emission events to one reception event
should exist. This is an important theoretical fact meaning that each point of
the orbit at which the satellite emits or intercepts a signal can be connected
to a null geodesics. Note that this concerns the case when one null geodesics
connects the signal-emitting and the signal-receiving points (i.e.
satellites).
\\  In the case investigated in this paper, when these points are moving and each
one of them is related to its own null cone, the situation will be quite
peculiar and a special condition will be derived. Since the world function is
one-half the square of the space-time distance between the points $P_{1}$ and
$P_{2}$, the fulfillment of the null - cone equation $ds^{2}=0$ is equivalent
to a null cone value $\Omega(P_{1},P_{2})=0$ of the world function. 

\subsection{Null cone equation, variational formalism for non-fixed boundary
points and arbitrary parametrization of the signal trajectory }

\label{sec:geodesic null}

\subsubsection{The standard variational approach for the geodesic lines in
gravitational physics with fixed boundary conditions}

In this paper the term "geodesic distance" further will be frequently used,
although the gravitational null cone equation will be used and not the
geodesic equation
\begin{equation}
\frac{d^{2}x^{\nu }}{d\lambda ^{2}}+\Gamma _{\alpha \beta }^{\nu }\frac{%
dx^{\alpha }}{d\lambda }\frac{dx^{\beta }}{d\lambda }=0\text{ \ \ \ }\alpha
,\beta ,\nu =0,1,2,3\text{ \ \ \ ,}  \label{HHH1}
\end{equation}%
where $\Gamma _{\alpha \beta }^{\nu }$ is the affine connection. It is
important that in the monograph by Fock \cite{Fock} the geodesic line is
defined as the line, connecting the two consequitive space-time points $%
x_{\alpha }^{(1)}$ and $x_{\alpha }^{(2)}$ of motion of a material body or
points along the trajectory of the light signal. The two points are
parametrized by the parameter $\lambda $ as
\begin{equation}
x_{\alpha }^{(1)}=\varphi ^{\alpha }(\lambda _{1})\text{ \ \ , \ \ \ }%
x_{\alpha }^{(2)}=\varphi ^{\alpha }(\lambda _{2})\text{ }  \label{HHH2}
\end{equation}%
and thus, the parameter $\lambda $ establishes a correspondence between the
space-time points and certain moments of time. The crucial fact here is that
this parameter is not obligatory to be chosen as a time variable, it can be
defined as an arbitrary parameter in the range $\lambda _{1}<\lambda
<\lambda _{2}$.

In the monograph by Moller \cite{Moller} it has been proved that the
geodesic equation (\ref{HHH1}) can be derived on the base of a variational
principle
\begin{equation}
\delta \int\limits_{\lambda _{1}}^{\lambda _{2}}Ld\lambda
=\int\limits_{\lambda _{1}}^{\lambda _{2}}\left[ \frac{\partial L}{\partial
x^{i}}\delta x^{i}(\lambda )+\frac{\partial L}{\partial \overset{.}{x^{i}}}%
\delta \overset{.}{x^{i}}(\lambda )\right] d\lambda \text{ \ ,}  \label{HHH3}
\end{equation}%
where for the concrete two-dimensional case $i=1,2$ and $\overset{.}{x^{i}}$
denotes the derivative with respect to the parameter $\lambda $, i.e. $%
\overset{.}{x^{i}}=\frac{dx^{i}}{d\lambda }$. In  \cite{Moller} it has been
assumed that the boundary conditions for the infinitesimal variations $%
\delta x^{i}(\lambda )$ are
\begin{equation}
\delta x^{i}(\lambda _{1})=\delta x^{i}(\lambda _{2})=0\text{ \ \ .}
\label{HHH4}
\end{equation}

\subsubsection{Variational formalism with non-fixed
boundary conditions}

Now we shall perform the variational principle without using this boundary
condition. The motivation is that a variational formalism with fixed
boundary conditions cannot be applied for the case, when a moving satellite
is sending a signal. The important point is that the null cone gravitational
equation is related to the variational formalism with fixed boundary
conditions, so it is necessary to check whether the result will be changed
for the case of non-fixed boundary conditions. For example, if the signal is
emitted by a moving along an elliptical orbit satellite (the elliptical
orbit is standardly described by the equations $x=a\cos E-ea$ and $y=a\sqrt{%
1-e^{2}}\sin E$), then the variations $\delta x$ and $\delta y$ are
dependent as
\begin{equation}
\delta y=-\sqrt{1-e^{2}}\cot gE\text{ }\delta x\text{ \ \ .}  \label{HHHL1}
\end{equation}%
Taking into account the simple equalities
\begin{equation}
\delta x^{i}(\lambda )=\overset{.}{x^{i}}(\lambda )\delta \lambda \text{ \ ,
\ \ }\delta \overset{.}{x^{i}}=\frac{d(\delta x^{i})}{d\lambda }\text{ \ \ \
\ }  \label{HHH5}
\end{equation}%
and performing an integration by parts of the second term in (\ref{HHH3}),
then this variational equation can be rewritten in the form
\begin{equation*}
\delta \int\limits_{\lambda _{1}}^{\lambda _{2}}Ld\lambda
=\int\limits_{\lambda _{1}}^{\lambda _{2}}\{\frac{\partial L}{\partial x^{i}%
}\overset{.}{x}^{i}(\lambda )-
\end{equation*}%
\begin{equation}
-\overset{.}{x^{i}}\frac{d}{d\lambda }\left( \frac{\partial L}{\partial
\overset{.}{x^{i}}}\right) \}\delta \lambda d\lambda +\left( \frac{\partial L%
}{\partial \overset{.}{x^{i}}}\overset{.}{x}^{i}\delta \lambda \right) \mid
_{\lambda _{1}}^{\lambda _{2}}\text{ \ .}  \label{HHH6}
\end{equation}%
If $L$ is a homogeneous function of the $n$-th degree of the variable $%
\overset{.}{x}^{i}$, then the following equality will be fulfilled
\begin{equation}
\frac{\partial L}{\partial \overset{.}{x^{i}}}\overset{.}{x}^{i}=nL\text{ \
\ \ .}  \label{HHH7}
\end{equation}%
Such a homogeneous function of the $2-$nd degree is with respect to the
variable $\overset{.}{x}^{i}$is $L=g_{ik}\overset{.}{x}^{i}\overset{.}{x}^{k}
$.

\subsubsection{The analogy with the electromagnetic case}

In particular, if
\begin{equation}
\omega (x,y,z,t)=0  \label{HHH8}
\end{equation}%
is the equation of the wave front of the signal, then the quadratic
polynomial with respect to the derivatives $\frac{\partial \omega }{\partial
x_{\alpha }}$
\begin{equation}
(\nabla \omega )^{2}=\sum\limits_{\alpha ,\beta =0}^{3}g^{\alpha \beta }%
\frac{\partial \omega }{\partial x_{\alpha }}\frac{\partial \omega }{%
\partial x_{\beta }}=0  \label{HHH9}
\end{equation}%
will represent the equation for the propagation of the wave front \cite{Fock}%
, in analogy with the electromagnetic theory equation for the propagation of
the wave front
\begin{equation}
\frac{1}{c^{2}}\left( \frac{\partial \omega }{\partial t}\right)
^{2}-\left\{ \left( \frac{\partial \omega }{\partial x}\right) ^{2}+\left(
\frac{\partial \omega }{\partial y}\right) ^{2}+\left( \frac{\partial \omega
}{\partial z}\right) ^{2}\right\} =0\text{ \ \ .}  \label{HHH10}
\end{equation}%
Consequently, if the condition (\ref{HHH7}) is substituted in the last term
of the variational equation (\ref{HHH6}), and then the second term is
integrated by parts, it can finally be obtained
\begin{equation}
\int\limits_{\lambda _{1}}^{\lambda _{2}}\left[ \left( \frac{\partial L}{%
\partial x^{i}}-\frac{d}{d\lambda }\left( \frac{\partial L}{\partial \overset%
{.}{x^{i}}}\right) \right) \overset{.}{x}^{i}d\lambda \right] \delta \lambda
+2L\delta \lambda \mid _{\lambda _{1}}^{\lambda _{2}}=0\text{ \ .}
\label{HHH11}
\end{equation}%
Provided that the variation along the parameter $\lambda $ is non-zero, i.e.
$\delta \lambda \mid _{\lambda _{1}}^{\lambda _{2}}\neq 0$, in its general
form this is a complicated integro-differential equation with respect to the
function $L$. Note that if the choice is made $\frac{\partial L}{\partial
x^{i}}-\frac{d}{d\lambda }\left( \frac{\partial L}{\partial \overset{.}{x^{i}%
}}\right) =0$, then its solution will be incompatible with $L=0$, and as a
consequence, (\ref{HHH11}) will not be  fulfilled. Therefore, a reasonable
choice is $L=g_{ik}\overset{.}{x}^{i}\overset{.}{x}^{k}=0$, which can be
written as
\begin{equation}
L(x,\overset{.}{x^{i}})=g_{ik}(x)\left( \frac{dx^{i}}{d\lambda }\right)
\left( \frac{dx^{k}}{d\lambda }\right) =0\text{ \ \ .}  \label{HHH12}
\end{equation}

\subsubsection{The gravitational null cone equation and the
propagation time }

In view of the fulfillment of the null cone equation, the parameter $\lambda
$ is not necessary to be chosen to coincide with the arclength along the
null geodesics. In the case, it will be chosen to be the eccentric anomaly
angle $E$, related to the motion of the satellite. Nevertheless, since the
signal is "propagating" on the null cone equation (\ref{HHH12}), if the time
variable is expressed from this equation and the integration of the
corresponding integral along the eccentric anomaly angle $E$ is performed,
then as a result the propagation time for the signal will be obtained. This
will be the essence of formulae (\ref{AAA4}) in Section  \ref{sec:propag
eccentr}. The important conclusion from the above considerations is that the
null cone equation (\ref{HHH12}) remains the basic theoretical formalism,
which has to be implemented, when considering the case of emission of a
signal from the moving along the elliptical orbit satellite. In fact, the
moving frame of the satellite serves as an initial boundary condition for
the propagation of the satellite.

In the framework of another formalism, based on the Hamilton-Jacobi
equations \cite{Fock}
\begin{equation}
\frac{dx_{k}}{dx_{0}}=\frac{\partial H}{\partial \omega _{k}}\text{ \ , \ \ }%
\frac{d\omega _{k}}{dx_{0}}=-\frac{\partial H}{\partial x_{k}}\text{ \ \ ,}
\label{HHH13}
\end{equation}%
it has been proved that for points on the wave front, the null cone equation
should be fulfilled. It should be noted that the proof is not related in any
way to the choice of a moving or a non-moving frame. 

In another publication it will be shown that if the functional for the propagation time 
is expressed from the null cone gravitational equation, then from the Fermat principle for the least time of travel for the signal, the 
signal (light ray or electromagnetic) trajectory can be found. Let us remind that according to this principle, the signal trajectory between the 
points of emission and reception is the line,  along which the signal travels for the least time. Then the application of the complicated variational formalisms for two non-fixed boundary points (corresponding to the signal-emitting and signal-receiving satellites) will clearly demonstrate that the 
trajectory of the signal will be different from the case of non-moving satellites.

\subsection{Shapiro time delay in VLBI radio
interferometry, instantaneous and variable Euclidean distance}
\label{sec:Shapiro delay}

The purpose of this section is to remind the basic assumptions, concerning the derivation
of the Shapiro delay formulae. Since the formulae is valid for fixed (non-moving) space points
of the emitter and the receiver, it will turn out that it is inappropriate to be used
with respect to emitters and receivers on moving satellites.

 Now for a moment we shall denote by $t$=TCG the Geocentric Coordinate Time (however, further for convenience the notation will be changed) and
we shall keep the notation $T$ for the propagation time of the signal between
two space points. If the coordinates of the emitter on the first satellite and of the receiver
on the second satellite are correspondingly $\mid x_{A}(t_{A}%
)\mid=r_{A}$ and $\mid x_{B}(t_{B})\mid=r_{B}$, and $R_{AB}=$ $\mid
x_{A}(t_{A})-x_{B}(t_{B})\mid$ is the Euclidean distance between the signal - emitting satellite
and the signal - receiving satellite, then from the null cone equation, the signal
propagation time $T_{AB}=T_{B}-T_{A}$ can be expressed by the known formulae
\cite{AAB21} (see also \cite{CA7A2B0} and also the review article
\cite{CA7A2B1} by Sovers, Fanselow and Jacobs on VLBI radio
interferometry)
\begin{equation}
T_{AB}=\frac{R_{AB}}{c}+\frac{2GM_{E}}{c^{3}}\ln\left(  \frac{r_{A}%
+r_{B}+R_{AB}}{r_{A}+r_{B}-R_{AB}}\right)  \text{ \ ,}\label{AA19}%
\end{equation}
where $GM_{E}$ is the geocentric gravitational constant and $M_{E}$ is the
Earth mass. Note the important moment that $R_{AB}$ is the Euclidean distance
between the space point of emission at the emission time $T_{A}$ and the point
of reception of the signal at the reception time $T_{B}$. In other words, this
is a distance, depending on two different moments of time. The second term in
formulae (\ref{AA19}) is the Shapiro time delay, \ accounting for the signal
delay due to the curved space-time. For low-orbit satellites, the Shapiro
delay is of the order of few picoseconds. However, since in real experiments
only the time of emission $T_{A}$ is known instead of the time of reception
$T_{B}$, the use of the Euclidean distance $R_{AB}$ in (\ref{AA19}) is not
very appropriate. Instead, as pointed out in the paper \cite{AAB21}, one may
use the s.c. \textit{"}instantaneous distance\textit{"} $D_{AB}=x_{A}%
(T_{A})-x_{B}(T_{A})$ (defined at the moment $T_{A}$ of emission of the
signal) and the Taylor decomposition of $x_{B}(t_{B})$ around the moment of
time $t_{A}$. The resulting expression, however, does not possess the symmetry
$A\Leftrightarrow B$ , i.e. station $A$ and satellite $B$ cannot be
interchanged \cite{AAB21} (see also the PhD thesis of Duchayne
\cite{CA7A3}). The lack of symmetry results in the fact that the relative
motion between the emitter (the station) and the receiver cannot be
accounted\textit{. }

It can be concluded that the dependence of the propagation time $T_{AB}$ on
the distance $R_{AB}$, determined at two different moments of time, leads to a
loss of accuracy if $R_{AB}$ is to be replaced by the instantaneous distance
$D_{AB}$. The complexity of the situation arises because in defining the
Euclidean distance $R_{AB}$, one has to keep account of the changes in the
location of the space-time points and also of the correspondence between these
space-points and the definite moments of time.

\subsection{Coordinate parametrizations for moving satellites and the
variable baseline distance $R_{AB}$}
\label{sec:parametriz moving}

From the above point of view, it will be very convenient to find some
new parametrization for the two space-time coordinates (defining the Euclidean
distance), so that these new parametrization variables will be related in a
prescribed way to a time variable, accounting for the motion of the satellites.
This is the essence of the approach, developed in this paper for the case of
two satellites moving along two elliptical orbits on one plane. For such a
case of plane motion, the most convenient variable turns out to be the
eccentric anomaly angle $E$, which parametrizes the two $x-y$ coordinates in
the framework of the standard Kepler motion in celestial mechanics. In this
approach, each satellite trajectory is parametrized by its own eccentric
anomaly, so two eccentric anomaly angles $E_{1}$ and $E_{2}$ are used. Their
values are taken at one and the same moment of time and since the eccentric
anomaly angles change with time, the Euclidean distance $R_{AB}$ also changes
with time. However, since further it will be shown that the space-time interval,
the condition for intersatellite communications and the geodesic distance do not
depend on the propagation time explicitly, then the two eccentric anomaly angles
$E_{1}$ and $E_{2}$ might be taken also at two different moments of time. In such a
way, the variable Euclidean distance can be accounted, because there will be an initial and
also a final value for the first eccentric anomaly angle, and an initial and a final
value for the second angle. It should be reminded that the parametrization of the satellites trajectories
in the framework of the Kepler problem is just a convenient tool for setting up the initial and final boundary
conditions for the propagation of the signal. These boundary conditions are related to the initial space point of emission of the signal and the final
space point of reception of the signal (by the second satellite), and these boundary points satisfy the corresponding two orbital equations, written in terms of the Kepler orbital elements for plane motion. However, the propagation time for the signal is found from the two gravitational
null cone equations, which constitute the basic ingredient of the General Relativity Theory description of the propagation of signals
in the gravitational field.
 \\  This theoretical approach is unlike the standard approach for
calculating the Shapiro delay in formulae (\ref{AA19}), where $R_{AB}$ has to
be constant so that the first term $\frac{R_{AB}}{c}$ will have the dimension
of time.

\subsection{The correspondence celestial time - eccentric anomaly
angle}
\label{sec:celest eccentric}

The time coordinate is chosen to be the celestial time $t_{cel}$, which is
related to the eccentric anomaly through the Kepler equation
\begin{equation}
E-e\sin E=n(t_{cel}-t_{p})=M\text{ \ \ ,}\label{AA11}%
\end{equation}
where $e$ is the eccentricity of the orbit, $n=\sqrt[.]{\frac{GM}{a^{3}}}$ is
the mean motion and $M$ is the mean anomaly. The mean anomaly $M$ is an
angular variable, which increases uniformly with time and changes by $360^{0}$
during one revolution. Let us remind also  the geometrical meaning of the mean
motion:  this is the motion of the satellite along an elliptical orbit,
projected onto an uniform motion along a circle with a radius equal to the
large axis of the ellipse. Usually $M$ is defined with respect to some
reference time - this is the time $t_{p}$ of perigee passage, where the
perigee is the point of minimal distance from the foci of the ellipse (the
Earth is presumed to be at the foci). Since the eccentric anomaly $E$ will play
an important role in the further calculation, let us also remind how this
notion is defined: if from a point on the ellipse a perpendicular is drawn
towards the large axis of the ellipse, then this perpendicular intersects a
circle with a centre $O$ at a point $P$. Then the
eccentric anomaly represents the angle between the joining line $OP$ and the
semi-major axis (the line of perigee passage).

In \cite{AA19A} $E$ is determined as an  auxiliary angular variable such
that $a-r=ae\cos E$, which has the following geometrical meaning with
respect to the ellipse: if $r_{p}$ and $r_{a}$ are the corresponding
radius-vectors at the perigee passage and at the apogee, then $E=0$ for $%
r=r_{p}$ and $E=\pi $ for $r=r_{a}$.

All these notions can be found as well in the standard textbooks on celestial
mechanics \cite{C8, C27, C17B, C28, C29, CA7A2, CA1, C17D} and in the books on
theoretical geodesy \cite{AAB15, AAB16, AAB17}. Extensive
knowledge about the most contemporary aspects of celestial mechanics can be
found in the recent textbook by Gurfil and Seidelmann \cite{C17DA1}. Note that
the geometrical meaning of the other three orbital parameters $(\Omega
,I,\omega)$ will also be outlined briefly in this paper, because they are important for
the space determination of the orbits. This will be necessary to be done, when
creating a theory of intersatellite communications between satellites on
different (space) Kepler orbits, characterized by the full set of six Kepler
parameters $(M,a,e,\Omega,I,\omega)$. For example, this might be a theory of
ISC between GPS, GLONASS and Galileo satellite constellations, situated
on different orbital planes.

Further, for concrete numerical values of the mean motion $M$ in the framework
of the numerical characteristics of the GPS orbit, the numerical values for
the eccentric anomaly $E$ will be calculated from the Kepler equation
(\ref{AA11}). The peculiar moment in such a consideration is that the
correspondence between $E$ and the celestial time $t_{cel}$ is not unique,
because $E~$\ is a solution of the transcendental Kepler equation
(\ref{AA11}). This means that for a given value of the celestial time, the
solution of (\ref{AA11}) with respect to $E$ is given in terms of an iterative
procedure. The more iterations are performed, then the more exact will be the
correspondence $E\Longleftrightarrow t_{cel}$. \

\subsection{The correspondence propagation time - eccentric anomaly
angle}
\label{sec:propag eccentr}

The second correspondence, which will be established is between the propagation
time $T$ and the eccentric anomaly angle $E$. This is not a trivial
correspondence because propagation time is an intrinsic characteristic of the
propagation of signals, which according to General Relativity Theory takes
place on the gravitational null cone $ds^{2}=0$ and the eccentric anomaly $E$
is a notion from celestial mechanics, based on the Newton equation. But it
turns out that these two characteristics are related - while moving along the
(two-dimensional, one-plane) elliptical orbit parametrized by the equations
\cite{CA1}
\begin{equation}
x=a(\cos E-e)\text{ \ \ \ \ , \ \ }y=a\sqrt[.]{1-e^{2}}\sin E\text{ \ \ ,
}\label{C4AA1}%
\end{equation}
the emitter of the first satellite emits a signal propagating on the
gravitational null cone
\begin{equation}
ds^{2}=0=g_{00}c^{2}dT^{2}+2g_{oj}cdTdx^{j}+g_{ij}dx^{i}dx^{j}\text{
\ \ .}\label{AAF3}%
\end{equation}
For the null-cone equation (\ref{AAF3}), the solution of this quadratic
algebraic equation  with respect to the differential $dT$ can be given as
\begin{equation*}
dT=\pm \frac{1}{c}\frac{1}{\sqrt[.]{-g_{00}}}\sqrt[.]{\left( g_{ij}+\frac{%
g_{0i}g_{0j}}{-g_{00}}\right) dx^{i}dx^{j}}+
\end{equation*}%
\begin{equation}
+\frac{1}{c}\left( \frac{g_{0j}}{-g_{00}}\right) dx^{j}\text{ \ \ ,}
\label{AA10}
\end{equation}%
where the metric tensor components are determined for an Earth
Reference System with an origin at the centre of the Earth. If the space
coordinates are parametrized in terms of the Keplerian (plane) elliptic orbital
parameters (semi-major axis $a$, eccentricity $e$ and eccentric anomaly
angle $E$), then formulae (\ref{AA10}) is the mathematical expression of the
correspondence eccentric anomaly angle $E\rightarrow $ propagation time.
After integrating, the propagation time can be found as
\begin{equation*}
T=\pm \frac{1}{c}\int\limits_{E_{0}}^{E_{1}}\frac{1}{\sqrt[.]{-g_{00}}}%
\sqrt[.]{\left( g_{ij}+\frac{g_{0i}g_{0j}}{-g_{00}}\right) \frac{dx^{i}}{dE}%
\frac{dx^{j}}{dE}}dE+
\end{equation*}%
\begin{equation*}
+\frac{1}{c}\int\limits_{E_{0}}^{E_{1}}\frac{g_{0j}}{-g_{00}}\frac{dx^{i}}{dE%
}dE\text{ }=
\end{equation*}%
\begin{equation}
=\int%
\limits_{E_{0}}^{E_{1}}M^{(1)}(e_{1},a_{1},g_{00}(x_{1},y_{1}),g_{0i}(x_{1},y_{1}),g_{ij}(x_{1},y_{1}))dE%
\text{\ \ \ .}  \label{AAA4}
\end{equation}
This is the time $T$ for propagation of the signal, while the eccentric
anomaly angle of the (first) satellite changes from some initial value $%
E=E_{init}=E_{0}$ (for example - at the initial time of perigee passage $%
t=t_{per}$) to the final value $E=E_{1}$. More about the determination of
the propagation time $T_{1}$ as an initial moment of time of emission of the
signal will be clarified in Section \ref{sec:propag null}. The second
propagation time $T_{2}$ can be written analogously, and the actual
propagation time for the signal to travel from the emitter of the first
satellite to the receiver of the second satellite is $T_{2}-T_{1}$.

\subsection{Two gravity null cones and the variable baseline
distance $R_{AB}$}
\label{sec:null vardist}

Note that the emission time $T$ is the time coordinate in this metric and the
space coordinates are in fact the parametrization equations (\ref{C4AA1}) for
the (first) elliptic orbit. Further, the signal is intercepted by the receiver
of the second satellite and this signal is propagating on a (second) null cone
$ds_{(2)}^{2}=0$, where the metric tensor components $g_{00}$, $g_{oj}$ and
$g_{ij}$ are determined at a second space point $x_{2}$, $y_{2}$, parametrized
again by the equations (\ref{C4AA1}) in terms of new orbital parameters
$a_{2}$, $e_{2}$ and $E_{2}$. The peculiar and very important feature of the
newly proposed formalism in this paper is that we have two gravity null cones
$ds_{(1)}^{2}=0$ and $ds_{(2)}^{2}=0$ for the emitted and the received signal
(with time of emission $T_{1}$ and time of reception $T_{2}$) with cone
origins at the points $(x_{1},y_{1},0)$ and $(x_{2},y_{2},0)$. Also an
equation about the differential of the square of the Euclidean distance
$dR_{AB}^{2}$ is written, which now is a variable quantity. Thus, it can be
noted that the two propagation times $T_{1}$ and $T_{2}$ are no longer treated
in the framework of just one null cone equation (as is the case with the known
equation (\ref{AA19}) for the Shapiro time delay), but in the framework of two
gravitational null cone equations. The derivation, the simultaneous solution
of these three equations and some physical consequences of the found solution
in terms of concrete numerical parameters for the GPS orbit are the main
objectives of this paper. It should be stressed that according to the theory
of propagation of electromagnetic signals in the gravitational field of
(moving) bodies developed by S. Kopeikin \cite{DOP1,DOP2,
DOP3, DOP4}, the interaction between the light and the
gravitational field should be described in terms of two null cones
\cite{DOP2} - the gravitational and the electromagnetic one.
In the approach of Kopeikin, the null gravitational
and light cones are situated at the light-deflecting body (Jupiter) and at the observer
on the Earth. Since the light ray originates from a very distant quasar, there is no
necessity of considering a second (moving) null gravitational or light cone with a
centre coinciding with the quasar. In our case, the relative motion between
the satellites is notable and is accounted by means of introducing two null
(gravitational) cones.

\subsection{Coordinate propagation time as a solution of the gravity
null cone equation}
\label{sec:propag null}

The propagation times $T_{1}$ and $T_{2}$ are coordinate times, which by
definition are determined for a region of space with a system of space-time
coordinates chosen arbitrary. The coordinate time is an independent variable
in the equations of motion for material bodies and in the equations for the
propagation of electromagnetic waves \cite{AAB9}. Examples of conventionally
defined coordinate times are the Terrestrial Time ($TT$) \cite{AAB5,
 AAB6}, the Geocentric Coordinate Time ($TCG$), defined for the space
around the Earth and also the Barycentric Coordinate Time ($TCB$), defined for
the region inside the Solar System. For the concrete case of the null-cone
equation (\ref{AAF3}) $ds^{2}=0$,  it   sets up a mathematical correspondence
between the eccentric anomaly $E$ and the propagation time $T$ and thus the
correspondence $T\Longleftarrow E$ is realized. This is so, because the metric
tensor components $g_{00}$, $g_{0j}$, $g_{ij}$ depend on the celestial
coordinates (\ref{C4AA1}), which in turn are expressed by the eccentric
anomaly angle $E$. The propagation time will depend on the eccentric anomaly
from some initial moment of time $t_{0}$ (of perigee passage) to some final
moment of time when $E=E_{fin}$.  After the initial moment of time, when
$E>E_{init}$, the emitted electromagnetic signal from the satellite is
completely decoupled from the motion of the satellite, but nevertheless it
"keeps track" of the position of the satellite along the elliptical orbit via
the eccentric anomaly angle\textit{ }$E$\textit{. } Let us note that the actual
emission of the signal (when the signal decouples from the motion of the
satellite) from the emitter of the first satellite is at the position on the
orbit with an eccentric anomaly angle $E_{1}$, but in order to determine
quantitatively this initial moment $T_{1}$ of emission, an additional
assumption is made that formulae (\ref{AAA4}) gives the propagation time $%
T_{1}$ of the signal during that period of time while the satellite changes
its position from the initial moment of perigee passage $t=t_{per}$ to the
moment when the satellite will have an eccentric anomaly angle $E_{1}$.
In such a way, together with the previously established correspondence
$E\Longleftrightarrow t_{cel}$, the correspondence between the celestial time
and the propagation time is realized.

But if there is such a correspondence, it is natural to ask whether there is some
standard (unit) for measuring both the celestial and the propagation times?
The answer is affirmative, and this standard is the proper time, which is the
actual reading of the atomic clock (the local time) in an inertial frame of
reference \cite{AAB9} at rest with respect to the reference frame, related to
the coordinate time. In this paper we shall present the basic algebraic geometry approach, related to
the atomic time, since it has much in common with the approach, developed with respect to the
propagation time. In more details, the atomic time shall be treated in subsequent publications.

\section{Propagation time and atomic time and the algebraic geometry approach - general outline, preliminary results
and perspectives}
\label{sec:atomic}

Atomic time is the time, registered by atomic clocks. It changes under the transportation of the atomic clocks,
i.e. it depends on the motion of the satellites along the elliptic orbit, since each satellite is equipped with at least
two atomic clocks. But if atomic time changes under transportation and the propagation time for the signal depends on the motion of both the satellites, it might naturally be conjectured that the propagation time and the atomic time are mutually related. This shall be proved explicitly further in this section. Most of the new achievements in this section are mentioned in \ref{sec:organization}, so here we shall outline only two of the new facts established: 1. The derivation for the general case of space-distributed orbits of the linear integral equation with respect to the square of the Euclidean distance, which confirms that the intersection of the two null four-dimensional cones with the hyperplane equation will provide another expression for the Euclidean distance, related to the space-time interval. 2. The strict mathematical proof that the Geocentric Coordinate Time in the defining formulae for the atomic time for the case of the applied diagonal metric is equal to the celestial time of motion for the satellite. This is a very useful and important result, enabling the practical calculation of the rate of change of the atomic time.

\subsection{Intersecting cones and hyperplanes and the propagation times of
emission and reception}
\label{sec:propag space0}

In this paper, the purpose will be to develop the approach about
intersecting null cones and the hyperplane only for the propagation time $T$%
, making use of the standard metric in the near-Earth space \
\begin{equation}
ds^{2}=-c^{2}\left( 1+\frac{2V}{c^{2}}\right) (dt)^{2}+\left( 1-\frac{2V_{.}%
}{c^{2}}\right) \left( (dx)^{2}+(dy)^{2}+(dz)^{2}\right) \text{ \ \ ,}
\label{DOP25}
\end{equation}%
where $V=\frac{G_{\oplus }M_{\oplus }}{r}$ is the standard gravitational
potential of the Earth, without taking into account any harmonics due to the
spherical form of the Earth since the GPS orbits are situated at a distance
more than $20000$ $km$. The null cones with origins at the signal-emitting
and signal-receiving satellites (i.e. at the space points $%
(x_{1},y_{1},z_{1})$ and $(x_{2},y_{2},z_{2})$ respectively) are
\begin{equation}
ds_{1}^{2}=0=-c^{2}\left( 1+\frac{2V_{1}}{c^{2}}\right) (dT_{1})^{2}+\left(
1-\frac{2V_{1.}}{c^{2}}\right) \left(
(dx_{1})^{2}+(dy_{1})^{2}+(dz_{1})^{2}\right) \text{ \ \ \ ,}
\label{DOP1F26}
\end{equation}%
\begin{equation}
ds_{2}^{2}=0=-c^{2}\left( 1+\frac{2V_{2}}{c^{2}}\right) (dT_{2})^{2}+\left(
1-\frac{2V_{2.}}{c^{2}}\right) \left(
(dx_{2})^{2}+(dy_{2})^{2}+(dz_{2})^{2}\right) \text{ \ \ }  \label{DOP2F26}
\end{equation}%
and they are intersecting, because are treated together with the hyperplane
equation
\begin{equation}
d\left[ (x_{1}-x_{2})^{2}\right] +d\left[ (y_{1}-y_{2})^{2}\right] +d\left[
(z_{1}-z_{2})^{2}\right] =dR_{AB}^{2}\text{ \ \ ,}  \label{DOPAB25}
\end{equation}%
obtained after differentiation of the formulae for the Euclidean distance.

The solution of the first equation (\ref{DOP1F26}) for the first propagation
time $T_{1}$ can be represented in the form
\begin{equation}
T_{1}=\frac{1}{c}\int_{E_{per}^{(1)}}^{\widetilde{E}_{1}}\sqrt{\widetilde{Q}%
_{1}(V_{1},t_{1},x_{1},y_{1},z_{1},E_{1})}dE_{1}+T_{1}^{init}\text{ \ ,}
\label{DOPC1}
\end{equation}%
where $\widetilde{Q}_{1}(V_{1},t_{1},x_{1},y_{1},z_{1},E_{1})$ is the
expression
\begin{equation}
\widetilde{Q}_{1}:=\frac{(c^{2}-2V_{1})}{(c^{2}+2V_{1})}\left(
(dx_{1})^{2}+(dy_{1})^{2}+(dz_{1})^{2}\right) \text{ \ \ \ .}  \label{DOPC2}
\end{equation}%
The expression (\ref{DOPC1}) is in fact identical with the propagation time of a signal,
emitted by a moving along an elliptical orbit satellite. This solution will
be derived in Section \ref{sec:proptime dimens} and in another paper it will be
proved that it can be represented as a sum of elliptic integrals of the first, second
and the third kind.

If from the hyperplane equation (\ref{DOPAB25}) the differential $dz_{2}$ is
expressed and substituted into the second equation (\ref{DOP2F26}) for $%
T_{2} $, then the solution can be written as
\begin{equation}
dT_{2}=\frac{1}{c}\sqrt{\widetilde{Q}%
_{2}(V_{1},t_{1},x_{1},y_{1},z_{1},E_{1},E_{2},x_{2},y_{2},z_{2},V_{2})}%
\text{ \ ,}  \label{DOPC3}
\end{equation}%
where the function $\widetilde{Q}_{2}(....)$ depends on the space
coordinates of both satellites. If these coordinates are parametrized by
means of the first and second eccentric anomaly angles respectively, then the
propagation time $T_{2}$ will also depend on both eccentric anomaly angles $%
\widetilde{E}_{1}$ and $\widetilde{E}_{2}$, i.e. $T_{2}=T_{2}\left(
\widetilde{E}_{1},\widetilde{E}_{2}\right) $.

\subsection{Finding the dependence of the second propagation time on the
first propagation time and on the variable Euclidean distance in the general
case of space-distributed orbits}
\label{sec:propag space}

We shall find this dependence only for the case when each  set $%
(x_{1},y_{1},z_{1})$ and $(x_{2},y_{2},z_{2})$ of the space coordinates
depends on the corresponding eccentric anomaly angles $E_{1}$ and $E_{2}$.
We shall omit the "wave sign" above $E_{1}$ and $E_{2}$, since the
expression for $T_{2}$ is obtained after the integration of the under -
integral expression $\frac{1}{c}\sqrt{\widetilde{Q}_{2}(...,E_{1},E_{2},...)}
$, which depends on $E_{1}$ and $E_{2}$.

Let us write the hyperplane equation (\ref{DOPAB25}) for the differential of
the square of the Euclidean distance as
\begin{equation}
2(z_{1}-z_{2})\widetilde{S}_{1}dE_{1}-2(z_{1}-z_{2})\widetilde{S}%
_{2}dE_{2}=dR_{AB}^{2}\text{ \ \ ,}  \label{DOPC4}
\end{equation}%
where $\widetilde{S}_{i}$ ($i=1,2$) is the introduced notation for
\begin{equation*}
\widetilde{S}_{i}:=\left( \frac{\partial z_{i}}{\partial E_{i}}\right) +%
\frac{(x_{1}-x_{2})}{(z_{1}-z_{2})}\frac{\partial x_{i}}{\partial E_{i}}+
\end{equation*}%
\begin{equation}
+\frac{(y_{1}-y_{2})}{(z_{1}-z_{2})}\frac{\partial y_{i}}{\partial E_{i}}%
\text{ \ \ .}  \label{DOPC5}
\end{equation}%
Expressing the differential $dE_{2}$ from (\ref{DOPC4}), substituting it
into expression (\ref{DOP2F26}) for the second propagation time, and
introducing also the notation ($i=1,2$)
\begin{equation}
\widetilde{R}_{i}:=\frac{1}{c}\sqrt[.]{\left( \frac{c^{2}-2V_{i}}{%
c^{2}+2V_{i}}\right) \left[ \left( \frac{\partial x_{i}}{\partial E_{i}}%
\right) ^{2}+\left( \frac{\partial y_{i}}{\partial E_{i}}\right) ^{2}+\left(
\frac{\partial z_{i}}{\partial E_{i}}\right) ^{2}\right] }\text{ \ \ ,}
\label{DOPC6}
\end{equation}%
one can represent expression (\ref{DOP2F26}) in the form
\begin{equation}
dT_{2}=\frac{1}{c}\sqrt[.]{\widetilde{A}_{1}(dT_{1})^{2}+\widetilde{A}%
_{2}(dT_{1})+\widetilde{A}_{3}}=\frac{1}{c}\sqrt[.]{\widetilde{Q}_{2}(....)}%
\text{ \ \ ,}  \label{DOPC7}
\end{equation}%
where $\widetilde{A}_{1}$, $\widetilde{A}_{2}$ and $\widetilde{A}_{3}$ are
complicated expressions, depending on the derivatives of the space
coordinates, on the expression (\ref{DOPC5}) and (\ref{DOPC6}) for $%
\widetilde{S}_{i}$ and $\widetilde{R}_{i}$ and also on the differential $%
dR_{AB}^{2}$ \ of the square of the Euclidean distance. The expressions for $%
\widetilde{A}_{1}$, $\widetilde{A}_{2}$ and $\widetilde{A}_{3}$ will be
given in Appendix G. It can be noted also that since from (\ref{DOPC1}) and (%
\ref{DOPC4}) the two differentials $dE_{1}$ and $dE_{2}$ can be expressed as
\begin{equation}
dE_{1}=\frac{dT_{1}}{\widetilde{R}_{1}}\text{ \ , }dE_{2}=\frac{\widetilde{S}%
_{1}}{\widetilde{S}_{2}}\frac{dT_{1}}{\widetilde{R}_{1}}-\frac{dR_{AB}^{2}}{%
2(z_{1}-z_{2})\widetilde{S}_{2}}\text{ \ \ }  \label{DOPC8}
\end{equation}%
and the differential $dE_{2}$ contains the unknown function $dR_{AB}^{2}$,
the differential equation in full derivatives cannot be used for the
determination of the second propagation time.

\subsection{Generalization of the formulae for the differential of the
second propagation time for the case of subsequent transmissions by means of a chain of $n$ satellites}
\label{sec:chain satell}

Suppose that a chain of $n$ satellites transmit a signal in one direction.
If the propagation time for the first satellite is $T_{1}$, the propagation
time for the second satellite will be $T_{2}=T_{2}(T_{1},\mathbf{r}_{1},%
\mathbf{r}_{2})$, where $\mathbf{r}_{1}=(x_{1},y_{1},z_{1})$ and $\mathbf{r}%
_{2}=(x_{2},y_{2},z_{2})$. Analogously, the $n-$th propagation time will be $%
T_{n}=T_{n}(T_{n-1},T_{n-2},.....T_{1},\mathbf{r}_{n},\mathbf{r}_{n-1\mathbf{%
\ }}......\mathbf{r}_{2},\mathbf{r}_{1})$ and will be dependent on all the
preceding propagation times and on the space coordinates of all the
satellites. Since the coefficient functions $\widetilde{A}_{1},\widetilde{A}%
_{2},\widetilde{A}_{3}$ in expression (\ref{DOPC7}) depend on $\mathbf{r}%
_{1},\mathbf{r}_{2}$ (see expressions (\ref{GGGG1}), (\ref{GGGG4}) and (\ref%
{GGGG5}) in Appendix G), the coefficient functions in the expression for the
differential of the $n-$th propagation time $dT_{n}$ will contain
coefficient functions, denoted as $\widetilde{A}_{n,n-1}^{(1)},\widetilde{A}%
_{n,n-1}^{(2)},\widetilde{A}_{n,n-1}^{(3)}$, where the lower subscripts $%
n,n-1$ denote the dependence of the functions on the $n-$th and $(n-1)-$th
space coordinates of the two satellites and the upper subscripts $%
(1),(2),(3) $ denote the order of appearance of the coefficient functions.
Consequently, the differential of the $n-$th propagation time can be
expressed as follows
\begin{equation}
dT_{n}=\frac{1}{c}\sqrt[.]{\widetilde{A}_{n,n-1}^{(1)}(dT_{n-1})^{2}+%
\widetilde{A}_{n,n-1}^{(2)}(dT_{n-1})+\widetilde{A}_{n,n-1}^{(3)}}\text{ \ \
,}  \label{DOPCC7}
\end{equation}%
where after applying the same formulae with respect to $dT_{n-1}$, the first
term will give terms with second inverse powers in $c$, i.e.
\begin{equation}
\widetilde{A}_{n,n-1}^{(1)}\frac{1}{c^{2}}\left[ \widetilde{A}%
_{n-1,n-2}^{(1)}(dT_{n-2})^{2}+\widetilde{A}_{n-1,n-2}^{(2)}(dT_{n-2})+%
\widetilde{A}_{n-1,n-2}^{(3)}\right] \text{ \ \ ,}  \label{DOPCC8}
\end{equation}%
and the second term $\widetilde{A}_{n,n-1}^{(2)}(dT_{n-1})$ will give an
embedded square root. The recurrent calculation of such expressions is a
very complicated mathematical problem. In order to find the sum of all the
differentials $dT_{n}$ from $1$ to $N$, one has to calculate the sum \
\begin{equation}
dT_{1}+\frac{1}{c}\sum\limits_{k=2}^{N}B_{k}\text{ \ \ ,}  \label{DOPCC9}
\end{equation}%
where
\begin{equation}
B_{k}:=\sqrt[.]{\widetilde{A}_{k,k-1}^{(1)}(dT_{k-1})^{2}+\widetilde{A}%
_{k,k-1}^{(2)}(dT_{k-1})+\widetilde{A}_{k,k-1}^{(3)}}\text{ \ \ .}
\label{DOPCC10}
\end{equation}%
The summation of such sums shall not be performed here.

\subsection{Derivation of the integral equation with respect to the square
of the Euclidean distance}
\label{sec:integral}

Up to the present moment, from the system of three equations (\ref{DOPAB25}%
), (\ref{DOP2F26}) and (\ref{DOP1F26}), only the first two have been used.
Consequently, one may combine the first and the third equation. For the
purpose, one may express the differential $dE_{1}$ from expression (\ref%
{DOPC4}) and substitute it in equation (\ref{DOP1F26}) for the first
propagation time $T_{1}$, then the following equation in full derivatives
will be obtained
\begin{equation*}
\left[ \frac{\partial T_{1}}{\partial E_{1}}-\frac{\widetilde{R}_{1}}{%
2(z_{1}-z_{2})\widetilde{S}_{1}}\frac{\partial R_{AB}^{2}}{\partial E_{1}}%
\right] dE_{1}+
\end{equation*}%
\begin{equation}
+\left[ -\frac{\widetilde{R}_{1}}{2(z_{1}-z_{2})\widetilde{S}_{1}}\frac{%
\partial R_{AB}^{2}}{\partial E_{2}}-\widetilde{R}_{1}\frac{\widetilde{S}_{2}%
}{\widetilde{S}_{1}}\right] dE_{2}=0\text{ \ \ \ . }  \label{DOPC9}
\end{equation}%
This equation will be solvable if it is of the type
\begin{equation}
\frac{\partial \widetilde{Q}_{1}}{\partial E_{1}}dE_{1}+\frac{\partial
\widetilde{Q}_{1}}{\partial E_{2}}dE_{2}=0\text{ }\Longrightarrow d%
\widetilde{Q}_{1}=0\Longrightarrow \text{\ }\widetilde{Q}_{1}=const\text{ \
\ \ .}  \label{DOPC10}
\end{equation}%
Setting up the first expression in the square brackets in (\ref{DOPC9})
equal to $\frac{\partial \widetilde{Q}_{1}}{\partial E_{1}}$ and performing
the integration with respect to $E_{1}$, the following expression is
obtained for $\widetilde{Q}_{1}$
\begin{equation}
\widetilde{Q}_{1}:=T_{1}-\frac{1}{2}\int \frac{\widetilde{R}_{1}}{%
(z_{1}-z_{2})\widetilde{S}_{1}}.\frac{\partial R_{AB}^{2}}{\partial E_{1}}%
dE_{1}+\varphi (E_{2})\text{ \ \ \ ,}  \label{DOPC11}
\end{equation}%
where an arbitrary function $\varphi (E_{2})$ has been added. If this
expression is differentiated by $E_{2}$ and is set up equal to the second
expression in the square brackets in front of $dE_{2}$ in (\ref{DOPC9}), the
derivative $\frac{\partial \varphi (E_{2})}{\partial E_{2}}$ can be
calculated to be
\begin{equation*}
\frac{\partial \varphi (E_{2})}{\partial E_{2}}=-\widetilde{R}_{1}\frac{%
\widetilde{S}_{2}}{\widetilde{S}_{1}}+\frac{1}{2}R_{AB}^{2}\frac{\partial }{%
\partial E_{2}}\left[ \frac{\widetilde{R}_{1}}{(z_{1}-z_{2})\widetilde{S}_{1}%
}\right] -
\end{equation*}%
\begin{equation*}
-\frac{1}{2}\int \frac{\partial R_{AB}^{2}}{\partial E_{2}}.\frac{\partial
}{\partial E_{1}}\left[ \frac{\widetilde{R}_{1}}{(z_{1}-z_{2})\widetilde{S}%
_{1}}\right] dE_{1}-
\end{equation*}%
\begin{equation}
-\frac{1}{2}\int R_{AB}^{2}\frac{\partial ^{2}}{\partial E_{1}\partial E_{2}%
}\left[ \frac{\widetilde{R}_{1}}{(z_{1}-z_{2})\widetilde{S}_{1}}\right]
dE_{1}\text{ \ \ \ .}  \label{DOPC12}
\end{equation}%
After performing the integration and substituting the derived expression for
$\varphi (E_{2})$ into (\ref{DOPC11}) for $\widetilde{Q}_{1}$, one can
obtain the following linear integral equation with respect to the square of
the Euclidean distance $R_{AB}^{2}=R_{AB}^{2}(E_{1},E_{2})$
\begin{equation*}
\frac{1}{2}\int R_{AB}^{2}\frac{\partial }{\partial E_{2}}\left( \frac{%
\widetilde{R}_{1}}{(z_{1}-z_{2})\widetilde{S}_{1}}\right) dE_{2}=\frac{1}{2}%
\frac{\widetilde{R}_{1}}{(z_{1}-z_{2})\widetilde{S}_{1}}R_{AB}^{2}-
\end{equation*}%
\begin{equation}
-T_{1}+\int \widetilde{R}_{1}\frac{\widetilde{S}_{2}}{\widetilde{S}_{1}}%
dE_{2}+const\text{ \ \ \ .}  \label{DOPC13}
\end{equation}%
This equation has an interesting interpretation, since the solution with
respect to the function $R_{AB}^{2}$ should not necessarily coincide with
the expression for the Euclidean distance. The reason is that the integral
equation has been obtained as a result of the intersection of two
four-dimensional null cones and a hyperplane equation. As a result, the
Euclidean distance will turn out to be related to the space-time distance.
Further in the text, equation (\ref{ABC36A}) in full derivatives in section %
\ref{sec:Deriv form} will be derived for the investigated two-dimensional
case of plane Keplerian motion. In a sense, equation (\ref{DOPC13}) is the
analogue of (\ref{ABC36A}), demonstrating that the two-dimensional case
offers some simplifications. \ \

\subsection{Intersecting cones and hyperplanes and comparison between atomic
clocks on satellites at a distance}
\label{sec:interscones atomic}

It may be noted that the same approach of intersecting cones and hyperplanes
may be developed also with respect to the proper time $\tau $. In a similar
way, the algebraic geometry approach of intersecting cones and hyperplanes
will give the mutual dependence of the readings of the atomic clocks on two distant
satellites, moving on one elliptical orbit or on two, space-distributed
orbits. The proper time, measured by the atomic clocks is the atomic time
and it is determined in a fixed, non-moving reference system ($d\overline{x}%
=d\overline{y}=d\overline{z}=0$) as
\begin{equation}
ds^{2}=-c^{2}d\tau ^{2}=-g_{\alpha \beta }dx^{\alpha }dx^{\beta }\text{ }\ \
\ (\alpha ,\beta =0,1,2,3)\text{\ \ .}  \label{DOP23}
\end{equation}%
Note that the interval $ds^{2}=-c^{2}d\tau ^{2}$ is defined with a minus
sign, although in some review papers \cite{C4}, \cite{C5} on GPS theory it
is defined with a positive sign.

\subsubsection{Clarification on the approach of Fock about
the atomic time definition}

The space-time interval is defined also in the monograph by Fock \cite%
{Fock} with a positive sign. In this monograph there is a statement that\ the
expression
\begin{equation}
\tau =\frac{1}{c}\int\limits_{0}^{t}\sqrt{g_{00}+2g_{0i}\overset{.}{x}%
_{i}+g_{ik}\overset{.}{x}_{i}\overset{.}{x}_{k}}dt  \label{DOP24}
\end{equation}%
should be used in the case of accelerating motion of the observer and should
replace the usual expression
\begin{equation}
\tau =\int\limits_{0}^{t}\left[ 1-\frac{1}{c^{2}}\left( \frac{1}{2}%
v^{2}+V\right) \right] dt  \label{DOPAA24}
\end{equation}%
for the proper time in case of uniform,non-accelerating motion, provided
that the integral (\ref{DOP24}) is invariant under arbitrary changes of the
space and time coordinates. In the present case, there is no such
non-invariance, since the parametrization of the space coordinates is fixed,
and as it will be shown, there will be no indeterminacy with respect to the
time coordinate. Another important remark in the monograph \cite{Fock} is
that "it is difficult to predict what will be the behaviour of these clocks,
when subjected to sudden or accelerated motion". From expression (\ref{C5A27}%
)$~$\ in Section \ref{sec:proptime dimens} for the Kepler velocity $\mathbf{v=}\sqrt{v_{x}^{2}+v_{y}^{2}}=\frac{na%
}{\left( 1-e\cos E\right) }\sqrt[.]{1-e^{2}\cos ^{2}E}$ it can be calculated that
the satellite has a constant accelerating motion because
\begin{equation}
a=\frac{\partial v}{\partial E}=-\frac{nae\sin E}{\left( 1-e\cos E\right)
\sqrt{1-e^{2}\cos ^{2}E}}\neq 0\text{ \ \ .}  \label{DOPBB24}
\end{equation}%
The remark by Fock should be properly understood not in the sense that it is
impossible to apply formulaes (\ref{DOP23}) and (\ref{DOP24}) for the case
of accelerated motion, but in the sense that these formulaes will not
provide exact numerical values for the atomic time, because they represent
complicated expressions, almost impossible to be integrated exactly. It will
be proved also in a future publication that if the metric (\ref{DOP25}) is
used, then the definition with a negative sign will not lead to any
contradiction.

\subsubsection{Two satellites and two intersecting
five-dimensional cones with the hyperplane equation for the two atomic times}

For the case of two satellites, the intersecting cones (considered together
with the hyperplane equation) will be five dimensional ones for the two
satellites with atomic times $\tau _{1}$ and $\tau _{2}$ respectively
\begin{equation}
-c^{2}d\tau _{1}^{2}=-c^{2}\left( 1+\frac{2V_{1}}{c^{2}}\right) (d\overline{T%
}_{1}^{.})^{2}+\left( 1-\frac{2V_{1.}}{c^{2}}\right) \left(
(dx_{1}^{.})^{2}+(dy_{1}^{.})^{2}+(dz_{1}^{.})^{2}\right) \text{ \ \ ,}
\label{DOPA25}
\end{equation}%
\begin{equation}
-c^{2}d\tau _{2}^{2}=-c^{2}\left( 1+\frac{2V_{2}}{c^{2}}\right) (d\overline{T%
}_{2}^{.})^{2}+\left( 1-\frac{2V_{2.}}{c^{2}}\right) \left(
(dx_{2}^{.})^{2}+(dy_{2}^{.})^{2}+(dz_{2}^{.})^{2}\right) \text{ \ \ .}
\label{DOPB25}
\end{equation}%
In these equations, $\overline{T}_{1}$ and $\overline{T}_{2}$ are no longer
propagation times, but they are the Geocentric Coordinate Time (TCG)
coordinates of the Geocentric Celestial Reference System (GCRS), defined for
the space around the Earth. Further it shall be proved that the TCG can be
identified with the celestial times of motion $t_{1}$ and $t_{2}$ for the
first and the second satellite, which are determined from the Kepler
equation $E-e\sin E=n(t-t_{per})$. The atomic times of the clock on the
first or the second satellite, when they are at positions on the orbit,
characterized by the corresponding eccentric anomaly angles $\widetilde{E}%
_{1}$ or $\widetilde{E}_{2}$, are given after performing the integrations of
the above equations, provided that the parametrizations of the two elliptic
orbits are known and also that $\overline{T}_{1}^{.}=t_{1}$ and $\overline{T}%
_{2}^{.}=t_{2}$, in view of the identification of the geocentric time with
the celestial time, which shall be proved for the case of the diagonal
metric (\ref{DOP25}). If equations (\ref{DOPA25}) and (\ref{DOPB25}) are
treated together with equation (\ref{DOPAB25}), then it will turn out that
the first atomic time will depend on the eccentric anomaly angle $\widetilde{%
E}_{1}$ because $\tau _{1}$ from (\ref{DOPA25}) can be expressed as
\begin{equation}
\tau _{1}=\frac{1}{c}\int_{E_{per}^{(1)}}^{\widetilde{E}_{1}}\sqrt{%
\widetilde{P}_{1}(V_{1},t_{1},x_{1},y_{1},z_{1},E_{1})}dE_{1}+\tau
_{1}^{init}\text{ \ .}  \label{DOP6B25}
\end{equation}%
In the above expression
\begin{equation*}
\widetilde{P}_{1}=-c^{2}\left( 1+\frac{2V_{1}}{c^{2}}\right) \left( \frac{%
\partial t_{1}}{\partial E_{1}}\right) ^{2}+
\end{equation*}%
\begin{equation}
+\left( 1-\frac{2V_{1}}{c^{2}}\right) \left[ \left( \frac{\partial x_{1}}{%
\partial E_{1}}\right) ^{2}+\left( \frac{\partial y_{1}}{\partial E_{1}}%
\right) ^{2}+\left( \frac{\partial z_{1}}{\partial E_{1}}\right) ^{2}\right]
\text{ \ .}  \label{DOP7B25}
\end{equation}%
In such a mathematical setting, the first atomic time $\tau _{1}$ depends
only on the first eccentric anomaly angle. However, if the second equation (%
\ref{DOPB25}) is considered together with the hyperplane equation (\ref%
{DOPAB25}), then the second atomic time $\tau _{2}$ will be expressed by a
formulae, analogous to (\ref{DOP6B25}), where the function $\widetilde{P}%
_{2} $ under the square is expressed by $\widetilde{P}%
_{2}(V_{2},t_{2},x_{2},y_{2},x_{1},y_{1},z_{1},E_{1},E_{2})$. Consequently,
the second atomic time $\tau _{2}$ depends on both eccentric anomaly angles $%
\widetilde{E}_{1}$ and $\widetilde{E}_{2}$. However, in case of two- way
transfer between satellites (which is not the case of time transfer between
two stations and a geostationary satellite, described in \cite{AAB21}), the
first atomic time can also depend on both eccentric anomaly angles.

A similar situation will arise further in this paper, when each of the
propagation times $T_{1}$ and $T_{2}$ will depend on $E_{1}$ or on both $%
E_{1}$ and $E_{2}$. Further the non-contradictory definition of the two
propagation times, defined respectively at the moment of emission and
reception of the signal, will also be explained, since it might seem that
only one propagation time is sufficient. In fact, the idea about two
propagation times has some apparent advantages, enabling the derivation of a
cubic algebraic equation, relating the differentials of the propagation and
the atomic times. Also, an idea will be given how the two eccentric anomaly
angles $\widetilde{E}_{1}$ and $\widetilde{E}_{2}$ are to be obtained from
the coupled system of nonlinear integral equations for the equality of the
celestial and the propagation times.

\subsection{Equality of celestial and propagation times as
conditions for emission of the signal by the first satellite and reception
of the signal by the second satellite - first possible definition}
\label{sec:celest propag1}

If the two celestial times are equal correspondingly to the first and the
second propagation time $T_{1}$ and $T_{2}$, i.e. if
\begin{equation}
t_{1}=T_{1}\text{ \ \ , \ \ \ }t_{2}=T_{2}\text{ \ \ \ ,}  \label{DOPC25}
\end{equation}%
then since the propagation time is expressed from the null cone equation $%
ds^{2}=0$, the right-hand sides of the defining equations (\ref{DOPA25}) and
(\ref{DOPB25}) will be equal to zero. In such a way, the two equalities in (%
\ref{DOPC25}) give the boundary conditions $c^{2}d\tau _{1}^{2}=0$ and $%
c^{2}d\tau _{2}^{2}=0$, i.e.$\tau _{1}^{.}=const_{1}$ and $\tau
_{2}^{.}=const_{2}$, meaning that at the moments of emission and reception
of the signal, the atomic times $\tau _{1}$ and $\tau _{2}$ cannot be
determined from the two systems of equations (\ref{DOP1F26}) and (\ref%
{DOP2F26}) for the two propagation times and (\ref{DOPA25}), (\ref{DOPB25})
for the two atomic times. This should be expected, because the change of the
atomic time is related to the transportation of the atomic clock (i.e. the
trajectory of the satellite) and not to the path of the signal. It is
important to mention this fact, because in the next section the advantage of
another definition will be shown, based on the difference between the first propagation
time of emission and the second propagation time of reception.

Furthermore, the notions about the first and second propagation times $T_{1}$
and $T_{2}$ are not different from the standard notion about the propagation
time, considered as the difference between the times of reception and
emission of the signal, if these times are counted from some initial moment,
coinciding with the moment of perigee passage. So the new definition about
the two propagation times $T_{1}$ and $T_{2}$, explained in details in
Section \ref{sec:proptime 1satel} implies that each propagation time is a
difference between the time of emission (or reception) of the signal and a
"fictitious" initial moment of emission (or reception) of the signal. However,
the definition of two propagation times (one for emission and one for
reception of the signal) gives the opportunity to count these times with
respect to two different initial moments of perigee passage for the first
and the second satellite.

\subsection{Second possible definition for the pair of celestial times,
equal to the difference in the two coordinate times of reception and
emission}
\label{sec:celest propag2}

The representation of the propagation time as a difference between two
coordinate times of reception and emission provides the opportunity to
define the celestial times $t_{1}$ and $t_{2}$
\begin{equation}
t_{1}=t_{2}=T_{2}-T_{1}\text{ \ \ \ .}  \label{DOPD26}
\end{equation}%
This means that the first celestial time $t_{1}$ is determined as the time
of motion (counted from the perigee passage) of the first satellite from the
moment of emission of the signal to the moment of reception of the signal by
the second satellite. Note that from $t_{1}=t_{2}$ it does not necessarily
follow that $E_{1}=E_{2}$ since the time of perigee passage for the two
satellites might be different and moreover, the characteristics of the two
orbits may be different. Similarly the second celestial time can be defined.
Thus, substituting the condition (\ref{DOPD26}) into (\ref{DOPA25}) and (\ref%
{DOPB25}), and taking into account the null cone equations $ds_{1}^{2}=0$
and $ds_{2}^{2}=0$ for the signal propagation, one can obtain the following
pair of equations for both the atomic and propagation times
\begin{equation*}
-c^{2}(d\tau _{1})^{2}=-c^{2}(1+\frac{2V_{1}}{c^{2}})(dT_{2})^{2}+
\end{equation*}%
\begin{equation}
+2c^{2}(1+\frac{2V_{1}}{c^{2}})(dT_{1})(dT_{2})\text{ \ \ \ ,}
\label{DOP1B25}
\end{equation}%
\begin{equation*}
-c^{2}(d\tau _{2})^{2}=-c^{2}(1+\frac{2V_{2}}{c^{2}})(dT_{1})^{2}+
\end{equation*}%
\begin{equation}
+2c^{2}(1+\frac{2V_{2}}{c^{2}})(dT_{1})(dT_{2})\text{ \ \ \ .}
\label{DOP2B25}
\end{equation}%
Simple algebraic manipulations can prove that the above two quadratic
four-dimensional algebraic surfaces in terms of the variables $dT_{1}$, $%
dT_{2}$, $d\tau _{1}$, $d\tau _{2}$ have an intersecting variety, which is a
four-dimensional cubic algebraic surface
\begin{equation}
(d\tau _{2})^{2}(1+\frac{2V_{1}}{c^{2}})(dT_{2})+(d\tau _{1})^{2}(1+\frac{%
2V_{2}}{c^{2}})(dT_{1})=0\text{ \ \ .}  \label{DOP3B25}
\end{equation}%
This formulae illustrates how important is the calculation of the
propagation time  also with respect to the determination of the atomic
time. However, the above relation acquires a more complicated form if the
hyperplane equation (\ref{DOPAB25}) is taken into account. It should be
noted that the hyperplane equation might be required to intersect either the
system of (algebraic) equations for the propagation time or the equations
for the atomic time, or both system of equations for the propagation and
atomic times. However, although the two propagation times and the two atomic times are
related by (\ref{DOP3B25}), in principle they have a separate physical meaning. Consequently, the
more correct approach will be to consider the intersection of the hyperplane equation with the
two algebraic equations (five-dimensional cones in the general case) for the atomic times.

\subsection{Synchronization of atomic clocks on moving satellites and the
navigation message}
\label{sec:navig message}

The above theoretical model, for the moment exposed in its general form, has
some important consequences, concerning the clock correction in the
navigation message, which should ensure the synchronization of clocks. The
synchronization is not between an atomic clock on an Earth station and a
satellite \cite{ElRabbany}, but between atomic clocks on two satellites. For
the purpose, the navigation message should contain the clock correction for
the second, signal-receiving satellite at the moment of reception, the clock
correction for the first satellite at the moment of emission of the signal
and also at the moment of reception of the signal by the second satellite.
The last correction should be calculated in the framework of the proposed
formalism. In such a way, the two clock corrections for the first and the
second satellite can be compared at the moment of reception of the signal
with the purpose of synchronizing the clocks. In other words, at the moment
of reception the signal has to keep memory of its initial characteristics at
the moment of emission by the first satellite, determined by its motion
along the orbit. In the strict mathematical sense, this will be closely
related to the fact that the first propagation time will depend only on the
first eccentric anomaly angle, while the second propagation time will depend
on both eccentric anomaly angles.

Formulaes (\ref{DOP6B25}), (\ref{DOP7B25}) and the corresponding ones for $%
\tau _{2}$ within the formalism of intersecting null cones, as well as the four-dimensional cubic
algebraic surface (\ref{DOP3B25}), give the opportunity to treat the problem for synchronization of moving atomic
clocks. In the current literature, the problem for synchronization of an
moving atomic clock and a non-moving one has been studied in \cite{AAB10}.
The basic assumption in this treatment was that the coordinate intervals $%
dT_{m}$ for the moving and the non-moving one $dT$ should be equal, i.e. $%
dT_{m}=dT$. A moving atomic clock on the Earth surface is for example the
s.c. "transportable hydrogen quantum clock" (THQC) Sapfire \cite{AAB11}. In
the present case of two moving one with respect to another atomic clocks,
the equality of the atomic intervals should be replaced by the complicated
system of equations (\ref{DOP1F26}) and (\ref{DOP2F26}) for the
differentials of the two propagation times.

\subsection{The case of propagation of a signal, emitted by moving along an
elliptical orbit satellite}

The propagation time can be calculated also for the case of a signal,
emitted by a moving satellite. This case is treated also in this paper in
Section \ref{sec:one satellite} up to the derivation of the formulae for the
propagation time with the purpose of demonstrating that it has the proper
dimension of seconds. Since further the numerical calculation of the
corresponding elliptic integrals of the first, second and third kind
requires the application of the methods of complex analysis, this will be
postponed for another publication. The numerical analysis will demonstrate
one important feature - for equal eccentric anomaly angles, the propagation
time for the signal between moving satellites will turn out to be greater
than the propagation time for a signal emitted by a single satellite, when
there will be no need to implement the formalism of two null cones.

In this paper this problem is investigated for the case of the metric element (%
\ref{C5A19}), which neglects any tidal effects due to the Moon and the Sun,
orbital perturbations due to the harmonics of the Earth gravitational
potential. The basic assumption for derivation of the formulae for the
propagation time (\ref{C5A40}) in Section \ref{sec:proptime 1satel} is the
smallness of the parameter $\beta =$ $\frac{2G_{\oplus }M_{\oplus }}{c^{2}a}%
=0.334.10^{-9}\ll 1$, which means that terms smaller than this value are
neglected. Under such assumptions, the analytical formulae (\ref{C5A40}) for
the first (the $O(\frac{1}{c})$ correction) and the second (the $O(\frac{1}{%
c^{3}})$ correction) in the propagation time have been derived. In another
publication it will be shown that the numerical value of the first term is $%
I_{1}^{(E_{(3)})}=0.0281341332790419$ $\left[ \sec \right] $, where $E_{(3)}$
is the third iteration of the solution of the Kepler equation as a
transcendental equation with respect to the eccentric anomaly angle.

\subsection{The rate of change of the atomic time with respect to the
propagation time for the case of a single satellite, emitting a signal while
moving along an elliptical orbit - two equivalent representations for the
propagation time}
\label{sec:atom prop}

It would be interesting to compare the rate of change of the propagation
time  with respect to the atomic time. This means that a dependence of the
propagation time with respect to the atomic time should exist,i.e. $T=T(\tau
)$. Such a dependence is nontrivial because atomic time is a characteristic
of the atomic clock on the moving satellite, while the propagation time is a
characteristic of the propagating signal after its decoupling from the
motion of the satellite. So these two times in principle might not be
comparable. Let us perform a simple calculation, based on the known
experimental fact \ \cite{HandbookGNSS} \ that at an altitude of $20184$ $km$%
, due to the difference in the gravitational potential, the satellite atomic
time runs faster by $45$ $\mu \sec /d$ (microseconds per day). Consequently,
for one second the atomic time will run faster by $0.5208333.10^{-9}$ $[\sec
]$, which for the interval of propagation time $dT=0.0281341332790419$ $%
\left[ \sec \right] $ will give an interval of atomic time $d\tau
=0.0146531934.10^{-9}$ $[\sec ]$. The ratio of the two time intervals is
equal to
\begin{equation}
\frac{d\tau }{dT}=\frac{0.0146531934.10^{-9}}{0.0281341332790419}%
=0.52083329721.10^{-9}\text{ .}  \label{KLM0}
\end{equation}%
The very small atomic time interval compared to the propagation interval
means that the atomic time can serve as a standard for measuring the
propagation time, because it will be able to detect changes even at the
nanosecond level. From a theoretical point of view, the ratio $\frac{d\tau }{%
dT}$ can be found from the simple relation
\begin{equation}
\frac{d\tau }{dT}.\frac{dT}{dE}.\frac{dE}{dt_{cel}}=\frac{d\tau }{dt_{cel}}%
\text{ \ ,}  \label{KLM1}
\end{equation}%
where $\frac{dT}{dE}$ will be found from expression (\ref{C5A40}) for the
propagation time, calculated from the null cone equation (\ref{C5A21}) for
the plane motion of the satellite, $\frac{dE}{dt_{cel}}$ can be calculated
from the Kepler equation and $\frac{d\tau }{dt_{cel}}$ - from the two
equivalent representations for the propagation time. For the purpose, let us
write the null cone equation in the form
\begin{equation}
\ ds^{2}=g_{00}c^{2}(dT)^{2}+2g_{0j}cdTdx^{j}+g_{ij}dx^{i}dx^{j}=0\text{ \ \
.}  \label{KLM2}
\end{equation}%
\ From here the differential of the propagation time $dT$ can be expressed
as a solution of the quadratic equation in the form
\begin{equation*}
dT=\pm \frac{1}{c}\frac{1}{\sqrt[.]{-g_{00}}}\sqrt[.]{\left( g_{ij}+\frac{%
g_{0i}g_{0j}}{-g_{00}}\right) dx^{i}dx^{j}}+
\end{equation*}%
\begin{equation}
+\frac{1}{c}\left( \frac{g_{0j}}{-g_{00}}\right) dx^{j}\text{ \ \ .}
\label{KLM3}
\end{equation}

\subsubsection{First representation of the propagation time}

After integrating, dividing and multiplying by $d\tau $, one can obtain the
first representation for the propagation time as
\begin{equation*}
T=\pm \frac{1}{c}\frac{1}{\sqrt[.]{-g_{00}}}\int \sqrt[.]{\left( g_{ij}+%
\frac{g_{0i}g_{0j}}{-g_{00}}\right) \frac{dx^{i}}{d\tau }\frac{dx^{j}}{d\tau
}}d\tau +
\end{equation*}%
\begin{equation}
+\frac{1}{c}\int \left( \frac{g_{0j}}{-g_{00}}\right) \frac{dx^{j}}{d\tau }%
\text{ }d\tau \text{\ \ .}  \label{KLM4}
\end{equation}%
The differential of the atomic time $d\tau $ can be found from the defining
expression (\ref{DOP23}), multiplying and dividing by $dt_{cel}$
\begin{equation}
d\tau =\frac{1}{c}dt_{cel}\sqrt[.]{g_{00}c^{2}+2g_{0j}c\frac{dx^{j}}{dt_{cel}%
}+g_{ij}\frac{dx^{i}}{dt_{cel}}\frac{dx^{j}}{dt_{cel}}}\text{ \ \ ,}
\label{KLM5}
\end{equation}%
where $t_{cel\text{ }}$is the celestial time in the Kepler equation, counted
from the moment of perigee passage and $\frac{dx^{i}}{dt_{cel}}=v^{i}$ is
the $i-$th component of the velocity.

Expressing the differential of the celestial time $dt_{cel}$ from the Kepler
equation as
\begin{equation}
dt_{cel}=\frac{1}{n}(1-e\cos E)dE\text{ \ \ ,}  \label{KLM5A}
\end{equation}
substituting in the above expression and introducing the notation
\begin{equation}
\widetilde{N}:=g_{00}c^{2}+2g_{0k}c\frac{dx^{k}}{dt_{cel}}+g_{kl}\frac{dx^{k}%
}{dt_{cel}}\frac{dx^{l}}{dt_{cel}}\text{ \ ,}  \label{KLM6}
\end{equation}%
one can rewrite the differential $d\tau $ (\ref{KLM5}) as%
\begin{equation}
d\tau =\frac{1}{c}\frac{\left( 1-e\cos E\right) }{n}\sqrt[.]{\widetilde{N}}%
dE=\frac{1}{c}\sqrt[.]{\widetilde{N}}dt_{cel}\text{ \ \ .}  \label{KLM7}
\end{equation}%
Using the simple relations
\begin{equation}
\frac{dx^{i}}{d\tau }=\frac{dx^{i}}{dt_{cel}}\frac{dt_{cel}}{d\tau }=v^{i}%
\frac{dt_{cel}}{d\tau }\text{ \ \ }  \label{KLM8}
\end{equation}
and substituting the differential $d\tau $ (\ref{KLM7}) in expression (\ref%
{KLM4}), one can obtain the first representation for the propagation time $T$
\begin{equation*}
T=\mp \frac{1}{c^{2}}\int \frac{1}{\sqrt[.]{-g_{00}}}\sqrt[.]{\left( g_{ij}+%
\frac{g_{0i}g_{0j}}{-g_{00}}\right) v^{i}v^{j}\left( \frac{dt_{cel}}{d\tau }%
\right) ^{2}\widetilde{N}}dt_{cel}+
\end{equation*}%
\begin{equation}
+\frac{1}{c^{2}}\int \left( \frac{g_{oj}}{-g_{00}}\right) v^{j}\left( \frac{%
dt_{cel}}{d\tau }\right) \sqrt[.]{\widetilde{N}}dt_{cel}\text{ \ \ \ .}
\label{KLM9}
\end{equation}

\subsubsection{Second representation of the propagation time}

The second representation is derived after dividing and multiplying (\ref%
{KLM4}) by $dt_{cel}$
\begin{equation*}
T=\mp \frac{1}{c}\int \frac{1}{\sqrt[.]{-g_{00}}}\sqrt[.]{\left( g_{ij}+%
\frac{g_{0i}g_{0j}}{-g_{00}}\right) v^{i}v^{j}}dt_{cel}+
\end{equation*}%
\begin{equation}
+\frac{1}{c}\int \left( \frac{g_{oj}}{-g_{00}}\right) v^{j}dt_{cel}\text{ \
\ \ .}  \label{KLM10}
\end{equation}%
Comparing the two representations, the following simple relation is obtained
\begin{equation}
\frac{d\tau }{dt_{cel}}=\frac{1}{c}\sqrt[.]{\widetilde{N}}\text{ \ \ .}
\label{KLM11}
\end{equation}%
Taking \ into account that $\frac{d\tau }{dT}=\frac{d\tau }{dE}.\frac{dE}{dT}
$, formulae (\ref{KLM1}) can be rewritten as
\begin{equation}
\frac{d\tau }{dE}=\left( \frac{d\tau }{dt_{cel}}\right) \left( \frac{dt_{cel}%
}{dE}\right) =\text{ \ \ }  \label{KLM12}
\end{equation}
\begin{equation}
=\frac{1}{nc}(1-e\cos E)\sqrt[.]{\widetilde{N}}=  \label{KLM13}
\end{equation}%
\begin{equation}
=\frac{1}{n}(1-e\cos E)\sqrt{g_{00}c^{2}+2g_{0k}cv^{k}+g_{kl}v^{k}v^{l}}%
\text{ \ ,}  \label{KLM14}
\end{equation}%
where expressions (\ref{KLM11}), (\ref{KLM5A}) and (\ref{KLM6}) have been
used. For the diagonal metric (\ref{DOP25}), the above formulae can be
written as
\begin{equation}
\frac{d\tau }{dE}=\frac{c}{n}(1-e\cos E)\sqrt{g_{00}+\frac{1}{c^{2}}%
g_{kk}(v^{k})^{2}}\text{ \ \ .}  \label{KLM15}
\end{equation}%
In order to see whether the ratio $\frac{d\tau }{dT}$ of the atomic time
differential and the propagation time differential (\ref{KLM0}) is a very
small number, one has to compute
\begin{equation}
\frac{d\tau }{dT}=\frac{1}{nc}\frac{(1-e\cos E)\sqrt[.]{\widetilde{N}}}{%
\frac{dT}{dE}}=  \label{KLM15A}
\end{equation}%
\begin{equation}
=\frac{1}{n}\frac{(1-e\cos E)\sqrt{g_{00}+\frac{2g_{0k}v^{k}}{c}+\frac{%
g_{kl}v^{k}v^{l}}{c^{2}}}}{\frac{a}{c^{2}}\sqrt[.]{1-e^{2}\cos ^{2}E}\text{ }%
-\frac{2G_{\oplus }M_{\oplus }}{c^{4}}\sqrt[.]{\frac{1+e\cos E}{1-e\cos E}}}%
\text{ \ \ \ ,}  \label{KLM15B}
\end{equation}%
where formulaes (\ref{KLM13}), (\ref{KLM14}) and formulae (\ref{C5A40}) for
the propagation time $T$ (to be derived in the following sections) have been
used. The key problem for establishing whether $\frac{d\tau }{dT}$ will be
of the order of $10^{-9}$ will be to check whether the assumed approximation
\begin{equation}
\frac{2V}{c^{2}}=\frac{2GM}{c^{2}a(1-e\cos E)}=\frac{\beta }{(1-e\cos E)}=%
\frac{0.334.10^{-9}}{(1-e\cos E)}\ll 1  \label{KLM15C}
\end{equation}
and the resulting neglect of the terms, smaller than the numerical value $%
0.334.10^{-9}$ , will enable the desired numerical accuracy for $\frac{d\tau
}{dT}$. If not, the propagation time in (\ref{C5A40}) has to be calculated
beyond the approximation (\ref{KLM15C}) and consequently the expression in
the denominator in (\ref{KLM15B}) will be another.

\subsubsection{Comparison with similar formulaes and some important comments}

\ \ \ \ \ It is interesting to note that assuming the $g_{00}$ metric tensor
component to be equal to $g_{00}=$\ $1-\frac{V}{c^{2}}$, the above formulae
resembles very much the equation (see the monograph \cite{C25})
\begin{equation}
\frac{d\tau }{d\overline{T}}=\left[ 1-\frac{2V}{c^{2}}-\frac{v^{2}}{c^{2}}%
\right] ^{\frac{1}{2}}\approx 1-\frac{V}{c^{2}}-\frac{v^{2}}{2c^{2}}\text{ \
\ \ ,}  \label{KLM16}
\end{equation}%
obtained from the equality of the space-time intervals
\begin{equation}
ds^{2}=-c^{2}d\tau ^{2}=\left( 1-\frac{2V}{c^{2}}\right) c^{2}d\overline{T}%
^{2}-d\mathbf{X}^{2}  \label{KLM17}
\end{equation}%
after dividing by $(d\overline{T})^{2}$, where $\overline{T}=TCG$ is the
Geocentric Coordinate Time. However, the Geocentric Coordinate Time is not
the propagation time $T$, for which the space-time interval in (\ref{KLM16})
would have been zero. Expression (\ref{KLM17}) is derived by assuming that $%
\frac{V}{c^{2}}\ll 1$, $\frac{v^{2}}{c^{2}}$ $\ll 1$ and also $(1-x)^{\frac{1%
}{2}}\approx 1-\frac{x}{2}+\frac{x^{2}}{4}+.....$. In our case, these
approximations have not been used in the derivation of the relation (\ref%
{KLM15})\ , and the potential term in the space component of the metric $%
g_{kk}=1+\frac{2V}{c^{2}}$ couples to the square of the Keplerian
velocity, unlike in the simplified case for the formulae (\ref{KLM16}). Also,
in the present case the derivation is based on formulae (\ref{KLM5}) for the
atomic time in its more general form and also on the equality of the two
different representations (\ref{KLM9}) and (\ref{KLM10}) for the propagation
time. These two representations are based on the initial assumption that
there is a correspondence between the atomic and the propagation times, i.e.
there exists a parametrization of the algebraic equation for the propagation
time $T$ in terms of the atomic time $\tau $.

\subsection{ Two representations for the atomic time of the second satellite
and a proof for the equality of the geocentric time with the celestial time}
\label{sec:equal geoccelest}

\subsubsection{General notes on the correspondence between the Geocentric
Coordinate Time and the atomic time}

Now a remarkable fact shall be proved about the possibility to identify the
Geocentric Coordinate Times $\overline{T}_{1}$ and $\overline{T}_{2}$ in the
formulaes (\ref{DOPA25}) and (\ref{DOPB25}) with the corresponding celestial
times, found from the Kepler equation. In the next section, a correspondence
will be established between the position of the satellite on the orbit and
the propagation time of a signal, as a result of which the integral formulae
(\ref{C5A40}) will be derived. But since the atomic clock readings also
depend on the transportation of the clock and consequently on the motion of
the satellite along the orbit, it is natural to expect that there will be
also a correspondence between the atomic time and the propagation time. This
correspondence in both directions (in the sense that the dependence of the
propagation time on the atomic time is invertible) was proved in the preceding sections.
However, there is an important fact - propagation times cannot be used in
the equations (\ref{DOPA25}) and (\ref{DOPB25}) for the atomic time, because
the expressions for $d\tau _{1}^{2}$ and $d\tau _{2}^{2} $ will equal to
zero. The time coordinate in these equations will be the Geocentric
Coordinate Time (TCG), which is the time of the Geocentric Celestial
Reference System (GCRS).

Now let us remember the definition for synchronous atomic clocks in \cite%
{C25}, showing proper atomic times $\tau _{1}$ and $\tau _{2}$: The two
atomic clocks are called synchronous if their corresponding TCG values
agree, i.e.
\begin{equation}
TCG^{(1)}(\tau _{1},X_{1})=TCG^{(2)}(\tau _{2},X_{2})\text{ \ \ \ ,}
\label{KLMN1}
\end{equation}%
where $X_{1}$ and $X_{2}$ denote the positions of the atomic clock. In a
more general theoretical setting, one can assume that the second atomic time
$\tau _{2}$ depends on the second geocentric time $\overline{T}_{2}$ and on
the first atomic time $\tau _{1}$, while $\tau _{1}$ depends only on the
first geocentric time $\overline{T}_{1}$
\begin{equation}
\tau _{1}=\tau _{1}(\overline{T}_{1})\text{ \ \ \ , \ \ \ \ }\tau _{2}=\tau
_{2}(\tau _{1},\overline{T}_{2})\text{ \ \ .}  \label{KLMN2}
\end{equation}
In fact, the above assumption is a sort of "inverse" formulation of the definition (\ref{KLMN1}) of synchronous
atomic clocks, namely: what happens if the first atomic time depends on the first Geocentric Coordinate Time
(TCG) $\overline{T}_{1}$, but the second atomic time depends also on the first TCG $\overline{T}_{1}$, in accord with the
above assumption (\ref{KLMN2})? Although the next step will be to assume that the second atomic time depends on both Geocentric Coordinate
Times, the above assumption will allow to obtain an interesting result.

\subsubsection{The first representation for the atomic time
in terms of the Geocentric Coordinate Time}

Now let us consider the two analytical representations of the second atomic
time (i.e. the time of the atomic clock on the second satellite). Taking
into account the dependance of the second atomic time on the first
geocentric time $\overline{T}_{1}$, the first representation for the second
atomic time is obtained after dividing and multiplying the under-square
expression by the differential $d\overline{T}_{1}$ of the first geocentric
time\textit{\ }%
\begin{equation}
\tau _{2}^{(1)}=\frac{1}{c^{.}}\int d\overline{T}_{1}\sqrt[.]{\widetilde{g}%
_{00}c^{2}+2\widetilde{g}_{0j}c\frac{d\widetilde{x}^{j}}{d\overline{T}_{1}}+%
\widetilde{g}_{ij}\frac{d\widetilde{x}^{i}}{d\overline{T}_{1}}\frac{d%
\widetilde{x}^{j}}{d\overline{T}_{1}}}\text{ \ \ ,}  \label{BBF1}
\end{equation}%
where the subscript "$(1)$" in $\tau _{2}^{(1)}$ denotes the first
representation, $\widetilde{x}^{j}$ denote the space coordinates of the
position of the second satellite (only the integration variable $\overline{T}%
_{1}$ for the geocentric time changes from $0$ to $\overline{T}_{1}$) and
the tilda signs above the metric tensor components \ \ $\widetilde{g}_{00}$,
$\widetilde{g}_{0j}$, $\widetilde{g}_{ij}$ mean that they are taken at the
space points $\widetilde{x}^{j}=(x_{2},y_{2},$ $z_{2})$. Since the
derivatives $\frac{d\widetilde{x}^{j}}{dT_{1}}$ do not have any physical
meaning related to velocities, our further aim will be to find their
relation to the celestial velocities $\frac{d\widetilde{x}^{j}}{dt_{1}}$
along the orbit. For the purpose, the following simple representation will
be used
\begin{equation}
\frac{d\widetilde{x}^{j}}{d\overline{T}_{1}}=\frac{d\widetilde{x}^{j}}{dt_{2}%
}\frac{dt_{2}}{dt_{1}}\frac{dt_{1}}{d\overline{T}_{1}}=  \label{BBF2A}
\end{equation}%
\begin{equation}
=\widetilde{v}^{j}\frac{n_{1}}{n_{2}}\frac{\left( 1-e_{2}\cos E_{2}\right) }{%
\left( 1-e_{1}\cos E_{1}\right) }\frac{dE_{2}}{dE_{1}}\frac{dt_{1}}{d%
\overline{T}_{1}}\text{ \ .}  \label{BBF2B}
\end{equation}%
Note that the elliptical orbits of the two satellites are in one and the
same plane, but they have different eccentricities, semi-major axis and
different eccentric anomalies. At least, this is the construction of the
theoretical model.

Substituting (\ref{BBF2B}) and also the simple representation for the
differential $d\overline{T}_{1}$
\begin{equation}
d\overline{T}_{1}=\frac{d\overline{T}_{1}}{dE_{1}}\frac{dE_{1}}{dE_{2}}dE_{2}
\label{BBF3}
\end{equation}%
into the initial expression (\ref{BBF1}) for the second atomic time, one can
represent the expression as an integral over the second eccentric anomaly angle
\begin{equation}
\tau _{2}^{(1)}=-\frac{1}{c^{2}}\int dE_{2}\sqrt[.]{A(E_{1},E_{2})\left(
\frac{dE_{1}}{dE_{2}}\right) ^{2}+B(E_{1},E_{2})\left( \frac{dE_{1}}{dE_{2}}%
\right) +\widetilde{D}(E_{2})}\text{ \ \ \ ,}  \label{BBF4}
\end{equation}%
where the coefficient functions $A(E_{1},E_{2})$, $B(E_{1},E_{2})$ and $%
\widetilde{D}(E_{2})$ are factorizable expressions, i.e.
\begin{equation}
A(E_{1},E_{2}):=A_{1}(E_{1})A_{2}(E_{2})\text{ \ \ , \ \ \ \ \ }%
B(E_{1},E_{2}):=B_{1}(E_{1})B_{2}(E_{2})\text{ \ \ .}  \label{BBF5}
\end{equation}%
All the coefficient functions are given by the following expressions
\begin{equation}
A_{1}(E_{1}):=\frac{\left( 1-e_{1}\cos E_{1}\right) ^{2}}{n_{1}^{2}}\left(
\frac{d\overline{T}_{1}}{dt_{1}}\right) ^{2}\text{ \ \ , \ }%
A_{2}(E_{2}):=-c^{2}\widetilde{g}_{00}\text{\ \ \ ,}  \label{BBF6}
\end{equation}%
\begin{equation}
B_{1}(E_{1}):=\frac{1}{n_{1}}\left( 1-e_{1}\cos E_{1}\right) \left( \frac{d%
\overline{T}_{1}}{dt_{1}}\right) \text{ \ \ \ ,}  \label{BBF7}
\end{equation}%
\begin{equation}
B_{2}(E_{2}):=\frac{1}{n_{2}}\left( 1-e_{2}\cos E_{2}\right) 2\widetilde{g}%
_{0j}c\widetilde{v}^{j}\text{ \ \ ,}  \label{BBF8}
\end{equation}%
\begin{equation}
\widetilde{D}(E_{2}):=\frac{\left( 1-e_{2}\cos E_{2}\right) ^{2}}{%
n_{2}^{2}e_{2}}\widetilde{g}_{ij}\widetilde{v}^{i}\widetilde{v}^{j}\text{ \
\ \ .}  \label{BBF9}
\end{equation}

\subsubsection{The second representation for the atomic
time in terms of the celestial time}

The second representation for the atomic time of the second clock is based
on the integration over the celestial time $t_{2}$ for the second satellite
\begin{equation}
\tau _{2}^{(2)}=\frac{1}{c^{.}}\int dt_{2}\sqrt[.]{\widetilde{g}_{00}c^{2}+2%
\widetilde{g}_{0j}c\widetilde{v}^{j}+\widetilde{g}_{ij}\widetilde{v}^{i}%
\widetilde{v}^{j}}\text{ \ \ .}  \label{BBF10}
\end{equation}

Taking into consideration the simple relation
\begin{equation}
dt_{2}=\frac{\left( 1-e_{2}\cos E_{2}\right) }{n_{2}}dE_{2}  \label{BBF11}
\end{equation}%
and setting up equal the under-integral expressions (after squaring) in the
two representations (\ref{BBF4}) and (\ref{BBF10}) for the atomic time $\tau
_{2}$, one can obtain
\begin{equation}
\frac{\left( 1-e_{2}\cos E_{2}\right) ^{2}}{n_{2}^{2}}\left( \widetilde{g}%
_{00}c^{2}+2\widetilde{g}_{0j}c\widetilde{v}^{j}+\widetilde{g}_{ij}%
\widetilde{v}^{i}\widetilde{v}^{j}\right) =  \label{BBF12}
\end{equation}%
\begin{equation}
=A(E_{1},E_{2})\left( \frac{dE_{1}}{dE_{2}}\right) ^{2}+B(E_{1},E_{2})\left(
\frac{dE_{1}}{dE_{2}}\right) +\frac{\left( 1-e_{2}\cos E_{2}\right) ^{2}}{%
n_{2}^{2}e}\widetilde{g}_{ij}\widetilde{v}^{i}\widetilde{v}^{j}\text{ \ \ .}
\label{BBF13}
\end{equation}%
The third terms on both sides cancel and as a result the following second -
order algebraic equation is obtained
\begin{equation}
A(E_{1},E_{2})\left( \frac{dE_{1}}{dE_{2}}\right) ^{2}+B(E_{1},E_{2})\left(
\frac{dE_{1}}{dE_{2}}\right) -D(E_{2})=0\text{ \ ,}  \label{BBF14}
\end{equation}%
where $D(E_{2})$ is given by the expression
\begin{equation}
D(E_{2}):=\frac{\left( 1-e_{2}\cos E_{2}\right) ^{2}}{n_{2}^{2}}\left(
\widetilde{g}_{00}c^{2}+2\widetilde{g}_{0j}c\widetilde{v}^{j}\right)
\label{BBF15}
\end{equation}%
and $A(E_{1},E_{2})$, $B(E_{1},E_{2})$ are given by expressions (\ref{BBF5})
- (\ref{BBF8}). The solution of the quadratic equation (\ref{BBF14}) with
respect to $\frac{dE_{1}}{dE_{2}}$ in the general case is%
\begin{equation}
\frac{dE_{1}}{dE_{2}}=-\frac{B}{2A}\pm \sqrt[.]{\frac{D}{A}+\left( \frac{B}{%
2A}\right) ^{2}}\text{ \ \ ,}  \label{BBF16}
\end{equation}%
but for the partial case of a diagonal metric of the type (\ref{DOP25}), the
solution is
\begin{equation}
\frac{dE_{1}}{dE_{2}}=\pm \sqrt[.]{\frac{D}{A}}=\pm \frac{n_{1}}{n_{2}}.%
\frac{\left( 1-e_{2}\cos E_{2}\right) }{\left( 1-e_{1}\cos E_{1}\right) }.%
\frac{1}{\frac{d\overline{T_{1}}}{dt_{1}}}\text{ \ \ .}  \label{BBF17}
\end{equation}%
This expression can be written in the form of a differential relation
\begin{equation}
\frac{\left( 1-e_{1}\cos E_{1}\right) }{n_{1}}dE_{1}.\frac{d\overline{T_{1}}%
}{dt_{1}}=\frac{\left( 1-e_{2}\cos E_{2}\right) }{n_{2}}dE_{2}\text{ \ \ .}
\label{BBF18}
\end{equation}%
Taking into account (\ref{BBF11}), written for the indices $1$ and $2$, the
last relation can be presented as
\begin{equation}
d\overline{T}_{1}=dt_{2}\text{ \ }\Longrightarrow \text{ \ }\overline{T_{1}}%
=t_{2}\text{ \ \ .}  \label{BBF19}
\end{equation}%
The integration constant is set up to zero, assuming that at some initial
moment the celestial time and the geocentric time are equal to zero. In a
similar way, if we assume that the second satellite is not moving, but the
first is moving, one can obtain he case%
\begin{equation}
\overline{T}_{2}=t_{1}\text{ \ \ .}  \label{BBF20}
\end{equation}
The crucial moment here is that the Geocentric Time $\overline{T}_{1}$ for
the case, when the first satellite is moving and emits the signal (while the
second one is non-moving) is equal to the second Geocentric Time $\overline{T%
}_{2}$ for the case, when the signal is emitted by the second (moving)
satellite and the first satellite is non-moving. This can be noted also from
the symmetry of the pair of equations (\ref{DOPA25}) and (\ref{DOPB25}) for
the atomic times with respect to the change of indices $1\leftrightarrow 2$.
It is clear also that this symmetry will not be valid, when both satellites
are moving and the distance between them is changing, meaning that the two
equations have to be intersected by the hyperplane equation (\ref{DOPAB25}).
Consequently, from the equality $\overline{T}_{1}=\overline{T}_{2}$ and ( %
\ref{BBF19}) and (\ref{BBF20}) it will follow that $\overline{T}_{1}=t_{1}$,
which proves that the Geocentric Time in (\ref{DOPA25}) and (\ref{DOPB25})
can be identified with the corresponding celestial times.

After this identification is performed, one can still consider the case of
moving satellites and perform the intersection of these equations with the
hyperplane equation-then the asymmetry will be displayed in the final
expressions for the two atomic times $\tau _{1}$ and $\tau _{2}$. However,
one can perform the same calculations after the second equation (\ref{DOPB25}%
) is intersected with the hyperplane equation in order to check whether
again the Geocentric time can be identified with the celestial time. It
should be kept in mind that the equality of the two representations for the
atomic time presupposes also the fulfillment of the definition (\ref{KLMN1})
for synchronous atomic clocks, which does not contain the case, when the
comparison of the atomic clock readings is performed during the motion of
the two satellites. Yet, the condition for synchronization (i.e. equality)
of the two atomic times can also be required for such a case, when the
distance between the two satellites is changing. This will be performed in a
future publication.

\subsection{Moving atomic clock and the rate of change of the atomic time
with respect to the Geocentric Coordinate Time (TCG)}
\label{sec:atom geoc}

The consideration in this section will be restricted to one moving atomic
clock and will essentially use the established relation (\ref{KLM11}) $\frac{%
d\tau }{dt_{cel}}=\frac{1}{c}\sqrt[.]{\widetilde{N}}$ (where $\widetilde{N}$
is given by (\ref{KLM14})), obtained after setting up equal the two
representations (\ref{KLM9}) and (\ref{KLM10}) for the propagation time.
From the initial formulae (\ref{DOPA25}) for the atomic time, let us divide
and multiply by the differential of the Geocentric time $d\overline{T}$ to
obtain
\begin{equation}
\frac{d\tau }{d\overline{T}}=\frac{1}{c}\sqrt[.]{g_{00}c^{2}+2g_{0j}c\frac{%
dx^{j}}{d\overline{T}}+g_{ij}\frac{dx^{i}}{d\overline{T}}\frac{dx^{j}}{d%
\overline{T}}}\text{ \ \ \ .}  \label{5B.4}
\end{equation}%
Now one can represent $\frac{dx^{j}}{d\overline{T}}$ as
\begin{equation*}
\frac{dx^{j}}{d\overline{T}}=\frac{dx^{i}}{dt_{cel}}\frac{dt_{cel}}{d%
\overline{T}}=v^{i}\frac{dt_{cel}}{d\overline{T}}=
\end{equation*}%
\begin{equation}
=v^{i}\frac{dt_{cel}}{d\tau }\frac{d\tau }{d\overline{T}}=v^{i}\frac{c}{\sqrt%
[.]{\widetilde{N}}}\frac{d\tau }{dT}\text{ \ \ ,}  \label{5B.3}
\end{equation}%
where we have used the derived equality (\ref{KLM11}). Substituting (\ref%
{5B.3}) into (\ref{5B.4}) and introducing the notation
\begin{equation}
\widetilde{F}=1-\frac{g_{ij}v^{i}v^{j}}{\widetilde{N}}\text{ \ \ ,}
\label{5B.7}
\end{equation}%
one can obtain from (\ref{5B.4}) the following quadratic equation with
respect to $\frac{d\tau }{d\overline{T}}$
\begin{equation}
\left( \frac{d\tau }{d\overline{T}}\right) ^{2}-2\frac{g_{0j}v^{j}}{%
\widetilde{F}\sqrt[.]{\widetilde{N}}}\left( \frac{d\tau }{d\overline{T}}%
\right) -g_{00}=0\text{ \ \ .}  \label{5B.8}
\end{equation}%
This simple algebraic equation has the solution
\begin{equation}
\frac{d\tau }{d\overline{T}}=\frac{g_{0j}v^{j}}{\widetilde{F}\sqrt[.]{%
\widetilde{N}}}\pm \sqrt[.]{g_{00}+\left( \frac{g_{0j}v^{j}}{\widetilde{F}%
\sqrt[.]{\widetilde{N}}}\right) ^{2}}\text{ \ \ ,}  \label{5B.10}
\end{equation}%
where the expression for $\widetilde{F}\sqrt[.]{\widetilde{N}}$, taking into
account (\ref{KLM14}) and (\ref{5B.7}), is the following
\begin{equation*}
\widetilde{F}\sqrt[.]{\widetilde{N}}:=\sqrt[.]{\widetilde{N}}-\frac{%
g_{ij}v^{i}v^{j}}{\sqrt[.]{\widetilde{N}}}=
\end{equation*}%
\begin{equation}
=\frac{c\left( g_{00}+2g_{0k}v^{k}\right) }{\sqrt[.]{g_{00}+2g_{0k}\frac{%
v_{.}^{k}}{c}+g_{kl}\frac{v_{.}^{k}v_{.}^{l}}{c^{2}}}}\text{ \ \ .}
\label{5B.11}
\end{equation}%
For a diagonal metric of the type (\ref{DOP25}) expression (\ref{5B.10})
considerably simplifies
\begin{equation}
\frac{d\tau }{d\overline{T}}=\pm \sqrt[.]{g_{00}}=\pm \sqrt[.]{-(c^{2}+2V)}%
=\pm i\sqrt[.]{(c^{2}+2V)}\text{ \ .}  \label{5B.12}
\end{equation}

\subsubsection{Physical interpretation of the obtained formulae}
\label{sec:phys interpr}

Now an important conclusion can be made, in view of the appearance of the
imaginary sign $i$ in the above expression. Combining it with the previously
defined interval in (\ref{DOP23}) as $ds=icd\tau $, one can see that the
interval becomes non-imaginary in terms of the geocentric time interval $d%
\overline{T}$, as it should be, i.e.
\begin{equation}
ds=\mp cd\overline{T}\sqrt[.]{(c^{2}+2V)}\text{ \ \ \ .}  \label{5B.13}
\end{equation}%
Let us take two space-time points, at which the corresponding atomic time
intervals are $d\tau _{1}$ and $d\tau _{2}$ and let us assume that these
points belong to a curve, on which all points have a constant differential
of the geocentric time, i.e. $d\overline{T}_{1}=d\overline{T}_{2}=d\overline{%
T}=const$ . Then the ratio of the two atomic time intervals will be
\begin{equation}
\frac{d\tau _{1}}{d\tau _{2}}=\frac{\sqrt[.]{g_{00}}^{(1)}}{\sqrt[.]{g_{00}}%
^{(2)}}\text{ \ \ \ .}  \label{5B.14}
\end{equation}%
This formulae should be compared to the known formulae (see for example the monograph \cite{C5AB2}) for
the ratio of the gravitational frequency shift for a photon, where the
source and the receiver on the path of the photon are assumed to be at rest.
  For the case of a diagonal metric, expression (\ref{5B.14}) will not be modified by the presence of a second term $g_{kk}(v^{k})^{2}$. At the same time, it can be noticed that the ratio of the differential atomic time to the differential celestial time in (\ref{KLM14}) for the case of
motion of the satellite depends on the additional term $g_{kk}(v^{k})^{2}$. This is a substantial difference between the two formulaes
(\ref{5B.14}) and (\ref{KLM14}). Also it is evident that if two satellites with equal eccentric anomaly angles and mean motion
are taken, then the ratio of the differentials of the two atomic times is
 \begin{equation}
\frac{d\tau _{1}}{d\tau _{2}}=\frac{\sqrt[.]{c^{2}g_{00}+g_{kk}(v^{k})^{2}}%
^{(1)}}{\sqrt[.]{c^{2}g_{00}+g_{kk}(v^{k})^{2}}^{(2)}}\text{ \ \ ,}
\label{BBF21}
\end{equation}%
where the subscripts $(1)$ and $(2)$ denote that the components of the
metric tensor and the velocities in the under-integral expressions are taken
at the first and the second space-time points correspondingly.
This corresponds to the case, when the frequencies $\nu _{2}$ and $\nu _{1}$ should be taken at the
space-time points $(1)$ and $(2)$, where the ratios $\left( \frac{d\tau }{dt_{cel}}\right)
_{2}$ and $\left( \frac{d\tau }{dt_{cel}}\right) _{1}$ are defined.

From a physical point of view, the two formulaes (\ref{5B.14}) and (\ref{BBF21}) can be understood as follows: the
first formulae is obtained by making use of an expression, containing the Geocentric Coordinate Time, which is irrelevant to
the motion of the satellites, while the second formulae is obtained from (\ref{KLM14}), which through the celestial time is
related to the motion of the satellites along the two orbits, consequently the formulae naturally is modified by the velocity term.

\subsection{Finding the eccentric anomaly angles of emission and reception
of the signal for the two satellites}
\label{sec:eccentranom}

Formulaes (\ref{DOPC25}) and (\ref{DOPD26}) for the equality of the
celestial times with the propagation time  as a condition for emission and
reception of the signal give the opportunity to find the eccentric anomaly
angles $\widetilde{E}_{1}$ and $\widetilde{E}_{2}$ on the two corresponding
points on the orbits, at which the emission and reception of the signal
takes place. These two formulaes represent a coupled system of nonlinear
integral equations with respect to the two eccentric anomaly angles
\begin{equation}
\frac{1}{n_{1}}(\widetilde{E}_{1}-e_{1}\sin \widetilde{E}_{1})=\int%
\limits_{E_{per}^{(1)}}^{\widetilde{E}_{1}}\sqrt{%
W_{1}(V_{1},x_{1}(E_{1}),y_{1}(E_{1}),z_{1}(E_{1}))}dE_{1}\text{ \ \ ,}
\label{DOP3BB25}
\end{equation}%
\begin{equation}
\frac{1}{n_{2}}(\widetilde{E}_{2}-e_{2}\sin \widetilde{E}_{2})=\int%
\limits_{E_{per}^{(2)}}^{\widetilde{E}_{2}}\sqrt{%
W_{2}(V_{1},x_{2},y_{2},z_{2},dR_{AB}^{2},x_{1},y_{1},E_{per}^{(1)},%
\widetilde{E}_{1})}dE_{1}\text{ \ \ ,}  \label{DOP4BB25}
\end{equation}%
and $W_{1}(V_{1},x_{1}(E_{1}),y_{1}(E_{1}),z_{1}(E_{1}))$ is the expression
for $T_{1}$ from (\ref{DOP1F26})
\begin{equation}
W_{1}=\frac{1}{c^{2}}\frac{(c^{2}-2V_{1})}{(c^{2}+2V_{1})}\left( \left(
dx_{1}\right) ^{2}+\left( dy_{1}\right) ^{2}+\left( dz_{1}\right)
^{2}\right) \text{ \ .}  \label{DOP4BBB25}
\end{equation}%
If $\widetilde{E}_{1}$ is found as a solution of (\ref{DOP3BB25}) and is
substituted into (\ref{DOP4BB25}), then this integral equation can be solved
with respect to $\widetilde{E}_{2}$. In (\ref{DOP4BB25}) the dependance of $%
W_{2}$ on $x_{1},y_{1},E_{per}^{(1)},\widetilde{E}_{1}$ arises because
equation (\ref{DOP2F26}) has been combined with the hyperplane equation (\ref%
{DOPAB25}).

After solving the nonlinear integral equation (\ref{DOP4BB25}), finding $%
\widetilde{E}_{2}$ and substituting $\widetilde{E}_{1}$ and $\widetilde{E}%
_{2}$ into the equations (\ref{DOP1F26}) and (\ref{DOP2F26}) for the
propagation time and equations (\ref{DOPA25}) and (\ref{DOPB25}) for the
atomic time, the two propagation times and the two atomic times can be
found. In the present case, the eccentric anomaly angles will be taken from
the concrete data for the GPS orbits, \ given in the thesis \cite{C4}.

\subsection{The advantage of the second definition for the equality of the
celestial times to the difference between the two propagation times}

Since further two propagation times will be used due to the two null cone
equations (\ref{DOP1F26}) and (\ref{DOP2F26}), related to the hyperplane
equation (\ref{DOPAB25}), \ some argumentation will be presented about the
advantages to deal with such theoretical setting:

1. It is much more convenient to make use of the equation (\ref{DOP3B25}) in
cubic differentials instead of solving the nonlinear integral equations
(\ref{DOP3BB25}) and (\ref{DOP4BB25}).

2. If one considers the case, when the first propagation time $T_{1}$ and the
atomic time $\tau _{1}$ depend only the first eccentric anomaly angle $%
\widetilde{E}_{1}$, i.e. $T_{1}=T_{1}(\widetilde{E}_{1})$ and $\tau
_{1}=\tau _{1}(\widetilde{E}_{1})$, and the second propagation time $T_{2}$
and the second atomic time $\tau _{2}$ - on the two eccentric anomaly angles
$\widetilde{E}_{1}$ and $\widetilde{E}_{2}$, i.e. $T_{2}=T_{2}(\widetilde{E}%
_{1},\widetilde{E}_{2})$ and $\tau _{2}=\tau _{2}(\widetilde{E}_{1},%
\widetilde{E}_{2})$, then expressing the differentials in the cubic equation
(\ref{DOP3B25}), one can obtain a cubic algebraic equation with respect to
the ratio $\frac{d\widetilde{E}_{2}}{d\widetilde{E}_{1}}$. A similar cubic
equation (\ref{ABC19}) will be derived in Section \ref{sec:Sign prop},
where the coefficient functions will depend on the derivatives of the
propagation times. In the present case, the equation can be solved as an
algebraic one with respect to $\frac{d\widetilde{E}_{2}}{d\widetilde{E}_{1}}$
and further a solution of the obtained first-order nonlinear equation $\frac{%
d\widetilde{E}_{2}}{d\widetilde{E}_{1}}=f(\widetilde{E}_{1},\widetilde{E}%
_{2})$ can be derived.

\section{SIGNAL PROPAGATION TIME FOR THE CASE OF ONE SATELLITE, EMITTING A SIGNAL WHILE MOVING
ALONG AN ELLIPTICAL ORBIT}
\label{sec:one satellite}

\subsection{Conventions about positive and negative gravitational potentials in classical physics,
celestial mechanics and relativistic theories of gravity}
\label{sec:Convent GravPotential}

In equations (\ref{ABC1}) and (\ref{ABC2}) $V_{1}$ and $V_{2}$ are the
potentials of the gravitational field at the space points $%
r_{1}=(x_{1},y_{1},0)$ and $r_{2}=(x_{2},y_{2},0)$ respectively and the
potential $V$ (omitting the indices) is defined as
\begin{equation}
V=\frac{GM}{r}=\frac{GM}{a(1-e\cos E)}\text{ \ \ \ \ .}  \label{C5ABCD}
\end{equation}%
In the next subsection it shall be explained why such a choice of a positive
gravitational potential is admissable and non-contradictory, because the
conventions in celestial mechanics and in the theory of relativistic
reference systems are not identical with those from classical physical
theory \cite{KKK1}, \cite{KKK2}. \

If the center of the Earth is at the point $O$ and it is considered to be
also the center of the reference system, then the definition (\ref{C5ABCD})
with a positive sign of the gravitational potential is consistent with the
negative force in the equation of motion
\begin{equation}
m\overset{..}{\overrightarrow{r}}=\overrightarrow{F}=-\frac{GM}{r^{2}}%
\overrightarrow{r}=\nabla V\text{ \ ,}  \label{C5A1}
\end{equation}%
the negative sign meaning that this is a force of attraction, i.e. in the
"inward direction", towards the Earth center. As noted in chapter $3.2$ in
the monograph of Murray, Dermott \cite{CA1}, the choice of a positive
gravitational potential like in (\ref{C5ABCD}) is a standard convention in
celestial mechanics, since $V$ is considered to be a scalar function, which
in principle might contain many other terms - centrifugal terms, also terms
related to the multipole expansion of the gravitational field, which are
necessary to be accounted also in the theory of relativistic reference
frames (see the monograph by Kopeikin, Efroimsky, Kaplan \cite{CA7A2}). Such
a definition is consistent with the physical idea that the gravitational
potential is generated by masses, distributed continuously over some volume,
i.e. $V=G\int \frac{\rho (s)d^{3}s}{\mid r-s\mid }$ (see the monographs on
celestial mechanics \cite{C27} and \cite{C29}), where the density $\rho (s)$
may be replaced by the relativistic mass density $\sigma (t,x)$ \cite{AAB63}%
, \cite{AAB64}. However, this definition of the potential energy is not the
same as the definition of this energy in the sense of the work, done by conservative
forces $W_{12}=V_{1}-V_{2}=\int\limits_{1}^{2}Fdr$, equal to the
diminishing of the potential energy under a transition of the body from a
space point $1$ to a space point $2$ \cite{KKK1}. If the transition of a
body with a mass $m$ is from infinity to a final position $r$ due to action
of the gravitational force $F=G\frac{Mm}{r^{2}}$ and the gravitational field
is created by a non-moving body of mass $M_{\oplus }$ (in the case this will
be the Earth) , then the above formulae can be written as
\begin{equation}
W_{(\infty \rightarrow r)}=V_{\infty }-V_{r}=\int\limits_{\infty
}^{r}Fdr=\int\limits_{\infty }^{r}\frac{GM_{\oplus }m}{r^{2}}dr\text{ \ \ .}
\label{C5A2}
\end{equation}%
It means that a material body is created by performing work for moving its
elementary volumes from infinity to a space point at a distance $r$. Now
imagine that a body is being lifted from the Earth surface (let the Earth
has a radius $R_{E}$) to a point at a distance $h$ above the Earth surface.
Since the force acting on the body has to overcome the gravitational force
of attraction and has the opposite direction in comparison with the
preceding force, the potential difference between the points $r_{1}=R_{E}$
and $r_{2}=R_{E}+h$ is defined as
\begin{equation}
V_{R+h}-V_{R}=-\int\limits_{R_{E}}^{R_{E}+h}\frac{GM_{\oplus }m}{r^{2}}dr=-%
\frac{GM_{\oplus }m}{R_{E}}+\frac{GM_{\oplus }m}{R_{E}+h}\text{ \ \ .}
\label{C5A3}
\end{equation}%
Note that the potential difference is negative and for small \ $h=\Delta h$
is equal to
\begin{equation}
\Delta V\approx -\frac{GM_{\oplus }m}{R_{E}^{2}}\Delta h\text{ \ \ ,}
\label{C5A4}
\end{equation}%
if the corresponding potentials $V_{R_{E}}$ and $V_{R_{E}+h}$ are defined
with a positive sign, i.e. $V_{R_{E}}=\frac{GM_{\oplus }m}{R_{E}}$ and $%
V_{R_{E}+h}=\frac{GM_{\oplus }m}{R_{E}+h}$. Similarly, from the positive
potential (\ref{C5A}) for small changes in the eccentric anomaly angle $%
\Delta E$ one can find the small negative potential difference
\begin{equation}
\Delta V=-\frac{GM_{\oplus }e\sin E}{a(1-e\cos E)^{2}}\Delta E\text{ \ .}
\label{C5A5}
\end{equation}%
In \cite{KKK3} and \cite{KKK4} for values of the mean radius of the
Earth $R_{E}\approx 6371$ $km$ and Earth mass $M_{E}\approx 5.97\times
10^{24}$ $kg$ from formulae (\ref{C5A4}) it was obtained that a change of
the Earth potential $\Delta V\simeq -0.1$ $m^{2}s^{-2}$ corresponds to a
sensitivity of the geoid height $\Delta h\simeq 1$ $cm$.

\subsection{Change of the gravitational potential along an
elliptical orbit - numerical comparison with the potential change above the
geoid of the Earth}
\label{sec:changegrav potent}

Let us find whether there is a similar change of the potential difference
for the case of formulae (\ref{C5A5}). The numerical parameters for the
eccentric anomaly $E_{(3)}$ (third iteration - to be found as an iterative
solution of the Kepler equation in section \ \ref{sec:restr ellip}) is
\begin{equation}
E_{(3)}^{exact}=-0.31758547588467897473\text{ \ \ }[rad]\text{ \ \ \ }
\label{C5A6}
\end{equation}%
and the eccentricity $e$ of the $GPS$ - orbit is chosen to be%
\begin{equation}
e=0.01323881349526\text{ \ \ .}  \label{C5A7}
\end{equation}%
In performing the calculation for the eccentric anomaly $E_{(3)}$
as an iterative solution of the Kepler equation and denoting by $%
M=n(t_{cel}-t_{per})$ the mean anomaly ($n=\sqrt[.]{\frac{GM_{\oplus }}{a^{3}%
}}$ is the mean motion, $t_{cel}$ is the celestial time determined from the
Kepler equation, $t_{per}$ is the time of perigee passage, usually
considered as an initial time), the formulae for the third iteration
\begin{equation*}
E_{3}=M+e\sin (M+e\sin M)\cos (\frac{1}{2}e^{2}\sin 2M)+
\end{equation*}%
\begin{equation}
+e\cos (M+e\sin M)\sin (\frac{1}{2}e^{2}\sin 2M)\text{ \ \ \ }  \label{C5A8}
\end{equation}%
has been used without the approximation $\sin (\frac{1}{2}e^{2}\sin
2M)\approx \frac{1}{2}e^{2}\sin 2M$ \ up to the thirteenth digit after the
decimal dot and also the approximation $\cos \left( \frac{1}{2}e^{2}\sin
2M\right) \approx 1$ up to the twelveth digit. In the case when these
approximations are taken into account, one will obtain the approximate value
$E_{3}^{approx}=-0.31758547631968691128$ $[rad]$ for the eccentric anomaly
angle, which differs from the exact value (\ref{C5A6}) after the eight digit
after the decimal dot. However, if the result for $E_{(3)}^{exact}$ (\ref%
{C5A6}) is compared with the result for the second iteration $E_{(2)}^{exact}
$
\begin{equation}
E_{(2)}^{exact}=-0.31758482974937\text{ }[rad]\text{ \ \ ,\ \ }  \label{BB28}
\end{equation}%
then the two numbers will differ after the fifth digit after the decimal
dot. Consequently, the third iterative solution $E_{(3)}^{exact}$
considerably improves the accuracy of determination of the eccentric anomaly
angle, and additionally, the accuracy is improved by three digits after the
decimal dot if no approximations for $\sin (\frac{1}{2}e^{2}\sin 2M)$ and $%
\cos \left( \frac{1}{2}e^{2}\sin 2M\right) $ are being used. In performing
the above calculations, the numerical value for the mean anomaly $M$ was
taken from the PhD thesis of Gulklett \cite{C4} $M=-0.3134513508155$ $[rad]$%
, the minus sign meaning that the satellite encircles along the elliptical
orbit in the opposite direction of the chosen clockwise direction.

The importance of performing such precise calculations, depending on the
formulaes for the iterative solutions of the Kepler equation can become
clear if the potential difference (\ref{C5A5}) is calculated for a small
eccentric anomaly angle change $\Delta E=1$ $milliarcsecond=1$ $%
mas=10^{-3}$ $arcseconds $. Let us note first that if $\Delta E$ is
expressed in arcseconds, this will not change the dimensions of the formulae
(\ref{C5A5}), because arcseconds can be expressed as some part of the $360$
degrees. But since previously radians were used for the numerical value $%
E_{(3)}^{exact}$ in (\ref{C5A6}), one should be able to convert the angle
change $\Delta E$ from arcseconds (or milliarcseconds) to radians. For the
purpose, the definition for the radian is used: the radian corresponds to
that angle of the circle, in which the arclength is equal to the radius%
\textit{,} i.e.
\begin{equation}
1\text{ }rad=\frac{360}{2\pi }=\frac{180}{\pi }\text{ \ \ }[\deg ]=\frac{180%
}{\pi }\times 60\times 60\text{ \ \ }[arcsec]\text{ \ .}  \label{C5A9}
\end{equation}%
For $\Delta E=1$ $[arcsec]$ the calculated potential difference from (%
\ref{C5A5}) is equal to%
\begin{equation}
\Delta V=-0.636329\text{ \ }\left[ \frac{km^{2}}{s^{2}}\right] \text{ \ \ \
\ ,}  \label{C5A10}
\end{equation}%
which is much higher than the change of the Earth potential $\Delta V\simeq
-0.1$ $[m^{2}s^{-2}]$. However, if $\Delta E=1$ $milliarcsecond=1$ $%
mas=10^{-3}$ $arcsec$ and the inverse conversion formulae (see (\ref%
{C5A9})) from arcseconds to radians is used, the potential change acquires a
value, which is the same magnitude as $\Delta V\simeq -0.1$ $[m^{2}s^{-2}]$,
i.e.
\begin{equation}
\Delta \widetilde{V}=\frac{\Delta V\pi }{180}\frac{1}{3600}10^{-3}=0.308501%
\text{ \ }\left[ \frac{m^{2}}{s^{2}}\right] \text{ \ .}  \label{C5A11}
\end{equation}%
The change $\Delta r$ of the linear distance along the ellipse,
corresponding to a change $\Delta E=1$ $mas=10^{-3}$ $arcsec$ of the
eccentric anomaly can be found by means of calculating the difference $%
(r+\Delta r)-r$ from the simple formulae $\ r=a(1-e\cos E)$
\begin{equation}
(r+\Delta r)-r=a(1-e\cos (E_{3}+\Delta E))-a(1-e\cos (E_{3}))=  \label{C5A12}
\end{equation}%
\begin{equation}
=ae(\cos (E_{3})-\cos (E_{3}+\Delta E))\text{ \ \ .}  \label{C5A13}
\end{equation}%
Since the third iteration $E_{3}$ is expressed from the Kepler equation in
radians, $\Delta E=$ $1$ $milliarcsec$ should also be converted to
radians by means of the inverted formulae (\ref{C5A9})
\begin{equation}
\Delta E=\frac{\pi }{36\times 18}\times 10^{-6}\text{ }\left[ rad\right]
=0.4848136811095359935899141\times 10^{-8}\text{ }\left[ rad\right] \text{ \
.}  \label{C5A14}
\end{equation}%
Two separate cases should be considered - when the change of the eccentric
anomaly $\Delta E$ is performed in the clockwise direction, and the second
case - when the change is performed in the anticlockwise direction. In the
first case, $\Delta E$ should be added to $E_{3}$
\begin{equation}
E_{3}+\Delta E=-0.31758418317360994037  \label{C5A15}
\end{equation}%
and in the second case - substracted from $E_{3}$
\begin{equation}
E_{3}-\Delta E=0.31758418802174675146\text{ \ .}  \label{C5A16}
\end{equation}%
Both numbers are identical up to the ninth digit after the decimal
point. The calculation of $\Delta r$ according to formulae (\ref{C5A13})
gives
\begin{equation}
\Delta r=-0.532362094952630\text{ }\left[ mm\right] \text{ \ for \ }%
E_{3}+\Delta E  \label{C5A17}
\end{equation}%
and
\begin{equation}
\Delta r=0.53236209474111\text{ \ \ }\left[ mm\right] \text{ \ for \ }%
E_{3}-\Delta E\text{ \ \ \ .}  \label{C5A18}
\end{equation}%
Consequently, the change of the potential is much more rapid in comparison
with the change of the potential due to height variations near the geoid,
calculated in \cite{KKK3} and \cite{KKK4}.

\subsection{Propagation time for a signal, emitted by a moving along an
elliptical orbit satellite - general considerations and comparison with the
case of two moving satellites}
\label{sec:proptime 1satel}

Let us consider the standard case of propagation of the signal, emitted by
an emitter of a satellite, moving along an elliptical orbit. There are
several problems, related to this issue, but the most important
are:

1. How can the propagation time $T$ be defined uniquely by means of
establishing an one-to-one correspondence of the time $T$ with the position
of the satellite on the orbit, characterized by the eccentric anomaly angle?
In other words, the propagating signal "keeps track" of the position of the
satellite. This problem will be discussed in this section and in the next
one. It will be proved that the propagation time can be expressed by a
combination of elliptic integrals of the first, second and third kind, which
can be exactly calculated. This numerical calculation will be performed in a future publication.
Thus, the correspondence eccentric anomaly angle
- propagation time will be proved analytically.

2. Since the propagation time is obtained as a solution of the null cone
equation
\begin{equation}
ds^{2}=0=-(c^{2}+2V)(dT)^{2}+(1-\frac{2V}{c^{2}})((dx)^{2}+(dy)^{2}+(dz)^{2})%
\text{ \ }  \label{C5A19}
\end{equation}%
with an origin at the space-time point $(T_{1},x_{1},y_{1})$, parametrized
by the Kepler equations (for this concrete case we omit the indice $1$)
\begin{equation}
x=a(\cos E-e)\text{ \ \ , \ \ \ \ }y=a\sqrt[.]{1-e^{2}}\sin E\text{ \ \ ,}
\label{C5A20}
\end{equation}%
will it be evident that it is positive and non-imaginary? From the solution
of (\ref{C5A19}), found after the parametrization equations (\ref{C5A20})
are substituted into (\ref{C5A19}) (for plane motion $z=0$)
\begin{equation}
dT=\frac{1}{c}\sqrt{\frac{c^{2}-2V}{c^{2}+2V}}((dx)^{2}+(dy)^{2})\text{ \ \
\ ,}  \label{C5A21}
\end{equation}%
it will be shown that this requirement for real-valuedness is fulfilled. Also, it is important
to mention that the solution of (\ref{C5A21}) is non-contradictory from the
viewpoint of dimensional analysis, providing a clear evidence
that the obtained expression for $T$ is expressed in seconds. This will turn
out to be the case.

3. Another important point is whether a positive or negative gravitational
potential will lead to a consistent expression for the propagation time. It
will be shown that in both cases, the propagation time is consistently
defined, but both cases will differ in the sign of the third $O(c^{-3})$
(third inverse power in the velocity of light $c$).

All these three important topics for the standard case of one null cone
equation (\ref{C5A19}) should be kept in mind further, when further the more
complicated case of two null cone equations is considered.

After the integration of (\ref{C5A21}), the solution will depend on the
eccentric anomaly angle $E$. So it is natural to assume that the propagation
time will also depend on the angle $E$. Nevertheless, according to the
terminology in the papers \cite{AAB9} and \cite{AAB6}, the coordinate time
of propagation of a signal in an Earth Centered Inertial (ECI) coordinate
system is $\Delta T\approx \frac{1}{c}\int\limits_{path}\sqrt{g_{ij}dx^{i}dx^{j}}$,
which is the Euclidean path length, divided by $c$. It is understood that
the path is along the points of the trajectory of the signal. In the present
case, the path integral will not be over the trajectory of the signal, but
will be from some initial value of the eccentric anomaly angle $E_{init}$
(corresponding to the moment of emission of the signal) to some final value $%
E_{fin}$, corresponding to the final moment of time, when the propagation
time is determined. This turns out to be possible due to the unique
correspondence between the eccentric anomaly angle $E$ and the propagation
time $T$, which was discussed previously in section \ref{sec:propag eccentr}%
. The path integral over the eccentric anomaly angle can be determined if
some initial value $E_{init}$ is fixed - it will be assumed that this
initial value is zero at the moment of perigee passage of the satellite (the
perigee is the point on the orbit of smallest distance of the satellite from
the Earth). So this is how the propagation time of the signal is defined as
an integral over the eccentric anomaly variable $E$ from some initial point $%
E_{init}$ to some final point $E_{fin}$
\begin{equation}
T=\frac{1}{c}\int\limits_{E_{init}}^{E_{fin}}dE\sqrt{\frac{c^{2}-2V}{%
c^{2}+2V}}.\sqrt{1-e^{2}\cos ^{2}E}\text{ }+C\text{\ ,}  \label{C5A22}
\end{equation}
where the constant $C$ will be zero due to the zero initial value $%
E_{init}=0 $ for the eccentric anomaly angle since for $E_{fin}=0$ the
propagation time should be zero. Note that this definition allows one to
determine an initial propagation time $T_{1}$, coinciding with the moment of
emission of the signal at the point on the orbit $E_{fin}=E_{1}$. This means
that this initial propagation time $T_{1}=\frac{1}{c}\int%
\limits_{E_{init}=0}^{E_{1}}dE\sqrt{\frac{c^{2}-2V}{c^{2}+2V}}.\sqrt{%
1-e^{2}\cos ^{2}E}$ represents a time difference, corresponding to a
"fictitious" propagation of the signal from the point $E_{init}=0$ on the
orbit to the point $E_{fin}=E_{1}$. Having determined this initial
propagation time $T_{1}$, any "real" (i.e. corresponding to a real
propagation of the signal) propagation time difference $\Delta T=T_{2}-T_{1}$
can be determined according to the previous formulae as
\begin{equation}
\Delta T=T_{2}-T_{1}=\frac{1}{c}\int\limits_{E_{1}}^{E_{2}}dE\left( \sqrt{%
\frac{c^{2}-2V}{c^{2}+2V}}.\sqrt{1-e^{2}\cos ^{2}E}\right) =\text{ }
\label{C5A23}
\end{equation}%
\begin{equation}
=\frac{1}{c}\int\limits_{0}^{E_{2}}dE\left( ...\right) -\frac{1}{c}%
\int\limits_{0}^{E_{1}}dE\left( ...\right) \text{ \ \ \ .}  \label{C5A24}
\end{equation}

So since for this case to each moment of propagation time corresponds the
position of the first (the only) satellite, the determination of the
propagation time $\Delta T=T_{2}-T_{1}$, corresponding to the two positions
of the satellite at the two eccentric anomaly angles $E_{1}$ and $E_{2}$
does not depend on the initial eccentric anomaly $E_{init}$. The situation
will change considerably, when further the more generalized setting of signal
transmission between two moving satellites is considered. Then due to the
different moments of time for the perigee passage for the two satellites,
the propagation time $T$ will depend on both eccentric anomaly angles $E_{1}$
and $E_{2}$ and the representation (\ref{C5A24}) will no longer be valid.
Further this will become evident when solving the differential equation (%
\ref{ABC36A}) in full derivatives in section \ref{sec:Deriv form}. The
solution will be found after performing an indefinite integration of the two
differential equations (\ref{ABC39}) and (\ref{ABC42}). However, the
indefinite integrals can be replaced by definite integrals with integration
boundaries correspondingly from $E_{init}^{(1)}$ to $E_{fin}^{(1)}$ and from
$E_{init}^{(2)}$ to $E_{fin}^{(2)}$. Thus, it can be seen that there is a
qualitative difference between the approaches for calculating the
propagation time for the two cases - the first case for a signal, emitted by
a satellite, moving along an elliptical orbit and the second case for a pair
of satellites, moving with respect to one another on one elliptical orbit.
In both cases, the trajectory of the signal due to the action of the gravitational field will
be a curved one. However, in the second case due to the motion of the second
satellite it can be expected that it will take more time for the signal to
arrive at the space point of reception in comparison with the first case,
when the initial and final moments of propagation of the signal correspond
to positions of the satellite at $E_{init}=0$ and $%
E_{fin}=E_{(3)}=-0.31758547588467897473$ \ \ $[rad]$ (see the value (\ref%
{C5A6})). A partial second case is when the two satellites move along one
and the same orbit (i.e. $a_{1}=a_{2}=a$ and $e_{1}=e_{2}=e$), their initial
times of perigee passage are different, but the difference between the
celestial times $t_{cel}^{(1)}$ and $t_{cel}^{(2)}$ at two different
positions $1$ and $2$ of the two satellites and their initial times $%
t_{init}^{(1)}$ and $t_{init}^{(2)}$ will be assumed to be constant. Since
this means that
\begin{equation}
t_{cel}^{(1)}-t_{init}^{(1)}=t_{cel}^{(2)}-t_{init}^{(2)}\text{ \ \ ,}
\label{C5AB1}
\end{equation}%
then from the corresponding two Kepler equations
\begin{equation}
E_{1}-e\sin E_{1}=n(t_{cel}^{(1)}-t_{init}^{(1)})  \label{C5AB2}
\end{equation}%
and
\begin{equation}
E_{2}-e\sin E_{2}=n(t_{cel}^{(2)}-t_{init}^{(2)})  \label{C5AB3}
\end{equation}%
it will follow that $E_{1}=E_{2}=E$. Conversely, from the equality of the
eccentric anomaly angles, the equality (\ref{C5AB1}) will be fulfilled. All
these general considerations are important for understanding the results in
Section \ref{sec:Sign ellip}.

 \subsection{Analytical expression for the propagation time of the signal,
the neglect of relativistic corrections and consistency of the dimensions of
the two coefficients}
\label{sec:proptime dimens}

This section has the purpose to find the expression for the propagation time
and in such way, to prove that the coefficients in front
of the two constituent expressions will have dimensions  $[\sec ]$. Thus it
will be proved that the propagation time will be correctly defined from the
null cone equation. Further in a subsequent publication it will be shown
that the resulting expression in fact is a sum of three elliptic integrals
of the first, second and third kind, which  can be exactly calculated
numerically.

It can be noted that if the right-hand side of (\ref{C5A21}) is divided and
multiplied by $(dt)^{2}$, where $t$ is the celestial time from the Kepler
equation, one can obtain
\begin{equation}
(c^{2}+2V)\left( dT\right) ^{2}=(1-\frac{2V}{c^{2}})\mathbf{v}^{2}\text{ \ ,}
\label{C5A25}
\end{equation}%
where $\mathbf{v}^{2}$ is the velocity of the satellite along the orbit
\begin{equation}
\mathbf{v}^{2}=v_{x}^{2}+v_{y}^{2}+v_{z}^{2}=\left( \frac{dx}{dt}\right)
^{2}+\left( \frac{dy}{dt}\right) ^{2}+\left( \frac{dz}{dt}\right) ^{2}\text{
\ .}  \label{C5A26}
\end{equation}%
Taking into account the Kepler parametrization (\ref{C5A20}), the above
expression for the velocity for the case of plane motion ($z=0$) can be
rewritten as
\begin{equation}
\mathbf{v=}\sqrt{v_{x}^{2}+v_{y}^{2}}=\frac{na}{\left( 1-e\cos E\right) }%
\sqrt[.]{1-e^{2}\cos ^{2}E}\text{ \ \ .}  \label{C5A27}
\end{equation}%
Substituting this formulae into (\ref{C5A25}) with account also of formulae (%
\ref{C5A}) for the gravitational potential, the following expression for the
propagation time, equivalent to (\ref{C5A22}) is obtained after performing
the integration over the eccentric anomaly angle $E$
\begin{equation}
T=\int \frac{\mathbf{v}}{c}\sqrt[.]{\frac{(c^{2}-2V)}{(c^{2}+2V)}}\text{ }%
dE+C=  \label{C5A28}
\end{equation}%
\begin{equation}
=\frac{a}{c}\int \sqrt[.]{\frac{(1-e^{2}\cos ^{2}E)\left[ a\left( 1-e\cos
E\right) -\beta \right] }{\left[ a\left( 1-e\cos E\right) +\beta \right] }}%
\text{ }dE+C\text{\ .}  \label{C5A29}
\end{equation}%
The constant $\beta $ is determined as
\begin{equation}
\beta =\frac{2GM_{\oplus }}{ac^{2}}\text{ \ . }  \label{C5A30}
\end{equation}%
The other constant $C$ can be determined as a result of the integration of
the null cone equation from some initial condition - for example, the
propagation time $T$ should be equal to zero, when $E=0$ (if $E_{init}=0$).
Further, making the substitution
\begin{equation}
1-e\cos E=y\text{ \ \ \ }\Rightarrow \text{ \ \ }dE=\frac{dy}{e\sin E}
\label{C5A31}
\end{equation}%
in the integral (\ref{C5A29}) and denoting $\overline{\beta }=\frac{\beta }{a%
}$, one can represent (\ref{C5A29}) as
\begin{equation}
T=\frac{a}{c}\int \sqrt[.]{\frac{y(2-y)(y-\overline{\beta })}{(y+\overline{%
\beta })\left[ e^{2}-\left( 1-y\right) ^{2}\right] }}dy\text{ \ \ \ .}
\label{C5A32}
\end{equation}%
This integral is of the form
\begin{equation}
T=\frac{a}{c}\int \sqrt[.]{\frac{a_{1}y^{3}+a_{2}y^{2}+a_{3}y}{%
b_{1}y^{3}+b_{2}y^{2}+b_{3}y+b_{4}}}dy\text{ \ \ \ ,}  \label{C5A33}
\end{equation}%
where the numerical constants $a_{1}$, $a_{2}$, $a_{3}$, $b_{1}$, $b_{2}$, $%
b_{3}$, $b_{4}$ can easily be calculated. If the square root, representing
an irrational function of the variable $y$, is denoted as
\begin{equation}
x=\sqrt[.]{\frac{a_{1}y^{3}+a_{2}y^{2}+a_{3}y}{%
b_{1}y^{3}+b_{2}y^{2}+b_{3}y+b_{4}}}\text{ }=\sqrt[.]{P(y)}\text{ \ \ \ \ ,}
\label{C5A34}
\end{equation}%
then the integral
\begin{equation}
\int xdy=\int \sqrt[.]{P(y)}dy  \label{C5A35}
\end{equation}%
is a partial case of a class of integrals (often denoted in mathematical
literature as $\int R(x,y)dy$). They are known in mathematics as abelian
integrals (see the monograph by Prasolov and Solovyev \cite{C3}), related
to the curve
\begin{equation}
F(x,y):=x^{2}-P(y)=0\text{ \ \ \ \ .}  \label{C5A36}
\end{equation}%
Such integrals is not possible to solve analytically, but from a physical
point of view this would not be necessary to be performed. The motivation
for this is that we are interested in the case
\begin{equation}
\frac{2V}{c^{2}}=\frac{2G_{\oplus }M_{\oplus }}{c^{2}a(1-e\cos E)}\ll 1\text{
\ \ , }  \label{C5A37}
\end{equation}%
which can \ be assumed to be fulfilled, because $\beta =$ $\frac{2G_{\oplus
}M_{\oplus }}{c^{2}a}\ll 1$. For the parameters of a $GPS$ orbit, the
constant $\beta $ can exactly be calculated to be $0.33.10^{-9}$, which
justifies the above strong inequality. In the strict mathematical sense, the
above inequality (\ref{C5A37}) takes place when $\cos E\leq \frac{1}{e}-%
\frac{2G_{\oplus }M_{\oplus }}{c^{2}ae}$, which is always fulfilled, because
the first term is a large number of the order $\frac{1}{13}.10^{3}$, while
the second term is a very small number of the order $1.4.10^{-8}$.  Another
values for the parameter $\beta $ have been obtained in the literature. For
example, in the review article by J. Pascual Sanchez \cite{CA8}, it was
obtained for the approximate radius of the satellite orbit $r_{s}=26561$ $[km]$
\begin{equation}
\frac{G_{\oplus }M_{\oplus }}{r_{s}c^{2}}=0.167.10^{-9}\text{ \ \ .}
\label{C5A38}
\end{equation}%
If multiplied by $2$ (in order to obtain the constant $\beta =$ $\frac{%
2G_{\oplus }M_{\oplus }}{c^{2}a}$), this will give $0.334.10^{-9}$, which is
in full accord (up to the second digit) with the estimate in this paper $%
\beta =0.33.10^{-9}$. It can be noted also that the first post-Newtonian
(PN) level of Einstein theory is formulated in terms of the two potentials $w
$ and $w^{i}$ \cite{CA8A}, where the potential $w$ is a generalization of
the Newtonian potential $U$ (and agrees with $U$ in the limit $c\rightarrow
\infty $). So relativistic effects due to the field of the Earth are
determined from $\left( \frac{v}{c}\right) ^{2}\sim \frac{w}{c^{2}}$ (from
the virial theorem), which gives a numerical value about $10^{-9}$ in the
vicinity of the Earth surface. In our case, the small value for $\beta =$ $%
\frac{2G_{\oplus }M_{\oplus }}{c^{2}a}\ll 1$ means that similarly to the
neglect of relativistic effects near the Earth surface, relativistic effects
in the propagation of the signal smaller than $10^{-9}$ will also be
neglected.

The inequality (\ref{C5A37}) turns out to be the necessary condition for
calculating the integral (\ref{C5A28}), which can be represented as a sum of
$O(\frac{1}{c})$ term (first inverse power in $c$) and an $O(\frac{1}{c^{3}}%
) $ term (third inverse power in $c$)
\begin{equation}
T=\int \frac{\mathbf{v}}{c}\sqrt[.]{\frac{(1-\frac{2V}{c^{2}})}{(1+\frac{2V}{%
c^{2}})}}\text{ }dt\approx \int \frac{\mathbf{v}}{c}(1-\frac{2V}{c^{2}}%
)dt=I_{1}+I_{2}=  \label{C5A39}
\end{equation}%
\begin{equation}
=\frac{a}{c}\int \sqrt[.]{1-e^{2}\cos ^{2}E}\text{ }dE-\frac{2G_{\oplus
}M_{\oplus }}{c^{3}}\int \sqrt[.]{\frac{1+e\cos E}{1-e\cos E}}dE\text{ \ \ .}
\label{C5A40}
\end{equation}%
Note the following two consequences from the last two formulaes:
\\ 1. In case of a negative gravitational potential $\widetilde{V}=-V$, where $%
V $ is the potential (\ref{C5A}), the under - integral expression in (\ref%
{C5A39}) can be written as
\begin{equation}
\frac{\mathbf{v}}{c}\sqrt[.]{\frac{(1-\frac{2\widetilde{V}}{c^{2}})}{(1+%
\frac{2\widetilde{V}}{c^{2}})}}=\frac{\mathbf{v}}{c}\sqrt[.]{\frac{(1+\frac{%
2V}{c^{2}})}{(1-\frac{2V}{c^{2}})}}\approx \int \frac{\mathbf{v}}{c}(1+\frac{%
2V}{c^{2}})\text{ \ \ .}
\label{C5AB41}
\end{equation}%
This will lead to a positive sign in front of the second term in (\ref{C5A40}%
). Thus, depending on the sign of the gravitational potential, there will be
an increase or a decrease of the propagation time $T$. However, the numerical
analysis can show that the first term in (\ref{C5A40}) will be nearly $1000$
times greater than each of the terms in the second term. Consequently, the
sign of the potential up to a certain level of accuracy of the measurement
will not depend on the sign of the coefficient of the second term.
\\ 2. The coefficient $\frac{a}{c}$ as a ratio of the large semi-major axis of
the orbit and the velocity of light $c=299792458$ \ $[\frac{m}{\sec }]$ will
have a dimension $[m/\frac{m}{\sec }]=[\sec ]$, as it should be. The first
integral, as well as the second integral in (\ref{C5A40}) can be represented
as a sum of elliptic integrals. They can be exactly calculable and will be
represented by numbers (for the given numerical values of the parameters of the elliptical orbit
and also the eccentric anomaly angle), but since the numerical calculation requires the
application of the methods from the theory of analytic functions, this will not
be performed in this publication.
\\ The second coefficient $\frac{2G_{\oplus }M_{\oplus }}{c^{3}}$ is twice the
ratio of the s.c. {"}geocentric gravitational constant {"} $%
G_{\oplus }M_{\oplus }$ $=3986005\times 10^{8}$ \ $[\frac{m^{3}}{\sec ^{2}}]$
(obtained after multiplying the gravitational constant of the Earth $%
G_{\oplus }$ with the Earth mass $M_{\oplus }$) and the third power of the
velocity of light. The corresponding dimension is $[\frac{m^{3}}{\sec ^{2}}:%
\frac{m^{3}}{\sec ^{3}}]=[\sec ]$, which clearly proves that formulae (\ref%
{C5A40}) has the proper dimensions.

\section{SIGNAL  PROPAGATION TIMES  FROM THE INTERSECTION OF THE TWO NULL  GRAVITATIONAL  CONES - THE CASE OF
TWO MOVING SATELLITES}
\label{sec:Signal propag}

\subsection{General considerations about intersecting algebraic equations and the correspondence eccentric anomaly angles - propagation times}
\label{sec:Gen consid}

Having in mind the argumentation in the previous section, it is important to
derive a formulae for the propagation time of a signal for the case, when both
the emitter and the receiver are moving. We shall consider in this section
only satellites moving along different elliptical orbits (different semi-major
axis $a_{1}$, $a_{2}$, eccentricity parameters $e_{1}$, $e_{2}$ and eccentric
anomalies $E_{1}$, $E_{2}$).

Let the gravitational null cone metric for the signal emitted by the first
satellite at the space point $(x_{1},y_{1},z_{1})$ is
\begin{equation}
ds_{1}^{2}=0=-(c^{2}+2V_{1})(dT_{1})^{2}+ \nonumber \\
\end{equation}
\begin{equation}
\\+(1-\frac{2V_{1}}{c^{2}})\left(
(dx_{1})^{2}+(dy_{1})^{2}+(dz_{1})^{2}\right)  \label{ABC1}%
\end{equation}
and the null cone metric for the second signal - receiving satellite is
\begin{equation}
ds_{2}^{2}=0=-(c^{2}+2V_{2})(dT_{2})^{2}+ \nonumber \\
\end{equation}
\begin{equation}
\\+(1-\frac{2V_{2}}{c^{2}})\left(
(dx_{2})^{2}+(dy_{2})^{2}+(dz_{2})^{2}\right)  \text{ \ \ \ .}\label{ABC2}%
\end{equation}
In equations (\ref{ABC1}) and (\ref{ABC2}) $V_{1}$ and $V_{2}$ are the
potentials of the gravitational field at the space points $%
r_{1}=(x_{1},y_{1},0)$ and $r_{2}=(x_{2},y_{2},0)$ respectively and the
potential $V$ (omitting the indices) is defined as
\begin{equation}
V=\frac{GM}{r}=\frac{GM}{a(1-e\cos E)}\text{ \ \ \ \ .}  \label{C5A}
\end{equation}%
The definition of the gravitational potential with a positive sign was given
 in section \ref{sec:Convent GravPotential} and in section \ref{sec:proptime dimens}
a motivation was presented that this definition is physically correct and moreover, a negative
definition of the potential will influence not the leading term (first inverse power in the velocity of light)
in the formulae for the propagation time, but the second (third inverse powers in the velocity of light) term,
which is considerably smaller than the first term.

Note that often in the literature the square of the space distance is written
as $dx^{2}+dy^{2}+dz^{2}$ and not in the way as are written in the above formulaes
$\left(  (dx_{1})^{2}+(dy_{1})^{2}+(dz_{1})^{2}\right)  $. In fact, the square
of the space-time distance is $ds^{2}=g_{\mu\nu}dx^{\mu}dx^{\nu}$ ($\mu
,\nu=0,1,2,3$), so this is an expression quadratic in the differentials, while
$dx^{2}+dy^{2}+dz^{2}$ might be understood as $2xdx+2ydy+2zdz$. For the use of
the notation in (\ref{ABC1}) and (\ref{ABC2}), the interested reader may
consult the monograph \cite{CA7A7}.

    The first system of coordinates $dT_{1}$, $dx_{1}$, $dy_{1}$, $dz_{1}$ for
the null cone (\ref{ABC1}) are related to the emission time $T_{1}$ of the
first satellite and the differential $dT_{2}$ in the second system of
coordinates of the null cone (\ref{ABC2}) - to the time of reception of the
signal by the second satellite. The space coordinates $dx_{2}$, $dy_{2}$, $%
dz_{2}$ of the second satellite are again determined at the initial moment
of emission of the first satellite. However, the evolution of the space
coordinates  $x_{2}$, $y_{2}$, $z_{2}$ up to the moment of reception of the
signal depends on the celestial motion of the satellite. Consequently,
during the propagation of the signal from the first to the second
satellite, the second satellite moves from the initial coordinates $x_{2}$, $%
y_{2}$, $z_{2}$ to the final coordinates  $x_{2}^{(fin)}$, $y_{2}^{(fin)}$, $%
z_{2}^{(fin)}$ of reception of the signal, and this evolution is governed by
the Kepler equation (\ref{AA11}). But at the same time, these space
coordinates enter the two gravitational null cone equations, which
physically means that the propagation times $dT_{1}$ and $dT_{2}$, found
from the intersection of these two null cones, also "keep track" of the
(constantly changing) positions of the two satellites. In other words, the
space coordinates in the null cone equations simply are parametrized in
terms of variables, related to the motion of the satellites. So this
parametrization does not have  relation with any "mixing up" of the
coordinates of the satellite and the space points of propagation of the
signal.

\ It is important to mention that  finding the differentials of the
propagation times $dT_{1}$ and $dT_{2}$ turns out to be a  complicated
problem from algebraic geometry. In fact, the changing positions (Euclidean
distance) between the satellites mean that the two four-dimensional null
cones have to be additionally intersected with the six-dimensional
hyperplane, which will depend also on the variable Euclidean distance
(meaning that $dR_{AB}^{2}\neq 0$). After finding the differentials $dT_{1}$
and $dT_{2}$ as solutions of the intersecting algebraic variety of algebraic
equations, the solutions of the obtained complicated differential equations
with respect to $dT_{1}$ and $dT_{2}$ will give the evolution of the
propagation time as a function of the changing positions of both satellites
(i.e. changing Euclidean distance). From the standard expression for the
Euclidean distance between the points $A$ and $B$
\begin{equation}
R_{AB}^{2}=(x_{1}-x_{2})^{2}+(y_{1}-y_{2})^{2}+(z_{1}-z_{2})^{2}
\label{ABC2A1}
\end{equation}%
one may obtain after differentiation the following hyperplane equation for
the case of variable Euclidean distance (i.e. $dR_{AB}^{2}\neq 0$)
\begin{equation}
dR_{AB}^{2}=2(x_{1}-x_{2})d(x_{1}-x_{2})+ \nonumber \\
\end{equation}
\begin{equation}
+2(y_{1}-y_{2})d(y_{1}-y_{2})+2(z_{1}-z_{2})d(z_{1}-z_{2})\text{ \ .}
\label{ABC2A2}
\end{equation}%
This is a $6-$dimensional hyperplane in terms of the variables $dx_{1}$, $%
dy_{1}$, $dz_{1}$, $dx_{2}$, $dy_{2}$, $dz_{2}$, which intersects the two
four-dimensional cones. Consequently, the notions of Euclidean distance $%
R_{AB}$ and the propagation times $T_{1}$ and $T_{2}$ are closely related to
the intersection variety of the hyperplane with the two gravitational cones.
In a subsequent paper, this complicated algebraic geometry approach \cite{AAB43B} will be applied
in the general case for the propagation of signals between satellites on
different space orbits, characterized by the full set of $6$ Keplerian
parameters. This problem is important in view of the operational interaction
(transmission of signals) between the satellites, belonging to different
satellite constellations - GPS, GLONASS and Galileo. Further in this paper we shall deal
only with the two-dimensional case of satellites on one and the same
elliptical orbit or on one - plane non-intersecting orbits.

\subsection{Two types of differentials and the equation in full derivatives with
respect to the square of the Euclidean distance}
\label{sec:Two diff}

Our further aim will be to find a relation between the Euclidean distance
$R_{AB}$ (a notion from Newtonian mechanics) and the variables in the null
cone equations. As previously, the two plane elliptical orbits are
parametrized by the  equations
\begin{equation}
x_{1}=a_{1}(\cos E_{1}-e_{1})\text{ \ \ , \ \ \ \ }y_{1}=a_{1}\sqrt[.]%
{1-e_{1}^{2}}\sin E_{1}\text{ \ \ ,}\label{ABC3}%
\end{equation}%
\begin{equation}
x_{2}=a_{2}(\cos E_{2}-e_{2})\text{ \ \ , \ \ \ \ }y_{2}=a_{2}\sqrt[.]%
{1-e_{2}^{2}}\sin E_{2}\text{ \ \ .}\label{ABC4}%
\end{equation}
In order to relate the differential $dR_{AB}$ to the space-coordinate
differentials in the null cone equations (\ref{ABC1}) and (\ref{ABC2}), let us
consider the differential
\[
d(x_{1}^{2}+x_{2}^{2})=d(x_{1}^{2}+x_{2}^{2}-2x_{1}x_{2}+2x_{1}x_{2})=
\]%
\begin{equation}
=d\left[  (x_{1}-x_{2})^{2}+2x_{1}x_{2}\right]  \text{ \ \ .}\label{ABC5}%
\end{equation}
Performing the same for the $y_{1},y_{2}$ coordinates and summing up with
(\ref{ABC5}), one can obtain
\[
d(x_{1}^{2}+x_{2}^{2}+y_{1}^{2}+y_{2}^{2})=d\left[  (x_{1}-x_{2})^{2}%
+(y_{1}-y_{2})^{2}\right]  +
\]%
\begin{equation}
+2\left[  x_{2}dx_{1}+x_{1}dx_{2}+y_{2}dy_{1}+y_{1}dy_{2}\right]  \text{
\ \ .}\label{ABC6}%
\end{equation}
Substituting the expressions (\ref{ABC3}) and (\ref{ABC4} for the elliptical coordinates in
the above formulae and keeping in mind that $R_{AB}^{2}=\left[  (x_{1}-x_{2})^{2}+(y_{1}-y_{2}%
)^{2}\right]  $, one can derive
\begin{equation}
d(x_{1}^{2}+x_{2}^{2}+y_{1}^{2}+y_{2}^{2})=dR_{AB}^{2}+ \nonumber  \\
\end{equation}
\begin{equation}
\\+S_{1}(E_{1}%
,E_{2})dE_{1}+S_{2}(E_{1},E_{2})dE_{2}\text{ \ ,}\label{ABC7}%
\end{equation}
where $S_{1}(E_{1},E_{2})$ denotes  the expression
\[
S_{1}(E_{1},E_{2}):=-2[a_{1}a_{2}\sqrt[.]{(1-e_{1}^{2})(1-e_{2}^{2})}\sin
E_{2}\cos E_{1}+
\]%
\begin{equation}
+a_{1}a_{2}\sin E_{1}\cos E_{2}-e_{2}a_{1}a_{2}\sin E_{1}]\text{
\ \ \ .}\label{ABC8}%
\end{equation}
 The expression for $S_{2}(E_{1},E_{2})$ is the same as the previous one,
but with interchanged $E_{1}\Longleftrightarrow E_{2}$, i.e. $S_{2}%
(E_{1},E_{2})=S_{1}(E_{2},E_{1})$.

The next step is to compare the differentials $(dx_{1})^{2}+(dy_{1})^{2}$ and
$d(x_{1}^{2}+y_{1}^{2})$, written in terms of the elliptical coordinates
(\ref{C4AA1}). Eliminating the term $\left[  a_{1}^{2}%
(1-e_{1}\cos E_{1})dE_{1}\right]  $ in the expressions for the two
differentials, one can obtain
\begin{equation}
d(x_{1}^{2}+y_{1}^{2})=\frac{2e_{1}\sin E_{1}}{(1+e_{1}\cos E_{1})}%
\frac{\left[  (dx_{1})^{2}+(dy_{1})^{2}\right]  }{dE_{1}}\text{ \ \ \ .}%
\label{ABC12}%
\end{equation}
Analogous expression can be obtained also for $d(x_{2}^{2}+y_{2}^{2})$. Now it
can be noted that $\left[  (dx_{1})^{2}+(dy_{1})^{2}\right]  $ and $\left[
(dx_{2})^{2}+(dy_{2})^{2}\right]  $ can be expressed also from the
gravitational null cone equations (\ref{ABC1}) and (\ref{ABC2})
\begin{equation}
(dx_{1})^{2}+(dy_{1})^{2}=\frac{c^{2}(c^{2}+2V_{1})}{(c^{2}-2V_{1})}%
(dT_{1})^{2}\text{ \ , \ }\label{ABC13A}%
\end{equation}
and the second expression is the same but with the indice "$1$" replaced by
"$2$". It should be kept in mind that the symmetry with respect to the interchange of the
indices means that the motion of the satellites can be reversed in the sense that the
signal can be send from the second satellite to the first one. Consequently, one can reverse the indices also in
formulae (\ref{DOPC7}) for the general case in section \ref{sec:propag space}, but this will
mean that the signal is being sent from the second satellite and the first satellite starts its motion from
its starting position. The reversal of the indices should not be performed in a two-way exchange of
signals, when after the second satellite has received the signal from the first one, the signal is sent back to the
first satellite, but during the time of propagation of the signal the first satellite has moved along the orbit and
will intercept the "backward" signal not at the initial point, but at another one. This "backward" point of reception of the signal is found
from the Kepler equation for the motion of the first satellite and this motion is realized during the propagation time of the signal from the
second satellite towards the first one. Such a model of "two-way" exchange of signals is very important to be created, but it will be treated in another publication. It is easy to guess that the model of "two-way" exchange of signals between moving satellites should be based on the formulae (\ref{DOPC7}) in Section \ref{sec:propag space} about the dependence of the differential of the second propagation time on the differential of the first propagation time and also its recurrent version (\ref{DOPCC7}) in Section \ref{sec:chain satell}. As an example, one may take the "two-way" exchange of signals between two or several satellites, moving  on one orbit within the GPS, GLONASS, Galileo satellite constellations.

Substituting the above expressions into (\ref{ABC12}) and into the analogous
formulae for $d(x_{2}^{2}+y_{2}^{2})$, one can obtain a second representation
for $d(x_{1}^{2}+x_{2}^{2}+y_{1}^{2}+y_{2}^{2})$
\begin{equation}
d(x_{1}^{2}+x_{2}^{2}+y_{1}^{2}+y_{2}^{2})=P_{1}(E_{1})\frac{(dT_{1})^{2}%
}{dE_{1}}+P_{2}(E_{2})\frac{(dT_{2})^{2}}{dE_{2}}\text{ \ ,}\label{ABC14}%
\end{equation}
where $P_{1}(E_{1})$ is the expression
\begin{equation}
P_{1}(E_{1}):=\frac{2e_{1}\sin E_{1}}{(1+e_{1}\cos E_{1})}\frac{c^{2}%
(c^{2}+2V_{1})}{(c^{2}-2V_{1})}\text{ \ \ \ ,}\label{ABC15}%
\end{equation}
and the expression for $P_{2}(E_{2})$ is the same, but with the indice "$1$"
replaced by "$2$".

Setting up equal the two equivalent representations (\ref{ABC7}) and
(\ref{ABC14}), one can find the following symmetrical relation
\[
dR_{AB}^{2}+S_{1}(E_{1},E_{2})dE_{1}+S_{2}(E_{1},E_{2})dE_{2}=
\]%
\begin{equation}
=P_{1}(E_{1})\frac{\left(  dT_{1}\right)  ^{2}}{dE_{1}}+P_{2}(E_{2}%
)\frac{\left(  dT_{2}\right)  ^{2}}{dE_{2}}\text{ \ \ \ \ \ .}\label{ABC17}%
\end{equation}

\subsection{Algebraic geometry meaning of the derived relation}
\label{sec:algeb geom}
Now it is important to clarify the result from the viewpoint of algebraic
geometry, after rewriting the above equation in the form of a
four-dimensional cubic algebraic surface (provided that $dE_{1}\neq 0$ and $%
dE_{2}\neq 0$) in terms of the variables $dE_{1}$, $dE_{2}$ and the two
differentials $dT_{1}$, $dT_{2}$ of the propagation time
\begin{equation*}
(dR_{AB}^{2})(dE_{1})(dE_{2})+S_{1}(E_{1},E_{2})(dE_{1})^{2}(dE_{2})+
\end{equation*}%
\begin{equation*}
+S_{2}(E_{1},E_{2})(dE_{1})(dE_{2})^{2}=
\end{equation*}%
\begin{equation}
=P_{1}(E_{1})\left( dT_{1}\right) ^{2}(dE_{2})+P_{2}(E_{2})\text{ }\left(
dT_{2}\right) ^{2}(dE_{1})\text{\ \ \ \ \ .}  \label{ABCD17}
\end{equation}%
In terms of the algebraic geometry terminology, this cubic surface is the
intersecting variety of the two gravitational null cone equations (\ref{ABC1}%
),  (\ref{ABC2}) and the hyperplane equation (\ref{ABC2A2}) in the previous
Section \ref{sec:Gen consid}. The last equation (\ref{ABC2A2})\ in the
present section is replaced by the relation (\ref{ABC6}) between the two
types of differentials. Although equation (\ref{ABCD17}) is a
multi-dimensional cubic surface, it can possess nontrivial solutions for the
differential of the two propagation times - in the papers \cite{BOG1} and
\cite{BOG2} it has been proved that multi-dimensional cubic algebraic
equations have solutions, depending (in a complicated way) on the
Weierstrass elliptic function. It should be reminded also that previously \
in Section \ref{sec:celest propag2} a four-dimensional cubic algebraic
surface (\ref{DOP3B25}) was obtained with respect to the differentials $%
dT_{1}$, $dT_{2}$ of the two propagation times and the differentials $d\tau
_{1}$, $d\tau _{2}$ of the two atomic times.

The four-dimensional cubic equation (\ref{ABCD17}) can be additionally
intersected by the two (quadratic) null cone equations (\ref{ABC1}), (\ref%
{ABC2}) after expressing the two differentials $dT_{1}$ and $dT_{2}$ and
substituting into (\ref{ABCD17}). The resulting intersecting variety is the
equation of a straight line (\ref{ABC36A}), derived in one of the following
sections \ref{sec:Deriv form} in terms of the differentials $dE_{1}$ and $%
dE_{2}$. This equation confirms the fact that the differential of the square
of the Euclidean distance is not given by the expressions (\ref{ABC6}) and (%
\ref{ABC7}), which do not take into account the null cone equations (\ref%
{ABC1}), (\ref{ABC2}), but by the differential equation in full derivatives (%
\ref{ABC36A}). The result of the integration of this equation with respect
to $R_{AB}^{2}$ will be the starting point for creating the theory about the
space-time distance in Section \ref{sec:SpaceTime Int} and about the
geodesic distance in Section  \ref{sec:geod dist}.

\subsection{The signal propagation times for the two satellites as
two-point time transfer functions of the two eccentric anomaly angles}
\label{sec:Sign prop}

From the last equality $dT_{2}$ can be expressed if $dT_{1}$ is known. It follows also that
if $T_{1}$ is a function only of the first eccentric anomaly angle, i.e.
$T_{1}=T_{1}(E_{1})$, then the second propagation time $T_{2}$ for the process
of signal propagation from the second satellite to the first one will depend on
both eccentric anomaly angles.

If the expression for the square of the differential $(dT_{2})^{2}$ in its
standard form
\[
\left(  dT_{2}\right)  ^{2}=\left(  \frac{\partial T_{2}(E_{1},E_{2}%
)}{\partial E_{1}}\right)  ^{2}(dE_{1})^{2}+
\]%
\begin{equation}
+2\frac{\partial T_{2}(E_{1},E_{2})}{\partial E_{1}}\frac{\partial T_{2}%
(E_{1},E_{2})}{\partial E_{2}}dE_{1}dE_{2}+ \nonumber \\
\end{equation}
\begin{equation}
\\+\left(  \frac{\partial T_{2}%
(E_{1},E_{2})}{\partial E_{2}}\right)  ^{2}(dE_{2})^{2}\label{ABC18}%
\end{equation}
is substituted into (\ref{ABC17}) and both sides are divided by $(dE_{1})^{2}%
$, then the following cubic polynomial with respect to $\frac{dE_{2}}{dE_{1}}$
is obtained
\begin{equation}
Q_{1}\left(  \frac{dE_{2}}{dE_{1}}\right)  ^{3}+Q_{2}\left(  \frac{dE_{2}%
}{dE_{1}}\right)  ^{2}+Q_{3}\left(  \frac{dE_{2}}{dE_{1}}\right)
+Q_{4}=0\text{ \ \ ,}\label{ABC19}%
\end{equation}
where the functions $Q_{1}(E_{1},E_{2})$, $Q_{2}(E_{1},E_{2})$, $Q_{3}%
(E_{1},E_{2})$ and $Q_{4}(E_{1},E_{2})$ are given in Appendix $A$.

Equation (\ref{ABC19}) can be transformed into an equation, depending on the
derivatives $\frac{\partial T_{1}}{\partial E_{1}}$ and $\frac{\partial T_{2}%
}{\partial E_{1}}$
\begin{equation}
\left[  \frac{\partial T_{2}}{\partial E_{1}}+\frac{P_{1}(E_{1})}{2P_{2}%
(E_{2})}\frac{\partial T_{1}}{\partial E_{1}}\left(  \frac{dE_{2}}{dE_{1}%
}\right)  \right]  ^{2}=K\text{ \ \ ,} \label{ABC21}%
\end{equation}
where $K$ can be represented as
\begin{equation}
K:=G_{1}(E_{1},E_{2})\left(  \frac{dE_{2}}{dE_{1}}\right)  ^{2}+G_{2}%
(E_{1},E_{2})\left(  \frac{dE_{2}}{dE_{1}}\right)  \text{ \ \ \ .}
\label{ABC22}%
\end{equation}
The functions $G_{1}(E_{1},E_{2})$ and $G_{2}(E_{1},E_{2})$ are given by
expressions (\ref{BBB5}) and (\ref{BBB6}) in Appendix $A$. Equation
(\ref{ABC21}) enables to express the unknown derivative $\frac{\partial T_{2}%
}{\partial E_{1}}$ as
\begin{equation}
\frac{\partial T_{2}}{\partial E_{1}}=-\frac{P_{1}}{2P_{2}}\frac{\partial
T_{1}}{\partial E_{1}}\frac{dE_{2}}{dE_{1}}+\epsilon\sqrt[.]{K}\text{ \ \ \ ,}
\label{ABC23}%
\end{equation}
where $\epsilon=\pm1$.

The interchange of the indices $1\Leftrightarrow2$ allows to find
\[
\frac{\partial T_{1}}{\partial E_{2}}=-\frac{P_{2}}{2P_{1}}\frac{\partial
T_{2}}{\partial E_{2}}\frac{dE_{1}}{dE_{2}}+
\]%
\begin{equation}
+\epsilon\sqrt[.]{\overline{G}_{1}(E_{2},E_{1})\left(  \frac{dE_{1}}{dE_{2}%
}\right)  ^{2}+\overline{G}_{2}(E_{2},E_{1})\left(  \frac{dE_{1}}{dE_{2}%
}\right)  }\text{ \ \ \ ,} \label{ABC24}%
\end{equation}
where $\overline{G}_{1}(E_{2},E_{1})$ (expression (\ref{BBB7}) in Appendix $A
$) and $\overline{G}_{2}(E_{2},E_{1})$ are the functions $G_{1}(E_{1},E_{2})$
and $G_{2}(E_{1},E_{2})$, but with interchanged indices $1$ and $2$. If the
derivative $\frac{\partial T_{1}}{\partial E_{2}}$ is known, then the
dependence of the second propagation time on $E_{2}$ can be found as
\begin{equation}
\frac{\partial T_{2}}{\partial E_{2}}=-\frac{1}{2}\left(  \frac{dE_{1}}%
{dE_{2}}\right)  \left(  1+4\frac{\partial T_{1}}{\partial E_{2}}\right)
+\epsilon\sqrt[.]{N}\text{ \ \ \ ,} \label{ABC25}%
\end{equation}
where the function $N$ is also given in Appendix $A$.

Thus, after integration of (\ref{ABC23}) and (\ref{ABC25}), it is possible to
find the dependence of the second propagation time $T_{2}$ on the
eccentric anomaly angles $E_{1}$ and $E_{2}$. The corresponding integrals however are
very complicated and not possible to be solved analytically. It is important
that both expressions depend on the derivatives of the square of the Euclidean
distance $R_{AB}^{2}$, which according to the parametrization equations
(\ref{ABC3}) and (\ref{ABC4}) is given by the formulae
\[
R_{AB}^{2}=\left[  \left(  a_{1}\cos E_{1}-a_{2}\cos E_{2}\right)  +\left(
a_{2}e_{2}-a_{1}e_{1}\right)  \right]  ^{2}+
\]%
\begin{equation}
+\left[  a_{1}\sqrt[.]{1-e_{1}^{2}}\sin E_{1}-a_{2}\sqrt[.]{1-e_{2}^{2}}\sin
E_{2}\right]  ^{2}\text{ \ \ .} \label{ABC27}%
\end{equation}

\subsection{Signal propagation times for satellites moving on
elliptical orbits with equal eccentric anomalies}
\label{sec:Sign ellip}

For orbits with equal eccentric anomalies (this notion will be clarified
further) there can be two cases.

First case: Equal eccentricities $e_{1}=e_{2}=e$ and semi-major axis
$a_{1}=a_{2}=a$.

 This is the case when two or more satellites move along one and the same
orbit. This corresponds to the satellite dispositions for the GLONASS, GPS
and Galileo constellations. For the GPS and GLONASS constellations, four
satellites are selected per plane \cite{AAB16A}, while in the Galileo
constellation nine satellites are equally spaced per plane.

Second case. Equal eccentric anomalies, but different eccentricities and
semi-major axis.

If one considers the first case then
\begin{equation}
P_{1}=P_{2}\text{ \ , \ }S_{1}=S_{2}\text{ \ \ ,\ \ }\frac{dE_{1}}{dE_{2}%
}=1\text{ \ , \ }\frac{\partial R_{AB}^{2}}{\partial E}=0\text{ \ \ \ .\ \ \ }%
\label{ABC28}%
\end{equation}
Consequently, expression (\ref{ABC23}) for $\frac{\partial T_{2}}{\partial E}$
can be rewritten as
\begin{equation}
\frac{\partial T_{2}}{\partial E}=-\frac{1}{2}\frac{\partial T_{1}}{\partial
E}+\varepsilon\sqrt[.]{G_{1}(E,E)+G_{2}(E,E)}\text{ \ \ \ ,}\label{ABC29}%
\end{equation}
where expressions $G_{1}(E,E)$ and $G_{2}(E,E)$ can be found from formulaes
(\ref{BBB5}) and (\ref{BBB6}) in Appendix $A$
\begin{equation}
G_{1}(E,E)+G_{2}(E,E)=\frac{1}{4}\left(  \frac{\partial T_{1}}{\partial
E}\right)  ^{2}+\frac{1}{2}\frac{\partial T_{1}}{\partial E}-\frac{S}{P}\text{
\ \ .}\label{ABC30}%
\end{equation}
After performing the integration in (\ref{ABC29}), one can obtain
\begin{equation}
T_{2}=-\frac{1}{2}T_{1}+\epsilon%
{\displaystyle\int}
dE\sqrt[.]{\left(  \frac{\partial T_{1}}{\partial E}\right)  ^{2}+\frac{1}%
{2}\left(  \frac{\partial T_{1}}{\partial E}\right)  -\frac{S}{P}}\text{
\ \ \ .}\label{ABC31}%
\end{equation}
From (\ref{ABC25}) for $E_{1}=E_{2}=E$ (however, it is not assumed that
$\frac{\partial T_{1}}{\partial E_{2}}=0$) one can also derive
\[
\frac{\partial T_{2}}{\partial E}=-\frac{1}{2}\left(  1+4\frac{\partial T_{1}%
}{\partial E}\right)  +
\]%
\begin{equation}
+\epsilon\sqrt[.]{2\left(  \frac{\partial T_{1}}{\partial E}\right)
^{2}+2\left(  \frac{\partial T_{1}}{\partial E}\right)  +\frac{2S}{P}+\frac
{1}{4}}\text{ \ \ \ \ .}\label{ABC32}%
\end{equation}
From the equality of the expressions (\ref{ABC28}) and (\ref{ABC32}) for
\ $\frac{\partial T_{2}}{\partial E}$, one can obtain the following quartic
algebraic equation with respect to $\frac{\partial T_{1}}{\partial E}$
\[
\frac{5}{4}\left(  \frac{\partial T_{1}}{\partial E}\right)  ^{4}+6\left(
\frac{\partial T_{1}}{\partial E}\right)  ^{3}+\left[  \frac{7}{4}+\frac
{6S}{P}\right]  \left(  \frac{\partial T_{1}}{\partial E}\right)  ^{2}+
\]%
\begin{equation}
+\left(  \frac{1}{2}-\frac{6S}{P}\right)  \frac{\partial T_{1}}{\partial
E}-\left(  \frac{S}{P}+9\frac{S^{2}}{P^{2}}\right)  =0\text{ \ \ \ .}%
\label{ABC33}%
\end{equation}
This equation can be solved as an algebraic equation with respect to
$\frac{\partial T_{1}}{\partial E}$ and then, after integration, the function
$T_{1}=T_{1}(E)$ can be found. However, if it is assumed that $\frac{\partial
T_{1}}{\partial E_{2}}=0$, then the function $T_{1}(E)$ can be found from the
following integral
\begin{equation}
T_{1}=%
{\displaystyle\int}
dE\frac{\left[  \frac{1}{2}+\frac{3S}{P}-\sqrt[.]{\frac{1}{4}+\frac{2S}{P}%
}\right]  }{\left[  1-\sqrt[.]{\frac{1}{4}+\frac{2S}{P}}\right]  }\text{
\ \ \ \ .}\label{ABC33A}%
\end{equation}
In both cases, the obtained integrals are rather complicated and not possible
to be solved analytically. It is interesting also to see from (\ref{ABC31})
the asymmetry and inequality between the two propagation times $T_{1}$ and
$T_{2}$. In the next section it will be shown how to determine the upper
integration boundary in the integral (\ref{ABC33A}) and in the integral
derived from (\ref{ABC33}).

\section{PHYSICAL\ \ AND\ \ MATHEMATICAL\ \ THEORY\ OF\ \ THE\ \ SPACE - TIME\
\ INTERVAL\ \ ON\ \ INTERSECTING\ \ GRAVITATIONAL\ \ NULL\ \ CONES\ \ FOR\ \ THE\
\ CASE\ OF\ NON-SPACE\ \ ORIENTED\ \ ORBITS}
\label{sec:SpaceTime Int}
\subsection{Derivation of the formulae for the space - time distance after integrating a
differential equation in full derivatives}
\label{sec:Deriv form}

Let us first express from the null cone equations (\ref{ABC1}) and
(\ref{ABC2}) the square of the differentials of the propagation times
$\left(  dT_{1}\right)  ^{2}$ and $\left(  dT_{2}\right)  ^{2}$ .
Combining these expressions with the relation (\ref{ABC15}) (also with the
one for the indice $2$) and making use of the parametrization equations (\ref{ABC3})
and (\ref{ABC4}) for the elliptical orbit, one can obtain the simple relation
\begin{equation}
\left(  dT_{1}\right)  ^{2}P_{1}(E_{1})=a_{1}^{2}\left(  1-e_{1}\cos
E_{1}\right)  2e_{1}\sin E_{1}\left(  dE_{1}\right)  ^{2}\text{ \ \ \ .}%
\label{ABC36}%
\end{equation}
If we substitute this relation and the analogous one for $\left(
dT_{2}\right)  ^{2}P_{2}(E_{2})$ \ into (\ref{ABC17}), then the following
equation in full differentials with respect to the (variable) square of the
Euclidean distance is obtained
\begin{equation}
dR_{AB}^{2}=F_{1}(E_{1},E_{2})dE_{1}+F_{2}(E_{1},E_{2})dE_{2}\text{
\ \ \ ,}\label{ABC36A}%
\end{equation}
where $F_{1}(E_{1},E_{2})$ and $F_{2}(E_{1},E_{2})$ are the expressions
\begin{equation}
F_{1}(E_{1},E_{2}):=2e_{1}a_{1}^{2}\left(  1-e_{1}\cos E_{1}\right)  \sin
E_{1}-S_{1}(E_{1},E_{2})\text{ \ \ ,}\label{ABC37}%
\end{equation}%
\begin{equation}
F_{2}(E_{1},E_{2})=2e_{2}a_{2}^{2}\left(  1-e_{2}\cos E_{2}\right)  \sin
E_{1}-S_{2}(E_{1},E_{2})\text{ \ \ }\label{ABC38}%
\end{equation}
and as previously, $S_{1}(E_{1},E_{2})$ is  given by (\ref{ABC8}),
$S_{2}(E_{1},E_{2})$ is the analogous expression, but with interchanged
indices. The conditions (\ref{ABC36A}) to be an equation in full differentials
are (see any textbook on differential equations, for example \cite{AAB16DF})
\begin{equation}
F_{1}(E_{1},E_{2})=\frac{\partial R_{AB}^{2}}{\partial E_{1}}\text{ \ \ ,
\ }F_{2}(E_{1},E_{2})=\frac{\partial R_{AB}^{2}}{\partial E_{2}}\text{
\ \ .}\label{ABC39}%
\end{equation}
If the first equation is integrated then
\begin{equation}
R_{AB}^{2}=%
{\displaystyle\int}
F_{1}(E_{1},E_{2})dE_{1}+\varphi(E_{2})=\label{ABC40}%
\end{equation}%
\[
=-2e_{1}a_{1}^{2}\cos E_{1}+\frac{1}{2}e_{1}^{2}a_{1}^{2}\cos(2E_{1})+
\]%
\[
+2a_{1}a_{2}\sqrt[.]{\left(  1-e_{1}^{2}\right)  \left(  1-e_{2}^{2}\right)
}\sin E_{1}\sin E_{2}-
\]%
\begin{equation}
-2a_{1}a_{2}\cos E_{1}\cos E_{2}
+2e_{2}a_{1}a_{2}\cos E_{1}+\varphi(E_{2})\text{ \ \ ,}\label{ABC41}%
\end{equation}
where $\varphi(E_{2})$ is a function, which has to be determined from the
second equation in (\ref{ABC39}). If from (\ref{ABC41}) the derivative
$\frac{\partial R_{AB}^{2}}{\partial E_{2}}$ is calculated and then is set up
equal to $F_{2}(E_{1},E_{2})$ given by expression (\ref{ABC38}), the following
simple differential equation for $\varphi(E_{2})$ can be obtained
\begin{equation}
\frac{\partial\varphi(E_{2})}{\partial E_{2}}=\left(  2e_{2}a_{2}^{2}%
-2e_{1}a_{1}a_{2}^{.}\right)  \sin E_{2}-e_{2}^{2}a_{2}^{2}\sin(2E_{2})\text{
\ \ .}\label{ABC42}%
\end{equation}
If the equation is integrated and the result is substituted into (\ref{ABC41}),
then the final expression for $R_{AB}^{2}$ is obtained
\[
R_{AB}^{2}=\left(  -2e_{1}a_{1}^{2}\cos E_{1}-2e_{2}a_{2}^{2}\cos
E_{2}\right)  +
\]%
\[
+\left(  2e_{2}a_{1}a_{2}\cos E_{1}+2e_{1}a_{1}a_{2}^{.}\cos E_{2}\right)  +
\]%
\[
+\frac{1}{2}\left(  e_{1}^{2}a_{1}^{2}\cos\left(  2E_{1}\right)  +e_{2}%
^{2}a_{2}^{2.}\cos\left(  2E_{2}\right)  \right)  -2a_{1}a_{2}\cos E_{1}\cos
E_{2}+
\]%
\begin{equation}
+2a_{1}a_{2}\sqrt[.]{\left(  1-e_{1}^{2}\right)  \left(  1-e_{2}^{2}\right)
}\sin E_{1}\sin E_{2}\text{ \ \ .}\label{ABC43}%
\end{equation}
Note that this expression is symmetrical and does not change under interchange
of the indices $1$ and $2$ , as it should be. This is in fact the second
representation for the square of the Euclidean distance, based on the equation
(\ref{ABC36}) in full differentials, when $R_{AB}^{2}$ is expressed according to
(\ref{ABC43}).

\subsection{General idea about the positive, negative or zero space - time
distance from the intersection of the two null cones}
\label{sec:Phys spacetime}

Relation (\ref{ABC43}) has been found from the intersection of the two null
cones (\ref{ABC1}) and (\ref{ABC2}) and consequently,
 represents a General Relativity Theory (GRT) notion.
It is more correct to call it "a space-time interval", which according to
GRT can be either positive, negative or equal to zero. In fact, an important clarification should be made:
GRT clearly defines what is a "gravitational null cone" and also a "space-time interval". But as
mentioned before, GRT and also Special Relativity Theory \cite{AAB16A1}
do not give an answer to the problem: will the intersection of two gravitational null cones again possess the
property of the space-time interval, i.e. can it be again positive, negative and zero? The investigation of the
"intersecting space-time interval" (\ref{ABC43}) in some partial simplified cases, but also in the general case will
give an affirmative answer to this problem. In other words, it will become evident that this "intersecting" interval
will again preserve the property of being positive, negative or null.

At the same time, the function $R_{AB}^{2}$ in (\ref{ABC43}) is the
Euclidean distance $R_{AB}^{2}=(x_{1}-x_{2})^{2}+(y_{1}-y_{2})^{2}$
in the initial formulae (\ref{ABC27}), which can be presented in a more
symmetrical way
\[
R_{AB}^{2}=a_{1}^{2}+a_{2}^{2}+(a_{2}e_{1}-a_{1}e_{2})^{2}-2a_{1}a_{2}\cos
E_{1}\cos E_{2}-
\]%
\[
-2a_{1}a_{2}\sqrt[.]{\left(  1-e_{1}^{2}\right)  \left(  1-e_{2}^{2}\right)
}\sin E_{1}\sin E_{2}-a_{1}^{2}e_{1}^{2}\sin^{2}E_{1}-
\]%
\[
-a_{2}^{2}e_{2}^{2}\sin^{2}E_{2}+2a_{1}a_{2}(e_{2}\cos E_{1}+e_{1}\cos E_{2})-
\]%
\begin{equation}
-2e_{1}a_{1}^{2}\cos E_{1}-2e_{2}a_{2}^{2}\cos E_{2}\text{ \ \ .}\label{ABC44}%
\end{equation}

\subsection{The compatibility condition for intersatellite communications}
\label{sec:Compat cond}

It turns out that $R_{AB}^{2}$ has two equivalent representations - the
first representation as a space-time interval (\ref{ABC43}), found from the
intersection of the null cones with the hyperplane equation, which can be positive, negative or zero, and
the second representation as an Euclidean distance (\ref{ABC44}), which
can be only positive. There is nothing strange that the Euclidean distance can be transformed into
a space-time interval, since it is found also from another
equations (the gravitational null cone equations). But in any case, they denote one and the
same function denoted as $R_{AB}^{2}$. The only possibility for the compatibility of the two
representations is they to be equal both to zero or to be both positive.
However, as further it shall be explained, it is not obligatory to impose the requirement for the compatibility of the
two representations - the space-time interval (\ref{ABC43}) can be treated as an independent
notion from the Euclidean distance. But in the case of light or signal propagation, when the two
distances have to be compatible because the signal travels a macroscopic distance,
the two representations (\ref{ABC43}) and (\ref{ABC44}) have to be
set up equal. Then from the equality of the two representations (\ref{ABC43}) and (\ref{ABC44}) for
$R_{AB}^{2}$, one can obtain the following simple \ relation between the
eccentric anomalies, semi-major axis and the eccentricities of the two orbits
\[
4a_{1}a_{2}\sqrt[.]{\left(  1-e_{1}^{2}\right)  \left(  1-e_{2}^{2}\right)
}\sin E_{1}\sin E_{2}=
\]%
\begin{equation}
=a_{1}^{2}+a_{2}^{2}+(a_{2}e_{2}-a_{1}e_{1})^{2}-\frac{1}{2}\left(  e_{1}%
^{2}a_{1}^{2}+e_{2}^{2}a_{2}^{2}\right)  \text{ \ \ \ .}\label{ABC45}%
\end{equation}

This relation can be conditionally called "a condition for intersatellite
communications between satellites on (one - plane) elliptical orbits". It is
obtained as a compatibility condition between the large-scale, Euclidean distance
(\ref{ABC44}) and the space-time interval (\ref{ABC43}).

\subsection{Positivity and negativity of the space-time interval for the
case of equal eccentric anomaly angles, eccentricities and semi-major axis
- consistency check of the calculations}
\label{sec:Posit negat}

The best way to understand the difference between the physical meaning of
the space-time interval and the Euclidean distance with account of the
condition (\ref{ABC45}) is to prove that the space-time interval for some
specific cases can be of any signs, while the situation will turn out to be
different for the geodesic distance.  It is remarkable that the
positivity of the geodesic distance will become evident when performing a
simple algebraic substitution of the condition (\ref{ABC45}) into formulae (%
\ref{ABC43}) and at the same time, this will be confirmed by the analysis of
a complicated algebraic equation of fourth degree.

Let us first write again the space-time interval (\ref{ABC43}) for the case
of equal eccentricities, semi-major axis and eccentric anomaly angles ($%
e_{1}=e_{2}=e$, $a_{1}=a_{2}=a$, $E_{1}=E_{2}=E$)
\begin{equation}
R_{AB}^{2}=4a^{2}\sin ^{2}E.(1-e^{2})+a^{2}(e^{2}-2)\text{ \ .}  \label{F1}
\end{equation}%
Now it is interesting to note that this space-time interval is positive for
\begin{equation}
\sin ^{2}E\geq \frac{2-e^{2}}{4(1-e^{2})}\text{ \ \ ,}  \label{F2}
\end{equation}%
but for
\begin{equation}
\sin ^{2}E\leq \frac{2-e^{2}}{4(1-e^{2})}\text{ \ \ }  \label{F3}
\end{equation}%
it can be also negative. This fact for the partial case suggests that this should be so also
for the general case of different one from another
eccentricities, large semi-major axis and eccentricity anomaly angles.
However, the general case will be much more complicated, because the
space-time distance (\ref{ABC43}) will turn out to be a fourth-degree
algebraic equation with respect to the variable $y=\sin E_{2}$. By means of
theorems from higher algebra and without solving this complicated equation,
it will further be proved that the space-time distance again can be zero,
negative or positive.

The lower bound for which $R_{AB}^{2}\geq 0$ for the case of a typical GPS\
orbit with eccentricity $e=0.01323881349526$ (see the PhD thesis \cite{C4}
of Gulklett) is given by the limiting value $E_{\lim }$ for the eccentric
anomaly angle
\begin{equation*}
E_{\lim }=\arcsin \left[ \frac{1}{2}\sqrt{\frac{2-e^{2}}{1-e^{2}}}\right] =
\end{equation*}%
\begin{equation}
=45.002510943228\text{ }[\deg ]\text{ \ \ .}  \label{F4}
\end{equation}%
Respectively, the upper bound $\frac{2-e^{2}}{4(1-e^{2})}$ in (\ref{F3}),
for which $R_{AB}^{2}\leq 0$, can be found from $E\leq E_{\lim }$. It is
curious to note that if the condition for intersatellite
communications (\ref{ABC45}) is taken into considerations, neither of
the two inequalities is realized. The reason is that for $e_{1}=e_{2}=e$, $%
a_{1}=a_{2}=a$, $E_{1}=E_{2}=E$ this condition gives the relation
\begin{equation}
4a^{2}(1-e^{2})\sin ^{2}E=2a^{2}-e^{2}a^{2}\text{ \ \ }\Longrightarrow \sin
E=\frac{1}{2}\sqrt[.]{\frac{\left( 2-e^{2}\right) }{\left( 1-e^{2}\right) }}%
\text{\ \ ,\ }  \label{ABC49A}
\end{equation}%
which, if substituted into the space-time interval (\ref{F1}), gives $%
R_{AB}^{2}=0$. This should be expected and in fact is a consistency check of
the calculations because for equal eccentricities, semi-major axis and
eccentric anomaly angles, the Euclidean distance (\ref{ABC44}) is equal to
zero. Then the compatibility condition (\ref{ABC49A}), when
substituted in the formulae for the space-time interval, should give also zero. The obtained result is fully
consistent with what should be expected. However, the equality to zero
of the space-time distance (but only for this specific case investigated) is ensured only when the
compatibility condition is applied, which justifies its name. Without the compatibility condition,
the space-time interval is different from zero for equal eccentricities, semi-major axis and
eccentric anomaly angles, while the Euclidean distance is equal to zero.

\subsection{Positive space-time interval from non-zero
Euclidean distance - the case of different eccentric anomaly angles}
\label{sec:Nonzero Euclid}

      Now we shall investigate the other case of non-zero Euclidean
distance, when two points on the two corresponding orbits do not coincide.
This will be the case of equal eccentricities and semi-major axis, but
different eccentric anomaly angles ($E_{1}\neq E_{2}$). Three important
facts will be proved:

1. The space-time interval can be positive.

2. The space-time interval can also be negative. This case shall be demonstrated
in the next subsection \ref{sec:Nonzero EuclidNeg}.

3. The space-time interval for the case of non-zero Euclidean distance can be equal to zero. This case will be
investigated in Section \ref{sec:High algeb}. Since the space-time interval will be represented by a fourth-degree
polynomial, the proof about a zero space-time interval will be equivalent to the proof that this fourth-degree polynomial
has roots within the unit circle. For the purpose, some more non-trivial theorems from higher algebra will be applied.

The space-time interval (\ref{ABC43}) can be written as
\begin{equation*}
R_{AB}^{2}=e^{2}a^{2}-e^{2}a^{2}(\sin E_{1}+\sin E_{2})^{2}-
\end{equation*}%
\begin{equation}
-2a^{2}\cos (E_{1}+E_{2})\text{ \ \ .}  \label{F6}
\end{equation}%
The space- time interval will be positive (i.e. $R_{AB}^{2}>0$), if the
following inequality is satisfied
\begin{equation*}
e^{2}-e^{2}(\sin ^{2}E_{1}+\sin ^{2}E_{2})+
\end{equation*}%
\begin{equation}
+2(1-e^{2})\sin E_{1}\sin E_{2}>2\cos E_{1}\cos E_{2}\text{ \ \ .}
\label{F7}
\end{equation}%
If we take into account the standard inequalities for the $\cos $-function
\begin{equation}
\cos E_{1}\leq 1\text{ \ \ , }\cos E_{2}\leq 1\text{ \ ,}  \label{F8}
\end{equation}%
then the first two terms on the first line of the above inequality can be
written as
\begin{equation*}
e^{2}-e^{2}(\sin ^{2}E_{1}+\sin ^{2}E_{2})=
\end{equation*}%
\begin{equation}
=e^{2}\cos ^{2}E_{1}-e^{2}+e^{2}\cos ^{2}E_{2}\leq e^{2}-e^{2}+e^{2}=e^{2}%
\text{ \ .}  \label{F9}
\end{equation}%
Substituting into inequality (\ref{F7}), it can be derived
\begin{equation}
2\cos E_{1}\cos E_{2}<e^{2}+2\sin E_{1}\sin E_{2}\text{ \ \ ,}  \label{F10}
\end{equation}%
which can be represented also as
\begin{equation}
\cos (E_{1}+E_{2})<\frac{e^{2}}{2}\text{ \ }\Longrightarrow \text{ \ \ \ \ }%
E_{1}+E_{2}>\arccos (\frac{e^{2}}{2})\text{ \ \ .}  \label{F11}
\end{equation}%
For the typical value of the eccentricity of the GPS orbit, it can be
obtained
\begin{equation}
E_{1}+E_{2}>89.994978993712\text{ }[\deg ]\text{ . }  \label{F12}
\end{equation}%
Note that the sign is greater because $\cos $ is a decreasing function
with the increase of the angle. This is valid for the first and the second quadrant, but for the
third and the fourth quadrant $\cos $ is an increasing function and the sign should be
the reverse one. Also the sign (for the angle within the first quadrant) should be the
reverse one to the sign in (\ref{F12}), if the space-time interval is negative, i.e. (i.e. $R_{AB}^{2}<0$).
For the moment, we shall investigate the case when the eccentric anomaly angle is
in the first and the second quadrant. If one sets up $E_{1}=E_{2}=E$ in (\ref{F10}%
), then (let us take again the case of positive space-time interval)
\begin{equation}
\sin ^{2}E>\frac{1}{2}(1-\frac{e^{2}}{2})\text{ \ }\Longrightarrow \text{ }E>%
\overline{E}=\arcsin \frac{\sqrt{2-e^{2}}}{2}\text{ \ ,}  \label{F13}
\end{equation}%
where the numerical result for $\overline{E}$ is twice as smaller than (\ref%
{F12})
\begin{equation}
\overline{E}=44.997489496856\text{ \ }[\deg ]\text{ \ .}  \label{F14}
\end{equation}
It should be clarified that this numerical value is a little lower that the
limiting value $45.002510943228\text{ }[\deg ]$ (\ref{F4}) in the preceding section \label{sec:Posit negat},
because in the case the property (\ref{F8}) of the trigonometric functions has been used. In the previous section,
the limiting value has been obtained as an exact value. Comparison between these two values will be performed in the
Discussion part of this paper.

\subsection{Negative space-time interval from non-zero
Euclidean distance - the case of different eccentric anomaly angles}
\label{sec:Nonzero EuclidNeg}

The space-time interval $R_{AB}^{2}$ will be negative, i.e. $R_{AB}^{2}<0$,
when the following inequality is fulfilled
\begin{equation}
e^{2}<e^{2}(\sin E_{1}+\sin E_{2})^{2}+2\cos (E_{1}+E_{2})\text{ \ \ .}
\label{PPP1}
\end{equation}%
Taking again into account that
\begin{equation}
\sin E_{1}\leq 1\text{ \ \ \ \ , \ \ \ \ }\sin E_{2}\leq 1\text{ \ \ ,}
\label{PPP1A}
\end{equation}%
it can be obtained
\begin{equation}
(\sin E_{1}+\sin E_{2})^{2}\leq 4\text{ \ \ .}  \label{PPP2}
\end{equation}%
Thus, inequality (\ref{PPP1}) can be represented as
\begin{equation}
e^{2}<4e^{2}+2\cos (E_{1}+E_{2})\text{ \ \ ,}  \label{PPP3}
\end{equation}%
from where it follows that
\begin{equation}
\cos (E_{1}+E_{2})>-\frac{3e^{2}}{2}\text{ }\Rightarrow E_{1}+E_{2}<\arccos
(-\frac{3e^{2}}{2})\text{\ \ .}  \label{PPP4}
\end{equation}%
For the value of the GPS orbit $e=0.01323881349526$ one can obtain $-\frac{%
3e^{2}}{2}=-0.00026289927414342$. It is evident that $\cos
(E_{1}+E_{2})$ is in the second quadrant, where it is a decreasing function.
Consequently, the sign in the inequality for $E_{1}+E_{2}$ is reversed and
one can obtain
\begin{equation}
E_{1}+E_{2}<90.015063019019\text{ }[deg]\text{ \ \ .}  \label{PPP5}
\end{equation}%
Together with the previously found inequality $E_{1}+E_{2}>89.994978993712$ $%
[\deg ]$ in (\ref{F12}), the sum of the two eccentric anomaly angles, found
by means of simple inequality estimates, is in the interval
\begin{equation}
89.994978993712[\deg ]<E_{1}+E_{2}<90.015063019019\text{ }[deg]\text{ \ \ ,}
\label{PPP6}
\end{equation}%
which is well within the interval (\ref{GGG3}), which will be predicted in Section \ref{sec:interv zero} after the application of the  Schur
theorem. Note also that twice the limiting value (\ref{F4}) $E_{\lim
}=45.0025109432281$ $[\deg ]$, calculated from the condition for
intersatellite communications (\ref{ABC49A}) for equal eccentric anomaly
angles, eccentricities and semi-major axis (and coinciding with the value,
above which the space-time interval is positive and below which it is
negative), also falls within the extremely small interval (\ref{PPP6}). This
proves that the trigonometric inequalities (\ref{F8}) and (\ref{PPP1A})
might provide precise information about the location of one of the roots.  \

\subsection{Zero space-time interval from non-zero
Euclidean distance - analysis of fourth-degree algebraic equations by means
of higher algebra theorems}
\label{sec:High algeb}

Now it remains to establish when the space-time interval can be zero
for the case of non-zero Euclidean distance. For the purpose,
expression (\ref{F6}) can be written as
\begin{equation*}
2\sqrt{(1-\sin ^{2}E_{1})(1-\sin ^{2}E_{2})}=
\end{equation*}%
\begin{equation*}
=e^{2}-e^{2}(\sin ^{2}E_{1}+\sin ^{2}E_{2})+
\end{equation*}%
\begin{equation}
+2(1-e^{2})\sin E_{1}\sin E_{2}\text{ \ \ .}  \label{F15}
\end{equation}%
After some transformations and introducing the notation $\sin ^{2}E_{1}=y$,
the above expression can be presented in the form of a quartic
(fourth-degree) algebraic equation
\begin{equation}
y^{4}+a_{1}y^{3}+a_{2}y^{2}+a_{3}y+a_{4}=0\text{ \ \ .}  \label{F16}
\end{equation}%
The coefficient functions of this equation will be given in Appendix C.
Consequently, the problem about finding those values of the eccentric
anomaly angle $E_{1}$ for which the space-time interval (\ref{F6}) is zero
is equivalent to the algebraic problem of finding all the roots of the above
quartic (fourth-order) algebraic equation, which are within the circle $\mid y\mid =\mid \sin
^{2}E_{1}\mid <1$ (we exclude the boundary points $y=\sin ^{2}E_{1}=1$).
It is well-known that an algebraic equation of fourth degree will always possess roots. The
problem is that these roots should be within the circle $\mid y\mid <1$.

\subsection{General overview of some higher algebra theorems about the
existence of roots within the unit circle}
\label{sec:Unit circle}

In order to prove that equation (\ref{F16}) has roots within the circle $%
\mid y\mid <1$, the following theorem of Enestrom-Kakeya from higher algebra
\cite{AAB16AB} shall be used:

\begin{theorem}
If for a $n-$th degree polynomial
\begin{equation}
f(y)=a_{0}y^{n}+a_{1}y^{n-1}+a_{2}y^{n-2}+.....+a_{n}  \label{F17}
\end{equation}%
one has
\begin{equation}
a_{0}>a_{1}>a_{2}>a_{3}>......a_{n}>0\text{ \ , }  \label{F18}
\end{equation}%
then the roots of the polynomial $f(y)$ are situated within the circle $\mid y\mid <1$%
.
\end{theorem}

In the monograph of Prasolov \cite{AAB16ABC} this theorem has a slightly
different formulation:

\begin{theorem}
If all the coefficients of the polynomial (\ref{F17}) are positive, then for
every root $\zeta $ of this polynomial the following estimate is valid
\begin{equation}
\min_{1\leq i\leq n}\{\frac{a_{i}}{a_{i-1}}\}=\delta \leq \mid \zeta \mid
\leq \gamma =\max_{1\leq i\leq n}\{\frac{a_{i}}{a_{i-1}}\}\text{ \ \ .}
\label{F19}
\end{equation}
\end{theorem}

\bigskip Since the Enestrom-Kakeya theorem shall not be used in this paper
due to reasons which are given below, its proof will not be given in
Appendix B. In Appendix C the coefficient functions $a_{1}$, $a_{2}$, $a_{3}$%
, $a_{4}$ of the polynomial (\ref{F16}) will be presented.

The theorem of Enestrom - Kakeya has two major shortcomings:

1. It requires the fulfillment of the "chain" of inequalities $%
a_{0}>a_{1}>a_{2}>a_{3}>a_{4}>0$. The last means that all coefficient
functions should be positive and moreover, beginning from the free term (the
coefficient $a_{4}$), each subsequent coefficient should be greater than the
preceding one. This is a serious restriction, since in the present case the
coefficient functions $a_{0}$, $a_{1}$, $a_{2}$, $a_{3}$, $a_{4}$ are not
positive.

2. The theorem represents only a necessary condition. This means that the
theorem is appropriate to be used if the condition (\ref{F18}) (the chain of
inequalities) can be proved to be valid and thus it will follow that the
roots of the polynomial $f(y)$ are within the circle $\mid y\mid <1$.
However, since the theorem does not represent a necessary and sufficient
condition, the non-fulfillment of the condition (\ref{F18}) does not guarantee
that the polynomial $f(y)$ will not have any roots. Some other condition may exist so that
the polynomial still might possess roots.

Another approach may be proposed with the aim to eliminate these
shortcomings by means of combining the theorem of Enestrom - Kakeya with
some other theorems. For example, for the first case those coefficient
functions, which are negative shall be taken with a negative sign (so that they will become positive), and in
this way a new chain of coefficients $b_{0}$, $b_{1}$, $b_{2}$, $b_{3}$, $%
b_{4}$ shall be obtained, and further they shall be arranged in a definite
order
\begin{equation}
b_{4}>b_{2}>b_{3}>b_{1\text{ \ \ \ \ \ .}}  \label{F20}
\end{equation}%
Introducing the new notations
\begin{equation}
b_{0}^{\ast }>b_{1}^{\ast }>b_{2}^{\ast }>b_{3}^{\ast }>b_{4}^{\ast }\text{
\ ,}  \label{F21}
\end{equation}%
so that
\begin{equation}
b_{0}^{\ast }=b_{4}\text{ \ , }b_{1}^{\ast }=b_{2}\text{ \ , \ }b_{2}^{\ast
}=b_{3}\text{ \ , \ }  \label{F22}
\end{equation}%
the following polynomial can be constructed
\begin{equation}
B(y):=b_{0}^{\ast }y^{4}+b_{1}^{\ast }y^{3}+b_{2}^{\ast }y^{2}+b_{3}^{\ast
}y+b_{4}^{\ast }\text{ \ ,}  \label{F23}
\end{equation}%
where the coefficient function $b_{4}^{\ast }$ shall be determined later
from a special condition. The polynomial $B(y)$ shall be defined also in
another representation as
\begin{equation}
B(y):=\overline{b}_{0}+\left(
\begin{array}{c}
4 \\
1%
\end{array}%
\right) \overline{b}_{1}y+\left(
\begin{array}{c}
4 \\
2%
\end{array}%
\right) \overline{b}_{2}y^{2}+\left(
\begin{array}{c}
4 \\
3%
\end{array}%
\right) \overline{b}_{3}y^{3}+\left(
\begin{array}{c}
4 \\
4%
\end{array}%
\right) \overline{b}_{4}y^{4}\text{ \ \ \ .}  \label{F24}
\end{equation}%
At the same time, the original polynomial (\ref{F17}) shall be represented
in another form
\begin{equation*}
f(y)=A(y):=\overline{a}_{0}+\left(
\begin{array}{c}
4 \\
1%
\end{array}%
\right) \overline{a}_{1}y+\left(
\begin{array}{c}
4 \\
2%
\end{array}%
\right) \overline{a}_{2}y^{2}+
\end{equation*}
\begin{equation}
+\left(
\begin{array}{c}
4 \\
3%
\end{array}%
\right) \overline{a}_{3}y^{3}+\left(
\begin{array}{c}
4 \\
4%
\end{array}%
\right) \overline{a}_{4}y^{4}\text{ \ \ \ ,}  \label{F25}
\end{equation}
which requires the calculation of the coefficient functions with the "bar"
sign above. Then the assertion of the s.c. Grace theorem (see again Obreshkoff monograph \cite{AAB16AB}) is

\begin{theorem}
 If the roots of the polynomial $B(y)$ (\ref{F24}) are within the circle $\mid y\mid <1$ (which
again should be checked by means of the Enestrom - Kakeya theorem), then the
polynomial $f(y)=A(y)$ (\ref{F17}) (or (\ref{F25})) has at least one root
within the circle $\ \mid y\mid <1$, provided also that the following
relation holds between the coefficient functions $\overline{a}_{0}$, $%
\overline{a}_{1}$, $\overline{a}_{2}$, $\overline{a}_{3}$, $\overline{a}_{4}$
and $\overline{b}_{0}$, $\overline{b}_{1}$, $\overline{b}_{2}$, $\overline{b}%
_{3}$, $\overline{b}_{4}$ of the two polynomials \
\begin{equation}
\overline{a}_{0}\overline{b}_{4}-4\overline{a}_{1}\overline{b}_{3}+\frac{4.3%
}{2}\overline{a}_{2}\overline{b}_{2}-\frac{4.3.2}{3.2}\overline{a}_{3}%
\overline{b}_{1}+\overline{a}_{4}\overline{b}_{0}=0\text{ \ \ .}  \label{F26}
\end{equation}
\end{theorem}

From the last relation, the coefficient function $b_{0}=\overline{b}_{0}$ of
the newly constructed polynomial $B(y)$ may be expressed. In the
mathematical literature \cite{AAB16AB} and \cite{AAB16ABC}, polynomials $%
B(y) $ (\ref{F24}) and $A(y)$ (\ref{F25}), satisfying the condition (\ref%
{F26}) are called apolar polynomials.

This approach shall not be developed in this paper, because it requires
rather tedious calculations, which at the end will not result in an equality
or non-equality, giving the opportunity to make the conclusion
whether the given polynomial has roots or not within the circle $\mid y\mid
<1$. The more serious reason is again in the lack of a necessary and
sufficient condition in the formulation of the Grace theorem - this means
that if the inequalities (\ref{F20}) $b_{4}>b_{2}>b_{3}>b_{1}$ of the
Enestrom-Kakeya theorem or the relation (\ref{F26}) are not fulfilled , then
this by itself does not guarantee that the polynomial $A(y)$ will not have a
root within the unit circle. In other words, the absence of a sufficient condition
means that some other necessary condition instead of the inequalities
(\ref{F20}) $b_{4}>b_{2}>b_{3}>b_{1}$ may exist, so that the polynomial will have
again roots within the unit circle. The formulation of the Schur theorem confirms
this conclusion.

\subsection{Schur theorem as a basic mathematical instrument for proving
the existence of roots within the unit circle for the space-time algebraic
equation}
\label{sec:Shur theor}

In this paper a preference is given to a theorem, which has a necessary and
sufficient condition. This is the Schur theorem \cite{AAB16AB} (originally published in 1918 in \cite{Schur1}),
which for the general $n-$ dimensional case has the following formulation:

\begin{theorem}
(Schur) The necessary and sufficient conditions for the polynomial of $n-$th
degree
\begin{equation}
f(y)=a_{0}y^{n}+a_{1}y^{n-1}+....+a_{n-2}y^{2}+a_{n-1}y+a_{n}  \label{F27}
\end{equation}%
to have roots only in the circle $\mid y\mid <1$ are the the following ones:

1. The fulfillment of the inequality
\begin{equation}
\mid a_{0}\mid >\mid a_{n}\mid .  \label{F28}
\end{equation}

2. The roots of the polynomial of $(n-1)$ degree
\begin{equation}
f_{1}(y)=\frac{1}{y}\left[ a_{0}f(y)-a_{n}f^{\ast }(y)\right] \text{ \ \ }
\label{F29}
\end{equation}%
should be contained in the circle $\mid y\mid <1$, where $f^{\ast }(y)$ is
the s.c. "inverse polynomial", defined as
\begin{equation}
f^{\ast }(y)=y^{n}f(\frac{1}{y}%
)=a_{n}y^{n}+a_{n-1}y^{n-1}+....+a_{2}y^{2}+a_{1}y+a_{0}\text{ \ \ .}
\label{F30}
\end{equation}%
In case of fulfillment of the inverse inequality%
\begin{equation}
\mid a_{0}\mid <\mid a_{n}\mid  \label{F31}
\end{equation}%
the $(n-1)$ degree polynomial $f_{1}(y)$ (again with the requirement the
roots to remain within the circle $\mid y\mid <1$) is given by the
expression
\begin{equation}
f_{1}(y)=a_{n}f(y)-a_{0}f^{\ast }(y)\text{ \ \ .}  \label{F32}
\end{equation}
\end{theorem}

The proof of this theorem, taken from the Obreshkoff monograph \cite{AAB16AB},
will be presented in Appendix B. The interested reader can see also the proof in the
monograph \cite{Schur2}. Concerning the necessary conditions, the Schur
theorem has one another advantage - if the condition (\ref{F29})
(or (\ref{F32})) about the roots of the polynomial $%
f_{1}(y)$ is not fulfilled, then the polynomial $f(y)$ will not have any
roots within the circle. This allows one to apply the theorem not only with
respect to the space-time interval algebraic equation (which shall be proved
to have roots in Appendix C), but also with respect to the geodesic
equation, which should not have any roots within the circle $\mid y\mid <1$
(this shall be proved in Appendix E). The last fact shall be confirmed
by independent calculations, since it shall be proved in the next sections
that the geodesic distance is greater than the Euclidean distance, so it
cannot become zero (for non-zero Euclidean distance). This is fully consistent from a physical point of view,
since it is not occasional that light or signal propagation is related to
the geodesic distance and not to the space-time interval, which can also be
equal to zero or even become negative. In this aspect, it is really amazing
how the physical interpretation is consistent with the mathematical results
about these two algebraic equations. It is important to mention that these
conclusions and mostly the mathematical proof for the general case of different eccentric anomaly angles
are valid in view of the fact that the eccentricity $e$ is very
small (for the GPS orbits, the typical eccentricity is of the order of $0.01$), and on the
base of this it is possible to compare terms with \ inverse powers in $e$ in
the corresponding inequalities - the higher inverse powers in $e$ will lead to a \ larger
number. For example, a term of the order of $\frac{1}{e^{2}}$ will give
a number of the order of $10000$, but as it will be shown, there will be terms
proportional to $\frac{1}{e^{10}%
}$, $\frac{1}{e^{12}}$ and even  $\frac{1}{e^{14}}$, which are extremely
large numbers. It is important that terms which differ by two orders in
inverse powers of $e$ will have greatly different numerical values.

On the base of such analysis, the Schur theorem gives the opportunity not
only to prove the existence of roots for the space-time interval equation
(without solving this equation), but also to predict the numerical interval
for the eccentric anomaly angle $E_{2}$, where the space-time interval can
become zero. This interval is
\begin{equation}
15.64\text{ }[\deg ]<E_{2}<56.88\text{ }[\deg ]\text{ \ .}  \label{F33}
\end{equation}%
In the Discussion part it will be explained that the restriction (from the properties
of trigonometric functions) on compatibility condition
for intersatellite communications will give a higher lower bound for the above inequality,
thus confirming the difference between the space-time interval and the geodesic distance, which will
be derived by means of the compatibility condition. Since all the expressions are symmetric with respect to the two angles $%
E_{1} $ and $E_{2}$, the same interval is valid also for $E_{1}$.
  \\ It is important to mention one peculiarity of the Schur theorem, which made possible
the derivation of the above result. This is the fact that the polynomial of $%
(n-1)$ degree (\ref{F29}) is a sufficient condition for the existence of
roots within the unit circle of the initial polynomial of $n-$th degree. But
then, if a new polynomial of $(n-2)$ degree is constructed according to
formulae (\ref{F29}) or (\ref{F32}), then this polynomial can become a
sufficient condition for the roots of the $(n-1)$ degree polynomial. In such
a way, a chain of lower-degree polynomials is constructed - each polynomial
represents a necessary and at the same time a sufficient condition for the
construction of a lower degree polynomial. The last constructed polynomial
will be of first order, and from it the condition for the roots to be
contained in the unit circle can easily be found. Note the important role of
the necessary and sufficient condition - if from the linear polynomial the
condition for the roots is found, then it will be a  sufficient
condition for the second-degree polynomial, further this polynomial will be a necessary
and sufficient condition for the third-degree polynomial and etc. In such a way, the
first-order polynomial will turn out to be a sufficient
condition for the roots to remain within the unit circle with respect
to the initial $n$-th degree polynomial, provided
also that for each polynomial the corresponding inequalities between the
coefficient functions are fulfilled. It can be claimed that this "chain"
of lower-degree polynomials, together with the corresponding inequalities between the
coefficient functions, represents a modified version of the Schur theorem. So from the point of
view of pure mathematics, such a modified version without any doubt is interesting, and the peculiar moment is
that the physical information (availability of roots with respect to the space - time equation and absence of
any roots with respect to the geodesic equation) is very important for the confirmation of such a modified version
of the theorem from a physical point of view. Of course, the proof is limited for the investigated case of polynomials of fourth degree.

\subsection{New definition of the Euclidean distance by means of
intersecting null cones - geometrical importance of the new result}
\label{sec:Dist cones}

The main result in this paper concerns the intersection of two
four-dimensional gravitational null cones (\ref{ABC1}) and (\ref{ABC2}) and
also with the hyperplane equation (\ref{ABC2A2}), which in fact
defines a variable distance (\ref{ABC43}) on the intersection of these null
cones. From Special Relativity Theory it is known that a distance on the
null cone can be either positive, negative or null. So the main nontrivial
result in this paper is that on the intersecting variety of these null
cones and the hyperplane equation, the distance again preserves this
important characteristics and can be positive, negative or null.

The fact that this distance can be positive (see (\ref{F2})) or negative
(see (\ref{F3})) for the partial case of equal semi-major axis,
eccentricities and eccentric anomaly angles

1. confirms the correctness of the interpretation of the formulae (\ref{F1})
as the square of the space-time interval, which can be either positive,
negative or zero.

2. raises up the important question whether this is a result only for the
partial case or also for the more general case of different semi-major axis,
eccentricities and eccentric anomaly angles. This is the case of equation (%
\ref{F6}) and (\ref{F15}), which is a fourth-degree algebraic equation with
respect to the variable $y=\sin ^{2}E_{1}$. The implementation of the Schur
theorem proves that this equation has roots within the circle $\mid y\mid <1$%
, which means that the space-time distance (\ref{ABC43}) can become zero,
meaning also that it can be also positive or negative.

Negative distances are not prohibited by geometry - these are the s.c.
hyperbolic geometries, known also as Lobachevsky geometries with negative
scalar curvature. So these three-dimensional hyperbolic geometries are
obtained as an intersection of four-dimensional null cones - this is an
interesting fact from  mathematical point of view, not
studied yet in the literature. In fact, because of the assumption for plane
orbital motion, the hyperbolic geometries will be two-dimensional ones.

It should be remembered also that the starting point for the calculations of the
space-time distance $R_{AB}$ (\ref{ABC43}) was the definition (\ref{ABC27})
of this function as the Euclidean distance. That is why, the Euclidean
distance can be affirmed to represent a partial case of a more general case,
related to the space-time distance. Thus, one can define the Euclidean
distance as a positive space-time distance, measured along the intersection
of a hyperplane equation and two null four-dimensional gravitational null cones,
attached to two moving observers (on the emitting - signal satellite and on the
receiving - signal satellite). Up to now the proof was given for the case of planar orbits.
It will be interesting to see whether such a physical interpretation will be valid also for the more
general case of space-distributed satellite orbits.

\section{IMPORTANT\ \ PHYSICAL\ \ CONSEQUENCES\ \ FROM\ \ THE \ CONDITION
\ FOR\ \ INTERSATELLITE\ \ COMMUNICATIONS}
\label{sec:Phys conseq}

\subsection{Satellites on one orbit and the restriction on the GPS - orbit eccentric
anomaly angle $E$}
\label{sec:orbit anomaly}

Let us begin with one important physical consequence from the condition for
intersatellite communications (\ref{ABC45}), which gives
\ restriction on the parameters of the orbit. More physical and numerical
consequences shall be given in the second (forthcoming) part of the paper.

Let us  calculate the found relation (\ref{ABC45}) for the case of
elliptical orbits with equal semi-major axis $a_{1}=a_{2}=a$, equal
eccentricities $e_{1}=e_{2}=e$ and equal eccentric anomalies $E_{1}=E_{2}=E$.
In order to understand properly the meaning of "equal eccentric anomalies", let
us remember the definition for the eccentric anomaly angle. Let us denote by $O$ the
center of the ellipse and from the position of the satellite on the elliptical
orbit, a perpendicular is drawn towards the large semi-major axis. If this
perpendicular at the point $M$ intersects the circle with a radius equal to
the semi-major axis, then the angle between the semi-major axis and the line
$OM$ is called the eccentric anomaly angle\textit{ }$E$\textit{. }Since the
 equal eccentricities and semi-major axis correspond to the case of
several satellites on one orbit, the notion "equal eccentric anomalies" means,
that for a fixed interval of time (counted from the moment of perigee
passage), the satellites encircle a distance along the orbit corresponding to
equal eccentric anomaly angles. However, when distances between satellites on
one orbit are calculated, the eccentric anomaly angles of the two satellites
should be different depending on their different, non-coinciding positions on
the orbit.

Let us assume again the value $e=0.01323881349526$ for the eccentricity of the
orbit for a GPS satellite, which is taken from the PhD thesis \cite{C4} of
Gulklett. From (\ref{ABC45}) it follows that the sine of the angle $E$ does
not depend on the semi-major axis
\begin{equation}
4a^{2}(1-e^{2})\sin^{2}E=2a^{2}-e^{2}a^{2}\text{ \ \ }\Longrightarrow\sin
E=\frac{1}{2}\sqrt[.]{\frac{\left(  2-e^{2}\right)  }{\left(  1-e^{2}\right)
}}\text{\ \ .\ } \label{ABC49A}%
\end{equation}
For the given eccentricity, the eccentric anomaly angle can be found to be
$E=45.002510943228$ \ \ $[\deg]$ or in radians $E=0.785441987624$ \ $[rad]$.

\subsection{The restriction on the ellipticity of the orbit}
\label{sec:restr ellip}

Let us compare this value with the one obtained as an iterative solution of
the Kepler equation (\ref{AA11}). For the purpose, the initial (zero)
approximation $E_{0}$ is taken from the dissertation \cite{C4} to be equal
to the mean anomaly $M$
\begin{equation}
E_{0}=M=-0.3134513508155\text{ \ \ }[rad]\text{ \ \ \ \ .}\label{BB26A}%
\end{equation}
However, since the mean anomaly $M$ is related to a projected uniform motion
along a circle, the more realistic angular characteristics is the eccentric
anomaly $E$, which can be found as an iterative solution of the transcendental
Kepler equation (\ref{AA11}). The iterative solution, described for example
in the monograph \cite{CA1}, is performed according to the formulae
\begin{equation}
E_{i+1}=M+e\sin E_{i}\text{ \ \ \ , \ \ }i=0,1,2,.....\text{ \ .}\label{BB22A}%
\end{equation}
Consequently, the first three iterative solutions are given according to the
following formulaes:
\begin{equation}
E_{1}=M+e\sin M\text{ \ ,}\label{BB23A}%
\end{equation}%
\begin{equation}
E_{2}=M+e\sin E_{1}=M+e\sin(M+e\sin M)\text{ \ \ ,}\label{BB24A}%
\end{equation}%
\begin{equation}
E_{3}=M+e\sin E_{2}=M+e\sin[M+e\sin(M+e\sin M)]\text{ \ .}\label{BB25A}%
\end{equation}
The third iteration gives the value
\begin{equation}
E_{3}^{.}=M+e\sin E_{2}=-0.31758547588467897473\text{ \ }[rad]\text{
\ ,}\label{BB50A}%
\end{equation}
The above value for $E$ is considerably lower than the calculated according to
(\ref{ABC49A}) value, which might only mean that this (initial) eccentric
anomaly angle is not very favourable for intersatellite
communications.

Since $\sin E\leq1$, one should have also
\begin{equation}
\sin E=\frac{1}{2}\sqrt[.]{\frac{\left(  2-e^{2}\right)  }{\left(
1-e^{2}\right)  }}\leq1\text{ \ \ ,} \label{ABC53A}%
\end{equation}
which is fulfilled for
\begin{equation}
e^{2}\leq\frac{2}{3}\text{ \ \ \ or \ }e\leq0.816496580927726\text{ \ \ .}
\label{ABC54A}%
\end{equation}
Surprisingly, highly eccentric orbits (i.e. with the ratio $e=\frac
{\sqrt[.]{a^{2}-b^{2}}}{a}$ tending to one, where $a$ and $b$ are the great
and small axis of the ellipse), are not favourable for intersatellite
communications. For GPS\ satellites, which have very small eccentricity orbits
(of the order $0.01$) and for communication satellites on circular orbits
($e=0$), intersatellite communications between moving satellites can be
practically achieved.

As mentioned in the Introduction, the RadioAstron space mission with a large
semi-major axis of $a\approx2\times10^{8}$ $m$ has a variable orbital
eccentricity ranging from $e=0.59$ to\textit{\ }the large value $e=0.966$,
which is higher than the value $0.816496580927726$. So for eccentricity in the
interval $0.59<e<0.816$, intersatellite communications of RadioAstron with
another satellites on the same orbit will be possible, but this will not be
possible for eccentricities in the interval $0.816<e<0.966$. Again, it should be reminded that by "intersatellite
communications" it is meant that the signal trajectory should take into account not only the space-time curvature, but also the
additional distance, which the signal has to travel so that it is intercepted by the second, moving satellite.

\section{POSSIBLE\ \ EXTENSIONS\ \ OF\ \ THE\ \ NEW\ \ RESULTS\ \ FOR\
THE\ \ SPACE-TIME\ \ INTERVAL\ \ TO\ \ THE\ CASE\ OF\ \ SPATIALLY-ORIENTED\ \ ORBITS}
\label{sec:Spat orient}
\subsection{GPS, GLONASS and Galileo satellite constellations and exchange
of signals between satellites on space-oriented orbits}
\label{sec:GPS GLONASS}

This section has the aim only to point out a number of research topics,
related to the problem about the generalization of the developed approach
for planar orbits to the case of space-oriented orbits. This
case is much more complicated and shall not be investigated in this paper.
Nevertheless, the purpose of the section will be to outline the basic
principles and equations to be used further for the construction of such a
more general theory. In particular, newly derived will be equations (\ref{BB4C11})
and (\ref{BB4C13}) in the subsequent sections, where the modified version of the
Kepler equation will be presented.

One of the main motivations for the idea for constructing an extension of
the theory for propagation of signals between moving satellites is the
requirement that the Global Navigation Satellite System ($GNSS$),
consisting of $30$ satellites and orbiting the Earth at a height of $23616$
km, should be interoperable with the other two navigational systems $GPS$
and $GLONASS$ \cite{C25}. This means that satellites on different orbital
planes should be able to exchange signals between each other. The
construction of such a theory is possible, and the main prerequisite for
this is the knowledge of the full set of six Keplerian elements $%
(M,a,e,\Omega ,I,\omega )$. For example, the satellites of the Galileo
constellation are situated on three orbital planes with nine-equally spaced
operational satellites in each plane. The Galileo satellites are in nearly
circular orbits with semi-major axis of $29600$\ km and a period of about $%
14 $\ hours \cite{C26} and an inclination of the orbital planes $56$
degrees. For comparison, the Russian Global Navigation Satellite System $%
GLONASS$, managed by the Russian Space Forces and launched in $1982$,
consists of $21$ satellites in three orbital planes (with three non-orbit
spares). Each satellite operates in nearly circular orbits with semi-major
axis of $25510$ km, and the satellites within the same orbital plane are
equally spaced by $45$ degrees. Each orbital plane has an inclination angle
of $64.8$ degrees, which is more than the inclination angle $56$ degrees of
the orbital planes of the Galileo satellites. Moreover, a $GLONASS$
satellite completes an orbit in approximately $11$ hours $16$ minutes - less
than the period of $14$ hours for the Galileo satellite. Consequently, the
three characteristic angles of rotation - the eccentric anomaly $E$, the
mean anomaly $M=n(\tau -t)$ and the true anomaly $f$ should be different for
the two satellites.

Different from $GLONASS$ orbital parameters have also the satellites of the $%
GPS$ satellite constellation, consisting of $24$ operational satellites,
deployed in six evenly spaced planes ($A$ to $F$) with $4$ satellites per
plane and an inclination of the orbit $55$ degrees \cite{C27}.

\subsection{The spatially oriented orbits and their orbital
characteristics}
\label{sec:Non planar}

The more interesting and complicated case is the one for space-oriented Keplerian orbits,
when the orbit is parametrized by the full set of six orbital elements $(M,a,e,\Omega ,I,\omega )$, where the first
three ones are characteristics of the planar motion and have been previously
defined. The next three Keplerian elements $(\Omega ,I,\omega )$ characterize
the spatial orientation of the orbit and are defined in the framework of the
geocentric equatorial coordinate system, in which the $x-$axis is aligned
with the vernal equinox ($\Upsilon $), the $z$ $\ \ $- axis points to the
north pole and the origin of the system is at the center of the Earth \cite%
{C29}. The vernal equinox describes the direction of the Sun as seen from
the Earth at the beginning \ of the spring season, which is equivalent to
considering the intersection of the equatorial plane with the Earth's
orbital plane. The direction of the $z-$axis clearly shows that the
precession and nutation of the Earth are neglected. Therefore, the polar
motion is not taken into account and a mean pole of the Earth rotation%
\textit{\ }is chosen, representing the average of all the changes in the
direction of the true rotational axis of the Earth \cite{C17D}. Such a
mean pole is called also Conventional International Origin (CIO).
  \\ In order to define the inclination of the orbit, first one should define the
line of nodes. This is the line of intersection of the orbital plane
with the equatorial plane (for Earth satellites and for celestial bodies in the
Solar system,  this reference plane will be the ecliptic) \cite{C17A}. The
line of intersection contains two ending points - the ascending node and
the descending node. Besides the inclination, the second measure to orient
the orbital plane is the right ascension of the ascending node\textit{\ }$%
\Omega $. It denotes the point where the satellite moves from the southern
hemisphere of the Earth to the northern hemisphere \cite{C17A}, \cite%
{C17B}. The angle $\Omega $\ of the longitude of the ascending node is the
angle between the ascending node and the $x-$axis (oriented towards the
vernal equinox). The argument of perigee (periapsis) $\omega $ is the angle
within the orbital plane from the ascending node to perigee in the direction
of the satellite motion $(0\leq \omega \leq 360^{0})$.
  \\ Very often, another variable is used - the argument of latitude $u=\omega +f$%
, being defined as the sum of the argument of perigee $\omega $ and the true anomaly $f
$ and geometrically representing the angle between the line of nodes and the
position vector $r$. The argument of latitude will appear further in the
calculation of the propagation time in terms of the celestial coordinates.\
A similar additive angular variable is the eccentric longitude $F=E+\omega
+\Omega $ (equinoctial orbital characteristic), representing the sum of the
eccentric anomaly $E$, the right ascension of the ascending node $\Omega $
and the argument of the perigee $\omega $.

\subsection{Intersecting null cones of observers on spatially oriented
orbits - the possible generalization of the approach}
\label{sec:Null space}

If a space-time interval is obtained as a result of the intersection of null cones
for the case of planar orbits, then it is reasonable to ask whether the
intersection of null cones at the space points of the signal-emitting and signal receiving satellites
on spatially oriented orbits (parametrized by $(M,a,e,\Omega ,I,\omega )$) will again produce a
space-time structure with positive, negative and null distance? This general
case will require the coordinate transformation from the orbital coordinates
$(M,a,e,\Omega ,I,\omega )$ to the cartesian coordinates $(x$, $y$, $z)$ of
the geocentric equatorial coordinate system (see the monograph of Brauer,
Clemence \cite{C28})
\begin{equation}
x=r\cos ^{2}\frac{I}{2}\cos (f+\Omega +\omega )+r\sin ^{2}\frac{I}{2}\cos
(f+\omega -\Omega )\text{ \ \ \ ,}  \label{EE5}
\end{equation}%
\begin{equation}
y=r\cos ^{2}\frac{I}{2}\sin (f+\Omega +\omega )-r\sin ^{2}\frac{I}{2}\sin
(f+\omega -\Omega )\text{ \ \ ,}  \label{EE6}
\end{equation}%
\begin{equation}
z=r\sin I\sin (f+\omega )\text{ \ \ ,}  \label{EE7}
\end{equation}%
where the true anomaly $f$ is expressed through the eccentric anomaly $E$
and the eccentricity parameter $e$ by means of the formulae
\begin{equation}
\tan \frac{f}{2}=\sqrt[.]{\frac{1+e}{1-e}}.\tan \frac{E}{2}\Longrightarrow
f=2\arctan \left[ \sqrt[.]{\frac{1+e}{1-e}}.\tan \frac{E}{2}\right] \text{ \
\ .}  \label{EE8}
\end{equation}

\bigskip In other words, if instead of the plane parametrization of the
orbit (\ref{C4AA1}) the parametrization (\ref{EE5})-(\ref{EE7}) is applied,
then an analogous formulae to (\ref{ABC43}) for the space-time distance can
be obtained. However, it can be expected that the corresponding equation
will not be an algebraic one, since the radius-vector $r$ is in the orbital
plane and is expressed by the true anomaly by means of the standard formulae
\begin{equation}
r=\frac{a(1-e^{2})}{1+e\cos f}\text{ \ \ .}  \label{EE8A}
\end{equation}
Consequently, the resemblance between the plane transformation (\ref{C4AA1})
in matrix notations
\begin{equation}
\left(
\begin{array}{c}
x \\
y%
\end{array}%
\right) =\left(
\begin{array}{c}
-ae \\
0%
\end{array}%
\right) +\left(
\begin{array}{cc}
a & 0 \\
0 & a\sqrt[.]{1-e^{2}}%
\end{array}%
\right) \left(
\begin{array}{c}
\cos E \\
\sin E%
\end{array}%
\right)  \label{EE8B}
\end{equation}%
and the non-planar transformations (\ref{EE5})-(\ref{EE7}) represented as%
\begin{equation}
\left(
\begin{array}{c}
x \\
y \\
z%
\end{array}%
\right) =R_{z}(-\Omega )R_{x}(-I)R_{z}(-\omega )\left(
\begin{array}{c}
r\cos f \\
r\sin f \\
0%
\end{array}%
\right) \text{ \ \ ,}  \label{EE9}
\end{equation}%
is only at first glance. The meaning of the above formulae is that
expressions (\ref{EE5}) - (\ref{EE7}) can be obtained after performing three
successive rotations $R_{z}(-\omega )$, $R_{x}(-I)$ and $R_{z}(-\Omega )$
with respect to the orbital vector $(r\cos f$, $r\cos f$, $0)^{T}$ (the
transponed vector to the vector-column in (\ref{EE9})), where $R_{z}(-\omega
) $ is the matrix of rotation at an angle $(-\omega )$ in the
counterclockwise direction around the $z$ axis, $R_{x}(-I)$ is the matrix of
rotation at an angle $(-I)$ around the $x-$axis, $R_{z}(-\Omega )$ is the
matrix of rotation at an angle $(-\Omega )$ around the $z-$axis \cite{C29}%
, \cite{C17D}. So the corresponding (non-algebraic and nonlinear) equation
for the space-time distance shall be derived after substituting the
transformations (\ref{EE9}) (with the corresponding indices $1$ and $2$) in
the null cone equations $ds_{(1)}^{2}=0$ and $ds_{(2)}^{2}=0$ for the
emitted and the received signal. The cone origins will be at the space
points $(x_{1},y_{1},z_{1})$ and $(x_{2},y_{2},z_{2})$.

It should be stressed that the transformations (\ref{EE5}) - (\ref{EE7}) for
the general case do not depend explicitly on the eccentricity $e$ because
the true anomaly $f$ \ depends on both the eccentricity $e$ and on the
eccentric anomaly angle $E$. This is the substantial difference from the
"planar orbit transformations" (\ref{EE8B}) used in this paper, which
depended directly on the eccentricity, and further the coefficient functions
of the fourth - order \ algebraic equations (\ref{C25}) for the space-time
distance and (\ref{E1}) for the geodesic distance exibited dependence on the
eccentricity. The algebraic proof that first equation (\ref{C25}) has the
property of the space-time distance was based substantially on the smallness
of the eccentricity.

However, if in the general case there will be no dependence of the
coefficient functions of the algebraic equation on the eccentricity
parameter $e$, then an interesting problem arises: \ will the algebraic proof
\ of the space-time distance property (of being positive, negative or zero)
be again possible? In other words, is the smallness of the eccentricity $e$
a very important ingredient of the mathematical proof? The smallness of $e$
can be taken into account after a series decomposition of (\ref{EE8}). In
this aspect, the following important problem arises: is it accidental that
the eccentricities of the orbits of the celestial bodies in the Solar
system are very small? For example, for Venus the eccentricity of the orbit
is $0.01$, for the Earth - $0.02$, for Mars - $0.09$, for Jupiter - $0.05$,
for Saturn - $0.06$, for Uranius - $0.05$, for Neptune - $0.01$. For
artificial body such as the RadioAstron SRT (Space Radio Telescope), the
range of the changing eccentricity of the orbit is between $0.59$ and $0.966$%
, but evidently such large eccentricities are not favoured by Nature. But this is
valid inside our Solar system, outside the Solar system some distant stars can have
relatively large eccentricities of their orbits of the order $0.4-0.6$.

\subsection{The true anomaly $f$ and the Runge-Lentz-Laplace vector}
\label{sec:Runge Lentz}

The dependence of the true anomaly $f$ on both the eccentricity and the
eccentric anomaly angle $E$ is a more peculiar feature of the orbital
characteristics. The true anomaly is defined as the geometric angle in the
plane of the ellipse between periapsis (the closest approach to the central
body) and the position of the orbiting satellite at any given time \cite%
{C17B}. There is also another more "mathematical" definition, which makes
use of the \ s.c. "Runge-Lentz" (or Laplace) vector $A_{L}$ \cite{C29},
which lies in the orbital plane and thus is orthogonal to the
angular-momentum vector $J=\mathbf{r}\times \overset{.}{\mathbf{r}}$ (called
also areal velocity $\overline{h}=\mathbf{r}\times \overset{.}{\mathbf{r}}=const=J$)
\begin{equation}
\overset{.}{\mathbf{A}_{L}:=\mathbf{r}}\times \left( \mathbf{r}\times
\overset{.}{\mathbf{r}}\right) -G_{\oplus }M_{\oplus }\frac{\mathbf{r}}{r}%
\overset{.}{=\mathbf{r}}\times \mathbf{J}-G_{\oplus }M_{\oplus }\frac{%
\mathbf{r}}{r}\text{ \ \ .}  \label{BB4A}
\end{equation}
Thus, the true anomaly $f$ is the angle between the Runge-Lentz vector $%
\mathbf{A}_{L}$ and the position vector $\mathbf{r}$. The vector $A_{L}$ appears
as an additive constant after integrating the equation
\begin{equation}
\mathbf{h}\times \overset{..}{\mathbf{r}}=-G_{\oplus }M_{\oplus }\frac{d}{dt}%
\left( \frac{\mathbf{r}}{r}\right) \text{ \ \ ,}  \label{BB4B}
\end{equation}
and $\Delta A=\frac{1}{2}\mid \mathbf{r}\times \overset{.}{\mathbf{r}}\Delta
t\mid =\frac{1}{2}\mid \overline{h}\mid \Delta t$ is the area, swept by radius-vector $%
r $ during the time $\Delta t$. The square of the Runge-Lentz-Laplace vector
can be calculated to be \cite{C29}
\begin{equation}
\mathbf{A}_{L}^{2}=G_{\oplus }M_{\oplus }^{2}+2J^{2}(\frac{1}{2}\overset{.}{r%
}^{2}-\frac{G_{\oplus }M_{\oplus }}{r})=G_{\oplus }M_{\oplus
}^{2}+2J^{2}\digamma \text{ \ ,}  \label{BB4BB1}
\end{equation}
where $\digamma $ is the conserved energy per unit mass. Thus, since the
magnitude and direction of the vector $A_{L}$ are conserved, the number of
the independent integrals of motion of the reduced two-body problem is
increased by one. It can be calculated that
\begin{equation*}
\overset{.}{r}^{2}=\frac{n^{2}a^{2}}{(1-e^{2})}[1+e^{2}(1-\frac{1}{2}\sin
(2\Omega )\sin (2\omega )(1-\cos i))+
\end{equation*}%
\begin{equation*}
+2e(\sin \omega \sin (\omega +f)+\cos \omega \cos ^{2}i\cos (\omega +f)+
\end{equation*}%
\begin{equation}
+\cos \omega \cos (\omega +f)\sin i)]\text{ \ .}  \label{BB4B2}
\end{equation}%
Consequently, after the decomposition of the pre-factor $\frac{n^{2}a^{2}}{%
(1-e^{2})}$ into an infinite sum of terms, depending on the small eccentricity parameter $e$ and
combining (\ref{BB4BB1}) and (\ref{BB4B2}), it can be seen that the conserved
energy of unit mass $\digamma $ and the orbital parameters $\omega $, $f$
, $i$ \ depend on the geocentric gravitational constant $G_{\oplus
}M_{\oplus }=3986205.266\times $ $10^{8}$ $\ [\frac{m^{3}}{\sec ^{2}}]$.
This numerical value however is not strictly determined - for example, the
value for $G_{\oplus }M_{\oplus }$ obtained from the analysis of laser
distance measurements of artificial Earth satellites is\textit{\ }
\begin{equation}
G_{\oplus }M_{\oplus }=\left( 3986004.405\pm 1\right) \times 10^{8}\ [\frac{%
m^{3}}{\sec ^{2}}]\text{ \ \ .}  \label{BB10}
\end{equation}%
In the review papers \cite{C17F}, \cite{C17E} by P. Mohr, B. Taylor and
D. B. Newell, where the value of $G$ (experimentally determined by means of
different experiments, performed by different groups) ranges from
\begin{equation}
G_{\oplus }M_{\oplus }=3986056.75236\times 10^{8}\ [\frac{m^{3}}{\sec ^{2}}]%
\text{ \ \ ,}  \label{BB11}
\end{equation}%
to
\begin{equation}
G_{\oplus }M_{\oplus }=3987999.07898\times 10^{8}\ [\frac{m^{3}}{\sec ^{2}}]%
\text{ \ \ .}  \label{BB12}
\end{equation}%

\subsection{Generalized Kepler equation for space-oriented orbits}
\label{sec:Gener Kepler}

The eccentric anomaly angle $E$ and the true anomaly $f$ are by definition
plane characteristics of the orbit, related to one another by means of
the differential relation
\begin{equation}
dE=\frac{\sqrt[.]{1-e^{2}}}{1+e\cos f}df\text{ \ \ \ or \ \ \ }ndt=\frac{%
\left( 1-e^{2}\right) ^{\frac{3}{2}}}{\left( 1+e\cos f\right) ^{2}}df\text{
\ \ \ \ \ .}  \label{BB4C1}
\end{equation}%
However, for space-oriented orbits the radius vector can be determined by
the formulae \cite{C29}, \cite{C29A1}
\begin{equation}
r=a(1-k\cos F-h\sin F)\text{ \ \ \ ,}  \label{BB4C}
\end{equation}
which is a generalization of the usual plane - orbit expression (\ref{C4AA1}%
) $x=a(\cos E-e)$ \ \ , \ $y=a\sqrt{1-e^{2}}.\sin E$. Correspondingly,
instead of the Kepler equation (\ref{AA11}) for the eccentric anomaly $E$,
the evolution of the eccentric longitude $F=E+\omega +\Omega $ for the case
of spatial orbit is governed by the modified Kepler equation \cite{C29}
\begin{equation}
F-k\sin F+h\cos F=l\text{ \ , }  \label{BB4C2}
\end{equation}%
where $l=M+\omega +\Omega $ is the mean longitude and $k$ and $h$ are the
trigonometric functions
\begin{equation}
k=e\cos (\omega +\Omega )\text{ \ \ , \ }h=e\sin (\omega +\Omega )\text{ \ \
\ .}  \label{BB4C3}
\end{equation}%
In fact, the functions $k$ and $h$ specify the orientation of the orbital
plane after a rotation at an angle $\omega +\Omega $. It is seen also that
for the case $\omega =0$ and $\Omega =0$ the modified Kepler equation (\ref%
{BB4C2}) transforms in the usual Kepler equation. The problem is: if the
above two orbital parameters are kept constant, then will the Kepler equation
also preserve its form for such a case? At first glance, it might seem that
this will happen. However, one should bear in mind that although the
eccentric anomaly $E$ and the true anomaly $f$ are plane characteristics,
their value is being accounted by means of the radius-vector. In the case of
space orbit with non-zero, but constant $\omega $ and $\Omega $, the radius
vector may not lie in the orbital plane so there will be an angle between
the orbital plane and the radius vector. So it might be expected that the
standard Kepler equation will be modified.

One more comparison may be performed between the plane-orbit and the
space-orbit cases. Previously, it was mentioned that the Kepler equation
establishes a correspondence between the eccentric anomaly $E$ and the
celestial time $t_{cel}$, i.e. $E\Longrightarrow t_{cel.}$. In fact, if the
celestial time $t_{cel}$ is known, then the iterative solution of the Kepler
equation establishes an approximate correspondence $t_{cel}\Longrightarrow E$%
. In the case of space orbits, due to the complicated integral
\begin{equation}
t_{cel}=\frac{1}{n}\int \frac{\left( 1-e^{2}\right) ^{\frac{3}{2}}}{\left(
1+e\cos f\right) ^{2}}df  \label{DOP20}
\end{equation}
resulting from the differential relation (\ref{BB4C1}) for the true anomaly $%
f$ and the celestial time $t_{cel}$, it might seem that a correspondence
between the true anomaly $f$ and the celestial time $f\Longrightarrow
t_{cel.}$ is not possible. In fact, this can be proved to be not true since
the above integral can be exactly calculated
\begin{equation*}
t_{cel}=\frac{\sqrt[.]{1-e^{2}}}{n}[-e\frac{\sin f}{(1+e\cos f)}+
\end{equation*}

\begin{equation}
+\frac{2}{\sin \delta }\arctan \left( \cot an\frac{\delta }{2}\tan \frac{f}{2%
}\right) ]\text{ \ \ ,}  \label{DOP21}
\end{equation}
where $\delta $ is the following numerical parameter
\begin{equation}
\delta =\arccos e\text{ \ \ \ .}  \label{DOP22}
\end{equation}%
However, an approximate correspondence $t_{cel}\Longrightarrow f$ in this
case cannot be established since $f$ cannot be expressed from (\ref{DOP21})
if $t_{cel}$ is known. It should be noted that in most monographs on celestial
mechanics only approximate solutions of the integral (\ref{DOP20}) are given. An
integral of the kind (\ref{DOP21}) will appear also in another problems, related to the
change of the proper time of an atomic clock, when transported along a given orbit. The analytical
techniques for finding the exact value of this integral will be presented in another paper.

\subsection{Modified Kepler equation only in terms of the eccentric anomaly angle}
\label{sec:Modif Kepler}

Now it shall be proved that such a modification of the Kepler equation will
really take place but most strangely, this modification will include terms
with the eccentric anomaly $E$ only.

For constant $\omega $ and $\Omega $, from (\ref{BB4C2}) it can be found
\begin{equation}
(-k\cos F-h\sin F)dE=Edt\text{ \ \ .}  \label{BB4C4}
\end{equation}%
But on the other hand, making use of the second formulae in (\ref{BB4C1}),
the last formulae can be rewritten as
\begin{equation}
\frac{dE}{df}=\frac{(1-e^{2})^{\frac{3}{2}}}{n(1+e\cos f)}.\frac{E}{(-k\cos
F-h\sin F)}\text{ \ \ .}  \label{BB4C5}
\end{equation}%
Since this determination of $E$ and $f$ should be compatible with the
plane-orbit relation (\ref{BB4C1}), both relations (\ref{BB4C1}) and (\ref%
{BB4C5}) should be fulfilled. This compatibility gives
\begin{equation}
1=\frac{(1-e^{2})^{.}}{n(1+e\cos f)}.\frac{E}{(-k\cos F-h\sin F)}\text{ \ \
\ .}  \label{BB4C6}
\end{equation}%
The second expression in the denominator can be written as
\begin{equation}
-k\cos (E+\omega +\Omega )-h\sin (E+\omega +\Omega )=  \label{BB4C7}
\end{equation}%
\begin{equation*}
=-e\cos (\omega +\Omega )\left[ \cos E\cos (\omega +\Omega )-\sin E\sin
(\omega +\Omega )\right] -
\end{equation*}%
\begin{equation*}
-e\sin (\omega +\Omega )[\sin E\cos (\omega +\Omega )+
\end{equation*}%
\begin{equation}
+\cos E\sin (\omega +\Omega )]=-e\cos E\text{ \ \ .}  \label{BB4C8}
\end{equation}
Substitution of this expression into (\ref{BB4C6}) gives a formulae for the
eccentric anomaly $E$ not dependent on the orbital parameters $\omega $ and $%
\Omega $
\begin{equation}
E=-\frac{ne(e+\cos f)}{1-e^{2}}\text{ \ \ \ .}  \label{BB4C9}
\end{equation}%
Note that this expression is not identical with the relation between $E$ and
$f$ for the case of planar orbits
\begin{equation}
\cos E=\frac{e+\cos f}{1+e\cos f}\text{ \ \ .}  \label{BB4C10}
\end{equation}%
Denoting $\widetilde{q}=\cos f$ and combining the last two expressions, the
following transcendental equation with respect to $\ \widetilde{q}$ can be
obtained%
\begin{equation}
\cos \left[ \frac{ne(e+\widetilde{q})}{1-e^{2}}\right] =\frac{e+\widetilde{q}%
}{1+e\widetilde{q}}\text{ \ \ \ .}  \label{BB4C11}
\end{equation}%
In terms of the eccentric anomaly $E$, this equation can be written also as
\begin{equation}
\cos X=\cos E\text{ \ \ , \ }X=\frac{ne\cos E}{1-e\cos E}\text{ \ \ ,}
\label{BB4C12}
\end{equation}%
or, taking into account that $X=E+2k\pi $, the following modified version of
the Kepler equation can be obtained
\begin{equation*}
E-eE\cos E=ne\cos E+
\end{equation*}%
\begin{equation}
+2k\pi -2k\pi e\cos E\text{ \ \ .}  \label{BB4C13}
\end{equation}%
This equation is second order in $E$, unlike the standard Kepler equation.

\subsection{Space orbits and the nontrivial problem for small
eccentricities}
\label{sec:small eccentr}

For the case of space orbits, the smallness of the eccentricity of the GPS
orbit creates an additional problem since, as mentioned in the monograph
\cite{C29}, for small $e$ and nearly circular orbits, the argument of
perigee $\omega $ is not a well-defined orbital element. The reason is that
small changes of the orbit may change the perigee location significantly.
For such a case, instead of the usual full set of Kepler parameters $%
a_{\alpha }=(a,e,I,M,\omega ,\Omega )$, another set of parameters $p_{\alpha
}=(a,l,h,k,\widetilde{p},\widetilde{q})$ had been implemented in the papers
by Broucke, Cefola \cite{C29A1} and Deprit, Rom \cite{C29A2}, where $l$ is
the mean longitude, $h$ and $k$ are given by (\ref{BB4C3}) and $\widetilde{p}
$ and $\widetilde{q}$ are the expressions
\begin{equation}
\widetilde{p}=\tan \frac{I}{2}\sin \Omega \text{ \ \ , \ }\widetilde{q}=\tan
\frac{I}{2}\cos \Omega \text{ \ \ .}  \label{BB4C14}
\end{equation}%
All these definitions enable the determination of the derivatives $\frac{%
\partial x}{\partial p}$ and $\ \frac{\partial x}{\partial q}$ in such a way,
so that they are consistent even when $e=0$.

\section{PHYSICAL\ \ AND\ \ MATHEMATICAL\ \ THEORY\ OF\ \ THE \ \ GEODESIC\ \ DISTANCE\ \ FOR\
 \ THE\ \ CASE\ OF\ NON-SPACE\ \ ORIENTED\ \ ORBITS}
\label{sec:geod dist}

\subsection{Geodesic distance as a result of the compatibility between the condition for intersatellite
communications and the space-time interval}
\label{sec:compat intersat}

We shall begin with a simple explanation, concerning how the formulae for the geodesic distance is obtained.
If (\ref{ABC45}) is substituted into expression (\ref{ABC43}) for $R_{AB}^{2}$
and the simple formulae $\cos(2E)=1-2\sin^{2}E$ is used, then expression
(\ref{ABC43}) can be written as
\[
\widetilde{R}_{AB}^{2}=\frac{1}{2}(a_{1}^{2}+a_{2}^{2})+\frac{1}{2}\left(
a_{2}e_{2}-a_{1}e_{1}\right)  ^{2}+\frac{1}{4}\left(  a_{1}^{2}e_{1}^{2}%
+a_{2}^{2}e_{2}^{2}\right)  -
\]%
\[
-\left(  2e_{1}a_{1}^{2}\cos E_{1}+2e_{2}a_{2}^{2}\cos E_{2}\right)  -
\]%
\[
-\left(  e_{1}^{2}a_{1}^{2}\sin^{2}E_{1}+e_{2}^{2}a_{2}^{2}\sin^{2}%
E_{2}\right)  -2a_{1}a_{2}\cos E_{1}\cos E_{2}+
\]%
\begin{equation}
+2a_{1}a_{2}\left(  e_{2}\cos E_{1}+e_{1}\cos E_{2}\right)  \text{
\ \ \ .}\label{ABC46}%
\end{equation}
The square $\widetilde{R}_{AB}^{2}$ of the Euclidean distance, when it is in
the form of the condition for intersatellite communications (transmission of
signals) is a two-point function, depending on the semi-major axis $a_{1}$,
$a_{2}$, eccentricities $e_{1}$, $e_{2}$ and eccentric anomaly angles $E_{1}$,
$E_{2}$ (which represent the variables in the investigated problem) of the two satellites. It is denoted with the tilde sign
$\widetilde{R}_{AB}^{2}$ in order to distinguish it from the usual expression
(\ref{ABC44}) for the Euclidean distance $R_{AB}^{2}=(x_{1}-x_{2})^{2}%
+(y_{1}-y_{2})^{2}$, expressed in the orbital elliptic coordinates. In the
paper \cite{AAB64}, when the propagation of signals from one satellite to
another is investigated, the distance travelled by light is called "geodesic
distance". So in analogy with this paper, we shall make a distinction between
the Euclidean distance $R_{AB}$ and the geodesic distance $\widetilde{R}%
_{AB}^{2}$. The difference between the two distances can be found by
substracting (\ref{ABC46}) from (\ref{ABC44})
\[
R_{AB}^{2}-\widetilde{R}_{AB}^{2}=\frac{1}{2}(a_{1}^{2}+a_{2}^{2})-e_{1}%
e_{2}a_{1}a_{2}+
\]%
\begin{equation}
+\frac{1}{4}(a_{1}^{2}e_{1}^{2}+a_{2}^{2}e_{2}^{2})-2a_{1}a_{2}\sqrt[.]%
{\left(  1-e_{1}^{2}\right)  \left(  1-e_{2}^{2}\right)  }\text{
\ .}\label{ABC47}%
\end{equation}

\subsection{Consistency of the calculations - geodesic and Euclidean distances
for equal eccentricities, semi-major axis and eccentric anomaly angles}
\label{sec:all equal1}

Note also another consistency of the calculations - from (\ref{ABC45}) for the
case $e_{1}=e_{2}=e$, $a_{1}=a_{2}=a$ and $E_{1}=E_{2}=E$ it can be derived
that $\sin^{2}E=\frac{2-e^{2}}{4(1-e^{2})}$ . If substituted into expressions
(\ref{ABC44}) for $R_{AB}^{2}$ and (\ref{ABC46}) for $\widetilde{R}_{AB}^{2}$,
it can be obtained that both these expressions are equal to zero. This should be
so - for coinciding positions of the satellites ($R_{AB}^{2}=0$), the geodesic
distance should also equal zero, i. e. $\widetilde{R}_{AB}^{2}=0$. For
equal eccentricities, semi-major axis and eccentric anomaly angles and without
taking into account the compatibility condition (\ref{ABC45}), the geodesic distance
(\ref{ABC43}) is different from zero. However, since the geodesic distance is derived by
using the compatibility condition, it should always be taken into account.

Now let us demonstrate that the boundary value
$E_{\lim }=45.002510943228\text{ }[\deg ]$ (\ref{F4}) is essential for the consistency between the
space-time interval, the geodesic distance and the Euclidean distance for this particular case, but is irrelevant for the
definition of the geodesic distance itself. For the purpose, let us write formulae (\ref{ABC46}) for the geodesic
distance again  for the case of equal eccentricities, equal semi-major axis
and equal eccentric anomaly angles
\begin{equation}
\widetilde{R}_{AB}^{2}=-a^{2}+\frac{1}{2}a^{2}e^{2}+2a^{2}(1-e^{2})\sin ^{2}E%
\text{ \ \ .}  \label{L1}
\end{equation}%
Then for
\begin{equation}
\sin ^{2}E\geq \sin ^{2}E_{\lim }=\frac{1}{4}\frac{(2-e^{2})}{(1-e^{2}}\text{
\ \ ,}  \label{L2}
\end{equation}%
substituting the above inequality in the expression (\ref{L1}), it can be
derived
\begin{equation}
\widetilde{R}_{AB}^{2}\geq -a^{2}+\frac{1}{2}a^{2}e^{2}+2a^{2}\frac{1}{4}%
(2-e^{2})=0\text{ \ \ .}  \label{L3}
\end{equation}
However, this cannot be considered as a proof of the positivity of the
geodesic distance, since from the condition for intersatellite
communications (\ref{ABC45}) it follows that the equality sign in (\ref{L2})
should be fulfilled, i.e. $\sin ^{2}E=\sin ^{2}E_{\lim }$. Consequently,
from (\ref{L3}) $\widetilde{R}_{AB}^{2}=0$. For the value  (\ref{L2}) of $E$%
, taken to be equal to $E_{\lim }$ and also identical with the value (\ref%
{ABC49A}), both the space-time interval and the geodesic distance are equal
to zero. Therefore, this  supports the consistency of  their  defining
formulaes (\ref{ABC43}) in Section \ref{sec:SpaceTime Int} and (\ref{ABC46}) in
Section \ref{sec:compat intersat}.

Now it can also be understood why the choice $E_{1}=E_{2}=E_{\lim }$ \ for
the inverse inequality to (\ref{F12}), when the space-time interval will be
negative in Section \ref{sec:Nonzero Euclid}, will not be acceptable. Let us
put
\begin{equation}
E_{1}=E_{2}=E=E_{\lim }=45.002510943228\ [\deg ]  \label{L4}
\end{equation}%
in the inverse inequality \
\begin{equation}
E_{1}+E_{2}<89.994978993712\text{ }[\deg ]\text{ . }  \label{L5}
\end{equation}%
Then it will follow
\begin{equation}
E<44.997489496\text{ }[\deg ]<45.002510943228\ [\deg ]\text{ \ .}  \label{L6}
\end{equation}%
However, the value (\ref{L4}) for the space-time interval has to be
compatible with the geodesic distance, which means that the geodesic
distance $\widetilde{R}_{AB}^{2}$ according to (\ref{L3}) should be equal to
zero. This is not possible for the inequality sign in (\ref{L6}),
consequently the choice $E=45.002510943228\ [\deg ]$ in (\ref{L4}) is
incompatible with the inequality (\ref{L5}).

\subsection{Compatibility condition and positive geodesic distance - the
case of different eccentric anomaly angles but equal eccentricities and
semi-major axis}
\label{sec:all equal2}

It is instructive to investigate the case for non-zero Euclidean distance
(given by formulae (\ref{ABC44})) and to compare it with the geodesic
distance (\ref{ABC46}). The Euclidean distance will be non-zero when the
eccentric anomaly angles $E_{1}$ and $E_{2}$ are different. Note however
that the difference $R_{AB}^{2}-$ $\widetilde{R}_{AB}^{2}$ in (\ref{ABC47})
does not depend on the eccentric anomaly angles. So for $e_{1}=e_{2}=e$ and
for $a_{1}=a_{2}=a$ one can represent (\ref{ABC47}) as
\begin{equation}
\widetilde{R}_{AB}=\sqrt{R_{AB}^{2}+a^{2}(1-\frac{3}{2}e^{2})}\text{ \ \ \ .}
\label{ABC55A1}
\end{equation}%
Taking into account the restriction (\ref{ABC54A}) $e^{2}\leq \frac{2}{3}$
on the value of the ellipticity of the orbit, the second term under the
square root in (\ref{ABC55A1}) is positive. Due to this
\begin{equation}
\widetilde{R}_{AB}\geq R_{AB}\text{ \ \ ,}  \label{ABC55A2}
\end{equation}%
which means that the geodesic distance, travelled by the signal is greater
than the Euclidean distance. This simple result, obtained by applying the
formalism of two intersecting null cones is a formal proof of the validity
of the Shapiro time delay formulae for the case of moving emitters and
receptors of the signals. Due to the larger geodesic distance, any signal in
the presence of a gravitational field will travel a greater distance and
thus will be additionally delayed.

From a formal point of view, an equality sign in (\ref{ABC55A2}) is possible
when $e^{2}=\frac{2}{3}$. This may take place if $\sin E=1$. This would mean
that $E=\frac{\pi }{2}+2k\pi $, which will contradict the initial assumption
about arbitrary values of the eccentric anomaly angle, not dependent on the ellipticity restriction (\ref{ABC54A}) $e^{2}\leq \frac{2}{3}$. Therefore, the geodesic distance should be considered strictly greater than the Euclidean one.

\subsection{Positivity of the geodesic distance in the general case of different plane orbital elements}
\label{sec:posit gener}

It is natural to expect that the geodesic distance will be greater than the
Euclidean one also in the general case of different eccentricities of the
two orbits, different semi-major axis and eccentric anomaly angles.

Let us first write the condition for intersatellite communications (\ref%
{ABC45}) as
\begin{equation}
\sin E_{1}\sin E_{2}=p\text{ \ \ \ ,}  \label{ABC55A3}
\end{equation}%
where $p$ is the introduced notation for
\begin{equation}
p=\frac{\overline{P}_{1}(e_{1},a_{1};e_{2},a_{2})}{\overline{Q}%
_{1}(e_{1},a_{1};e_{2},a_{2})}\text{ \ \ ,}  \label{ABC55A4}
\end{equation}
 $\overline{P}_{1}(e_{1},a_{1};e_{2},a_{2})$ and $\ \overline{Q}%
_{1}(e_{1},a_{1};e_{2},a_{2})$
for given values of the two eccentricities and the semi-major axis are the
numerical parameters
\begin{equation}
\overline{P}%
_{1}(e_{1},a_{1};e_{2},a_{2}):=a_{1}^{2}+a_{2}^{2}+(a_{2}e_{2}-a_{1}e_{1})^{2}-%
\frac{1}{2}(e_{1}^{2}a_{1}^{2}+e_{2}^{2}a_{2}^{2})\text{ ,}  \label{ABC55A5}
\end{equation}%
\begin{equation}
\overline{Q}_{1}(e_{1},a_{1};e_{2},a_{2}):=4a_{1}a_{2}\sqrt{%
(1-e_{1}^{2})(1-e_{2}^{2})}\text{ \ \ \ \ .}  \label{ABC55A6}
\end{equation}
Since
\begin{equation}
\sin E_{1}\sin E_{2}\leq 1\text{ \ ,}  \label{ABC55A7}
\end{equation}%
from the preceding relations it can be obtained
\begin{equation*}
-\frac{1}{2}(a_{1}^{2}+a_{2}^{2})+a_{1}a_{2}e_{1}e_{2}\geq
\end{equation*}%
\begin{equation*}
\geq -\frac{1}{4}(e_{1}^{2}a_{1}^{2}+e_{2}^{2}a_{2}^{2})+\frac{1}{2}%
(e_{1}^{2}a_{1}^{2}+e_{2}^{2}a_{2}^{2})-
\end{equation*}%
\begin{equation}
-2a_{1}a_{2}\sqrt{(1-e_{1}^{2})(1-e_{2}^{2})}\text{ \ \ \ .}  \label{ABC55A8}
\end{equation}%
Substituting the terms in the left-hand side of the above inequality in the
expression (\ref{ABC47}) for $\widetilde{R}_{AB}^{2}-R_{AB}^{2}$, it can be obtained
\begin{equation*}
\widetilde{R}_{AB}^{2}-R_{AB}^{2}\geq -\frac{1}{4}%
(e_{1}^{2}a_{1}^{2}+e_{2}^{2}a_{2}^{2})+\frac{1}{2}%
(e_{1}^{2}a_{1}^{2}+e_{2}^{2}a_{2}^{2})-
\end{equation*}%
\begin{equation*}
-2a_{1}a_{2}\sqrt{(1-e_{1}^{2})(1-e_{2}^{2})}-\frac{1}{4}%
(e_{1}^{2}a_{1}^{2}+e_{2}^{2}a_{2}^{2})+
\end{equation*}%
\begin{equation}
+2a_{1}a_{2}\sqrt{(1-e_{1}^{2})(1-e_{2}^{2})}\text{ \ \ \ .}  \label{ABC55A9}
\end{equation}%
All the terms in the right-hand side of the above inequality cancel, so one
obtains
\begin{equation}
\widetilde{R}_{AB}^{2}\geq R_{AB}^{2}\text{ \ \ .}  \label{ABC55A10}
\end{equation}%
Note the interesting fact that for different eccentricities and semi-major
axis the equality sign is fully legitimate. So the Euclidean distance
becomes equal to the geodesic distance when
\begin{equation}
\sin E_{1}=\sin E_{2}=1\text{ \ .}  \label{ABC55A11}
\end{equation}%
Since then%
\begin{equation}
E_{1}=E_{2}=\frac{\pi }{2}+2k\pi  \label{ABC55A12}
\end{equation}%
in a coordinate system, in which the small axis of the ellipses coincides
with the $y$ - axis, the two satellites should be situated one above
another (of course, remaining on the plane orbit). From the relation (\ref{ABC45}) or
(\ref{ABC47}), it can be found that this may happen, when the ratio $m=\frac{a_{1}}{a_{2}}$ of the two
semi-major axis satisfies the quadratic algebraic equation
\begin{equation*}
\frac{1}{4}(1+2e_{1}^{2})m+\frac{1}{4}(1+2e_{2}^{2})\frac{1}{m}-
\end{equation*}%
\begin{equation}
-e_{1}e_{2}=2\sqrt{(1-e_{1}^{2})(1-e_{2}^{2})}\text{ \ \ .}  \label{ABC55A13}
\end{equation}

\subsection{Geodesic distance in terms only of the first eccentric
anomaly angle and for equal eccentricities and equal semi-major axis}
\label{sec:geod anom}

Let us substitute $\sin E_{2}$ from the condition for intersatellite
communications $\sin E_{1}\sin E_{2}=p$ (\ref{ABC55A3}) in expression (\ref%
{F6}) for the square of the geodesic distance, which can be rewritten as
\begin{equation*}
\widetilde{R}_{AB}^{2}=2pa^{2}(1-e^{2})+e^{2}a^{2}(\cos ^{2}E_{1}-
\end{equation*}%
\begin{equation}
-\frac{p^{2}}{\sin ^{2}E_{1}})-2a^{2}\cos E_{1}\sqrt{1-\frac{p^{2}}{\sin
^{2}E_{1}}}\text{ \ \ \ .}  \label{ABC55A14}
\end{equation}%
Note that expression (\ref{F6}) was obtained after applying (\ref{ABC55A3})
to the expression (\ref{ABC44}) for the Euclidean distance. Consequently,
the above formulae is obtained after a subsequent application of (\ref%
{ABC55A3}) with respect to (\ref{F6}), assuming again the equality of the two eccentricities and
of the two semi-major axis. Since (\ref{ABC55A14}) depends only
on the first eccentric anomaly $E_{1}$, it is not fully equivalent to (\ref%
{ABC46}) but only in the sense that (\ref{ABC46}) depends on both eccentric
anomaly angles.

Remembering the numerical value for $p=\frac{2-e^{2}}{4(1-e^{2})}$, one can
write
\begin{equation}
2pa^{2}(1-e^{2})+e^{2}a^{2}=a^{2}+\frac{3}{2}e^{2}a^{2}\text{ \ \ .}
\label{ABC55A15}
\end{equation}%
By means of this expression and considering the geodesic distance $%
\widetilde{R}_{AB}^{2}$ to be positive, from (\ref{ABC55A14}) one can obtain
the inequality
\begin{equation*}
2\sqrt{\left( 1-\sin ^{2}E_{1}\right) \left( 1-\frac{p^{2}}{\sin ^{2}E_{1}}%
\right) }<
\end{equation*}%
\begin{equation}
<\left( 1+\frac{3}{2}e^{2}\right) -e^{2}\left( \sin ^{2}E_{1}+\frac{p^{2}}{%
\sin ^{2}E_{1}}\right) \text{ \ .}  \label{ABC55A16}
\end{equation}

\subsection{Restrictions on the lower bound of the
eccentric anomaly angle}
\label{sec:restr lower}

Since the left-hand side of the above inequality is positive, the right-hand
side should also be positive. It can be written as
\begin{equation}
\frac{\sin ^{4}E_{1}+p^{2}}{\sin ^{2}E_{1}}<\frac{3}{2}+\frac{1}{e^{2}}\text{
\ .}  \label{ABC55A17}
\end{equation}%
For small eccentricities $e\sim 0.01$, the number $\frac{1}{e^{2}}\sim 10000$
in the right-hand side is much greater than the numerical value in the
left-hand side. So the inequality is fulfilled, and since it is a
consequence of the previous one, based on the inequality (\ref{ABC55A2}) $%
\widetilde{R}_{AB}\geq R_{AB}$, it should be considered as another indirect
confirmation of the positiveness of the geodesic distance. In a rough
approximation, the above inequality will be fulfilled, if $\sin E_{1}>e$,
which for the typical eccentricity of the GPS orbit $e=0.01323881349526$
gives
\begin{equation*}
E_{1}>\arcsin e=0.013238426779\text{ }[rad]=
\end{equation*}%
\begin{equation}
=0.000231060882\text{ }[\deg ]\text{ \ \ .}  \label{ABC55A18}
\end{equation}%
So the eccentric anomaly angle should not be smaller than the above value,
which is an extremely small number. Note also that up to the sixth digit
after the decimal dot the simple relation $\sin E\approx E$ ($\arcsin
E\approx E$) is fulfilled. This is evident if the angle $E_{1}$ is expressed
in radians.

However, the lower bound on $E_{1}$ should be more stringent because from
the condition (\ref{ABC55A3}) and from the requirement to define properly
the second eccentric anomaly angle $E_{2}$, it can be obtained
\begin{equation}
\sin E_{2}=\frac{p}{\sin E_{1}}<1\text{ \ \ \ .}  \label{ABC55A18B1}
\end{equation}%
From here and for the numerical value $p=0.50004382422659548$ for the case
of equal eccentricities and semi-major axis it follows that
\begin{equation*}
E_{1}>\arcsin p=30.00289942985\text{ }[\deg ]=
\end{equation*}
\begin{equation}
=0.523649380196\text{ }[rad]\text{ \ \ \ .}  \label{ABC55A18B2}
\end{equation}%
It should be reminded why such a restriction appears on the eccentric anomaly angle, entering the condition
for intersatellite communications (\ref{ABC55A3}), which might seem to be in contradiction with the idea of
propagation of the signal between any space points. The problem in the case is that the mathematical formalism for the
propagation of the signal should take into account not only the curved trajectory of the signal due to the
delayed action of the gravitational field, but also the curved trajectory at that moment of time, when the signal will be intercepted by the
second, also moving satellite. That is why, because of the movement of the second satellite, such a lower bound on the value of the
eccentric anomaly angle is quite natural to exist. In fact, it confirms our physical intuition that due to the motion of the second satellite,
the length of the curved path of the signal should be greater in comparison with the case of a non-moving (second) satellite.
 Of course, the calculated numerical value for $p$ is just for one specific case, it might be in
principle another for other cases of different eccentricities and semi-major
axis, but in any case the eccentric anomaly angles should not be small. The value
(\ref{ABC55A18B2}) turns out to be higher than the lower bound (\ref{F33}) (from formulae (\ref{C101A1})
in Appendix C) for the space-time interval. But since the geodesic distance is positive, while the space-time
interval can be of any signs, it is natural to expect that the range of values for the eccentric anomaly angle, related to the
definition of the geodesic distance will be more restrictive in comparison with the range of values of this angle, related to the
space-time interval. This conclusion is confirmed by the numerical analysis.

\subsection{Fourth-order algebraic equation for the
geodesic distance without any roots}
\label{sec:fourthgeod roots}

Taking the square of inequality (\ref{ABC55A16}), it can be written as
\begin{equation*}
(1+\frac{9}{4}e^{4}+3e^{2}+2e^{4}p^{2}-4-4p^{2})+
\end{equation*}%
\begin{equation*}
+e^{4}(\sin ^{4}E_{1}+\frac{p^{4}}{\sin ^{4}E_{1}})-e^{2}(2+3e^{2})\sin
^{2}E_{1}-
\end{equation*}%
\begin{equation}
-\frac{e^{2}p^{2}}{\sin ^{2}E_{1}}(2+3e^{2})+4(\frac{p^{2}}{\sin ^{2}E_{1}}%
+\sin ^{2}E_{1})>0\text{ \ \ .}  \label{ABC55A19}
\end{equation}%
The most significant contributions will be given by the terms with the
smallest powers in $E$. Now one can realize the importance of the fact that
the angle $E_{1}$ should not be a small one - a term with $\sin E_{1}$ ($%
\sin ^{2}E_{1}$ or $\sin ^{4}E_{1}$) in the denominator will give a very
large value. But in view of the lower bound (\ref{ABC55A18B2})
in the previous section, the numerical value of the
eccentric anomaly angle cannot be a small number, since the geodesic distance is
ultimately related to the restrictions imposed by the condition for intersatellite
communications.

That is why, another choice is made in this paper. By denoting $y=\sin
^{2}E_{1}$, the above expression has been presented in the form of a
fourth-degree algebraic equation (\ref{E1}) \ $g(y)=\overline{a}_{0}y^{4}+%
\overline{a}_{1}y^{3}+\overline{a}_{2}y^{2}+\overline{a}_{3}y+\overline{a}%
_{4}$ with coefficient functions, some of which depend on the inverse powers
of the eccentricity parameter $e$. The equation (\ref{E1}) is investigated
in Appendix E again by applying the Schur theorem. Since the final polynomial
from the chain of polynomials will not satisfy the sufficient condition of
the theorem for small eccentricities, it will follow that the necessary
condition of the theorem for the roots to be within the unit circle $\mid
y\mid <1$ will not be fulfilled.  Thus, the polynomial will not have any
roots and so it cannot be equal to zero. This is consistent with the previous proof
about the positivity of the geodesic distance.

\section{DISCUSSION}
\label{sec:Discus}

In this paper we have presented a theoretical model of intersatellite
communications based on two gravitational null cones, the origin of each one
of them situated at the emitter and at the receiver of the corresponding
satellites. The satellites are assumed to move on (one-plane) elliptic
orbits with different eccentricities $e_{1}$, $e_{2}$ and semi-major axis $%
a_{1}$, $a_{2}$. The standard formulae (\ref{AA19}) for the Shapiro time
delay, used in VLBI radio interferometry \cite{CA7A2B1} (see also the
textbook \cite{CA7A3D} on GR theory) presumes that the coordinates $%
r_{A}=\mid x_{A}(t_{A})\mid$ and $r_{B}=\mid x_{B}(t_{B})\mid$ are varying
in such a way so that the Euclidean distance $R_{AB}=\mid
x_{A}(t_{A})-x_{B}(t_{B})\mid$ is not changing. In such a formalism, it is
natural to derive the signal propagation time from one gravitational null cone equation.
It should be noted also that the coordinates $r_{A}$ and $r_{B}$ enter
formulae (\ref{AA19}) in a symmetrical manner.

The situation changes, when two satellites are moving with respect to one
another and with respect to the origin of the chosen reference system. For
such a case, two different parametrizations of the space coordinates are
used, corresponding to the two different (uncorrelated) motions of the
satellites. This means that if in terms of the first parametrization the
null cone gravitational equation $ds^{2}=0$ is fulfilled, then it might not
be fulfilled with respect to the other parametrization. Consequently, this
is the motivation for making use of the two gravitational null cone
equations (\ref{ABC1}) and (\ref{ABC2}). The positions of the satellites are
characterized by the two eccentric anomaly angles $E_{1}$ and $E_{2}$ and by
the semi-major axis $a_{1}$, $a_{2}$ and eccentricities $e_{1}$, $e_{2}$,
and all these parameters satisfy the s.c.\textit{"}condition for
intersatellite communications\textit{" }(\ref{ABC45}), derived for the first
time in this paper. On the base of this expression the subsequent formulae (%
\ref{ABC46}) for the difference between the squares of the geodesic distance
$\widetilde{R}_{AB}^{2}$ and the Euclidean distance $R_{AB}^{2}$ is
obtained. Since light and radio signals travel along null geodesics, the
geodesic distance will be different from the Euclidean distance, and this is
exactly proved mathematically by formulae (\ref{ABC47}) for the case $%
a_{1}=a_{2}=a$ and $e_{1}=e_{2}=e$.
One of the main purposes of this paper is to build up a consistent physical and
mathematical theory of the two notions about the space-time interval and the geodesic distance,
which are closely related one to another. From a conceptual point of view,
the most significant contribution in this paper is the conclusion that the
intersection of two space-time intervals can be related to the macroscopic Euclidean distance.
There is nothing strange in this conception since General Relativity approaches, for example
the motion of a body in a spherically-symmetric gravitational field of a Schwarzschild
metric, turns out to be the key for understanding a celestial-mechanics effect - the
precession of the perihelion of an orbit (see the monograph by T.Padmanabhan
\cite{CA7A3D1}). Moreover, the perihelion shift of the orbit of Mercury could not
be found by the methods of celestial mechanics \cite{CA7A3D}. Likewise, the space-time
distance and the geodesic distance in this paper provide interesting and new information
about the values of some celestial mechanics parameters, which should be so that the
transmission of signals between moving satellites can be achieved.

Now let us summarize the physical consequences of the condition for
intersatellite communications (\ref{ABC45}), which constitutes one of the most
important results in this paper. The first important consequence is that
unlike formulae (\ref{AA19}), where the symmetry between the coordinates $%
r_{A}$ and $r_{B}$ results in the "reversibility" of the propagation time
difference, i.e. $T_{2}-T_{1}=-(T_{1}-T_{2})$, there is no time
reversibility for the case of moving gravitational null cones. This can be
seen from expression (\ref{ABC31}) $T_{2}=-\frac{1}{2}T_{1}+\epsilon \int dE%
\sqrt[.]{\left( \frac{\partial T_{1}}{\partial E}\right) ^{2}+\frac{1}{2}%
\left( \frac{\partial T_{1}}{\partial E}\right) -\frac{S}{P}}$ (for the case
$E_{1}=E_{2}=E$) where $T_{2}$ and $T_{1}$ are not symmetrical since $T_{2}$
depends not only on $T_{1}$, but also on $\frac{\partial T_{1}}{\partial E}$%
. In view of the motion of the satellites while transmitting the signals
this means that the "forward" and "backward" optical paths (but only when the satellites are moving) will be
different, i.e. $\widetilde{R}_{A_{1}B_{1}}\neq \widetilde{R}_{B_{1}A_{2}}$,
where the optical paths will be in fact the geodesic distances $\widetilde{R}%
_{A_{1}B_{1}}$ and $\widetilde{R}_{B_{1}A_{2}}$, given by formulae (\ref%
{ABC46}). The idea that the "forward" and "backward" optical paths for the case of moving satellites
should not be equal was proposed by Klioner in \cite{AAB33}. The difference
of the optical paths however has nothing to do with the "reversibility" of
the numeration of points - one may choose the space point $2$ as the
"initial" point from where the signal is being sent in the direction of the
second receiving-signal satellite, situated at the space-point $1$. Then
formulae (\ref{ABC31}) should be rewritten as (the indices $1$ and $2$ are
interchanged)
\begin{equation}
T_{1}=-\frac{1}{2}T_{2}+\epsilon \int dE\sqrt[.]{\left( \frac{\partial T_{2}%
}{\partial E}\right) ^{2}+\frac{1}{2}\left( \frac{\partial T_{2}}{\partial E}%
\right) -\frac{S}{P}}\text{ \ \ .}  \label{GGG1}
\end{equation}%
The equations (\ref{ABC31}) and (\ref{GGG1}) give a solution for the
propagation time $T_{2}$ in the form of a complicated integro-differential
equation.

The second consequence concerns the derived relation (\ref{ABC49A}) $\sin E=%
\frac{1}{2}\sqrt[.]{\frac{\left( 2-e^{2}\right) }{\left( 1-e^{2}\right) }}$,
which does not depend on the semi-major axis of the orbit and is derived
from the condition for intersatellite communications (\ref{ABC45}) for the
case of equal semi-major axis, eccentricities and eccentric anomaly angles
(the case of zero Euclidean distance). For the typical eccentricity of the
GPS-orbit $\ e=0.01323881349526$, the obtained value for the eccentric
anomaly angle from equality (\ref{ABC49A}) is $E=45.00251094$ $[\deg ]$,
which is surprisingly close to the value $f=45.541436900412$ $[\deg ]$ for
the true anomaly angle $f$ calculated according to formulae $\cos f=\frac{%
\cos E-e}{1-e\cos E}$ \cite{CA7A2}, \cite{CA1}. In fact, these numerical
values are the only ones for the case, when the satellites may move along one
and the same orbit situated equidistantly one from another (without
colliding with each other). The value $E=45.00251094$ $[\deg ]$
was proved to be the same as the limiting value
(\ref{F4}) $E_{\lim }=45.00251094$ $[\deg ]$, above which the
space-time interval (\ref{F1}) is positive and below which the space-time
interval is negative. So this fact suggests the interpretation that
the condition for intersatellite communication represents a boundary value
above which the space-time interval is positive or below which this interval
is negative.

In (\ref{ABC49A}) the condition for intersatellite communications was
written for a partial case but in (\ref{ABC45}) it was represented in the
general case. In deriving the formulae (\ref{ABC55A19}) (taken with the
positive sign) in Section VII G, the $\sin E_{2}$
function was expressed from the condition for intersatellite communications (%
\ref{ABC55A3}) and was substituted into the geodesic equation (\ref{ABC46}).
Consequently, the positivity of the geodesic distance was established with
respect to equation (\ref{ABC55A19}) and not with respect to the initial
equation (\ref{ABC46}). Nevertheless, the obtained result about the absence of
any roots of this equation is mathematically correct. Note also that in
deriving both the algebraic equations for the space-time distance and the
geodesic distance, for simplicity the case of equal semi-major axis and
eccentricities was considered. However, the algebraic structure of these
equations depends on the eccentric anomaly angles \ and not on the
eccentricities and semi-major axis because the important fact is that
different eccentric anomaly angles give different Euclidean distances.
Concretely for the algebraic treatment, the other important fact is the smallness
of the eccentricities of the GPS orbits - the proofs based on higher algebra
theorems are valid only for such a case.

Some facts from experimental point of view may be pointed out which
might be related to the obtained in this paper value $%
E=45.00251094 $ $[\deg ]$ for the eccentric anomaly angle. For example, in
the GLONASS constellation the satellites within one and the same plane are
equally spaced at $45$ degrees. Eight satellites can be situated in this
way. The eccentricity of the orbit for the GLONASS constellation is $e=0.02$
(close to the eccentricity of the GPS orbit), so the value for the eccentric
anomaly angle according to (\ref{ABC49A}) is obtained to be $E=45.00573$ $%
[\deg ]$. This is surprisingly close to the angle of equal spacing for the
GPS satellites within one plane. For the Galileo constellation, \ the satellites are $9$
per one plane, thus equally spaced at $40$ degrees. Of course, from a formal
point of view the coincidence between the angle of equal spacing with the
eccentric anomaly angle from the formalism of two null gravitational null
cones might seem to be accidental but yet the question remains: what is the
role of the angle $45$ $[\deg ]$ in the \ GPS (or GLONASS) intersatellite
communications?

Now some evidence shall be presented for the mutual consistency of the
obtained numerical results. When considering the case of non-zero Euclidean distance and the positive or
negative space-time distance, the numerical inequality (\ref{F12}) $%
E_{1}+E_{2}>89.994978993712$ $[\deg ]$ was derived for the lower bound of
the sum of the two eccentric anomaly angles. But since this case is more
general, the derived formulae (\ref{F12}) should be valid also for the
partial case of equal eccentric anomaly angles when the Euclidean and the
space-time distances are equal to zero - as a consequence formulae (\ref%
{ABC49A}) was derived and also the equivalent formulae\ (\ref{F4}) for the
limiting value. This means that inequality (\ref{F12}) should be fulfilled
for the partial case of equal eccentric anomaly angles, given by $E=E_{\lim
}=45.00251094$ $[\deg ]$ according to formulaes (\ref{ABC49A}) and (\ref{F4}%
). Indeed, twice the value of $E=E_{\lim }$ is greater than the number $%
89.994978993712$ $[\deg ]$ in the right-hand side of inequality (\ref{F12}).

Another evidence for a consistent result is the derived formulae (\ref{F33})
$15.64$ $[\deg ]<E_{2}<56.88$ $[\deg ]$ in Section D3
of Appendix C for the numerical interval for the angle $E_{2}$, when the
space-time interval can become zero. The last means that the four possible
roots of the space-time algebraic equation (\ref{C25}) in Section C of Appendix C
 are expected to be found in this numerical interval.
Since in all formulaes there is a symmetry with respect to the angles $E_{1}$
and $E_{2}$, the same inequality as (\ref{F33}) should be valid with respect
to $E_{1}$, i.e.
\begin{equation}
15.64 \ [\deg ]<E_{1}<56.88 \ [\deg ]\text{ \ \ .}  \label{GGG2}
\end{equation}%
If (\ref{F33}) and (\ref{GGG2}) are summed up, then one can obtain
\begin{equation}
31.28 \ [\deg ]<E_{1}+E_{2}<113.76 \ [\deg ]\text{ \ \ .}  \label{GGG3}
\end{equation}%
Since in deriving the formulae \ (\ref{GGG2}) in Appendix C one of the basic
assumptions was about the smallness of the eccentricity $e$, the lower bound
is in a "broader range" (i.e. considerably smaller) in comparison with the
lower bound $89.994978993712$ $[\deg ]$ from inequality (\ref{F12}), derived
under the assumption that trigonometric functions are less or equal to $1$,
i.e. $\cos E_{1}\leq 1$, $\cos E_{2}\leq 1$. So in fact (\ref{GGG3}) should
be written as
\begin{equation}
89.994978993712\text{ }[\deg ]<E_{1}+E_{2}<113.76\text{ }[\deg ]\text{ \ \ .}
\label{GGG4}
\end{equation}%
It is really amazing that twice the value of $E=E_{\lim }$ (the partial case
for equal eccentric anomaly angles) from (\ref{ABC49A}) and (\ref{F4}) \
remains within this interval! Moreover, the value $E=E_{\lim }$ is obtained
for a partial case, while the upper bound $113.76\ [\deg ]$ in (\ref{GGG4}) is
a result for the general case of different eccentric anomaly angles, derived
after the application of the Schur theorem.

Another interesting information is contained in the inequality (\ref{PPP6}),
related to the negativity of the space-time distance and derived on the base
of trigonometric inequalities. This very small interval
\begin{equation}
89.994978993712[\deg ]<E_{1}+E_{2}<90.015063019019\text{ }[deg]\text{ \ \ ,}
\label{GGGH4}
\end{equation}%
also falls within the interval (\ref{GGG4}), the upper boundary of which is predicted by the
Schur theorem, and the lower bound is a consequence both of the Schur theorem and the trigonometric
inequalities. It can be noted also that twice the limiting value (\ref{F4}) $E_{\lim
}=45.0025109432281$ $[\deg ]$ also is within this very tiny interval. Since the limiting value
(\ref{F4}) is that value of the eccentric anomaly angle, for which the space-time interval is zero, this is a
confirmation of the fact that one of the roots of the space-time algebraic equation (\ref{C25}) is within the
interval (\ref{GGG3}), determined by the Schur theorem.

There is one more curious and interesting fact. The lower bound $15.64\ [\deg
] $ in (\ref{GGG2}) is related to the numerical interval for the eccentric
anomaly angle, when the space-time distance can become zero. However, if the
restriction (\ref{ABC55A18B1}) $\sin E_{2}=\frac{p}{\sin E_{1}}<1$ from the
condition for intersatellite communications (\ref{ABC55A3}) is taken into
account, then the allowed lower bound of $E_{1}$ (or $E_{2}$) $E_{1}>\arcsin
p=30.002899$ $[\deg ]$ (\ref{ABC55A18B2}) turns out to be higher than the
lower bound in $15.64$ $ [\deg ]<E_{1}<56.88$ $ [\deg ]$ (\ref{GGG2}). \
Consequently, there is an interval
\begin{equation}
15.64 \ [\deg ]<E_{1}<30.002899 \ [\deg ]\text{ \ ,}  \label{GGG5}
\end{equation}%
where the space-time interval can exist (and can have zeroes, or can be negative or positive), but the
condition for intersatellite communications and the resulting from it
geodesic distance equation (\ref{ABC46}) cannot be defined. This confirms
the conclusion  that the space-time
distance is a more broader notion and has a more general meaning in
comparison with the Euclidean distance and the geodesic distance. This also
means that the notion of space-time distance can be defined independently
from the geodesic distance. In fact, this was the logical consequence of
derivation of these equations - first  the
space-time distance equation (\ref{ABC43}) was derived, then after setting
up equal (\ref{ABC43}) with the Euclidean distance (\ref{ABC44}), the
condition for intersatellite connections (\ref{ABC45}) was derived and
finally - formulae (\ref{ABC44}) for the geodesic distance. It should be clear
also that the space-time interval is defined  outside the interval (\ref{GGG5}) as well,
where the space-time algebraic equation does not have roots (zeroes in terms of the
chosen variable $y=\sin ^{2}E_{2}$), but can have either positive or negative values.
Whether the sign of the space-time algebraic polynomial outside the interval
(\ref{C101A}) will be positive or negative depends on the values of the
polynomial $f(y)$ at the endpoints $y=0$ and $y=1$. However, in order to
ensure the fulfillment of the substitution theorem for the availability of
an even or odd number of roots inside the given interval, at these endpoints
the values of the polynomial should be determined by the conditions (\ref{D1}%
) and (\ref{D3}), defined by the inequalities $f(y=0)<0$ and $f(y=1)<0$
respectively. These conditions, investigated in details in Appendix D again
under the realistic assumption about smallness of the eccentricity of the
orbit, guarantee that there will be an even number of roots (two or four)
of the investigated fourth-degree space-time algebraic polynomial. The
second inequality $f(y=1)<0$ is proved to be fulfilled if the inequality (%
\ref{D9}) $E_{2}<2\arctan (-2)=-126.869897645844$ $[\deg ]$ is valid, which
in terms of the variable $y=\sin ^{2}E_{2}$ and the fact that $\sin (...)$
in the third quadrant is a decreasing function, can be rewritten as
\begin{equation}
y>\sin ^{2}(-126.869897645844)=0.64=y_{0}\text{ \ \ .}  \label{KKK1}
\end{equation}%
Now it is interesting to compare this result from the application of the
substitution theorem with the final inequality (\ref{C101A}), derived after
the application of the Schur theorem in Appendix C. In terms of the chosen
variable $y$ the inequality assumes the form
\begin{equation}
\sin ^{2}(15.64)\leq y\leq \sin ^{2}(56.88)\text{ \ \ }  \label{KKK2}
\end{equation}%
or, taking into account the numerical values, it can be written as
\begin{equation}
0.07267993\leq y\leq 0.701453\text{ \ \  .}  \label{KKK3}
\end{equation}%
It is very interesting to note that the value $y_{0}=0.64$ from (\ref{KKK1})
falls within the interval (\ref{KKK3}), which is an evidence about the
consistency between the two theorems. On the base of this consistency, it
can be asserted that in the interval
\begin{equation}
0.64<y\leq 0.701453\text{ \ \  }  \label{KKK4}
\end{equation}%
there should be at least two (i.e. two or four) roots of the space-time
algebraic equation, and in the other interval $0.07267993\leq y<0.64$ -
either two roots, or no roots at all (in case if all the roots fall within
the interval (\ref{KKK4})).

  Note also that the numerical boundaries of the interval (\ref{GGG2}) are determined by the
inequality (\ref{C73}) after comparing the terms $\frac{2.16^{2}\sin
^{2}E_{2}}{e^{14}}$ in (\ref{C68}) and $\frac{16^{2}\sin E_{2}}{e^{14}}-%
\frac{16^{2}}{e^{14}}\sin E_{2}\cos E_{2}$ in (\ref{C71}), which represent
the highest inverse powers of the eccentricity $e$. Since the next inverse
powers of $e$ are proportional to $\sim \frac{1}{e^{12}}$, they will be $%
10000$ times smaller than the leading terms and therefore will not give any
substantial contributions to inequality (\ref{C73}), from where the interval (%
\ref{GGG2}) is obtained. Consequently, this numerical estimate (although
approximate) can be trusted.  \\
  \  The third consequence follows from relation (\ref{ABC53A}) and since $\sin
E\leq 1$, it places a restriction on the value of the eccentricity of the
orbit. Hence, it can be derived that $e\leq 0.816496580927726$ ($e^{2}\leq
\frac{2}{3}$). Since this value is too high and GPS-orbits have a very low
eccentricity, the above restriction is of no importance for the GPS-
intersatellite communications. It is of importance for the RadioAstron space
project, where the eccentricity of the orbit varies in a wide range. In this
paper it is established that intersatellite communications of RadioAstron
with another satellites on the same orbit will be possible in the range%
\textit{\ } $0.59<e<0.816$ , but this will not be possible for
eccentricities in the interval $0.816<e<0.966$. In other words, the
formalism of the two gravitational null cones will not be valid in this
range. In the Introduction in Section \ref{sec:multi-range curved} it was pointed out
(see the monograph \cite{Liu}) that the current and future technologies will allow the construction of
satellites, which can have varying eccentricities in their movement along the orbit.  \\
    However, the eccentricity restriction plays an important role for the
greatness of the geodesic distance $\widetilde{R}_{AB}$ in comparison with
the Euclidean distance $R_{AB}$ in formulae (\ref{ABC55A1}) $\widetilde{R}%
_{AB}=\sqrt{R_{AB}^{2}+a^{2}(1-\frac{3}{2}e^{2})}$. This again confirms the
fact about the mutual consistency between the different numerical values \
obtained in the framework of the formalism.

One more topic in this paper, which for the moment is not the central one and will be developed in details in
forthcoming publications is the possibility to apply the algebraic geometry approach of intersecting cones with respect to the atomic time.
In the framework of this approach, the analysis of the corresponding equations confirmed the relation of the atomic time to the propagation time of
the signal. From a physical point of view this should also be so - the propagation time of the signal depends on the changing distance
between the satellites, but the atomic time of each of the satellites also depends on the trajectory of each of the satellites (i.e.
on the transportation of the atomic clocks along each orbit). Consequently, each atomic time will correspond to the initial moment of emission of the signal (the initial propagation time) and the final moment of reception of the signal (the final propagation time) for the case of moving satellites. Unlike the case about the propagation times, the case about the atomic times is more complicated from a conceptual and technical point of view, because the defining formulae for the atomic time (see formulaes (\ref{DOP23}) and (\ref{DOP24})) contains the Geocentric Coordinate Time (TCG), which is not a directly measurable quantity and by definition only enters the corresponding equations. However, by using two different representations for the atomic time, depending respectively on the propagation time and on the celestial time, it has been proved that for the given metric of the near-Earth space (\ref{DOP25}), the TCG can be identified with the celestial time (determined from the Kepler equation in celestial mechanics). This important proof in fact (contained in the derived formulaes (\ref{BBF18})-(\ref{BBF20}) ) makes possible the analytical and numerical calculation of the change of atomic time under the transportation of the atomic clock. However, the calculation is performed for the given metric and also for just one atomic time, which is the case of one moving satellite, not two moving one with respect to another satellites. This is the first important case, considered in the paper.

The second case is related to the fact that there are four different times in the investigated problem about the exchange of signals between moving satellites: the atomic time, the TCG (Geocentric Coordinate Time), the propagation time and the celestial motion of the satellite. One important problem is which one of the four times can be used for the measurement of all the other three times? No doubt this is the atomic time, because it fulfills the basic requirement of being very small with respect to all the other times. This has been proved by means of calculating (formulae (\ref{KLM0})) the ratio of the differential atomic time to the differential propagation time, based on the experimental data, taken from \cite{HandbookGNSS}. From a theoretical point of view, the calculation has been performed in (\ref{KLM15B}) by using two different representations for the propagation time - the first representation depends on the atomic time and the second one - on the celestial time.

The third case is when the atomic time depends on the Geocentric Time (TCG) and the other representation - when the atomic time depends on the propagation time. In such a way, by assuming the invertible dependence of the propagation time on the atomic one, the ratio of the differential atomic time to the differential Geocentric Time has been obtained to be equal to the square of the $g_{00}$ component of the metric tensor (see equality (\ref{5B.14})). Naturally, the right-hand side is not modified by any velocity term (related to the motion of the satellites), while the other obtained relation (\ref{BBF21}) is modified by the velocity term, because is obtained from relation (\ref{KLM14}), involving the celestial time.
\\  All the three cases can be subjected to substantial further development in the following way: 1. The obtained expressions for the ratios of differential quantities, for example the ratio (\ref{KLM15B}) between the differential atomic and the differential propagation times can be calculated for the more general case of space-distributed orbits, making use of the transformations (\ref{EE5}) - (\ref{EE7}), containing the full set of $6$ Keplerian parameters. 2. One can assume that each time depends simultaneously on the other two or even three times. For example, the propagation time can be assumed to depend simultaneously on the atomic time and on the celestial time. 3. In the framework of the algebraic geometry approach of intersecting null cones, one can assume two different propagation times at two different space points with two different representations for each one of them. The formalism becomes more complicated because the second propagation time depends not only on the second atomic and celestial times, but also on the differential of the first propagation time, according to formulae (\ref{DOPC7}). The analogous dependence of the second atomic time on the first one should exist due to the similar mathematical structure of the equations for the atomic time, but this has not been worked out in this paper.
  \\ The dependence of the second propagation time on the first one will stimulate further research in the framework of the mentioned in the Introduction new concept about "multi-ranging" in a curved space-time, the essence of which is the transmission of a signal through a "chain" of satellites instead of sending directly to the distant satellite. In order to answer the question whether it is more favourable to use the concept of "multi-ranging", taking into account the action of the gravitational field, one has to optimize the total geodesic length or the total sum of the initial (emission) and final (reception) propagation times. This can be performed on the base of the found recurrent formulae (\ref{DOPCC8}), from which becomes evident that it is important to find the first propagation time of emission, which in view of the fact that it depends only on the space coordinates of the first satellite, is the propagation time for the signal emitted by the moving first satellite. For the case of plane motion, the analytical formulaes (\ref{C5A39}) and (\ref{C5A40}) have been found. These expressions should be substituted in the formulae (\ref{DOPC3}) for the second propagation time of reception from the second satellite.

\section{APPENDIX A:\ SOME COEFFICIENT\ FUNCTIONS IN EQUATIONS
(\ref{ABC19}) AND (\ref{ABC21})}
\label{sec:AppA}

The coefficient functions $Q_{1}(E_{1},E_{2})$, $Q_{2}(E_{1},E_{2})$,
$Q_{3}(E_{1},E_{2})$ and $Q_{4}(E_{1},E_{2})$ in equation (\ref{ABC19}) have
the following form
\begin{equation}
Q_{1}(E_{1},E_{2}):=\frac{P_{1}(E_{1})}{P_{2}(E_{2})}\frac{\partial
T_{1}(E_{1})}{\partial E_{1}}\text{ \ \ \ ,} \label{BBB1}%
\end{equation}%
\begin{equation}
Q_{2}(E_{1},E_{2}):=\left(  \frac{\partial T_{2}(E_{1},E_{2})}{\partial E_{2}%
}\right)  ^{2}+ \nonumber \\
\end{equation}%
\begin{equation}
\\+2\frac{P_{1}(E_{1})}{P_{2}(E_{2})}\frac{\partial T_{1}(E_{1}%
)}{\partial E_{1}}\frac{\partial T_{2}(E_{1},E_{2})}{\partial E_{2}}- \nonumber \\
\end{equation}%
\begin{equation}
\\-\frac{1}{P_{2}(E_{2})}\frac{\partial R_{AB}^{2}}{\partial E_{2}}-\frac
{S_{2}(E_{1},E_{2})}{P_{2}(E_{2})}\text{ \ \ \ ,} \label{BBB2}%
\end{equation}%
\[
Q_{3}(E_{1},E_{2}):=\frac{P_{1}(E_{1})}{P_{2}(E_{2})}\left(  \frac{\partial
T_{1}(E_{1})}{\partial E_{1}}\right)  ^{2}-
\]%
\begin{equation}
-\frac{1}{P_{2}(E_{2})}\frac{\partial R_{AB}^{2}}{\partial E_{1}}-\frac
{S_{1}(E_{1},E_{2})}{P_{2}(E_{2})}\text{ \ \ \ ,} \label{BBB3}%
\end{equation}%
\begin{equation}
Q_{4}(E_{1},E_{2}):=\left(  \frac{\partial T_{2}(E_{1},E_{2})}{\partial E_{1}%
}\right)  ^{2}\text{ \ \ \ .} \label{BBB4}%
\end{equation}%

The coefficient functions $G_{1}(E_{1}.E_{2})$ and $G_{2}(E_{1},E_{2})$ in
(\ref{ABC22}) have the following form
\[
G_{1}(E_{1}.E_{2}):=\frac{1}{2}\frac{S_{2}(E_{1},E_{2})}{P_{2}(E_{2})}+
\]
\begin{equation}
+\left(  \frac{P_{1}(E_{1})}{2P_{2}(E_{2})}\right)  ^{2}\left(  \frac{\partial
T_{1}(E_{1})}{\partial E_{1}}\right)  ^{2}-\frac{P_{1}(E_{1})}{2P_{2}(E_{2}%
)}\frac{\partial T_{1}(E_{1})}{\partial E_{1}}\text{ \ ,} \label{BBB5}%
\end{equation}%
\[
G_{2}(E_{1},E_{2}):=\frac{1}{2}\frac{S_{1}(E_{1},E_{2})}{P_{2}(E_{2})}+
\]%
\begin{equation}
+\frac{1}{P_{2}(E_{2})}\frac{\partial R_{AB}^{2}}{\partial E_{1}}-\frac
{P_{1}(E_{1})}{2P_{2}(E_{2})}\left(  \frac{\partial T_{1}(E_{1})}{\partial
E_{1}}\right)  ^{2}\text{ \ \ .} \label{BBB6}%
\end{equation}
The function $\overline{G}_{1}(E_{1}.E_{2})$ in (\ref{ABC24}) is obtained from
(\ref{BBB5}) by interchanging the indices $1$ and $2$
\begin{equation*}
\overline{G}_{1}(E_{1}.E_{2}):=\frac{1}{2}\frac{S_{1}(E_{1},E_{2})}{%
P_{1}(E_{1})}+
\end{equation*}%
\begin{equation*}
+\left( \frac{P_{2}(E_{2})}{2P_{1}(E_{1})}\right) ^{2}\left( \frac{\partial
T_{2}(E_{1},E_{2})}{\partial E_{2}}\right) ^{2}-
\end{equation*}%
\begin{equation}
-\frac{P_{2}(E_{2})}{2P_{1}(E_{1})}\frac{\partial T_{2}(E_{1},E_{2})}{%
\partial E_{2}}\text{ \ .}  \label{BBB7}
\end{equation}
The function $\overline{G}_{2}(E_{1}.E_{2})$ can be obtained in an analogous
way from (\ref{BBB6}).

The function $N$ in (\ref{ABC25}) can be written as follows
\begin{equation*}
N:=\frac{1}{4}\left( \frac{dE_{1}}{dE_{2}}\right) ^{2}\left( 1+4\left( \frac{%
\partial T_{1}}{\partial E_{2}}\right) \right) ^{2}+
\end{equation*}%
\begin{equation*}
+\frac{S_{1}(E_{1},E_{2})}{P_{2}(E_{2})}\left( \frac{dE_{1}}{dE_{2}}\right)
^{2}+\frac{S_{2}(E_{1},E_{2})}{P_{2}(E_{2})}+
\end{equation*}%
\begin{equation}
+\frac{1}{P_{2}(E_{2})}\frac{\partial R_{AB}^{2}}{\partial E_{2}}-\frac{%
2P_{1}(E_{1})}{P_{2}(E_{2})}\frac{dE_{1}}{dE_{2}}\left( \frac{\partial
T_{1}(E_{1})}{\partial E_{1}}\right) ^{2}\text{ \ \ .}  \label{BBB8}
\end{equation}

\qquad\

\section{APPENDIX B: THREE\ THEOREMS\ FROM\ HIGHER\ ALGEBRA}
\label{sec:AppB}

This appendix does not present new material, but contains the proofs of three
theorems from higher algebra, which shall be extensively used for proving
 the existence of roots (within the unit circle) for the space-time
algebraic equation and for the non-existence of such roots for the geodesic
algebraic equation. These theorems are: the substitution theorem, the Rouche
theorem, and the Schur theoreom. All the proofs are taken from the Obreshkoff
monograph \cite{AAB16AB}.
 \\ Let us begin first with the formulation and the proof of the "substitution" theorem:

\begin{theorem}
\ If for two numbers $a$ and $b$ the polynomial $f(y)$ of
arbitrary degree has equal signs, then $f(y)$ has an even number of roots
(zeroes) in the interval $(a,b)$. If the signs of $f(y)$ at the endpoints $a$
and $b$ are different, then the polynomial $f(y)$ has an odd number of roots.
\end{theorem}

\textbf{Proof}: Let the roots of the polynomial in the interval $(a,b)$ are
\begin{equation}
a\leq \alpha _{1}\leq \alpha _{2}\leq .....\leq \alpha _{m}<b\text{ \ \ \ .}
\label{B7}
\end{equation}%
Since the polynomial has also roots outside the interval $(a,b)$, it can be
decomposed as
\begin{equation}
f(y)=(y-\alpha _{1})(y-\alpha _{2})....(y-\alpha _{n})\varphi (y)\text{ \ \
\ ,}  \label{B8}
\end{equation}%
where $\varphi (y)$ contains binomial multipliers of the kind $y-\beta $ ($%
\beta $ is a number outside the interval $(a,b)$), and also quadratic
multipliers of the kind $(y-\mu )^{2}+\nu ^{2}$, responsible for the
imaginary roots. Since $\beta >b>a$, one can write also
\begin{equation}
\frac{a-\beta }{b-\beta }>0\text{ \ }\Longrightarrow \text{ \ }\frac{\varphi
(a)}{\varphi (b)}>0\text{ \ \ .}  \label{B9}
\end{equation}%
But for each root $\alpha _{k}$ ($1\leq k\leq m$) inside the interval $(a,b)$,
the sign of $\frac{a-\alpha _{k}}{b-\alpha _{k}}$ is negative. Consequently,
if the decomposition (\ref{B8}) is used, then the following equality can be
written
\begin{equation}
\frac{f(a)}{f(b)}=\left[ \frac{(a-\alpha _{1})(a-\alpha _{2})......(a-\alpha
_{m})}{(b-\alpha _{1})(b-\alpha _{2})......(b-\alpha _{m})}\right] .\frac{%
\varphi (a)}{\varphi (b)}\text{ \ .}  \label{B10}
\end{equation}%
Due to the positivity of $\frac{\varphi (a)}{\varphi (b)}$, the sign of $%
\frac{f(a)}{f(b)}$ will be determined by the $m$-multipliers in the square
bracket. Since each one is with a negative sign, the overall sign of $\frac{%
f(a)}{f(b)}$ will be given by $(-1)^{m}$. So for $m$ even, one has $\frac{f(a)%
}{f(b)}>0$ and for $m$ odd it can be obtained $\frac{f(a)}{f(b)}<0$. This
proves the theorem.

The second theorem, which shall be presented in this Appendix is the Rouche theorem. It shall not be used in the concrete
calculations, but is an important ingredient of the proof of the Schur theorem. The Rouche theorem has the following formulation \cite{AAB16AB}:

\begin{theorem}
\ Let $f(x)$ and $\varphi (x)$ are two polynomials and $C$
is a closed curve on which these polynomials are defined. If on $C$ the
following inequality is defined
\begin{equation}
\mid f(x)\mid >\mid \varphi (x)\mid \text{ \ ,}  \label{B11}
\end{equation}%
then the two equations
\begin{equation}
f(x)=0\text{ \ \ \ , \ \ }f(x)+\varphi (x)=0\text{\ }  \label{B12}
\end{equation}%
have an equal number of roots inside $C$.
\end{theorem}

Proof: \ The theorem is valid in principle for the case of complex roots $%
\alpha _{k}$ of the polynomial $f(x)$, when for each root $\alpha _{k}$
within the curve $C$ it can be written
\begin{equation}
x-\alpha _{p}=r_{p}(\cos \varphi _{p}+i\sin \varphi _{p})\text{ \ , \ \ }%
p=1,2,.....k\text{ \ ,}  \label{B13}
\end{equation}%
and for each root $\beta _{s}$ outside the curve $C$, it can also be written
\begin{equation}
x-\beta _{s}=r_{s}(\cos \Psi _{s}+i\sin \Psi _{s})\text{ \ , \ \ }%
s=1,2,.....m\text{ \ .}  \label{B14}
\end{equation}%
Then the function $f(x)$ can be represented as
\begin{equation}
f(x)=R(\cos \Phi +i\sin \Phi )\text{ \ \ ,}  \label{B15}
\end{equation}%
where $R$ and $\Phi $ are correspondingly the module and the argument of $f$%
. When encircling along the curve, \ the argument $\Phi $ will change in one
or another direction, but one full encircling along the curve $C$ will
correspond to a change of the argument \ by $2k\pi $. It should be noted
that the argument $\Psi _{s}$ for the outside roots can increase or decrease
within certain limits, but $\Psi _{s}$ cannot change by $2\pi $.

Further, the following function is constructed
\begin{equation}
F(x)=f(x)+\varphi (x)=f(x)\left[ 1+\frac{\varphi (x)}{f(x)}\right] \text{ \
\ .}  \label{B16}
\end{equation}%
Let the first equation $f(x)$ has $p$ roots inside $C$. Then upon one full
encircling along $C$, the argument of $f(x)$ will increase by $2p\pi $. Since
$\mid \varphi (x)\mid <\mid f(x)\mid $, the point $u=\frac{\varphi (x)}{f(x)}
$ \ will remain within a circle with a radius smaller than $1$. Thus the
point
\begin{equation}
1+u=1+\frac{\varphi (x)}{f(x)}  \label{B17}
\end{equation}%
will be within a circle, centered at the point equal to $1$. Therefore, the
argument of $f(x)$ will return to its initial value upon one full encircling
along $C$. Since the argument of $F(x)$ is a sum of the arguments of the
functions $f(x)$ and $(1+u)$, this argument will have to increase by $2p\pi $%
. This means that the function $F(x)$ has $p$ roots, which precludes the
proof.

By means of the Rouche theorem, let us prove the Schur theorem, which further
will be the basic mathematical tool for investigation of the space-time algebraic
equation and the geodesic algebraic equation.

\begin{theorem}
\cite{AAB16AB} \ The necessary and sufficient conditions for the equation
\begin{equation}
f(y)=a_{0}y^{n}+a_{1}y^{n-1}+......+a_{n}=0\text{ \ , \ \ }a_{0}\neq 0
\label{B18}
\end{equation}%
to have roots within the circle $\mid y\mid <1$ are:

1. The inequality
\begin{equation}
\mid a_{0}\mid >\mid a_{n}\mid  \label{B19}
\end{equation}
should be fulfilled.

2. The polynomial of $(n-1)$-degree
\begin{equation}
f_{1}(y)=\frac{1}{y}\left[ \overline{a}_{0}f(y)-\overline{a}_{n}f^{\ast }(y)%
\right]  \label{B20}
\end{equation}%
should have roots only within the circle $\mid y\mid <1$. In (\ref{B20}) $%
\overline{a}_{0}$, $\overline{a}_{n}$ are the complex conjugated quantities
of the coefficients $a_{0}$, $a_{n}$ and $f^{\ast }(y)$ is the s.c.
"inverse" polynomial, \ defined as
\begin{equation}
f^{\ast }(y)=y^{n}\overline{f}(\frac{1}{y})=\overline{a}_{n}y^{n}+\overline{a%
}_{n-1}y^{n-1}+......+\overline{a}_{0}\text{ \ \ ,}  \label{B21}
\end{equation}%
where $\overline{a}_{n}$, $\overline{a}_{n-1}$,$......\overline{a}_{0}$ are
the complex conjugated coefficients (complex coefficient functions), related to the
coefficients (complex coefficient functions) $a_{n}$, $a_{n-1}$, $......a_{0}$. For
the present case, all the coefficient functions will be real, so there will
be no complex conjugated quantities and no "barred" coefficients, i.e.
\begin{equation}
f^{\ast }(y)=y^{n}f(\frac{1}{y})=a_{n}y^{n}+a_{n-1}y^{n-1}+......+a_{0}\text{
\ \ .}  \label{B22}
\end{equation}
\end{theorem}

Proof: Let us prove first the necessary condition, assuming that the roots of
the polynomial $f(y)=0$ are in the unit circle $\mid y\mid <1$. Since
according to the Wiet formulae the multiplication of all the roots $%
y_{1},y_{2},.....y_{n-1},y_{n}$ gives
\begin{equation}
(-1)^{n}\frac{a_{n}}{a_{0}}=y_{1}y_{2}......y_{n}\text{ \ \ }  \label{B23}
\end{equation}%
and also $\mid y\mid <1$, it follows that
\begin{equation}
\mid \frac{a_{n}}{a_{0}}\mid <1\text{ \ }\Longrightarrow \mid a_{0}\mid
>\mid a_{n}\mid \text{ \ .}  \label{B24}
\end{equation}%
Consequently, for $\mid y\mid =1$ one can write also
\begin{equation}
\mid a_{0}f(y)\mid >\mid a_{n}f^{\ast }(y)\mid \text{ \ .}  \label{B25}
\end{equation}%
Then from the Rouche theorem it follows that the polynomial $f(y)$ and the
polynomial of $(n-1)$-degree
\begin{equation}
yf_{1}(y)=a_{0}f(y)-a_{n}f^{\ast }(y)  \label{B26}
\end{equation}%
have $n$ roots within the circle $\mid y\mid <1$.

\section{APPENDIX C: THE\ SCHUR\ THEOREM\ AND\ THE\ PROOF\ THAT\ THE\ FOURTH-ORDER\ ALGEBRAIC\ EQUATION\ FOR\ THE\ SPACE-TIME\ INTERVAL\ HAS\ ROOTS\ WITHIN\ THE\ UNIT\ CIRCLE}
\label{sec:AppC}

\subsection{The general strategy for constructing a "chain" of lower-degree
polynomials}
\label{sec:strategy chain }

Since the Schur theorem is based on the construction of \ the $(n-1)$ degree
polynomial $f_{1}(y)$ with roots inside the circle $\mid y\mid <1$, it is
important to check whether the condition (\ref{B19}) $\mid a_{0}\mid >\mid
a_{4}\mid $(for the case of the four-dimensional polynomial (\ref{F16})) is
fulfilled. Then the condition the polynomial \ $f_{1}(y)$ to have roots
within the unit circle will be equivalent to the condition for the original
polynomial $f(y)$\ to have roots within the same circle.

Following the above mentioned algorithm and also formulae (\ref{B20}), another
polynomial of $(n-2)$ degree may be constructed. It will have roots inside
the circle \ $\mid y\mid <1$ only if condition (\ref{B19}) is valid with
respect to the coefficients $b_{0},b_{1}$, $b_{2}$.....$b_{n-1}$ of the
polynomial of $(n-1)$ degree, i.e. the inequality $\mid b_{0}\mid >\mid
b_{n-1}\mid $\ should be satisfied (for the $4-$dimensional case it will be $%
\mid b_{0}\mid >\mid b_{3}\mid $). Further, if the coefficients of the $%
(n-2) $ degree polynomial are $c_{0}$, $c_{1},....c_{n-2}$ and $\mid
c_{0}\mid >\mid c_{n-2}\mid $ is fulfilled, then another polynomial of $(n-3)$
degree can be constructed with roots within the circle $\mid y\mid <1$. In
such a way, a chain of polynomials of diminishing degrees can be constructed,
the final polynomial being a first order (linear) equation. It can easily be found
when its root by module is smaller than $1$. But then, since the Schur theorem contains
a necessary and sufficient condition, all the preceding polynomials of
second, third,....$(n-1)$, $n$ -th degree will have also roots within the
unit circle, provided that the chain of coefficient inequalities
\begin{equation}
\mid a_{0}\mid >\mid a_{n}\mid \text{ , }\mid b_{0}\mid >\mid b_{n-1}\mid
\text{, \ }\mid c_{0}\mid >\mid c_{n-2}\mid ........\text{\ \ }  \label{C1}
\end{equation}%
is fulfilled. In the following subsections, the algorithm will be developed
in details for the fourth-degree algebraic equation for the space-time
interval.

It should be kept in mind that the above coefficient inequalities might not
be fulfilled. For example, instead of $\mid a_{0}\mid >\mid a_{n}\mid $, the
inverse inequality $\mid a_{0}\mid <\mid a_{n}\mid $ might be fulfilled. For
such a case, instead of the formulae (\ref{B26}), the following formulae for
obtaining the $(n-1)$- degree polynomial should be used
\begin{equation}
a_{n}f(y)-a_{0}f^{\ast }(y)=f_{1}(y)\text{ \ .}  \label{C2}
\end{equation}%
Analogously, if for example another inverse inequality \ $\mid b_{0}\mid
<\mid b_{n-1}\mid $ is fulfilled, then the next $(n-2)$ degree polynomial $%
f_{2}(y)$ will be given by
\begin{equation}
b_{n-1}f_{1}(y)-b_{0}f_{1}^{\ast }(y)=f_{2}(y)\text{ \ .}  \label{C2B}
\end{equation}%
So some of the inequalities (\ref{C1}) might be fulfilled, but the
remaining  might be the inverse ones. Then both the formulaes of the type
(\ref{C2}) and (\ref{C2B}) should be applied.

\subsection{Calculation of the coefficient functions for the chain of
polynomials of diminishing degrees according to the Schur theorem}
\label{sec:coeff chain }

In order to derive the first-degree polynomial and to impose the requirement
$\mid y\mid <1$, one has to calculate the coefficient functions of all the
polynomials of $n$-th, $(n-1)$, $(n-2)$,$......2$ degree.

Let us first calculate the coefficient functions for the case, when
equalities (\ref{C1}) hold. For the case of the inequalities (\ref{C1}), after
applying formulae (\ref{B26}), one can obtain for the general case of the $%
n- $th degree polynomial
\begin{equation}
f(y)=a_{0}y^{n}+a_{1}y^{n-1}+...+a_{n-2}y^{2}+a_{n-1}y+a_{n}  \label{C3}
\end{equation}%
the following $(n-1)$ degree polynomial
\begin{equation*}
f_{1}(y)=\frac{1}{y}[a_{0}(a_{0}y^{n}+a_{1}y^{n-1}+...
\end{equation*}%
\begin{equation*}
+a_{n-2}y^{2}+a_{n-1}y+a_{n})-
\end{equation*}%
\begin{equation}
-a_{n}\left( a_{n}y^{n}+a_{n-1}y^{n-1}+......+a_{1}y+a_{0}\right) ]=
\label{C4}
\end{equation}%
\begin{equation}
=b_{0}y^{n-1}+b_{1}y^{n-2}+......+b_{n-2}y+b_{n-1}\text{ \ \ ,}  \label{C5}
\end{equation}
where the coefficients $b_{0}$, $b_{1}$,$......b_{n-2}$, $b_{n-1}$ are
obtained to be
\begin{equation}
b_{0}=a_{0}^{2}-a_{n}^{2}\text{ \ , \ }b_{1}=a_{0}a_{1}-a_{n-1}a_{n}\text{ ,
\ }b_{2}=a_{0}a_{2}-a_{n-2}a_{n}\text{ \ ,}  \label{C6}
\end{equation}%
\begin{equation}
b_{k-1}=a_{0}a_{k-1}-a_{n-k+1}a_{n}\text{ \ \ , \ \ }%
b_{n-1}=a_{0}a_{n-1}-a_{1}a_{n}\text{ \ \ .}  \label{C7}
\end{equation}%
In the same way, the $(n-2)$- degree polynomial
\begin{equation}
f_{2}(y)=\frac{1}{y}\left[ b_{0}f_{1}(y)-b_{n-1}f_{1}^{\ast }(y)\right] =
\label{C8}
\end{equation}%
\begin{equation*}
=c_{0}y^{n-2}+c_{1}y^{n-3}+c_{2}y^{n-4}+...
\end{equation*}%
\begin{equation}
+.....+c_{n-4}y^{2}+c_{n-3}y+c_{n-2}  \label{C9}
\end{equation}
has the coefficient functions
\begin{equation}
c_{0}=b_{0}^{2}-b_{n-2}^{2}=\left( a_{0}^{2}-a_{n}^{2}\right) ^{2}-\left(
a_{0}a_{n-1}-a_{1}a_{n}\right) ^{2}\text{ \ ,}  \label{C10}
\end{equation}%
\begin{equation}
c_{1}=b_{0}b_{1}-b_{n-2}b_{n-1}\text{ , \ }c_{2}=b_{0}b_{2}-b_{n-3}b_{n-1}%
\text{ \ \ ,}  \label{C11}
\end{equation}%
\begin{equation}
c_{n-3}=b_{0}b_{n-3}-b_{2}b_{n-1}\text{ \ , \ \ \ }%
c_{n-2}=b_{0}b_{n-2}-b_{1}b_{n-1}\text{ \ ,}  \label{C12}
\end{equation}%
\begin{equation}
c_{k-2}=b_{0}b_{k-2}-b_{n-k+1}b_{n-1}\text{ \ .}  \label{C13}
\end{equation}%
According to the Schur theorem, the polynomial $f_{2}(y)$ of $(n-2)$ degree has
roots within the circle $\mid y\mid <1$ if and only if the inequality $\mid
b_{0}\mid >\mid b_{n-1}\mid $ for the coefficient functions of the
polynomial $f_{1}(y)$ (\ref{C5}) is fulfilled. Taking into account (\ref%
{C6}) for $b_{0}$ and (\ref{C7}) for $b_{n-1}$, one can rewrite this
inequality as
\begin{equation}
\mid a_{0}^{2}-a_{n}^{2}\mid >\mid a_{0}a_{n-1}-a_{1}a_{n}\mid \text{ \ .}
\label{C14}
\end{equation}%
In the next subsection it will be shown that the non-fulfillment of the
initial inequality $\mid a_{0}\mid >\mid a_{n}\mid $ will lead to the
non-fulfillment of the above equality (\ref{C14}). Yet, a general rule cannot be
formulated.

Next, from the polynomial $f_{2}(y)$ (\ref{C9}) of $(n-2)$ degree one can
construct the $(n-3)$ degree polynomial
\begin{equation}
f_{3}(y)=\frac{1}{y}\left[ c_{0}f_{2}(y)-c_{n-2}f_{2}^{\ast }(y)\right]
\text{ \ \ .}  \label{C15}
\end{equation}%
For the initial polynomial (\ref{F16}) of fourth degree, $f_{2}(y)$ will be
the second-degree polynomial
\begin{equation}
f_{2}(y)=c_{0}y^{2}+c_{1}y+c_{2}\text{ \ \ , \ }f_{2}^{\ast
}(y)=c_{2}y^{2}+c_{1}y+c_{0}\text{ \ }  \label{C16}
\end{equation}%
and the polynomial $f_{3}(y)$ will be the linear polynomial
\begin{equation}
f_{3}(y)=d_{0}y+d_{1}=(c_{0}^{2}-c_{2}^{2})y+(c_{0}c_{1}-c_{1}c_{2})\text{ \
\ .}  \label{C17}
\end{equation}%
Taking into account expressions (\ref{C6}), (\ref{C7}) for $b_{0}$, ...$%
b_{n-1}$ and (\ref{C10}) - (\ref{C12}) for $c_{0}$, ...$c_{n-2}$, one can
represent the coefficient functions $c_{0}$, $c_{1}$, $c_{2}$ as
\begin{equation}
c_{0}=b_{0}^{2}-b_{3}^{2}=(a_{0}^{2}-a_{4}^{2})^{2}-(a_{0}a_{3}-a_{1}a_{4})^{2}%
\text{ ,}  \label{C18}
\end{equation}%
\begin{equation*}
c_{1}=b_{0}b_{1}-b_{2}b_{3}=(a_{0}^{2}-a_{4}^{2})(a_{0}a_{1}-a_{3}a_{4})-
\end{equation*}%
\begin{equation}
-(a_{0}a_{2}-a_{2}a_{4})(a_{0}a_{3}-a_{1}a_{4})\text{ \ ,}  \label{C19}
\end{equation}%
\begin{equation*}
c_{2}=b_{0}b_{2}-b_{1}b_{3}=(a_{0}^{2}-a_{4}^{2})(a_{0}a_{2}-a_{2}a_{4})-
\end{equation*}%
\begin{equation}
-(a_{0}a_{1}-a_{3}a_{4})(a_{0}a_{3}-a_{1}a_{4})\text{ \ \ .}  \label{C20}
\end{equation}%
The following expressions for the $b_{0}$, $b_{1}$, $b_{2}$, $b_{3}$
coefficient functions have been used
\begin{equation}
b_{0}=a_{0}^{2}-a_{4}^{2}\text{ \ , \ }b_{1}=a_{0}a_{1}-a_{3}a_{4}\text{ \ \
,}  \label{C21}
\end{equation}%
\begin{equation}
b_{2}=a_{0}a_{2}-a_{2}a_{4}\text{ \ , \ \ \ }b_{3}=a_{0}a_{3}-a_{1}a_{4}%
\text{ \ \ .}  \label{C22}
\end{equation}%
From the linear equation (\ref{C17}) \ one can find when the root is modulo
less than $1$
\begin{equation}
\mid y\mid =\mid -\frac{d_{1}}{d_{0}}\mid =\mid -\frac{%
(c_{0}c_{1}-c_{1}c_{2})}{(c_{0}^{2}-c_{2}^{2})}\mid =\mid -\frac{c_{1}}{%
c_{0}+c_{2}}\mid <1\text{ \ \ .}  \label{C23}
\end{equation}%
Taking into account the preceding expressions (\ref{C18}) - (\ref{C22}),
the inequality can be rewritten as
\begin{equation*}
\mid (a_{0}a_{2}-a_{2}a_{4})(a_{0}a_{3}-a_{1}a_{4})-
\end{equation*}%
\begin{equation*}
-(a_{0}^{2}-a_{4}^{2})(a_{0}a_{1}-a_{3}a_{4})\mid <\mid
(a_{0}^{2}-a_{4}^{2})+
\end{equation*}%
\begin{equation}
+(a_{0}-a_{4})\left[
a_{2}(a_{0}^{2}-a_{4}^{2})-(a_{0}a_{3}-a_{1}a_{4})(a_{1}+a_{3})\right] \mid
\text{ \ .}  \label{C24}
\end{equation}%
A convincing argument, demonstrating the validity of the Schur theorem for the
chain of algebraic equations with diminishing degrees is that the inequality
(\ref{C23}), derived from the linear equation (\ref{C17}) can also be
obtained from the quadratic equation (\ref{C16}) $%
f_{2}(y)=c_{0}y^{2}+c_{1}y+c_{2}$ after finding its roots and imposing the
restriction $\mid y\mid <1$. This simple calculation shall be performed in
the following subsections.

The above calculational scheme shall not be applied with respect to the
space-time interval algebraic equation since the basic calculational inequalities (\ref%
{C1}) will not be fulfilled. However, some of these inequalities will be
fulfilled with respect to the other fourth-degree algebraic equation called
in this paper "the geodesic equation". Again, by means of the Schur theorem
it will be proved that this equation will have no roots in the circle $\mid
y\mid <1$. From an algebraic point of view, it will be interesting to see
how the inequality (\ref{C24}) changes when the coefficient inequalities (%
\ref{C1}) are the inverse ones - all of them or some of them.

\subsection{Coefficient functions and inequalities for the chain of
polynomials derived from the space-time interval algebraic equation}
\label{sec:coeff inequal}

In this subsection the analogue of the inequality (\ref{C24}) for the case
of the space-time interval algebraic equation
\begin{equation}
f(y)=a_{0}y^{4}+a_{1}y^{3}+a_{2}y^{2}+a_{3}y+a_{4}=0  \label{C25}
\end{equation}%
will be derived. This equation has the following coefficient functions
\begin{equation}
a_{0}=1\text{ , }a_{1}=-\frac{4}{e^{2}}\left( 1-e^{2}\right) \sin E_{2}\text{
\ \ \ ,}  \label{C26}
\end{equation}%
\begin{equation}
a_{2}=\frac{1}{e^{4}}\left[ 4(1-e^{4})-6e^{2}(1-e^{2})\sin
^{2}E_{2}-2e^{2}\sin ^{2}E_{2}\right] \text{ \ ,}  \label{C27}
\end{equation}%
\begin{equation}
a_{3}=\frac{4}{e^{2}}(1-e^{2})\sin E_{2}\cos E_{2}\text{ \ ,}  \label{C28}
\end{equation}%
\begin{equation}
a_{4}=1-\frac{4}{e^{4}}-2\sin ^{2}E_{2}+\sin ^{4}E_{2}\text{ \ .}
\label{C29}
\end{equation}

\subsubsection{The first coefficient inequality}
\label{sec:first inequal }

Let us check first whether the inequality $\mid a_{0}\mid >\mid a_{4}\mid $ %
 is fulfilled, which can be written as
\begin{equation}
1>\mid 1-\frac{4}{e^{4}}-2\sin ^{2}E_{2}+\sin ^{4}E_{2}\mid \text{ .}
\label{C30}
\end{equation}
Since $e$ is the eccentricity of the orbit, which is approximately $0.01$,
its inverse powers will be very large numbers. It should be stressed that
 the proof that the space-time interval equation (\ref{C25}) \
has roots is based on the smallness of the eccentricity number $e$. Some
physical \ implication of this fact have been pointed out in the Discussion
part.

The term $(-\frac{4}{e^{4}})$ in the right-hand side of (\ref{C30}) is a
very large number equal to $-4.10^{8}$, which is predominant over the other
ones. Since $\mid x\mid =-x$, when $x<0$, (\ref{C30}) should be written as
\begin{equation}
1>-1+\frac{4}{e^{4}}+2\sin ^{2}E_{2}-\sin ^{4}E_{2}\text{ .}  \label{C31}
\end{equation}%
This inequality cannot be fulfilled because it is impossible for a large
positive number $\frac{4}{e^{4}}$ to be smaller than the number $1$. Thus it
is proved that the inverse inequality $\mid a_{0}\mid <\mid a_{4}\mid $
holds. Therefore, the  $(n-1)$ degree polynomial $\widetilde{f}_{1}(y)$
should be calculated according to formulae (\ref{C2}). For the fourth-degree
algebraic equation (\ref{C25}), the polynomial $\widetilde{f}_{1}(y)$ is
given by the formulae
\begin{equation}
\widetilde{f}_{1}(y)=a_{4}f(y)-f^{\ast }(y)=b_{0}^{\ast }y^{3}+b_{1}^{\ast
}y^{2}+b_{2}^{\ast }y+b_{3}^{\ast }\text{ \ ,}  \label{C32}
\end{equation}%
where the star "$\ast $" subscripts denote coefficient functions derived for
the case, when the inverse inequalities between the coefficient functions are
fulfilled. The coefficient functions $b_{0}^{\ast }$, $b_{1}^{\ast }$, $%
b_{2}^{\ast }$, $b_{3}^{\ast }$ can be expressed as follows
\begin{equation}
b_{0}^{\ast }=-b_{3}\text{ },\text{ \ }b_{1}^{\ast }=-b_{2}\text{ },\text{ \
}b_{2}^{\ast }=-b_{1}\text{ \ },b_{3}^{\ast }=-b_{0}\text{ \ ,}  \label{C33}
\end{equation}%
where the coefficients $b_{0}$, $b_{1}$, $b_{2}$, $b_{3}$ are given
according to formulaes (\ref{C21}) - (\ref{C22}).

\subsubsection{The second coefficient inequality}
\label{sec:second inequal }

Let us check whether the inequality $\mid b_{0}^{\ast }\mid >\mid
b_{3}^{\ast }\mid $ between the coefficient functions of the polynomial $%
\widetilde{f}_{1}(y)$ (\ref{C32}) is fulfilled. It can be rewritten as
\begin{equation}
\mid a_{1}a_{4}-a_{0}a_{3}\mid >\mid a_{4}^{2}-1\mid \text{ \ ,}  \label{C34}
\end{equation}%
which is in fact the inverse inequality of (\ref{C14}) for $a_{0}=1$. Taking
into account the expressions for the coefficient functions, it can be
rewritten as (keeping only the highest inverse powers of $e$, which
constitute the predominant contribution)
\begin{equation*}
\mid \frac{16\sin E_{2}}{e^{6}}-\frac{16\sin E_{2}}{e^{4}}+\frac{l_{1}}{e^{2}%
}+....+l_{2}\mid >
\end{equation*}%
\begin{equation}
>\mid \frac{16}{e^{8}}-\frac{\left[ 8(1+\sin ^{4}E_{2})\right] }{e^{4}}%
+....+l_{3}\mid \text{ \ \ ,}  \label{C35}
\end{equation}%
where $l_{1}$, $l_{2}$, $l_{3}$ are expressions, containing trigonometric
functions
\begin{equation}
l_{1}:=-4\sin E_{2}+8\sin ^{3}E_{2}-4\sin E_{2}-2\sin (2E_{2})\text{ \ ,}
\label{C36}
\end{equation}%
\begin{equation}
l_{2}:=8\sin E_{2}-8\sin ^{3}E_{2}+2\sin (2E_{2})\text{ \ ,}  \label{C37}
\end{equation}%
\begin{equation}
l_{3}:=\sin ^{8}E_{2}-4\sin ^{6}E_{2}+6\sin ^{4}E_{2}-4\sin ^{2}E_{2}\text{
\ .}  \label{C38}
\end{equation}%
The right-hand side of (\ref{C35}) contains a term, inversely proportional
to the eight power of the eccentricity $e$ of the orbit, while the left-hand
side contains a term inversely proportional to the sixth power of the
eccentricity. Therefore, the right-hand side is nearly $10000$ times greater
than the left-hand side, due to which inequality (\ref{C35}) cannot be
fulfilled. Moreover, the trigonometric terms in (\ref{C36})-(\ref{C38}) \
influence insignificantly both sides of the inequality. So since the inverse
inequality $\mid b_{0}^{\ast }\mid <\mid b_{3}^{\ast }\mid $ is valid, the
second degree polynomial $\widetilde{f}_{2}(y)$ should be given by an
analogous to (\ref{C2}) formulae
\begin{equation}
\widetilde{f}_{2}(y)=b_{3}^{\ast }\widetilde{f}_{1}(y)-b_{0}^{\ast }%
\widetilde{f}_{1}^{\ast }(y)=c_{0}^{\ast }y^{2}+c_{1}^{\ast }y+c_{2}^{\ast }%
\text{ \ ,}  \label{C39}
\end{equation}%
where $\widetilde{f}_{1}^{\ast }(y)$ is the inverse polynomial to (\ref{C32}%
)
\begin{equation}
\widetilde{f}_{1}^{\ast }(y)=b_{3}^{\ast }y^{3}+b_{2}^{\ast
}y^{2}+b_{1}^{\ast }y+b_{0}^{\ast }\text{ \ .}  \label{C40}
\end{equation}%
The coefficient functions $c_{0}^{\ast }$, $c_{1}^{\ast }$, $c_{2}^{\ast }$
are given by the expressions
\begin{equation}
c_{0}^{\ast }=b_{1}^{\ast }b_{3}^{\ast }-b_{0}^{\ast }b_{2}^{\ast
}=b_{2}b_{0}-b_{3}b_{1}=c_{2}\text{ \ \ ,}  \label{C41}
\end{equation}%
\begin{equation}
c_{1}^{\ast }=b_{2}^{\ast }b_{3}^{\ast }-b_{0}^{\ast }b_{1}^{\ast
}=b_{0}b_{1}-b_{2}b_{3}=c_{1}\text{ \ ,}  \label{C42}
\end{equation}%
\begin{equation}
c_{2}^{\ast }=b_{3}^{\ast 2}-b_{0}^{\ast 2}=b_{0}^{2}-b_{3}^{2}=c_{0}\text{
\ \ .}  \label{C43}
\end{equation}

\subsubsection{The third coefficient inequality}
\label{sec:third inequal }

It remains to check whether the inequality $\mid c_{0}^{\ast }\mid >\mid
c_{2}^{\ast }\mid $ is fulfilled. It can be rewritten as
\begin{equation}
\mid c_{2}\mid >\mid c_{0}\mid \text{ \ \ }\Longrightarrow \text{ }\mid
b_{0}b_{2}-b_{1}b_{3}\mid >\mid b_{0}^{2}-b_{3}^{2}\mid \text{ \ ,}
\label{C44}
\end{equation}%
or, in terms of the $a_{0}$, $a_{1}$, $a_{2}$, $a_{3}$, $a_{4}$ coefficient
functions, as
\begin{equation*}
\mid (1-a_{4}^{2})(a_{0}a_{2}-a_{2}a_{4})-
\end{equation*}%
\begin{equation*}
-(a_{0}a_{1}-a_{3}a_{4})(a_{0}a_{3}-a_{1}a_{4})\mid >
\end{equation*}%
\begin{equation}
>\mid (1-a_{4}^{2})^{2}-(a_{0}a_{3}-a_{1}a_{4})^{2}\mid \text{ .}
\label{C45}
\end{equation}%
Now let us write the two highest inverse powers of $e$ for the left-hand
side of the above inequality, which give the predominant contributions:
\begin{equation*}
(1-a_{4}^{2})(a_{0}a_{2}-a_{2}a_{4})\simeq
\end{equation*}%
\begin{equation}
\simeq -\frac{16^{2}}{e^{16}}+\frac{2.16^{2}\sin ^{2}E_{2}}{e^{14}}+....%
\text{ \ \ ,}  \label{C46}
\end{equation}%
\begin{equation*}
-(a_{0}a_{1}-a_{3}a_{4})(a_{0}a_{3}-a_{1}a_{4})\simeq
\end{equation*}%
\begin{equation}
\simeq \frac{16^{2}\sin ^{2}E_{2}\cos E_{2}}{e^{12}}-\frac{2.16^{2}\sin
^{2}E_{2}\cos E_{2}}{e^{10}}+.....\text{ \ .}  \label{C47}
\end{equation}%
Because of the presence of the large negative term $-\frac{16^{2}}{e^{16}}$
in (\ref{C46}) ($10000$ times larger than the next term $\simeq +\frac{%
2.16^{2}\sin ^{2}E_{2}}{e^{14}}$) and $10^{8}$ and $10^{12}$ times larger
than the corresponding terms in (\ref{C47}), the sum of the two terms (\ref%
{C46}) and (\ref{C47}) is negative. Thus with account of (\ref{C44}), it is
proved that $c_{2}<0$.

The largest terms on the right-hand side of (\ref{C45}) are
\begin{equation}
(1-a_{4}^{2})^{2}\simeq \frac{16^{2}}{e^{16}}-\frac{16^{2}(1-k_{1})}{e^{12}}+%
\frac{32f}{e^{8}}+.....\text{,}  \label{C48}
\end{equation}%
\begin{equation}
-(a_{0}a_{3}-a_{1}a_{4})^{2}\simeq -\frac{16^{2}\sin ^{2}E_{2}}{e^{12}}+%
\frac{2.16^{2}\sin ^{2}E_{2}}{e^{10}}-\frac{m_{3}}{e^{8}}\text{ \ ,}
\label{C49}
\end{equation}%
where $f$, $k_{1}$ and $m_{3}$ are combinations of trigonometric functions.
Both the left-hand and the right-hand side have terms $\sim \frac{16^{2}}{%
e^{16}}$, but in the left-hand side of (\ref{C46}) a term $\sim \frac{1}{%
e^{14}}$ is contained, which is not present in the right-hand sides of
expressions (\ref{C48}) and (\ref{C49}). Evidently the left-hand side of
inequality (\ref{C45}) is indeed greater that the right-hand side due to
which this inequality is fulfilled. It can be seen also that because of the
large positive term $\frac{16^{2}}{e^{16}}$ in (\ref{C48}) which is absent
in expression (\ref{C49}) for $-(a_{0}a_{3}-a_{1}a_{4})^{2}$, the calculated
according to (\ref{C18}) coefficient function $c_{0}$ is positive, i.e.
\begin{equation}
c_{0}=b_{0}^{2}-b_{3}^{2}=(1-a_{4}^{2})^{2}-(a_{0}a_{3}-a_{1}a_{4})^{2}>0%
\text{ \ .}  \label{C50}
\end{equation}%
From (\ref{C46}) - (\ref{C49}) it can easily be seen that
\begin{equation}
c_{0}+c_{2}>0\text{ \ ,}  \label{C51}
\end{equation}%
which is fully consistent with the inequality $\mid c_{2}\mid >\mid
c_{0}\mid $ because of the chain of inequalities
\begin{equation*}
c_{0}>-c_{2}\Longrightarrow c_{2}<-c_{0}\Longrightarrow -\overline{c}%
_{2}<-c_{0}\Longrightarrow
\end{equation*}

\begin{equation}
\Longrightarrow \overline{c}_{2}>c_{0}\Longrightarrow \mid c_{2}\mid >\mid
c_{0}\mid \text{ .}  \label{C52}
\end{equation}
In (\ref{C52}) $\overline{c}_{2}$ is the positive part of $c_{2}$ because $%
c_{2}$ is negative.

\subsubsection{Positive and negative coefficient function $c_{1}$}
\label{sec:posit negc }

We shall prove that the coefficient $c_{1}$ can be positive or
negative, depending on the value of the eccentricity angle $E_{2}$. This is
important in view of the fact that the restriction $\mid y\mid <1$ can be
proved either from the linear equation or equivalently, from the quadratic
equation (\ref{C39}).

The two parts of the expression (\ref{C19}) can be calculated by using
expressions (\ref{C26}) - (\ref{C29}) for the coefficient functions $a_{0}$,
$a_{1}$, $a_{2}$, $a_{3}$, taking into account only the highest inverse
powers of the eccentricity
\begin{equation*}
(a_{0}^{2}-a_{4}^{2})(a_{0}a_{1}-a_{3}a_{4})\simeq -\frac{16^{2}}{e^{14}}%
\sin E_{2}\cos E_{2}+
\end{equation*}
\begin{equation}
+\frac{16^{2}}{e^{12}}+\frac{64\left[ (1-k_{1})\sin (2E_{2})+n_{2}\right] }{%
e^{10}}+....\text{ \ ,}  \label{C53}
\end{equation}%
\begin{equation*}
-(a_{0}a_{2}-a_{2}a_{4})(a_{0}a_{3}-a_{1}a_{4})\simeq \frac{16^{2}}{e^{14}}%
\sin E_{2}-
\end{equation*}%
\begin{equation}
-\frac{16^{2}}{e^{14}}\sin E_{2}\cos E_{2}-\frac{2.16^{2}\sin ^{3}E_{2}}{%
e^{12}}+.....\text{ \ \ .}  \label{C54}
\end{equation}%
Again, the terms $\sim \frac{16^{2}}{e^{14}}$ are $10000$ times larger than
the terms $\sim \frac{1}{e^{12}}$ (we take into account the numerical value in the nominator), so
the sum only of the terms $\sim \frac{1%
}{e^{14}}$ in the above approximate equalities gives
\begin{equation}
c_{1}=\frac{16^{2}}{e^{14}}\sin (2E_{2})\left[ 2\sin E_{2}\sin (2E_{2})-%
\frac{1}{2}\right] \text{ \ \ .}  \label{C55}
\end{equation}%
If $\sin (2E_{2})>0$, i.e. $E_{2}\subset \left[ 0,\frac{\pi }{2}\right] $
and also if
\begin{equation}
2\sin E_{2}\sin (2E_{2})-\frac{1}{2}>0\text{ \ \ \ ,}  \label{C56}
\end{equation}%
then the coefficient function $c_{1}$ is positive. In terms of the notation $%
\sin ^{2}E_{2}=y$, the preceding inequality can be written as
\begin{equation}
y^{3}-y^{2}+\frac{1}{64}<0\text{ \ .}  \label{C57}
\end{equation}%
Conversely, if $\sin (2E_{2})<0$ (i.e. $E_{2}\subset \left[ \frac{\pi }{2},%
\frac{3\pi }{4}\right] \cup \left[ \frac{3\pi }{4},\pi \right] $), then the
inequality (\ref{C57}) has to be with a reversed sign so that $c_{1}$ is
again positive. In the same way, the case $c_{1}<0$ can be investigated. In
this paper, the above inequality will not be investigated because the final
inequality (quite similar to (\ref{C57})) for the localization of the roots
of the initial polynomial within the unit circle will be valid for both
positive and negative $c_{1}$.

\subsection{Finding the condition (in the form of an inequality) for localization of
the roots of the fourth-degree space-time interval equation within the unit
circle}
\label{sec:local roots }

Let us construct the linear equation
\begin{equation}
\widetilde{f}_{3}(y)=d_{0}^{\ast }y+d_{1}^{\ast }=c_{2}^{\ast }\text{ }%
\widetilde{f}_{2}(y)-c_{0}^{\ast }\text{ }\widetilde{f}_{2}^{\ast }(y)
\label{C58}
\end{equation}%
after applying the chain of higher-degree algebraic equations and
inequalities from the preceding subsections. Note also that the
first-order equation is derived on the base of the modified formulae (\ref%
{C39}) and not (\ref{C15}).

The coefficient functions $d_{0}^{\ast }$ $\ $and $d_{1}^{\ast }$ are given
by the following expressions
\begin{equation}
d_{0}^{\ast }=c_{1}^{\ast }c_{2}^{\ast }-c_{0}^{\ast }c_{1}^{\ast
}=c_{1}(c_{0}-c_{2})\text{ \ ,}  \label{C59}
\end{equation}%
\begin{equation}
d_{1}^{\ast }=c_{2}^{\ast 2}-c_{0}^{\ast 2}=c_{0}^{2}-c_{2}^{2}\text{ \ \ \ .%
}  \label{C60}
\end{equation}%
Consequently, the condition for the roots to remain within the unit circle
is
\begin{equation}
\mid y\mid =\mid -\frac{d_{1}^{\ast }}{d_{0}^{\ast }}\mid =\mid \frac{%
c_{2}^{2}-c_{0}^{2}}{c_{1}(c_{0}-c_{2})}\mid =\mid -\frac{(c_{0}+c_{2})}{%
c_{1}}\mid <1\text{ .}  \label{C61}
\end{equation}%
From the preceding subsection it follows that the following case is
fulfilled
\begin{equation}
c_{2}<0\text{ , \ \ }c_{0}>0\text{ \ , }c_{0}+c_{2}>0\text{ , \ }c_{1}>0%
\text{ or }c_{1}<0\text{ \ . }  \label{C62}
\end{equation}%
 From inequality (\ref{C61}) it can be written also that
\begin{equation}
c_{0}+c_{2}<c_{1}\text{ \ \ .}  \label{C63}
\end{equation}%
This in fact represents the condition for the roots of the linear equation
to remain within the unit circle. In view of the necessary and sufficient
conditions of the Schur theorem, it will represent also the condition for the
roots of all other higher-degree algebraic equations (second, third and the
final one - the fourth-degree) to remain within the circle $\mid y\mid <1$.

Let us demonstrate that the theorem is fulfilled for the first- and second-
degree algebraic equations. This means that the found relation (\ref{C63}) $%
c_{0}+c_{2}<c_{1}$ from the first-degree equation should be derived also
from the second-degree equation (\ref{C39}), which in view of (\ref{C41}) - (%
\ref{C43}) should be written as (see also (\ref{C39}))
\begin{equation}
\widetilde{f}_{2}(y)=c_{0}^{\ast }y^{2}+c_{1}^{\ast }y+c_{2}^{\ast
}=c_{2}y^{2}+c_{1}y+c_{0}\text{ \ .}  \label{C64}
\end{equation}%
From the roots of this quadratic equation and the condition $\mid y\mid <1$,
it follows
\begin{equation}
\mid y\mid =\mid -\frac{c_{1}}{2c_{2}}\pm \sqrt{\left( \frac{c_{1}}{2c_{2}}%
\right) ^{2}-\frac{c_{0}}{c_{2}}}\mid <1\text{ \ .}  \label{C65}
\end{equation}%
Let $c_{1}>0$ and let us assume a minus sign in front of the square root.
The expression inside the module will be negative (then $\mid x\mid =-x$ if $%
x<0$) and if $\overline{c}_{2}$ is the positive part of the negative
coefficient function $\overline{c}_{2}$, then it follows
\begin{equation}
\sqrt{\left( -\frac{c_{1}}{2\overline{c}_{2}}\right) ^{2}+\frac{c_{0}}{%
\overline{c}_{2}}}<1+\frac{c_{1}}{2\overline{c}_{2}}\text{ \ \ ,}
\label{C65B}
\end{equation}%
from where
\begin{equation}
\frac{c_{0}}{\overline{c}_{2}}<1+\frac{c_{1}}{\overline{c}_{2}}%
\Longrightarrow \frac{c_{1}+\overline{c}_{2}-c_{0}}{\overline{c}_{2}}%
>0\Longrightarrow c_{0}-\overline{c}_{2}<c_{1}\text{ ,}  \label{C66}
\end{equation}%
which is the same as (\ref{C63}). This condition with account of the
expressions (\ref{C18}) - (\ref{C20}) for $c_{0}$, $c_{1}$, $c_{2}$ can be
represented in the following way
\begin{equation*}
\mid \left( a_{0}^{2}-a_{4}^{2}\right) ^{2}-(a_{0}a_{3}-a_{1}a_{4})^{2}+
\end{equation*}%
\begin{equation*}
+(a_{0}^{2}-a_{4}^{2})a_{2}(a_{0}-a_{4})-
\end{equation*}%
\begin{equation*}
-(a_{0}a_{1}-a_{3}a_{4})(a_{0}a_{3}-a_{1}a_{4})\mid <
\end{equation*}%
\begin{equation*}
<\mid \left( a_{0}^{2}-a_{4}^{2}\right) (a_{0}a_{1}-a_{3}a_{4})-
\end{equation*}%
\begin{equation}
-a_{2}(a_{0}-a_{4})(a_{0}a_{3}-a_{1}a_{4})\mid \text{ .}  \label{C67}
\end{equation}%
The purpose further will be to derive the exact expression, using the
coefficient functions (\ref{C26}) - (\ref{C29}) and then to prove that the
right-hand side contains higher inverse powers of the eccentricity of the
orbit. Since it is a small number and the inverse powers represent very large
numbers, this would mean that the right-hand side will be greater that the
left-hand side and consequently, the inequality will be fulfilled.

\subsection{Final proof of the validity of the Schur theorem with respect
to the algebraic equation for the space-time interval}
\label{sec:final proofShur}

A lengthy calculation shows that the first three terms with highest inverse
powers of $e$ in the left-hand side of the preceding inequality (\ref{C67})
can be represented as
\begin{equation}
\frac{16^{2}}{e^{16}}-\frac{16^{2}}{e^{16}}+\frac{2.16^{2}\sin ^{2}E_{2}}{%
e^{14}}+\frac{p_{1}}{e^{12}}\text{ \ ,}  \label{C68}
\end{equation}%
where
\begin{equation*}
p_{1}=16^{2}\sin ^{2}E_{2}\cos E_{2}+64(2-3k_{1}-k_{2})-
\end{equation*}
\begin{equation}
-16^{2}\sin ^{2}E_{2}-16^{2}(1-k_{1})\text{ \ \ ,}  \label{C69}
\end{equation}%
where $k_{1}$ and $k_{2}$ are the trigonometric expressions
\begin{equation}
k_{1}:=2\sin ^{2}E_{2}-\sin ^{4}E_{2}\text{ \ , \ \ \ }k_{2}=6\sin
^{2}E_{2}-4\text{ \ \ \ .}  \label{C70}
\end{equation}%
Since terms $\sim $ $\frac{16^{2}}{e^{16}}$ cancel, the highest term is $%
\frac{2.16^{2}\sin ^{2}E_{2}}{e^{14}}$. The next term in (\ref{C68}) $\frac{%
p_{1}}{e^{12}}$ is $10000$ times smaller than the first one and moreover,
two of the constituent terms in $p_{1}$ are with a positive sign and two -
with a negative sign. So the term $\frac{p_{1}}{e^{12}}$ is really much
smaller than the preceding term.

Consequently, the largest term $\frac{2.16^{2}\sin ^{2}E_{2}}{e^{14}}$ from
the left-hand side of (\ref{C68}) has to be compared with the highest
inverse degree terms with the eccentricity parameter in the right-hand side
of \ (\ref{C67}). These terms  are
\begin{equation}
\frac{16^{2}\sin E_{2}}{e^{14}}-\frac{16^{2}}{e^{14}}\sin E_{2}\cos E_{2}%
\text{ \ ,}  \label{C71}
\end{equation}%
and they will be greater than the term in the left-hand side of (\ref{C68}%
), if the following inequality is fulfilled
\begin{equation}
\sin E_{2}\sin ^{2}(2E_{2})>\sin ^{2}E_{2}\text{ \ \ .}  \label{C72}
\end{equation}%
Introducing the notation $\overline{y}=\sin E_{2}$, this inequality can be
rewritten as
\begin{equation}
4\left[ \overline{y}^{3}-\overline{y}+\frac{1}{4}\right] >0\text{ \ \ \ .}
\label{C73}
\end{equation}%
Further a distinction should be made between the notations $\overline{y}%
=\sin E_{2}$ and the initially introduced notation $y=\sin ^{2}E_{1}$.

\subsubsection{Finding the solutions of the algebraic
inequality with respect to the function $\overline{y}=\sin E_{2}$}
\label{sec:solut sin }

Suppose that the cubic equation
\begin{equation}
\overline{y}^{3}-\overline{y}+\frac{1}{4}=0  \label{C73A}
\end{equation}%
has three roots $\overline{y}_{1}$, $\overline{y}_{2}$ and $\overline{y}_{3}$%
. It will be positive if the following inequalities are fulfilled
\begin{equation}
\overline{y}_{1}\leq \overline{y}\leq \overline{y}_{2}\text{ \ \ , \ \ }%
\overline{y}\geq \overline{y}_{3}\text{ \ \ .}  \label{C74}
\end{equation}%
However, one additional requirement should be taken into account - \ $\mid
\overline{y}\mid <1$. In view of the simplicity of the cubic equation, the
algebraic criteria for the localization of roots from the Schur and the
substitution theorems shall not be applied, but instead, the roots of the
cubic polynomial (\ref{C73A}) shall be found directly. It is known from
higher algebra that the roots of a cubic polynomial of the kind
\begin{equation}
\overline{y}^{3}+p\overline{y}+q=0\text{ }  \label{C75}
\end{equation}%
are found from the Kardano formulae
\begin{equation}
\overline{y}=\sqrt[3]{-\frac{q}{2}+\sqrt{\Delta }}+\sqrt[3]{-\frac{q}{2}-\sqrt{\Delta }}%
\text{ \ \ \ ,}  \label{C76}
\end{equation}%
where
\begin{equation}
\Delta =\frac{q^{2}}{4}+\frac{p^{3}}{27}  \label{C77}
\end{equation}%
is the discriminant of the cubic polynomial. In the present case $p=-1$, $q=%
\frac{1}{4}$, so the discriminant is
\begin{equation}
\Delta =\frac{1}{64}-\frac{1}{27}=\frac{27-64}{64.27}<0\text{ \ \ .}
\label{C78}
\end{equation}%
Thus the Kardano formulae becomes
\begin{equation}
\overline{y}=\sqrt[3]{-\frac{q}{2}+\sqrt{\widetilde{\Delta }}}+\sqrt[3]{-\frac{q}{2}-%
\sqrt{\widetilde{\Delta }}}\text{ \ \ \ ,}  \label{C79}
\end{equation}%
where $\widetilde{\Delta }$ is the positive part of $\Delta $. Denoting the
complex expression under the cubic root in (\ref{C79}) as
\begin{equation}
-\frac{q}{2}+\sqrt{\widetilde{\Delta }}=\rho (\cos \varphi +i\sin \varphi )
\label{C80}
\end{equation}%
and making equal the corresponding real and imaginary parts on both sides, one can find
\begin{equation}
\rho =\sqrt{\frac{q^{2}}{4}-\Delta }=\sqrt{-\frac{p^{3}}{27}}\text{ \ , \ }%
\cos \varphi =-\frac{q}{2\sqrt{-\frac{p^{3}}{27}}}=-\frac{q}{2\rho }\text{ \
\ .}  \label{C81}
\end{equation}%
For the numerical values of $p$ and $q$, the numerical values for $\rho $
and $\cos \varphi $ are
\begin{equation}
\rho =\sqrt{\frac{1}{27}}=0.192450089\text{ \ ,}  \label{C82}
\end{equation}%
\begin{equation}
\cos \varphi =-\frac{1}{8\rho }=-0.649519052\text{ \ \ ,}  \label{C83}
\end{equation}%
\begin{equation*}
\varphi =\arccos (-0.649519052)=
\end{equation*}%
\begin{equation}
=130.505350\text{ \ }[\deg ]\text{ \ \ \ .}  \label{C84}
\end{equation}%
Further, the two parts of the solution (\ref{C79}) can be represented as
\begin{equation}
\overline{y}=x+z=\sqrt[3]{\rho (\cos \varphi +i\sin \varphi )}+\sqrt[3]{\rho
(\cos \varphi -i\sin \varphi )}=  \label{C85}
\end{equation}%
\begin{equation*}
=\sqrt[3]{\rho }\left( \cos \frac{\varphi +2k_{1}\pi }{3}+i\sin \frac{%
\varphi +2k_{1}\pi }{3}\right) +
\end{equation*}%
\begin{equation}
+\sqrt[3]{\rho }\left( \cos \frac{\varphi +2k_{2}\pi }{3}+i\sin \frac{%
\varphi +2k_{2}\pi }{3}\right) \text{ \ \ .}  \label{C86}
\end{equation}%
On the other hand, substituting (\ref{C85}) $\overline{y}=x+z$ in the cubic
equation (\ref{C75}), one can derive
\begin{equation}
\overline{y}^{3}-3zx\overline{y}-(x^{3}+z^{3})=0\text{ \ \ ,}  \label{C87}
\end{equation}%
from where it follows
\begin{equation}
xz=-\frac{p}{3}\text{ \ \ , \ \ \ }x^{3}+z^{3}=q\text{ \ \ \ .}  \label{C88}
\end{equation}%
Making use of (\ref{C85}) and (\ref{C86}), one can obtain from the first
relation in (\ref{C88})
\begin{equation}
\rho ^{\frac{2}{3}}\left( \cos \frac{2(k_{1}-k_{2})\pi }{3}+i\sin \frac{%
2(k_{1}-k_{2})\pi }{3}\right) =-\frac{p}{3}\text{ \ \ ,}  \label{C89}
\end{equation}%
from where
\begin{equation}
k_{1}=k_{2}\text{ \ , \ \ \ }\rho =\left( i^{2}.\frac{p}{3}\right) ^{\frac{3%
}{2}}\text{ \ \ .}  \label{C90}
\end{equation}%
Consequently, the three roots of the cubic equation $\overline{y}^{3}-%
\overline{y}+\frac{1}{4}=0$ can be written as
\begin{equation}
\overline{y}_{m}=\frac{2}{\sqrt{3}}\cos \frac{\varphi +2k\pi }{3}\text{ \ ,
\ \ }k=0\text{, }1\text{, }2\text{ \ };\text{ \ \ \ }m=1\text{, }2\text{, }3%
\text{ \ \ \ \ \ \ \ .}  \label{C91}
\end{equation}%
Now let us assume that $0\leq E_{2}\leq \frac{\pi }{2}$, i.e. the eccentric
anomaly angle is in the first quadrant. Then in view of (\ref{C91}) and the
inequalities (\ref{C74}), it follows that
\begin{equation}
\frac{2}{\sqrt{3}}\cos \frac{\varphi }{3}\leq \sin E_{2}\leq \frac{2}{\sqrt{3%
}}\cos \frac{\varphi +2\pi }{3}\text{ \ \ \ ,}  \label{C92}
\end{equation}%
\begin{equation}
\sin E_{2}\geq \frac{2}{\sqrt{3}}\cos \frac{\varphi +2\pi }{3}\text{ \ \ \ \
.}  \label{C93}
\end{equation}%
For the first quadrant one can write the obvious relation
\begin{equation}
\sin E_{2}=\cos (\frac{\pi }{2}-E_{2})  \label{C94}
\end{equation}%
and since $\cos $ is a decreasing function, the substitution into (\ref{C92}%
) will give the inequality
\begin{equation*}
\frac{\pi }{2}+\arccos \left( \frac{2}{\sqrt{3}}\cos \frac{\varphi }{3}%
\right) \leq E_{2}\leq
\end{equation*}%
\begin{equation}
\leq \frac{\pi }{2}+\arccos \left( \frac{2}{\sqrt{3}}\cos \frac{\varphi
+2\pi }{3}\right) \text{ \ \ \ .}  \label{C95}
\end{equation}

\subsubsection{Finding the interval for the numerical values of the
eccentric anomaly angle, where the space-time interval can become zero}
\label{sec:interv zero}

Note that we have taken into account two important facts: \ with the
increase of the angle $\varphi $, $\cos \varphi $ is a decreasing function
in the first quadrant, but since $\arccos (...)$ is also a decreasing
function, $\arccos \left( \frac{2}{\sqrt{3}}\cos \frac{\varphi }{3}\right) $
will be an increasing function. However, (\ref{C95}) gives values for $E_{2}$
greater than $\frac{\pi }{2}$, in contradiction with the initial assumption.
The correct approach will be to take into account that $\cos (\frac{\pi }{2}%
-E_{2})=\cos (E_{2}-\frac{\pi }{2})$, then equality (\ref{C92}) can be
rewritten as
\begin{equation*}
\frac{\pi }{2}-\arccos \left( \frac{2}{\sqrt{3}}\cos \frac{\varphi +2\pi }{3}%
\right) \leq E_{2}\leq
\end{equation*}%
\begin{equation}
\leq \frac{\pi }{2}-\arccos \left( \frac{2}{\sqrt{3}}\cos \frac{\varphi }{3}%
\right) \text{ \ \ .}  \label{C96}
\end{equation}%
Now the concrete numerical value (\ref{C84}) for $\varphi =130.505$ $[\deg ]$
should be used in order to prove that the left-hand side of inequality (\ref%
{C96}) is undetermined, since it can be calculated
\begin{equation}
\frac{2}{\sqrt{3}}\cos \frac{\varphi +2\pi }{3}=-1.107159\text{ \ \ .}
\label{C97}
\end{equation}%
Consequently, the second root ($m=2$, $k=1$ in (\ref{C91})) is undefined
with respect to the angle $E_{2}$ (since $\mid \sin E_{2}\mid =\mid
\overline{y}\mid \leq 1$) and because of that, the function $\arccos \left(
\frac{2}{\sqrt{3}}\cos \frac{\varphi +2\pi }{3}\right) $ has an invalid
argument and the root $y_{2}$ is outside the circle $\mid y\mid <1$.

It is important however that for the value $\varphi =130.505$ $[\deg ]$ the
expression in the right-hand side of (\ref{C96}) can be exactly calculated.
So we have
\begin{equation}
\arccos \left( \frac{2}{\sqrt{3}}\cos \frac{\varphi }{3}\right)
=33.116078469887\text{ \ \ }[\deg ]  \label{C98}
\end{equation}%
and from (\ref{C96})
\begin{equation}
E_{2}\leq 56.883921530113\text{ \ }[\deg ]\text{ \ \ .}  \label{C99}
\end{equation}%
Let us now investigate the second inequality (\ref{C93}). It can be found
that
\begin{equation}
\frac{2}{\sqrt{3}}\cos \frac{\varphi +4\pi }{3}=0.2695944364053715\text{ \ \
\ .}  \label{C100}
\end{equation}%
Since in the first quadrant $\arcsin (...)$ is an increasing function, from (%
\ref{C93}) \ one can obtain
\begin{equation}
E_{2}\geq \arcsin \left( \frac{2}{\sqrt{3}}\cos \frac{\varphi +4\pi }{3}%
\right) =15.640134887147\text{ [}\deg ]\text{ \ \ . }  \label{C101}
\end{equation}%
Therefore, from the above analysis two important conclusions can be made:

1. The roots of the cubic equation (\ref{C73A}) (one of them is undetermined) are
not related to the number of roots of the original algebraic equation (\ref{C25}).

2. The two roots of the equation \ (\ref{C73A}) allow to determine the
possible range of values for the eccentric anomaly angle $E_{2}$, for which
the space-time interval can become zero. This is the range
\begin{equation}
15.64\text{ }[\deg ]\leq E_{2}\leq 56.88\text{ }[\deg ]\text{ \ \ .}
\label{C101A}
\end{equation}%
In fact, the lower bound should not be $15.64$ $[\deg ]$, because earlier in
(\ref{ABC55A18B2}) it was established that $E_{1}>\arcsin p=30.00289942985$ $%
[\deg ]$. Since the eccentric anomaly angles $E_{1}$ and $E_{2}$ enter all
formulaes symmetrically, the same bound should be valid also for the angle $%
E_{2}$. So if one would like to define properly both the space-time interval and
the geodesic distance, one should write the admissable numerical interval for the eccentric
anomaly angle as
\begin{equation}
30.002\text{ }[\deg ]\leq E_{2}\leq 56.88\text{ }[\deg ]\text{ \ \ .}
\label{C101A1}
\end{equation}

In the Discussion part of this paper it was however clarified that since the space-time
interval can be defined independently from the condition for intersatellite
communications and from the geodesic distance, then the numerical estimate (\ref{C101A}) can also
be accounted as denoting the boundaries of the interval, where the space-time interval can become
zero (in the first quadrant). It should be mentioned that the upper limit $56.88$ $[\deg ]$ in the above
equality may be obtained directly from (\ref{C92}), taking into account that
$\arcsin \left( \frac{2}{\sqrt{3}}\cos \frac{\varphi }{3}\right) $ in the
first quadrant is a decreasing function. Therefore, it can be written
\begin{equation*}
\arcsin \left( \frac{2}{\sqrt{3}}\cos \frac{\varphi +2\pi }{3}\right) \leq
E_{2}\leq
\end{equation*}%
\begin{equation}
\leq \arcsin \left( \frac{2}{\sqrt{3}}\cos \frac{\varphi }{3}\right) \text{
\ \ .}  \label{C102}
\end{equation}%
Direct calculation confirms that
\begin{equation}
\arcsin \left( \frac{2}{\sqrt{3}}\cos \frac{\varphi }{3}\right)
=56.883921530113\text{ \ }[\deg ]  \label{C103}
\end{equation}%
and consequently
\begin{equation}
\frac{\pi }{2}-\arccos (\frac{2}{\sqrt{3}}\cos \frac{\varphi }{3})=\arcsin
\left( \frac{2}{\sqrt{3}}\cos \frac{\varphi }{3}\right) \text{ \ \ \ .}
\label{C104}
\end{equation}%
In the same way, a similar analysis may be performed for the other three
quadrants. For example, for the second quadrant $E_{2}\in \lbrack \frac{\pi
}{2},2\pi ]$, \ where $\arccos (\frac{2}{\sqrt{3}}\cos \frac{\varphi }{3})$
is an increasing function, it can be obtained from inequality (\ref{C93})
for the third root that
\begin{equation}
-\cos E_{2}=\sin \left( \frac{\pi }{2}+E_{2}\right) \geq \frac{2}{\sqrt{3}}%
\cos \left( \frac{\varphi +4\pi }{3}\right) \text{ \ ,}  \label{C105}
\end{equation}%
from where
\begin{equation*}
E_{2}\leq \arccos (-\frac{2}{\sqrt{3}}\cos \frac{\varphi +4\pi }{3})=
\end{equation*}%
\begin{equation}
=105.640134887147\text{ }[\deg ]\text{ \ \ \ \ .}  \label{C106}
\end{equation}%
Therefore, $E_{2}$ is in the range $E_{2}\in \left[ \frac{\pi }{2},\text{ }105.64\right] $ $[\deg ]$.

\section{APPENDIX D: ROOTS\ OF\ THE\ SPACE-TIME\ INTERVAL\ ALGEBRAIC\ EQUATION\ FROM\ THE\ SUBSTITUTION\ THEOREM}
\label{sec:AppD}

In order to prove that an even or an odd number of roots of a polynomial
remain within a given interval, according to the substitution theorem one
has to compute the signs of the polynomial at the two endpoints of the
interval. Thus, in order to check whether the fourth-degree polynomial (\ref%
{C25}) $f(y)=a_{0}y^{4}+a_{1}y^{3}+a_{2}y^{2}+a_{3}y+a_{4}=0$ has roots
within the circle $\mid y\mid <1$, one has to determine the sign of the
polynomial (\ref{C25}) at the endpoints $y=0$ and $y=1$ .

Let us compute $f(0)$, keeping in mind that the coefficient functions of $%
f(y)$ are given by the expressions (\ref{C26})-(\ref{C29}). We have
\begin{equation}
f(0)=1-\frac{4}{e^{4}}-2\sin ^{2}E_{2}-\sin ^{4}E_{2}<0\text{ \ \ .}
\label{D1}
\end{equation}%
This is an expression with a negative sign because of the large negative term $-\frac{4}{%
e^{4}}=-4.10^{-8}$. The value of $f$ at $y=1$ is
\begin{equation*}
f(1)=a_{0}+a_{1}+a_{2}+a_{3}+a_{4}=1-\frac{4}{e^{2}}(1-e^{2})\sin E_{2}+
\end{equation*}%
\begin{equation*}
+\frac{1}{e^{4}}[4(1-e^{4})-6e^{2}(1-e^{2})\sin ^{2}E_{2}-2e^{2}\sin
^{2}E_{2}]+
\end{equation*}%
\begin{equation*}
+\frac{4}{e^{2}}(1-e^{2})\sin E_{2}\cos E_{2}+1-
\end{equation*}%
\begin{equation}
-\frac{4}{e^{4}}-2\sin ^{2}E_{2}-\sin ^{4}E_{2}\text{ \ \ \ .}  \label{D2}
\end{equation}%
Note that terms with $\frac{4}{e^{2}}$ cancel. According to the substitution
theorem, if at the two endpoints $f(0)<0$ and $f(1)<0$ (i.e. the polynomial
has equal \ signs), then the polynomial will have an even number of roots. In
view of the fact that the polynomial is of fourth degree, the even number of roots
might be two or four. The last seems more probable, so
the condition $f(1)$ to be negative can be written as
\begin{equation}
\frac{1}{e^{2}}\left[ -4\sin E_{2}-8\sin ^{2}E_{2}+2\sin (2E_{2})\right] <0%
\text{ \ \ \ ,}  \label{D3}
\end{equation}%
taking into account only the terms proportional to $\frac{1}{e^{2}}$, which
have the predominant contribution. Let us now find what is the consequence
from inequality (\ref{D3}), which can be rewritten as
\begin{equation}
\sin (2E_{2})-2\sin E_{2}-4\sin ^{2}E_{2}<0\text{ \ \ ,}  \label{D4}
\end{equation}%
or in an equivalent form
\begin{equation}
8\sin ^{2}\frac{E_{2}}{2}\cos ^{2}\frac{E_{2}}{2}\left( \tan \frac{E_{2}}{2}%
+2\right) >0\text{ \ \ \ \ .}  \label{D5}
\end{equation}%
The inequality is fulfilled if
\begin{equation}
E_{2}>2\arctan (-2)\text{ \ \ \ or \ \ \ \ }E_{2}<2\arctan (-2)\text{ \ \ \ .%
}  \label{D6}
\end{equation}%
The first case will take place if $\arctan ()$ is an increasing function and
the second case - if $\arctan ()$ is a decreasing function. Since
\begin{equation}
2\arctan (-2)\text{ }=-126.869897645844\text{ \ \ }[\deg ]\text{ \ \ ,}
\label{D7}
\end{equation}%
the choice of the first or the second option in (\ref{D6}) will depend on
whether the function $\arctan (...)$ in the third quadrant will be
increasing or decreasing. It should be clarified first that a minus sign of
the degrees is counted from the positive $x-$axis by rotation in the
anticlockwise direction. Thus $-126.86$ $[\deg ]$ is within the third
quadrant $E_{2}\subset \left[ \pi ,\frac{3\pi }{2}\right] $, \ where the
function $\tan (...)$ is a decreasing one. This can be established by
computing two arbitrary values, for example
\begin{equation}
\tan (-120)=1.73\text{ \ \ , \ \ \ }\tan (-150)=0.57\text{ \ \ \ \ .}
\label{D8}
\end{equation}%
Consequently, since the function $arc\tan (...)$ is also a decreasing one,
one should choose the second case in (\ref{D6}), i.e.
\begin{equation}
E_{2}<2\arctan (-2)=-126.869897645844\text{ \ \ }[\deg ]\text{ \ \ ,}
\label{D9}
\end{equation}%
which means that
\begin{equation}
E_{2}\subset (-126.86,-\frac{\pi }{2}]\cup \lbrack -\frac{\pi }{2}%
\text{,}0)\text{ \ \ .}  \label{D10}
\end{equation}

\section{APPENDIX E: PROOF\ BY\ MEANS\ OF\ THE\ SCHUR\ THEOREM\ THAT\ THE\ FOURTH\ DEGREE\
ALGEBRAIC\ EQUATION\ FOR\ THE\ GEODESIC\ DISTANCE\ DOES\ NOT\ HAVE\ ANY\ ROOTS\ WITHIN\ THE\ UNIT\ CIRCLE}
\label{sec:AppE}

\subsection{The geodesic algebraic equation and its coefficient functions}
\label{sec:geod coeff }

In this Section we shall follow again the algorithm, based on
the Schur theorem for construction of a chain of lower-degree polynomials,
beginning from the initial fourth-degree geodesic equation \
\begin{equation}
g(y)=\overline{a}_{0}y^{4}+\overline{a}_{1}y^{3}+\overline{a}_{2}y^{2}+%
\overline{a}_{3}y+\overline{a}_{4}=0\text{ \ \ \ .}  \label{E1}
\end{equation}%
It should be clarified that further in this Appendix the "barred" coefficients $\overline{a}_{0}, \overline{a}_{1}, \overline{a}_{2}, \overline{a}_{3}, \overline{a}_{4}$  will not be related to any complex conjugation, but will signify that they are different from the coefficients of the space-time interval algebraic equation (\ref{C25}) in Section \ref{sec:coeff inequal}. All other "barred" quantities will also be real functions or numbers and not complex conjugated ones. Further the aim will be to prove that the above equation  will not result in an equality
similar to (\ref{C67}) or (\ref{C24}%
), which is supposed to be fulfilled for small eccentricities. In view of the necessary and sufficient conditions of the Schur theorem, applied again with respect to a chain of polynomials of diminishing degrees and in this case not fulfilled, this would mean that the
polynomial shall have no roots within the circle $\mid y\mid <1$, as it should be
expected for the geodesic distance. It should be remembered also that by means of the condition for
intersatellite communications, the geodesic distance  was proved to be greater than the Euclidean
distance. The coefficient functions in the above polynomial are the
following ones
\begin{equation}
\overline{a}_{0}=1\text{ \ \ , \ \ }\overline{a_{1}}=-\frac{(2+3e^{2})}{e^{2}%
}\text{ \ \ \ , \ \ \ }\overline{a_{2}}=\frac{\widetilde{M}}{e^{4}}\text{ \
\ \ ,}  \label{E2}
\end{equation}%

\begin{equation}
\overline{a_{3}}=-\frac{p^{2}(2+3e^{2})}{e^{2}}\text{ \ \ \ , \ \ \ }%
\overline{a_{4}}=p^{4}\text{ \ \ \ \ \ ,}  \label{E3}
\end{equation}%
where $p$ and $\widetilde{M}$ are the numerical parameters
\begin{equation}
p:=\frac{2-e^{2}}{4(1-e^{2})}\text{ \ , \ }\widetilde{M}:=\frac{9}{4}%
e^{4}+3e^{2}+2e^{4}p^{2}-3-4p^{2}\text{ \ .\ \ }  \label{E4}
\end{equation}%
Earlier we have introduced the notation $p$ in (\ref{ABC55A4}) for the case
of different eccentricities $e_{1}$, $e_{2}$ and different semi-major axis $%
a_{1}$, $a_{2}$. Here we use the same notation because (\ref{E4}) is in fact
expression (\ref{ABC55A4}) for the case $e_{1}=$ $e_{2}$ and $a_{1}=a_{2}$.

\subsection{Construction of the chain of lower-degree polynomials and the
corresponding coefficient inequalities}
\label{sec:chain lower}

In order to construct the first third-degree polynomial $g_{1}(y)$ and its
inverse one $g_{1}^{\ast }(y)$
\begin{equation}
g_{1}(y)=\overline{b}_{0}y^{3}+\overline{b}_{1}y^{2}+\overline{b}_{2}y+%
\overline{b}_{3}\text{ \ , }g_{1}^{\ast }(y)=\overline{b}_{3}y^{3}+\overline{%
b}_{2}y^{2}+\overline{b}_{1}y+\overline{b}_{0}\text{\ }  \label{E5}
\end{equation}%
on the base of the defining polynomial (\ref{C4})-(\ref{C5}) or on the base
of the polynomial (\ref{C32}) with coefficient functions (\ref{C33}), one
has to check whether the inequality $\mid \overline{a}_{0}\mid >\mid
\overline{a}_{4}\mid $ is fulfilled. In the present case this inequality has
the simple form
\begin{equation}
1>p^{4}  \label{E5B}
\end{equation}%
and since $p$ can be represented as
\begin{equation}
p:=\frac{2-e^{2}}{4(1-e^{2})}=\frac{1}{4}+\frac{1}{4(1-e^{2})}<1\text{ \ \ ,}
\label{E6}
\end{equation}%
the inequality is fulfilled. Moreover, for the typical GPS eccentricity $e$,
the parameters $p$ and $p^{4}$ have the numerical values
\begin{equation}
p=0.50004382422651548\Longrightarrow p^{4}=0.062521\text{ \ \ .}  \label{E7}
\end{equation}%
Consequently, the coefficient functions $\overline{b}_{0}$, $\overline{b}%
_{1} $, $\overline{b}_{2}$, $\overline{b}_{3}$ of the third-degree
polynomial $g_{1}(y)$ (\ref{E5}) are given by the standard formulaes (\ref%
{C21}) - (\ref{C22}), but now with the "bar" coefficients $\overline{a}_{0}$%
, $\overline{a}_{1}$, $\overline{a}_{2}$, $\overline{a}_{3}$, $\overline{a}%
_{4}$ in (\ref{E2}) and (\ref{E3}).

In order to write the next second-degree polynomial from the chain of
polynomials, one has to check whether the inequality
\begin{equation}
\mid \overline{b}_{0}\mid >\mid \overline{b}_{3}\mid  \label{E9}
\end{equation}%
is fulfilled where $\overline{b}_{0}$ and $\overline{b}_{3}$ are given by
the expressions
\begin{equation}
\overline{b}_{0}=\overline{a}_{0}^{2}-\overline{a}_{4}^{2}\text{ \ , \ \ }%
\overline{b}_{3}=\overline{a}_{0}\overline{a}_{3}-\overline{a}_{1}\overline{a%
}_{4}\text{ \ \ .}  \label{E10}
\end{equation}%
The inequality (\ref{E9}) acquires the form
\begin{equation}
\mid 1-p^{8}\mid >\mid \frac{p^{2}(2+3e^{2})}{e^{2}}(p^{2}-1)\mid \text{ \ .}
\label{E11}
\end{equation}%
Denoting $p^{2}=\overline{p}<1$ and keeping in mind that $\mid p^{2}-1\mid
=1-p^{2}$ since $p^{2}-1<0$ ($\mid x\mid =-x$ if $x<0$), inequality (\ref%
{E11}) can be rewritten as
\begin{equation}
1+\overline{p}^{2}+\overline{p}^{3}-\overline{p}\frac{2(1+e^{2})}{e^{2}}>0%
\text{ \ \ .}  \label{E12}
\end{equation}%
The first three terms are not large, but they are positive, while the fourth
term $-\overline{p}\frac{2(1+e^{2})}{e^{2}}$ is a very large negative term
due to the second inverse power of the eccentricity $e$. Because of this,
the term will be proportional to $-10^{4}$. So the left-hand side of the above
inequality cannot be positive and inequalities (\ref{E11}) and (\ref{E9})
cannot be fulfilled. Consequently, since instead of (\ref{E9}) one has the
inequality $\mid \overline{b}_{0}\mid <\mid \overline{b}_{3}\mid $, the
second-order polynomial
\begin{equation}
g_{2}(y)=\widetilde{c}_{0}y^{2}+\widetilde{c}_{1}y+\widetilde{c}_{2}=%
\overline{b}_{3}g_{1}(y)-\overline{b}_{0}g_{1}^{\ast }(y)  \label{E13}
\end{equation}%
is constructed by means of formulaes (\ref{C2}) and (\ref{C32}). The "tilda"
signs of the coefficients $\widetilde{c}_{0}$, $\widetilde{c}_{1}$, $%
\widetilde{c}_{2}$ mean that these coefficients are not derived from
formulaes (\ref{C18})-(\ref{C20}), but from (\ref{C41})-(\ref{C43}). In the
present case, these expressions are
\begin{equation}
\widetilde{c}_{0}=\overline{b}_{1}\overline{b}_{3}-\overline{b}_{0}\overline{%
b}_{2}=-\overline{c}_{2}\text{ , \ \ }\widetilde{c}_{2}=\overline{b}_{3}^{2}-%
\overline{b}_{0}^{2}=-\overline{c}_{0}\text{ \ ,}  \label{E14}
\end{equation}%

\begin{equation}
\widetilde{c}_{1}=\overline{b}_{2}\overline{b}_{3}-\overline{b}_{0}\overline{%
b}_{1}=-\overline{c}_{1}\text{ \ \ .}  \label{E15}
\end{equation}%
Next, we have to check whether
\begin{equation}
\mid \widetilde{c}_{0}\mid >\mid \widetilde{c}_{2}\mid \text{ \ i.e. }\mid
\overline{c}_{2}\mid >\mid \overline{c}_{0}\mid \text{ \ .}  \label{E16}
\end{equation}%
Making use of the standard expressions (\ref{C18})-(\ref{C20}) (now with the
"barred" coefficients $\overline{a}_{0}$, $\overline{a}_{1}$, $\overline{a}%
_{2}$, $\overline{a}_{3}$, $\overline{a}_{4}$ instead of the "unbarred"
ones), the modulus $\mid \overline{c}_{2}\mid $ and $\mid \overline{c}%
_{0}\mid $ can be written as
\begin{equation*}
\mid \overline{c}_{2}\mid =\frac{(1-p^{2})}{e^{4}}.\mid \widetilde{M}%
(1-p^{8})(1+p^{2})-
\end{equation*}%
\begin{equation}
-(2+3e^{2})p^{2}(1-p^{6})\mid \text{ \ .}  \label{E17}
\end{equation}%

\begin{equation}
\mid \overline{c}_{0}\mid =(1-p^{2})^{2}.\left( \mid (1+p^{2})^{2}-\frac{%
p^{4}(2+3e^{2})^{2}}{e^{4}}\mid \right) \text{ \ .}  \label{E18}
\end{equation}%
Both sides of the inequality $\mid \overline{c}_{2}\mid >\mid \overline{c}%
_{0}\mid $ have denominators proportional to $\frac{1}{e^{4}}$. Let us
rewrite the inequality in the following form
\begin{equation*}
\mid \widetilde{M}.(1+p^{2})^{2}(1+p^{4})-
\end{equation*}%
\begin{equation*}
-(2+3e^{2})^{2}p^{2}(1+p^{2}+p^{4})\mid >
\end{equation*}%
\begin{equation}
>\mid e^{4}(1+p^{2})^{2}-p^{4}(2+3e^{2})^{2}\mid \text{ \ .}  \label{E19}
\end{equation}%
Terms proportional to $p$ (see also (\ref{E6})) and not to the eccentricity $%
e$ will be larger. So the largest terms in the left-hand side will be those
having the smallest powers in $p$. This is the term $\widetilde{M}p$, and
since $\widetilde{M}$ is given by (\ref{E4}), this will be the term $\mid
-3p\mid $, numerically equal to $1.50013$. In the right-hand side, the
largest will be the term not multiplied by powers of the eccentricity $e$ -
this will be the term $\mid -4p^{4}\mid $, equal to $0.25008$. Consequently,
inequality $\mid \overline{c}_{2}\mid >\mid \overline{c}_{0}\mid $ is
fulfilled.

\subsection{The last linear polynomial and the proof that the final
inequality is not satisfied}
\label{sec:last polynom}

From the polynomial (\ref{E13}) and formulae (\ref{C17}) from the preceding
appendix, the final condition for the roots of the linear polynomial
\begin{equation*}
g_{3}(y)=\overline{d}_{0}y+\overline{d}_{1}=
\end{equation*}%
\begin{equation}
=(\widetilde{c}_{0}^{2}-\widetilde{c}_{2}^{2})y+(\widetilde{c}_{0}\widetilde{%
c}_{1}-\widetilde{c}_{1}\widetilde{c}_{2})  \label{E20}
\end{equation}%
to be in the circle $\mid y\mid =\mid -\frac{\overline{d}_{1}}{\overline{d}%
_{0}}\mid <1$ can be expressed as
\begin{equation}
\mid -(\widetilde{c}_{0}\widetilde{c}_{1}-\widetilde{c}_{1}\widetilde{c}%
_{2})\mid <\mid (\widetilde{c}_{0}^{2}-\widetilde{c}_{2}^{2})\mid \text{ }%
\Longrightarrow \mid \overline{c}_{1}\mid <\mid \overline{c}_{2}+\overline{c}%
_{0}\mid  \label{E21}
\end{equation}%
In terms of the coefficient functions $\overline{a}_{0}$, $\overline{a}_{1}$%
, $\overline{a}_{2}$, $\overline{a}_{3}$, $\overline{a}_{4}$ (given by (\ref%
{E2}) - (\ref{E3})), the above inequality can be written as%
\begin{equation*}
\mid (\overline{a}_{0}^{2}-\overline{a}_{4}^{2})(\overline{a}_{0}\overline{a}%
_{1}-\overline{a}_{3}\overline{a}_{4})-
\end{equation*}%
\begin{equation}
-(\overline{a}_{0}\overline{a}_{2}-\overline{a}_{2}\overline{a}_{4})(%
\overline{a}_{0}\overline{a}_{3}-\overline{a}_{1}\overline{a}_{4})\mid <
\label{EE22}
\end{equation}%

\begin{equation*}
<\mid (\overline{a}_{0}^{2}-\overline{a}_{4}^{2})\overline{a}_{2}(\overline{a%
}_{0}-\overline{a}_{4})-
\end{equation*}%
\begin{equation*}
-(\overline{a}_{0}\overline{a}_{1}-\overline{a}_{3}\overline{a}_{4})(%
\overline{a}_{0}\overline{a}_{3}-\overline{a}_{1}\overline{a}_{4})+
\end{equation*}%
\begin{equation}
+(\overline{a}_{0}^{2}-\overline{a}_{4}^{2})^{2}-(\overline{a}_{0}\overline{a%
}_{3}-\overline{a}_{1}\overline{a}_{4})^{2}\mid \text{ .}  \label{EE23}
\end{equation}%
In a simple and compact form, this inequality can be represented as
\begin{equation}
\frac{p^{2}(2+3e^{2})}{e^{2}}.\mid (1+p^{2})(1+p^{4})-\frac{\widetilde{M}}{%
e^{4}}\mid <  \label{E24}
\end{equation}%

\begin{equation*}
<(1+p^{2}).\mid (1+p^{4})^{2}+\frac{\widetilde{M}(1+p^{4})}{e^{4}}-
\end{equation*}%
\begin{equation}
-\frac{p^{2}(2+3e^{2})^{2}}{e^{4}}\mid \text{ \ }  \label{E25}
\end{equation}%
and the numerical parameter $\widetilde{M}$ is given again by expression (\ref{E4}). Now
the largest terms in both sides of the inequality have to be compared - in
fact, these are the terms, proportional to the inverse powers of $e$. In the
left-hand side, this is the term
\begin{equation*}
\mid -\frac{2p^{2}\widetilde{M}}{e^{6}}\mid \sim \frac{\mid 6p^{2}\mid }{%
e^{6}}\text{ \ \ .}
\end{equation*}%
This is a large number with $12$ digits. On the other hand, in the
right-hand side the largest term is
\begin{equation*}
\simeq \frac{(1+p^{2})4p^{2}}{e^{4}}\text{ \ .}
\end{equation*}%
This is a number with $8$ digits. So, the right-hand side is impossible to be
larger than the left-hand side. Consequently, the Schur theorem cannot \ be
fulfilled, due to which the geodesic equation (\ref{E1}) cannot have any
roots in the interval $(0$, $1)$. This is in full accord with the fact that
the geodesic distance cannot be zero, since it is greater than the Euclidean
one.

\section{APPENDIX F: THE\ SUBSTITUTION\ THEOREM\ APPLIED\ TO\ THE\ GEODESIC\ EQUATION}
\label{sec:AppF}

The application of the substitution theorem to the geodesic equation (\ref%
{E1}) $g(y)=\overline{a}_{0}y^{4}+\overline{a}_{1}y^{3}+\overline{a}%
_{2}y^{2}+\overline{a}_{3}y+\overline{a}_{4}=0$ presumes that the signs of
the functions $g(0)$ and $g(1)$ have to be determined.

It can easily be found that
\begin{equation}
g(0)=p^{4}>0\text{ \ \ \ ,}  \label{FF1}
\end{equation}%
\begin{equation*}
g(1)=1-\frac{(2+3e^{2})}{e^{2}}+\frac{\widetilde{M}}{e^{4}}-
\end{equation*}%
\begin{equation}
-\frac{p^{2}(2+3e^{2})}{e^{2}}+p^{4}\text{ .}  \label{FF2}
\end{equation}%
It is seen that the largest term is $\frac{M}{e^{4}}$. More exactly, since $%
\widetilde{M}$ is given by (\ref{E4}), the largest term will be this part of $%
\widetilde{\text{ }M}$, which does not contain powers of $e$ in the nominator - they will
cancel with $e^{4}$ in the denominator and thus smaller inverse powers of $e$
will result in a smaller term\ $\frac{\widetilde{M}}{e^{4}}$. So the part of
$\widetilde{M}$, not containing $e$ is $\frac{-3-4p^{2}}{e^{4}}$,
consequently the predominant contribution in $f(1)$ is negative, i.e.
\begin{equation}
g(1)=-\frac{(3+4p^{2})}{e^{4}}<0\text{ \ \ . }  \label{FF3}
\end{equation}%
Since at the both endpoints, equal to the numbers zero and one, the polynomial has equal signs
(since $g(0)<0$ and $g(1)<0$ ), it should have an odd number of roots. The
odd number could be one or three. However, the polynomial is of fourth
degree, so it should have an even (four) number of roots. If these roots are
determined as $y=\sin ^{2}E_{1}$, then all of them should be situated in the
circle $\mid y\mid <1$. Since this is not what follows from the substitution
theorem, it can be concluded that there should be no roots at all in the
interval $(0,1)$, which confirms the conclusion in the preceding appendix. Let us also remember the formulation of the Schur
theorem, according to which all the roots of the algebraic equation should remain in the unit circle. Consequently, if one assumes that
the Schur theorem is fulfilled with respect to the geodesic algebraic equation (\ref{E1}) and thus complies with the substitution theorem,
then the case only one or three roots to remain within the unit circle cannot be fulfilled. Thus it is proved that the fourth-degree geodesic
equation cannot have any roots within the unit circle.

\section{Appendix G: Coefficient functions $\widetilde{A}_{1}$, $\widetilde{A%
}_{2}$ and $\widetilde{A}_{3\text{ }}$in the expression for the differential
of the second propagation time as a function of the differential of the
first propagation time}

The coefficient function $\widetilde{A}_{1}$ in the expression (\ref{DOPC7})
$dT_{2}=\frac{1}{c}\sqrt[.]{\widetilde{A}_{1}(dT_{1})^{2}+\widetilde{A}%
_{2}(dT_{1})+\widetilde{A}_{3}}$ \ for the differential $dT_{2}$ is of the
following form
\begin{equation*}
\widetilde{A}_{1}:=c^{2}\left( \frac{\widetilde{R}_{2}}{\widetilde{R}_{1}}%
\right) ^{2}.\left( \frac{\widetilde{S}_{1}}{\widetilde{S}_{2}}\right) ^{2}+
\end{equation*}%
\begin{equation*}
+\left( \frac{c^{2}-2V_{2}}{c^{2}+2V_{2}}\right) \left( \frac{\partial z_{2}%
}{\partial E_{1}}\right) ^{2}\left( \frac{1}{\widetilde{R}_{1}}\right) ^{2}+
\end{equation*}%
\begin{equation}
+2\left( \frac{c^{2}-2V_{2}}{c^{2}+2V_{2}}\right) \left( \frac{\partial z_{2}%
}{\partial E_{1}}\right) ^{2}\left( \frac{\partial z_{2}}{\partial E_{2}}%
\right) ^{2}\left( \frac{\widetilde{S}_{1}}{\widetilde{S}_{2}}\right) \left(
\frac{1}{\widetilde{R}_{1}}\right) ^{2}\text{ \ \ .}  \label{GGGG1}
\end{equation}%
   The dependence of the coordinate $z_{2}$ on both the eccentric anomaly
angles $E_{1}$ and $E_{2}$ can be seen, if from the hyperplane equation (\ref%
{DOPAB25}) $dz_{2}$ is expressed as
\begin{equation}
dz_{2}=dz_{1}+\frac{F}{2(z_{1}-z_{2})}\text{ \ \ \ ,}  \label{GGGG2}
\end{equation}%
where expression $F$ is
\begin{equation*}
F:=2(x_{1}-x_{2})dx_{1}-2(x_{1}-x_{2})dx_{2}+
\end{equation*}%
\begin{equation}
+2(y_{1}-y_{2})dy_{1}-2(y_{1}-y_{2})dy_{2}-dR_{AB}^{2}\text{ \ \ \ \ .}
\label{GGGG3}
\end{equation}%
The coefficient function $\widetilde{A}_{2}$ is given by the expression
\begin{equation*}
\widetilde{A}_{2}:=-c^{2}\left( \frac{\widetilde{R}_{2}}{\widetilde{S}_{2}}%
\right) ^{2}.\left( \frac{\widetilde{S}_{1}}{\widetilde{R}_{1}(z_{1}-z_{2})}%
\right) .dR_{AB}^{2}-
\end{equation*}%
\begin{equation}
-\left( \frac{c^{2}-2V_{2}}{c^{2}+2V_{2}}\right) .\left( \frac{\partial z_{2}%
}{\partial E_{1}}\right) .\left( \frac{\partial z_{2}}{\partial E_{2}}%
\right) \frac{dR_{AB}^{2}}{\widetilde{R}_{1}\widetilde{S}_{2}}\text{ \ .}
\label{GGGG4}
\end{equation}%
The coefficient function $\widetilde{A}_{3}$ is given by
\begin{equation}
\widetilde{A}_{3}:=c^{2}\left( \frac{\widetilde{R}_{2}}{\widetilde{S}_{2}}%
\right) ^{2}.\frac{\left( dR_{AB}^{2}\right) ^{2}}{4(z_{1}-z_{2})^{2}}\text{
\ \ \ .}  \label{GGGG5}
\end{equation}%

\qquad\

\begin{acknowledgments}
  The author is grateful to Prof. Vladimir Gerdjikov (Institute for Mathematics and Informatics (IMI),
  Bulgarian Academy of Sciences), to Assoc. Prof. Dr.
  Martin Ivanov (INRNE) and to  Dr. Stoyan Mishev (IAPS, New Bulgarian University, Sofia) for some advises,
  concerning the revtex style layout processing. Special gratitude also to Assoc. Prof. Dr. Ivan Chipchakov
  (Institute for Mathematics and Informatics, Bulgarian Academy of Sciences) for
  the organization of the seminar "Algebra and Logics" at IMI and also to the participants in this seminar.
  The author acknowledges small financial support from the Institute for Advanced Physical
  Studies (IAPS) for participation in the Jubilee Balkan Physical Conference in August 2018, when part of this work was reported.
\end{acknowledgments}

\end{document}